\documentclass[10pt]{article}

\usepackage{epsf,graphics,amssymb,epic,eepic}
\usepackage[normalsize]{caption}

\newcommand{\be}{\begin{eqnarray}}
\newcommand{\ee}{\end{eqnarray}}

\newcommand{\la}[1]{
\label{#1} }

\newcommand{\ba}{\begin{array}}
\newcommand{\ea}{\end{array}}

\newcommand\p{\psi }

\def\xslash{\rlap/{\mkern-1mu x}}

\newcommand{\pert}{{\cal D}}
\newcommand{\partialboth}{\stackrel{\leftrightarrow}{\nabla}}
\def\nn{\nonumber\\}
 
 \def\Tr{\mbox{Tr}}

 \def\Re{\mbox{Re}}

\def\Dirac1#1{#1\hskip-2pt/}
\def\Dirac#1{#1\hskip-6pt/}

\newcommand\pbar{\bar{\psi}}
\newcommand{\spinn}[1]{ \bar{N}(p^{\ \prime}){#1} N(p)}

\def\dd{\Dirac \partial}

\def\bra#1{{\langle#1\vert}}
 \def\ket#1{{\vert#1\rangle}}

 \def\mbig#1{\mbox{\rule[-2. mm]{0 mm}{6 mm}#1}}

\begin{document}

\title{Hard Exclusive Reactions and the Structure of Hadrons}
\author{K. Goeke$^a$, M.V. Polyakov$^{a,b}$, M. Vanderhaeghen$^c$}
\date{}
\maketitle

\vspace{-.75cm}
\begin{center}
{\it
$^a$ Institut f\"ur Theoretische Physik II, Ruhr-Universit\"at Bochum,
D-44780 Bochum, Germany \\
$^b$ Petersburg Nuclear Physics Institute, 188350, Gatchina, Russia \\
$^c$ Institut f\"ur Kernphysik, Johannes Gutenberg-Universit\"at,
D-55099 Mainz, Germany}
\end{center}

%
\vspace{2cm}

\begin{abstract}
We outline in detail the properties of generalized parton
distributions (GPDs), which contain new information on
the structure of hadrons and which
enter the description of hard exclusive reactions.
We highlight the physics content of the GPDs
and discuss the quark GPDs in the large $N_c$ limit and
within the context of the chiral quark-soliton model.
Guided by this physics, we then present a general parametrization for these
GPDs. Subsequently we discuss how these GPDs enter in
a wide variety of hard electroproduction processes and how they can be
accessed from them.
We consider in detail deeply virtual Compton scattering and the
hard electroproduction of mesons.
We identify a list of key observables which are sensitive
to the various hadron structure aspects contained in the GPDs
and which can be addressed by present and future experiments.
\end{abstract}
%
%
%


%
%

\vspace{6cm}
{\it Prog. Part. Nucl. Phys. {\bf 47} (2001) 401 - 515}

\newpage

\tableofcontents

\newpage

\section{INTRODUCTION}
\label{chap1}

The famous Lagrangian of Quantum Chromodynamics
(QCD):

\be
L= -\frac 14\ G_{\mu\nu}^a G^{a\mu\nu} +\sum_f \bar \psi_f
\biggl(i\Dirac{\partial} + g \Dirac{A}-m_f\biggr)\psi_f \, ,
\label{QCD}
 \ee
contains, in principle, all phenomena of hadronic and nuclear
physics ranging from the physics of pions to the properties of
heavy nuclei. The main difficulty and challenge in the derivation
of strong interaction phenomena from the QCD Lagrangian
(\ref{QCD}) is that the theory is formulated in terms of quark and
gluon degrees of freedom whereas phenomenologically one deals with
hadronic degrees of freedom. Understanding of how colorless
hadrons are built out of colored degrees of freedom (quarks and
gluons) would allow us to make predictions for strong interaction
phenomena directly from the Lagrangian (\ref{QCD}). The physics of
hadronization of quarks and gluons is governed by such phenomena
as confinement and spontaneous chiral symmetry breaking which are
in turn related to the non-trivial structure of the QCD vacuum. It
implies that the studies of hadronization processes provide us
with valuable information on the fundamental questions of the
vacuum structure of non-abelian gauge theories.

The quark and gluon structure of hadrons can be best revealed with the help
of weakly interacting probes, such as (provided by Nature) photons and $W$, $Z$
bosons. One needs probes which are
weakly coupled to quarks  in order to
``select" a well defined QCD operator expressed in terms of quark
and gluon degrees of freedom of the Lagrangian (\ref{QCD}).
By measuring the reaction of a hadron to such a probe, one measures the
matrix element of the well-defined quark-gluon operator over the hadron
state revealing the quark-gluon structure of the hadron.

Historically the famous experiment of Otto Stern~{\em et al.}
\cite{Fri33} measuring the anomalous magnetic moment of the proton
revealed for the first time the non-trivial (now we say quark-gluon)
structure of the proton. In modern language we can say that in
this experiment with the help of the low energy photon probe (weakly interacting to quarks)
one selects the QCD operator $\sum_q e_q \bar q \gamma^\mu q$,
then one couples it to the proton,
$\langle p^{\ \prime}|\sum_q e_q \bar q \gamma^\mu q|p\rangle$, and extracts
the structural information encoded in the values of hadron form
factors. The experimental observation of a large
Pauli form factor $F_2(0)$
\footnote{ The irony is that this form factor
carries the name of W.~Pauli who tried to talk out O.~Stern of his
experiment, because for all theoreticians at that time
there were no doubts that the proton is a point like particle. See the nice
historical overview in Ref.~\cite{DDre00}.} of the proton  \cite{Fri33}
has created the physics of hadrons as strongly interacting many body systems.

Are we limited to explore the structure of hadrons by QCD
operators created by photons and $W,Z$ bosons? Can we find other
weak coupling mechanism to select more sophisticated QCD operators to explore
the structure of hadrons? Actually weak interactions as such are
an inherent
property of
QCD due to the phenomenon of asymptotic freedom meaning that
at short distances the interactions between quarks and gluons
become weak. This implies that if one manages to create a small
size configuration of quarks and gluons it can be used as a new probe
of hadronic structure. The possibility to create  small size
configurations of quarks and gluons is provided by  hard
reactions, such as deep inelastic scattering (DIS), semi-inclusive
DIS, Drell-Yan processes, hard exclusive reactions, etc.

The common important feature of hard reactions is the possibility
to separate clearly the perturbative and nonperturbative stages of
the interactions,
this is the so-called factorization property. Qualitatively
speaking, the presence of a hard probe allows us to create
small size quark, antiquark and gluon configurations whose
interactions are described by means of perturbation theory due to
the
asymptotic freedom of QCD. The non-perturbative stage of
such a reaction describes how a given hadron reacts to this
configuration, or how this probe is transformed into
hadrons.

In the present work we deal with  hard exclusive reactions of
the type:
\be
\gamma^*(q) + T(p)\to \gamma(q^{\ \prime}) +T'(p^{\ \prime}) \, ,
\qquad \gamma^*_L(q) + T(p)\to M(q^{\ \prime})  +T'(p^{\ \prime})
\, , \la{reac} \ee in which a photon \(\gamma^*\)  with high
energy and large virtuality $-q^2=Q^2>0$ scatters off the hadronic
target $T$ and produces a meson $M$ (or a low mass mesonic
cluster) or a real photon $\gamma$, and a low-mass   hadronic
state $T'$ (where the mass is small compared to $Q^2$). The
all order factorization theorems for such kind of reactions have been proven
in Ref.~\cite{Col97} (meson production)
Refs.~\cite{Ji98a,Col99,Rad98} (real photon production) . These
theorems assert that the amplitude of the reactions (\ref{reac})
can be written in the form (we show this for the case of hard
meson production \cite{Col97}):
\begin{eqnarray}
   &&
   \sum _{i,j} \int _{0}^{1}dz  \int d x_1
   f_{i/p}(x_1 ,x_1 -x_B;t,\mu ) \,
   H_{ij}(Q^{2}x_1/x_B,Q^{2},z,\mu )
   \, \phi _{j}(z,\mu )
\nonumber\\
&&
   + \mbox{power-suppressed corrections} ,
\label{factorization}
\end{eqnarray}
where $f_{i/p}$ is a generalized parton distribution (GPD),
\cite{Bar82,Dit88,Mul94,Ji97b,Rad97,Col97}
(see details in Sec.~\ref{chap3}),
$x_1$ is the fraction of the  target momentum carried by
the  interacting parton,
$\phi_j$ is the distribution amplitude (DA) of the
meson, and $H_{ij}$ is a hard-scattering coefficient,
computable as a power series in the strong coupling constant $\alpha _{s}(Q)$.
The amplitude depends also on the Bjorken variable
$x_B=Q^2/2(pq)$, and on the
momentum transfer squared $t=(p-p^{\ \prime})^2$ which is
assumed to be much smaller than
the hard scale $Q^2$. In Eq.~(\ref{factorization}) the GPDs $f_{i/p}$
and the meson DAs $\phi_j$ contain the non-perturbative
physics and $H_{ij}$ the perturbative one.
The proof of cancellation of the soft gluon interactions in
the processes (\ref{reac})
is intimately
related to the fact that the final meson arises from a
quark-antiquark (gluon) pair generated by the hard interaction.
Thus the pair starts as a small size configuration and only
substantially later grows to a normal hadronic size, i.e. into
a meson.

\begin{figure}[h]
\epsfxsize=11 cm
\epsfysize=6 cm
\centerline{\hspace{-2cm} \epsffile{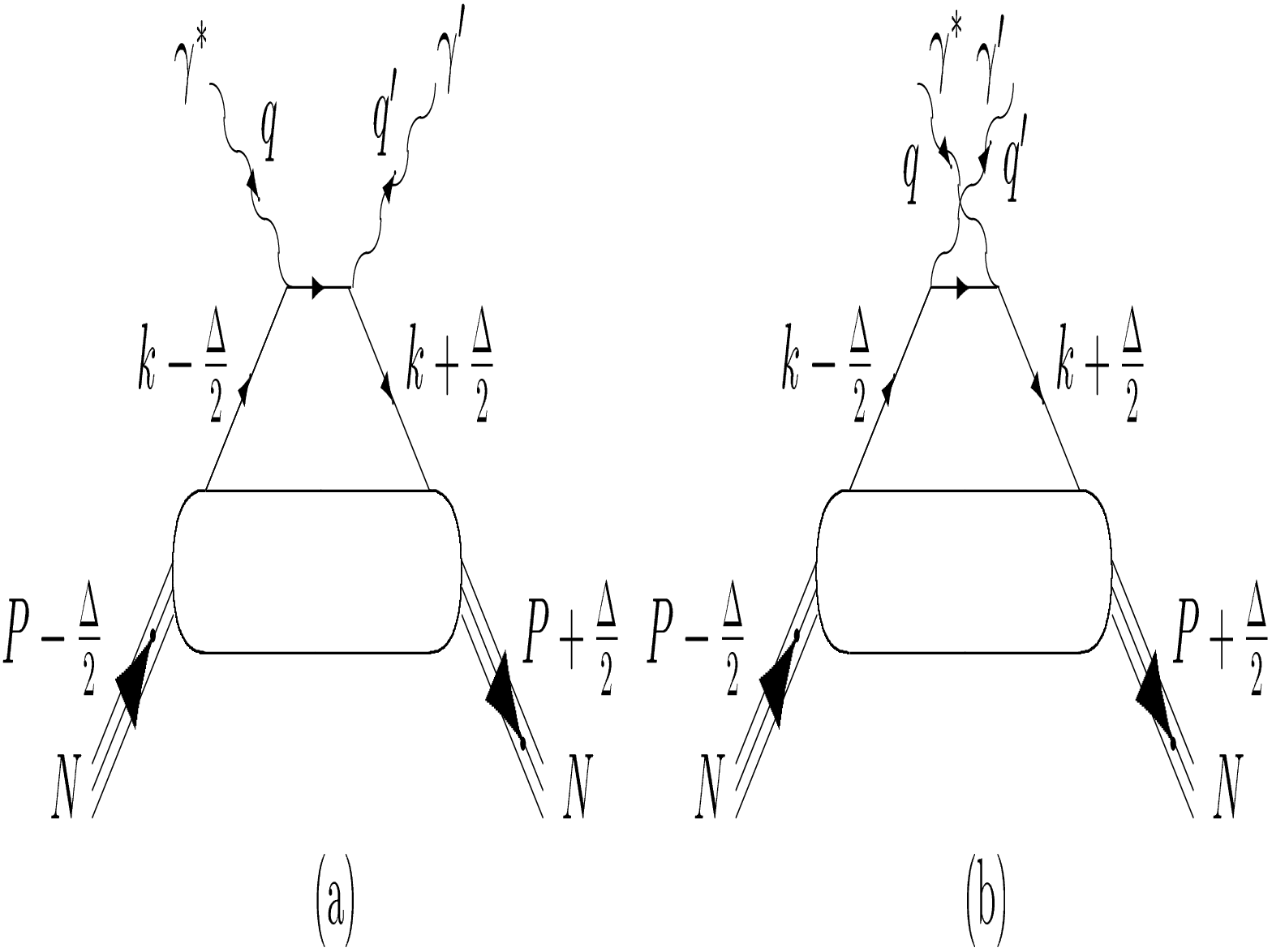}}
\vspace{-1.5cm}
\caption[]{\small ``Handbag'' diagrams for DVCS.}
\label{fig:handbags}
\end{figure}

\begin{figure}[h]
\vspace{1.5cm}
\epsfxsize=11 cm
\epsfysize=6. cm
\hspace{-2cm}\centerline{\epsffile{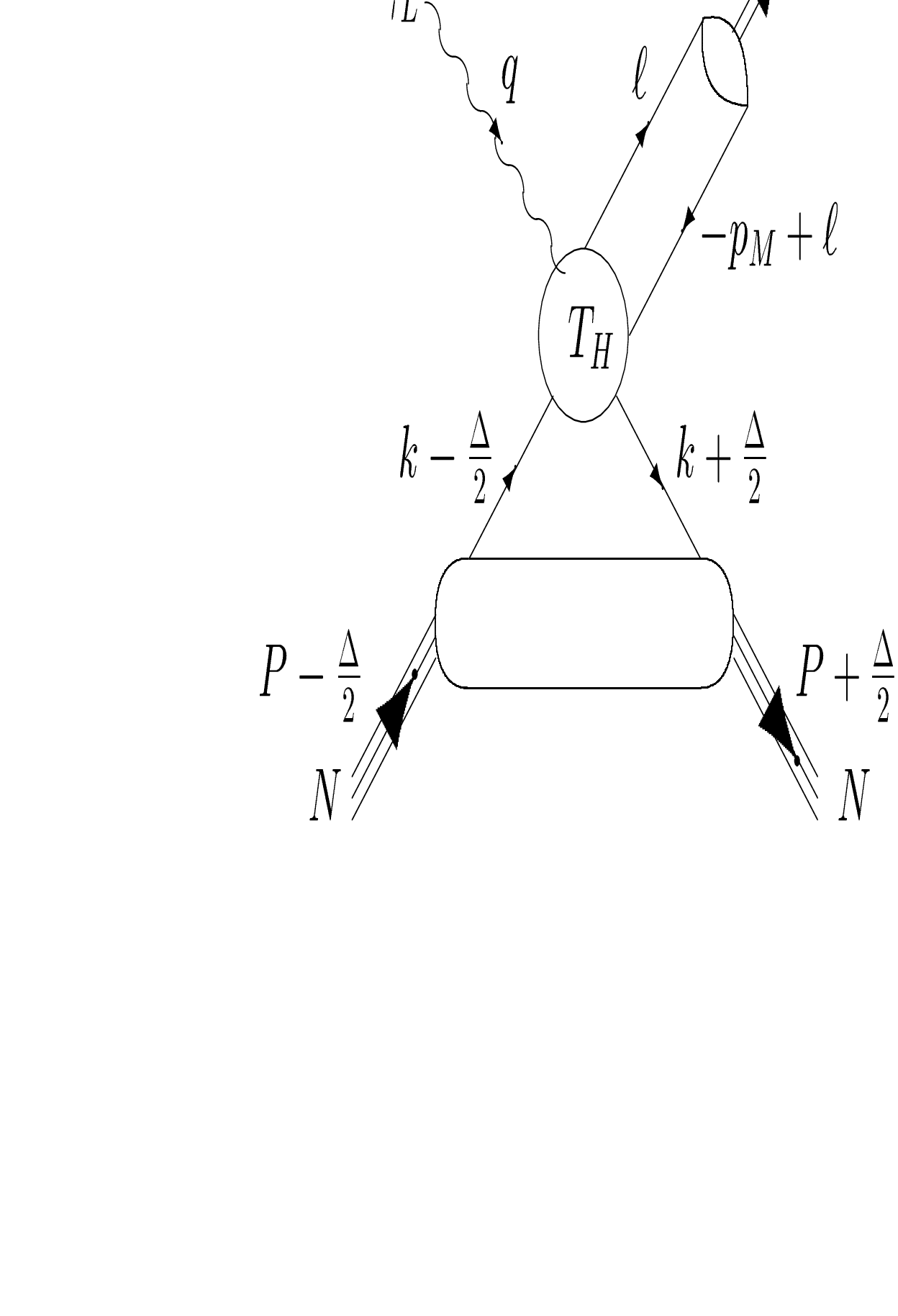}}
\vspace{-2.5cm}
\caption[]{\small Factorization for the leading order hard meson
electroproduction amplitude.}
\label{fig:factmeson}
\end{figure}

Qualitatively one can say that the reactions (\ref{reac}) allow
one to perform a ``microsurgery'' of a nucleon by removing in a
controlled way a quark of one flavor and spin and implanting
instead another quark (in general with a different flavor and
spin). It is illustrated in Fig.~\ref{fig:handbags} for the case
of the deeply virtual Compton scattering (DVCS) and in Fig.~\ref{fig:factmeson}

\noindent
for hard exclusive meson production (HMP). The lower blobs in both
figures correspond to GPDs. It is important that according to the
QCD factorization theorem in both types of processes
the same {\em universal} distribution functions enter. This allows
us to relate various hard processes to each other. Additionally
the detected final state can be used as a filter
for the spin, flavor, C-parity, etc. of the removed and  implanted
quarks.

In the processes as in Eq.~(\ref{reac})
the short distance stage of the reaction described by
$H_{ij}$ in Eq.~(\ref{factorization}) corresponds to the
interaction of a parton with a highly virtual photon. This stage is
described by perturbative QCD. The corresponding hard scattering
coefficients
have been computed to the NLO order in Refs.~\cite{Ji98a,Bel98b,Man98b}
for DVCS
and recently in Ref.~\cite{Bel01c} for the hard pion production.
Also the perturbative evolution of generalized parton
distributions is elaborated to the NLO order
\cite{Bel99,Bel00d,Bel00e}.
This shows that the perturbative QCD calculations of the hard
scattering coefficients and evolution kernels are under theoretical control.
In this work we shall not review the
corresponding perturbative
calculations, for a review see Refs.~\cite{Ji98b,Rad01b}, and
for our analysis we will
always stick to the LO expressions.

We can say that the hard scattering part (``handbag part" of Fig.~\ref{fig:handbags} and the
part denoted as $T_H$ in Fig.~\ref{fig:factmeson}) of the process
``creates" a well-defined QCD operator which is ``placed" into
the target nucleon and the outgoing meson.
We concentrate in present work mostly on studies of such
non-perturbative objects entering the factorization theorem
(\ref{factorization}). They can be generically described
by the following matrix elements:

\be
\langle B| \bar \psi_\alpha(0)\ {\rm P}e^{ig\int_0^z dx_\mu A^\mu}\
\psi_\beta (z) |A\rangle\, ,\qquad
\langle B| G_{\alpha \beta}^a(0)\ \Biggl[{\rm P}e^{ig\int_0^z dx_\mu A^\mu}\Biggr]^{ab}\
G_{\mu \nu}^{b}(z) |A\rangle\, ,
\label{AtoB}
\ee
where the operators are on the light-cone, i.e.
$z^2=0$.
Depending on the initial and the final states $A$ and $B$, the
following nomenclature of the non-perturbative matrix elements
exists:
\begin{itemize}
\item
\underline{$A=B=$~one particle state} -- parton distributions in the hadron $A$.
These objects enter the description of hard inclusive and semi-inclusive reactions.
\item
\underline{$B=$~vacuum, $A=$~one particle state} -- distribution amplitudes or
light-cone wave functions of a hadron $A$. They enter the
description of, say, hadronic form factors at large momentum
transfer as well as hard exclusive production of mesons.
\item
\underline{Both $A$ and $B$ are one particle states with different momenta},
i.e. $p(A)\neq p(B)$ -- generalized parton distributions.
They enter the description of hard exclusive production of mesons and
deeply virtual Compton scattering.
\item
\underline{$B=$~vacuum, $A$=~many particle state} -- generalized distribution
amplitudes. They enter the description of hard exclusive multi-meson
production or transition form factors between multi-meson states at large momentum
transfer.
\item
\underline{$A=$~one particle state, $B=$~many particle state}, e.g. $B=\pi N, KN$, etc.
 -- can be called
many body generalized parton distributions. They enter the description of
semi-exclusive production of mesons and semi-exclusive DVCS.
\item
\underline{$A=B=$~many particle state} -- can be called interference parton
distributions. An interesting object which has not yet been
considered in the literature.
\end{itemize}
Although all these objects have different physical meaning, their
unifying feature is that they describe the response of
hadronic states to quark-gluon configurations with small
transverse size, which have an extension along the light-cone
direction. This can also be rephrased by saying that one probes the transition
from one hadronic state to another by low-energy extended objects--``QCD
string operators" (\ref{AtoB}). From such a point of
view, the studies of hard exclusive reactions
allow us to extend considerably  the number of low-energy fundamental probes
that can be used to study hadrons.
An advantage of hard exclusive reactions is
that in this case the number of probes with different quantum
numbers is very large and also that these reactions
probe non-diagonal transitions in flavor and spin spaces.
To great extent,
the properties of these new probes  have not been
studied, especially such aspects as chiral perturbation theory for
these probes\footnote{See however works in this direction
\cite{Pol99a,Leh01,Tho00,Pob01,Che01,Arn01}
which we do not review here.},
the large $N_c$ limit, the influence of
spontaneous breaking of the chiral symmetry, etc.
In the present work we attempt to address these
questions.
\begin{table}[h]
{\centering
\begin{tabular}{|c|c|c|c|c|c|c|}
\hline
 & $0^{-+}$ & $0^{++}$  & $1^{--}$ & $1^{+-}$ & $1^{++}$ & $2^{++}$\\
\hline
 & $\eta(\eta')$ & $f_0$  & $\omega(\phi)$ & $h_1$ & $f_1$ & $f_2$\\
\hline
\hline
$e_f\ (q_f-\bar q_f)$  & &$\times$& & & & $\times$   \\
\hline
$e_f\ (q_f+\bar q_f)+\frac 34 G$  & & &$\times$&&&    \\
\hline
$e_f\ (\Delta q_f-\Delta \bar q_f)$  &$\times$& & &&$\times$&    \\
\hline
$e_f\ (\Delta q_f+\Delta \bar q_f)$  &&&&$\times$&&    \\
\hline
\end{tabular}
\caption{
{\small
Combinations of parton distributions entering the amplitude for
the
production of a
meson cluster with quantum numbers indicated in the first row
($J^{PC}$)
and with zero isospin.
In the second row we give typical meson representatives  for these
quantum numbers.}
}
\label{t1}}
\end{table}

As an illustration of the wider scope of hard exclusive reactions for
studies of hadronic structure we give here a qualitative discussion
to what type of usual parton distributions these reactions are
sensitive. We shall see that the hard exclusive reactions allow us
(leaving aside truly non-forward effects) to probe
flavor and C-parity combinations of usual parton distributions
which are not accessible in inclusive measurements.
The amplitude of a hard exclusive meson production on the nucleon
depends on the nucleon generalized parton distributions. In the
limiting case of zero ``skewedness" (i.e. when the momentum
transfer $\Delta\to 0$ in Figs.~\ref{fig:handbags},\ref{fig:factmeson})
the latter are reduced to usual
parton distributions in the nucleon.
In this sense it holds that
the
hard exclusive meson production is, among other things, sensitive to
the usual parton distributions in the nucleon.
In Tables~\ref{t1}-\ref{t3} \cite{Fra01}
we give examples of hard exclusive production processes of mesons with various
quantum numbers on a proton target and list the corresponding parton
distributions to which the amplitudes are sensitive.
In these tables the non empty entries mean that
the given combination of parton distributions
can be probed in the production of a
mesonic system with the given quantum numbers.
\begin{table}[h]
{\centering
\begin{tabular}{|c|c|c|c|c|c|c|}
\hline
 & $0^{-+}$ & $0^{++}$  & $1^{--}$ & $1^{+-}$ & $1^{++}$ & $2^{++}$\\
\hline
 & $\pi^0$ & $a_0$  & $\rho^0$ & $b_1$ & $a_1$ & $a_2$\\
\hline
\hline
$2 (u-\bar u)+ (d-\bar d)$  & &$\times$& & & & $\times$   \\
\hline
$2 (u+\bar u)+ (d+\bar d)+\frac 34 G$  & & &$\times$&&&    \\
\hline
$2 (\Delta u-\Delta \bar u)+
 (\Delta d-\Delta \bar d)
$  &$\times$& & &&$\times$&    \\
\hline
$2 (\Delta u+\Delta \bar u)+
 (\Delta d+\Delta \bar d)
$  &&&&$\times$&&    \\
\hline
\end{tabular}
\caption{ {\small
Combinations of parton distributions
entering the amplitude for the production amplitude of
a meson cluster with the quantum numbers indicated in the first row
($J^{PC}$)
and with isospin one and zero charge.
In the second row we give typical meson representatives  for these
quantum numbers.}}
\label{t2}}
\end{table}
\begin{table}[h]
{\centering
\begin{tabular}{|c|c|c|c|c|c|c|}
\hline
 & $0^{--}$ & $0^{++}$  & $1^{-+}$ & $1^{++}$ & $1^{+-}$ & $2^{++}$\\
\hline
 & $\pi^+$ & $a_0$  & $\rho^+$ & $b_1$ & $a_1$ & $a_2$\\
\hline
\hline
$ (2u-\bar u)- (2d-\bar d)$  & &$\times$& & & & $\times$   \\
\hline
$(2u+\bar u)-(2d+\bar d)$  & & &$\times$&&&    \\
\hline
$(2\Delta u-\Delta \bar u)-
(2\Delta d-\Delta \bar d)
$  &$\times$& & &&$\times$&    \\
\hline
$(2\Delta u+\Delta \bar u)-
(2\Delta d+\Delta \bar d)
$  &&&&$\times$&&    \\
\hline
\end{tabular}
\caption{ {\small
Combinations of  parton distributions
entering the amplitude for the production amplitude of
a meson cluster with the quantum numbers indicated in the first row
($J^{PG}$, instead of $C-$parity we indicate the $G$-parity)
and with charge +1.
In the second row we give typical meson representatives  for these
quantum numbers. }}
\label{t3}}
\end{table}
\noindent
The physics content of the GPDs is of course not reduced to the forward quark
distributions. In fact, the ``truly non-forward" effects play a crucial role
in the physics of hard exclusive processes (\ref{reac}), i.e. HMP as well
as DVCS.
These effects
also contain  new information about the structure of mesons and baryons
which is not accessible by standard hard inclusive reactions.

The paper is organized as follows. In section~\ref{chap3}
we review the formalism of generalized parton distributions.
We concentrate on such properties of quark GPDs which are
related to the structure of the nucleon leaving aside a discussion of the perturbative
evolution which has been extensively reviewed in
Refs.~\cite{Ji98b,Rad01b}.
Section~\ref{chap4} is devoted to studies of quark GPDs in the limit of
a large number of
colors $N_c$. In this chapter we also review and present new
calculations of GPDs in the chiral quark-soliton model.
The basic structures of GPDs discussed
in the sections~\ref{chap3} and \ref{chap4} are then
used in chapter~\ref{chap5} in order to construct a
phenomenological parametrization of the GPDs.
These parametrizations are then used to make predictions for a
wide variety of hard electroproduction processes. Since our main
objective is to investigate   how to reveal the basic features of the
nucleon structure in these reactions we restrict ourselves
in the calculation of the hard interaction kernels to the
leading perturbative order for observables. Studies of the NLO
effects on observables can be found in Refs.~\cite{Bel98,Bel00a,Bel01c}
\footnote{In these papers it was found that for some of observables
the NLO corrections can be rather large. This implies that more detail
studies of the NLO effects are needed.}.
In particular, we perform detailed studies for DVCS observables
in Sec.~\ref{chap5_3} at
twist-2 and twist-3 level. We demonstrate how various DVCS
observables are sensitive to parameters of GPDs, e.g. to the
quark total angular momentum
contribution to the nucleon spin. In
Sec.~\ref{chap5_5} we study hard meson production discussing in
detail for each specific process the information about nucleon structure one accesses.
In all calculations we restrict ourselves to the region of not
too small $x_{B}$ (valence region),
where one is mainly sensitive to the quark GPDs. Discussions of
various aspects of  hard exclusive meson
production at small $x_{B}$
can be found e.g. in Refs.~\cite{Bro94,Fra96,Abr97,Mar98b,Iva98,Iva99,Shu99,Cle00}
and of DVCS at small $x_{B}$ in Ref.~\cite{Fra98b,Fra99}.
We note that in the hard exclusive processes at small $x_B$
the GPDs can be related to the usual parton distributions in a
model independent way.

The GPDs have also been discussed within the context of wide angle
Compton scattering \cite{Rad98,Die99} where the GPDs are
related to soft overlap contributions to elastic and generalized
form factors (for a review of such processes see Ref.~\cite{Rad01b}).
Additionally the GPDs enter the description of the diffractive
photoproduction of heavy quarkonia, see e.g.
Refs.~\cite{Rys97,Fra98c,Fra99c,Mar99}, as well as in
diffractive $\gamma\to
2$jets \cite{Gol98} and in $\pi\to 2$jets processes
\cite{Fra93,Fra00b,Bra01,Cher01}. We do not review these
applications in the present work.

Finally, we summarize our main findings in section~\ref{chap6} and
list a number of key experiments to access the GPDs and to extract
their physics content.

\newpage

\section{GENERALIZED PARTON DISTRIBUTIONS {\normalsize (GPDs)}}
\label{chap3}

In this section we consider basic properties of the generalized
parton distributions (GPDs). We shall pay  special attention to
such properties of GPDs which are related to the non-perturbative
structure of hadrons and the QCD vacuum. We shall not discuss the
perturbative evolution of the GPDs, as this issue is excellently
reviewed in \cite{Ji98b,Rad01b}. Our aim will be to cover topics which
are complementary to the ones discussed in those reviews \cite{Ji98b,Rad01b}.


\subsection{Definitions, link with ordinary parton distributions
and nucleon form factors}
\label{chap3_1}

The factorization theorem for hard exclusive reactions allows us
to describe the wide class of such reactions in terms of {\em universal}
generalized parton distributions.
This nucleon structure information
can be parametrized, at leading twist--2 level, in
terms of four (quark chirality conserving) generalized structure
functions. These functions are the GPDs
denoted by $H, \tilde H, E, \tilde E$ (we do not consider chirally odd GPDs,
which are discussed in Refs.~\cite{Hoo98,Die01b})
which depend upon three variables : $x$, $\xi$ and $t$.
The light-cone momentum \footnote{using the definition
$a^{\pm} \equiv 1/\sqrt{2} (a^0 \pm a^3)$ for the light-cone
components} fraction $x$ is defined (see Figs.~\ref{fig:handbags} and
\ref{fig:factmeson}) by $k^+ = x\bar P^+$,
where $k$ is the quark loop momentum and
$\bar P$ is the average nucleon momentum ($\bar  P = (p + p^{\ \prime})/2$, where $p (p^{\ \prime})$
are the initial (final) nucleon four-momenta respectively).
The skewedness variable $\xi$ is
defined by $\Delta^+ = - 2 \xi \,\bar  P^+$, where $\Delta = p^{\ \prime} - p$ is the
overall momentum transfer in the process, and where
$2 \xi \rightarrow x_B/(1 - x_B/2)$ in the Bjorken limit.
Furthermore, the third variable entering the GPDs
is given by the Mandelstam invariant $t = \Delta^2$, being the total
squared momentum transfer to the nucleon.~\footnote{In what follows we shall
omit the $t$-dependence of GPDs if the corresponding quantity is
assumed to be defined at $t=0$.}
In a frame where the virtual photon momentum \( q^{\mu } \) and the average
nucleon momentum \(\bar  P^{\mu } \) are collinear
along the \( z \)-axis and in opposite direction, one can parametrize
the non-perturbative object in the lower blobs of
Figs.~\ref{fig:handbags} and \ref{fig:factmeson} as
(we follow Ref.~\cite{Ji97b})~\footnote{In all non-local
expressions we always assume the gauge link
${\rm Pexp}(ig\int dx^\mu A_\mu)$ ensuring the color gauge
invariance of expressions.}:
\begin{eqnarray}
&& {{\bar  P^{+}}\over {2\pi }}\, \int dy^{-}e^{ix\bar  P^{+}y^{-}}
\langle p^{'}|\bar{\psi }_{\beta }(-y/2) \psi _{\alpha}(y/2)
|p\rangle {\Bigg |}_{y^{+}=\vec{y}_{\perp }=0} \nonumber \\
&=& {1\over 4}\left\{ ({\gamma ^{-}})_{\alpha \beta }
\left[ H^{q}(x,\xi ,t)\; \bar{N}(p^{'})\gamma ^{+}N(p)\,
\right. \right.\nonumber\\
&&\hspace{2cm}\left. +\, E^{q}(x,\xi ,t)\; \bar{N}(p^{'})i\sigma ^{
+\kappa }{{\Delta _{\kappa }}\over {2m_{N}}}N(p)\right] \nonumber \\
&& \;+({\gamma _{5}\gamma ^{-}})_{\alpha \beta }
\left[ \tilde{H}^{q}(x,\xi ,t)\; \bar{N}(p^{'})\gamma ^{+}\gamma _{5}N(p)\,
\right.\nonumber\\
&&\left. \left. \hspace{2cm}+\, \tilde{E}^{q}(x,\xi ,t)\;
\bar{N}(p^{'})\gamma _{5}{{\Delta ^{+}}\over {2m_{N}}} N(p) \right]
\right\} , \;\;\;\;
\la{eq:qsplitting}
\end{eqnarray}
where \( \psi  \) is the quark field
of flavor $q$, \( N \) the nucleon spinor
and \( m_{N} \) the nucleon mass.
The {\it lhs} of Eq.~(\ref{eq:qsplitting}) can be interpreted as a Fourier
integral along the light-cone distance $y^-$ of a quark-quark
correlation function, representing the process where
a quark is taken out of the
initial nucleon (having momentum $p$) at the space-time point $y/2$, and
is put back in the final nucleon (having momentum $p^{\ \prime}$) at the space-time
point $-y/2$. This process takes place at equal light-cone time ($y^+
= 0$) and at zero transverse separation ($\vec y_\perp = 0$) between
the quarks. The resulting one-dimensional Fourier integral along the
light-cone distance $y^-$ is with respect to the quark light-cone
momentum $x \bar P^+$.
The {\it rhs} of Eq.~(\ref{eq:qsplitting}) parametrizes this
non-perturbative object in terms of four GPDs, according to whether
they correspond to a vector operator $(\gamma^-)_{\alpha \beta}$ or
an axial-vector operator $(\gamma_5 \gamma^-)_{\alpha \beta}$ at the
quark level. The vector operator corresponds at the nucleon side
to a vector transition (parametrized by the function $H^q$, for a quark
of flavor $q$) and
a tensor transition (parametrized by the function $E^q$).
The axial-vector operator corresponds at the nucleon side
to an axial-vector transition (function $\tilde H^q$)
and a pseudoscalar transition (function $\tilde E^q$).
\newline
\indent
In Fig.~\ref{fig:handbags}, the variable $x$ runs from -1 to 1.
Therefore, the momentum fractions ($x + \xi$ or $x - \xi$) of the
active quarks can either be positive or negative. Since positive
(negative) momentum fractions correspond to quarks (antiquarks), it
has been noted in \cite{Rad96a} that in this way, one can
identify two regions for the GPDs~:
when $x > \xi$ both partons represent quarks, whereas for
$x < - \xi$ both partons represent antiquarks. In these regions,
the GPDs are the generalizations of the usual parton distributions from
DIS. Actually, in the forward direction, the GPDs $H$ and $\tilde H$
reduce to the quark density distribution $q(x)$ and
quark helicity distribution $\Delta q(x)$ respectively, obtained from DIS~:
 \begin{eqnarray}
\la{eq:dislimit}
H^{q}(x,0,0)\,
&=& \left\{
\begin{array}{cr}
q(x),& \hspace{.5cm} x \; > \; 0\,, \\
- \bar q(-x),& \hspace{.5cm} x \; < \; 0 \,.
\end{array}
\right.\\
\hspace {0.5cm}
\la{eq:dislimitp}
\tilde{H}^{q}(x,0,0)\,
&=& \left\{
\begin{array}{cr}
\Delta q(x),& \hspace{.5cm} x \; > \; 0\,, \\
\Delta \bar q(-x),& \hspace{.5cm} x \; < \; 0 \,.
\end{array}
\right.
\end{eqnarray}
The functions $E$ and $\tilde E$ are not measurable
through DIS because the associated tensors
in Eq.~(\ref{eq:qsplitting}) vanish in the forward limit ($\Delta \to 0$).
Therefore, $E$ and $\tilde E$ are new leading twist functions, which
are accessible through the
hard exclusive electroproduction reactions, discussed in the following.
\newline
\indent
In the region $ -\xi < x < \xi$, one parton connected to the lower
blob in Fig.~\ref{fig:handbags} represents a
quark and the other one an antiquark. In this region, the GPDs
behave like a meson distribution amplitude and contain completely new
information about nucleon structure, because the region
$ -\xi < x < \xi$ is absent in DIS, which corresponds to the limit
$\xi \to 0$.
\newline
\indent
Besides coinciding with the quark distributions at vanishing momentum
transfer, the generalized parton distributions have interesting links with other
nucleon structure quantities. The first moments of the GPDs are related to
the elastic form factors
of the nucleon through model independent sum rules.
By integrating Eq.~(\ref{eq:qsplitting}) over \( x \), one
obtains for any $\xi$
the following relations for a particular quark flavor \cite{Ji97b} :
\begin{eqnarray}
\int_{-1}^{+1}dx\, H^{q}(x,\xi ,t)\, &=&\, F_{1}^{q}(t)\, ,
\la{eq:ffsumruleh}\\
\int _{-1}^{+1}dx\, E^{q}(x,\xi ,t)\, &=&\, F_{2}^{q}(t)\, ,
\la{eq:ffsumrulee}\\
\int_{-1}^{+1}dx\, \tilde{H}^{q}(x,\xi ,t)\, &=&\, g_{A}^{q}(t)\, ,
\la{eq:ffsumruleht}\\
\int _{-1}^{+1}dx\,\tilde{E}^{q}(x,\xi ,t)\, &=&\, h_{A}^{q}(t)\,.
\la{eq:ffsumruleet}
\end{eqnarray}
where $F_1^q(t)$ represents the elastic Dirac form factor for the
 quark flavor $q$ in the nucleon.
When referring to the quark form factors in the following,
 we understand them in our notation to be for the proton, e.g.
$F_1^u(t) \equiv F_{1}^{u/p}(t)$. In this notation, the $u$-quark form factor
is normalized at $t = 0$ as $F_{1}^{u}(0) = 2$
so as to yield the normalization of 2 for the
$u$-quark distribution in the proton, whereas the $d$-quark form
 factor is normalized at $t = 0$ as $F_{1}^{d}(0) = 1$
so as to yield the normalization of 1 for the
$d$-quark distribution in the proton.
These elastic form factors for one quark flavor in the proton,
are then related to the physical nucleon form factors
(restricting oneself to the \( u,d \) and \( s \) quark flavors), using
 $SU(2)$ isospin symmetry, as~:
\begin{eqnarray}
F_{1}^{u/p}\, &=&\, 2\,F_{1}^{p}\,+\,F_{1}^{n}\,+\,F_{1}^{s}\; , \nonumber\\
F_{1}^{d/p}\, &=&\, 2\,F_{1}^{n}\,+\,F_{1}^{p}\,+\,F_{1}^{s}\; ,
\la{eq:vecff}
\end{eqnarray}
where \( F_{1}^{p} \) and \( F_{1}^{n} \) are the proton and
neutron electromagnetic form factors respectively, with $F_1^p(0) = 1$
and $F_1^n(0) = 0$.
In Eq.~(\ref{eq:vecff}) \( F_{1}^{s} \)
is the strangeness form factor of the nucleon.
Relations similar to Eq.~(\ref{eq:vecff}) hold for the
Pauli form factors \( F_{2}^{q} \). For the axial vector form factors one uses
the isospin decomposition~:
\begin{equation}
\la{eq:axff}
g_{A}^{u}\, =\, {1\over 2}g_{A}+{1\over 2}g_{A}^{0}\; ,
\hspace {0.5cm}g_{A}^{d}\, =\, -{1\over 2}g_{A}+{1\over
  2}g_{A}^{0}\; ,
\end{equation}
where $g_A (g_A^0)$ are the isovector (isoscalar) axial form
factors of the nucleon
respectively. Similar relations exist for $h_A$.
The isovector axial form factor \( g_{A} \) is known from experiment, with
\( g_{A}(0)\approx 1.267 \). The induced pseudoscalar form factor
$h_A$ contains an important pion pole contribution, through the
partial conservation of the axial current (PCAC), as will be discussed in more details
below.

An interesting connection of GPDs to light-cone
wave functions of the nucleon was considered in
Refs.~\cite{Die01a,Bro01}. In these papers an
exact representation of generalized parton distributions for
unpolarised and polarized quarks and gluons as an overlap
of the nucleon light-cone wave functions has been constructed.
Using such an overlap representation of the GPDs model calculations
have been presented in Ref.~\cite{Cho01,Tib01,Burk01}.


\subsection{Polynomial conditions, D-term}
\label{chap3_2}

One of the non-trivial properties of the generalized parton
distributions is the polynomiality of their Mellin moments which
follows from the Lorentz invariance of nucleon matrix elements \cite{Ji97b}.
Indeed the $(N+1)$-th Mellin moment of GPDs corresponds to the nucleon matrix element
of the  twist-2, spin-$(N+1)$ {\it local} operator. Lorentz invariance then dictates that
the Mellin moments of GPDs should be polynomials maximally of the order $N+1$, {\it i.e.}
the polynomiality property implies that  \cite{Ji97b}\footnote{The general method
for counting of generalized form factors of twist-2 operators can
be found in Ref.~\cite{Ji01}}:

\be
\la{pc}
\int_{-1}^1 dx\ x^N\ H^q(x,\xi)&=&h_0^{q(N)}+h_2^{q(N)}\ \xi^2+\ldots+ h_{N+1}^{q(N)}\
\xi^{N+1}\, ,\\
\nonumber
\int_{-1}^1 dx\ x^N\ E^q(x,\xi)&=&e_0^{q(N)}+e_2^{q(N)}\ \xi^2+\ldots+ e_{N+1}^{q(N)}\
\xi^{N+1}\, .
\ee
Note that the corresponding polynomials contain only even powers of the skewedness parameter $\xi$.
This follows from the time reversal invariance,
see Ref.~\cite{Man98a,Ji98b}. This fact implies that the highest power
of $\xi$ is $N+1$ for odd $N$ (singlet GPDs ) and $N$ for even $N$ (nonsinglet GPDs).
Furthermore due to the fact that the nucleon has spin $1/2$,
the coefficients in front of the highest power of
$\xi$ for the singlet functions $H^q$ and $E^q$ are related to each other \cite{Ji97b,Ji98b}:
\be
\la{HE}
e_{N+1}^{q(N)}=-h_{N+1}^{q(N)}\, .
\ee
The polynomiality conditions (\ref{pc}) strongly restrict the
class of functions of two variables
$H^q(x,\xi)$ and $E^q(x,\xi)$. For example the conditions (\ref{pc})
imply that GPDs should satisfy the following integral constrains:
\be
\int_{-1}^1 \frac{dx}{x} \biggl[H^q(x,\xi+x z)-H^q(x,\xi)
\biggr]&=&
-\int_{-1}^1 \frac{dx}{x} \biggl[E^q(x,\xi+x z)-E^q(x,\xi)
\biggr]\nn&=&z \sum_{n=0}^\infty h_{n+1}^{q(n)}\ z^n \, .
\label{crit}
\ee
Note that the skewedness parameter $\xi$ enters the {\it lhs} of this
equation, whereas the {\it rhs} of the equation is $\xi$-independent.
Therefore this $\xi$-independence of the above integrals is a
criterion of whether the functions $H^q(x,\xi)$, $E^q(x,\xi)$ satisfy the
polynomiality conditions (\ref{pc}). Simultaneously these
integrals are generating functions for the highest coefficients
$h_{N+1}^{(N)}$. In addition, the condition (\ref{crit}) shows that there
are nontrivial functional relations between the functions $H^q(x,\xi)$
and $E^q(x,\xi)$.

An elegant possibility to implement the polynomiality conditions
(\ref{pc}) for the GPDs is to use the double distributions
\cite{Mul94,Rad96a,Rad97}. A detailed discussion of the double distributions
has been given in the
review of Ref.~\cite{Rad01b}.
In this case the generalized distributions
are obtained as a one--dimensional section of the two--variable
double distributions
$F^q(\beta,\alpha), K^q(\beta,\alpha)$ (Recently
in Refs.~\cite{Bel00c,Ter01}
the inversion formula
has been discussed):

\be
\la{dd2}
H^q_{DD}(x,\xi)=
\int_{-1}^{1}d\beta\
\int_{-1+|\beta|}^{1-|\beta|} d\alpha\
\delta(x-\beta-\alpha\xi)\  F^q(\beta,\alpha)\ \, ,
\ee
and an analogous formula for the GPD $E^q(x,\xi)$:
\be
\la{dd2e}
E^q_{DD}(x,\xi)=
\int_{-1}^{1}d\beta\
\int_{-1+|\beta|}^{1-|\beta|} d\alpha\
\delta(x-\beta-\alpha\xi)\  K^q(\beta,\alpha)\ \, .
\ee
Obviously, the double distribution function $F^q(\beta,\alpha)$
should satisfy the condition:
\be
\la{dd-fwd}
\int_{-1+|x|}^{1-|x|} d\alpha\
 F^q(x,\alpha)=q(x) \, ,
\ee
in order to reproduce the forward limit (\ref{eq:dislimit}) for
the GPD $H^q(x,\xi)$.
It is easy to check that the GPDs obtained by reduction from the
double distributions satisfy the polynomiality conditions
(\ref{pc}) but always lead to $h_{N+1)}^{q(N)}=e_{N+1}^{q(N)}=0$, {\em i.e.}
the highest power of $\xi$ for the singlet GPDs is absent. In other words the
parametrization of the singlet GPDs in terms of double distributions is not
complete. It can be completed by adding the so-called D-term to
Eq.~(\ref{dd2}) \cite{Pol99b}:

\be
H^q (x,\xi)&=&
\int_{-1}^{1}d\beta\
\int_{-1+|\beta|}^{1-|\beta|} d\alpha\
\delta(x-\beta-\alpha\xi)\  F^q(\beta,\alpha)+
\theta
\left[1-\frac{x^2}{\xi^2}\right]\ D^q\left(\frac{x}{\xi}\right) \, ,\nn
E^q(x,\xi)&=&
\int_{-1}^{1}d\beta\
\int_{-1+|\beta|}^{1-|\beta|} d\alpha\
\delta(x-\beta-\alpha\xi)\  K^q(\beta,\alpha)-
\theta\left[1-\frac{x^2}{\xi^2}\right]\ D^q\left(\frac{x}{\xi}\right) \, .
\la{addingDterm}
\ee
Here $D^q(z)$ is an odd function (as it contributes only to the singlet GPDs)
having a support $-1\le z\le 1$. In the Mellin moments, the D-term generates the highest power
of $\xi$:

\be
h_{N+1}^{q(N)}=-e_{N+1}^{q(N)}=\int_{-1}^1 dz\ z^N\ D^q(z)\, .
\label{D-term-generation}
\ee
Note that for both GPDs $H^q(x,\xi)$ and $E^q(x,\xi)$ the
absolute value of the D-term is
the same, it contributes to both functions with opposite sign.
The latter feature follows from the relation (\ref{HE}). We shall see
in the section~\ref{chap4_4} that estimates of the D-term in the
chiral quark--soliton model gives $D^u(z)\approx D^d(z)$. Therefore
in the calculations below we shall assume that:

\be
D^q(z)=\frac 1 {N_f} \ D(z)\, ,
\ee
where $N_f$ is the number of active flavors and $D(z)=\sum_q D^q(z)$
is the flavor singlet D-term.

The D-term evolves with the change of the renormalization scale
according to the ERBL evolution equation \cite{Efr80,Lep79}. Hence it is
useful to decompose the D-term in a Gegenbauer series (eigenfunctions
of the LO ERBL evolution equation):
\be
\la{dterm}
D(z)&=& (1-z^2) \biggl[ d_1 \ C_1^{3/2}(z)
+ d_3\ C_3^{3/2}(z) + d_5 \ C_5^{3/2}(z) + ...\biggr] \, ,\\
\la{dtermg}
D^g(z)&=& \frac 34 (1-z^2)^2 \biggl[ d_1^G
+ d_3^G\ C_2^{5/2}(z) + d_5^G \ C_4^{5/2}(z) + ...\biggr] \, ,
\ee
where $D^g(z)$ represents the gluon D-term.
Because the D-term is a singlet quantity therefore  the
quark D-term $D(z)$ is mixed under evolution
with the gluon D-term
$D^g(z)$. Asymptotically (for a renormalization scale $\mu\to\infty$) we obtain:

\be
D_{as}(z)&=&d\ \frac{N_f}{N_f+4 C_F}\  (1-z^2)\ C^{3/2}_1(z)\, ,\\
D_{as}^g(z)&=&d\ \frac{4 C_F}{N_f+4 C_F}\ \frac{3}{4}\ (1-z^2)^2\, ,
\ee
where $C_F=\frac{N_c^2-1}{2N_c}$. Furthermore, the scale independent constant $d$ is a
nonperturbative parameter characterizing the D-term at a low
normalization point, $d=d_1+d_1^G$ with $d_1$ from Eq.~(\ref{dterm}) and
$d_1^G$ being the first Gegenbauer coefficient of the gluon D-term (\ref{dtermg}).
We see that the D-term survives in the limit $\mu \to\infty$ and therefore the
complete form\footnote{In the literature, to the best of our knowledge,
only an incomplete form of the asymptotic
singlet GPDs was presented.} of the singlet quark GPDs at
an asymptotically large scale $\mu^2$ is the following:
\be
\sum_q H^{q}_{\rm
as}(x,\xi)&=& \frac{N_f}{N_f+4 C_F}
\left(1-\frac{x^2}{\xi^2}\right)C^{3/2}_1\left(\frac
x\xi\right)\
\left[
\frac 5{4\xi^2}
+
d \right] \theta\left(1-\frac{x^2}{\xi^2}\right)\, .
\ee
The asymptotic form of the
corresponding gluon GPD is the following:
\be
H_{\rm as}^g(x,\xi)&=& \frac{4C_F}{N_f+4 C_F}\xi\
\left(1-\frac{x^2}{\xi^2}\right)^2
\frac 3 4
\left[
\frac{5}{4\xi^2}+ d \right] \theta\left(1-\frac{x^2}{\xi^2}\right)\, .
\ee
Analogously, one can obtain the asymptotic form of the singlet GPD
$E(x,\xi)$:
\be
\sum_q E^{q}_{\rm as}(x,\xi)&=& -\frac{N_f}{N_f+4 C_F}\ d\
\left(1-\frac{x^2}{\xi^2}\right)C^{3/2}_1\left(\frac
x\xi\right) \theta\left(1-\frac{x^2}{\xi^2}\right)\, .
\ee
The asymptotic form of the
corresponding gluon GPD is the following:
\be
E_{\rm as}^g(x,\xi)&=&-\frac{4C_F}{N_f+4 C_F}\ \frac{3d}4\ \xi\
\left(1-\frac{x^2}{\xi^2}\right)^2
\theta\left(1-\frac{x^2}{\xi^2}\right)\, . \ee Furthermore, note
that asymptotically the GPD $E(x,\xi)$ is completely determined by
the D-term. Note that all expressions for the asymptotic singlet
GPDs depend on the scale independent (conserved) constant $d$.
From this point of view this constant is as fundamental as other
conserved characteristics of the nucleon, such as the total
momentum or total angular momentum.

Up to now we considered the polynomiality properties of the GPDs $H(x,\xi)$ and $E(x,\xi)$.
For the quark helicity dependent GPDs $\widetilde H(x,\xi)$ and $\widetilde E(x,\xi)$ the polynomiality
conditions are very similar to those for $H(x,\xi)$ and $E(x,\xi)$, see Eq.~(\ref{pc}).
The only difference is that
in the case of the quark helicity dependent GPDs the highest powers of the polynomial in $\xi$ is $N-1$ for
the singlet case and $N$ for the nonsinglet one. This implies that the D-term is absent for the GPDs
$\widetilde H(x,\xi)$ and $\widetilde E(x,\xi)$.

\subsection{Angular momentum sum rule}
\label{chap3_3}

The second Mellin moments of the quark helicity independent GPDs
are given by the nucleon form factors of the symmetric energy momentum tensor.
The quark part of the symmetric energy momentum tensor  is related to the quark
angular momentum operator by:
\be
J_{q}^i=\varepsilon^{i j k}\int d^3 x\ x^j\ T_q^{0k}\, .
\la{eq:JandT}
\ee
This relation implies that the forward nucleon matrix element
of the angular momentum operator can be related to the form factor of
the symmetric energy momentum tensor.
This results in
the sum rule relating the second Mellin moment of the GPDs to the angular
momentum carried by the quarks in the nucleon~\cite{Ji97b}:

\be
\la{eq:original_ji_sr}
\int_{-1}^1 dx\ x\ \left( H^q(x,\xi)+E^q(x,\xi)\right)=2
J^q\, .
\ee
In Eq.~(\ref{eq:original_ji_sr}) $J^q$ is the fraction of the nucleon angular momentum carried
by a quark of the flavor $q$ (i.e.
the sum of spin and orbital angular momentum). The closest analogy for this sum
rule is the relation between the magnetic moment of the nucleon
and the eletromagnetic (e.m.) form factors:

\be
\mu_N = F_1^N(0)+F_2^N(0)\, .
\ee
In this case the magnetic moment is given by the ``forward" form factor $F_1^N(0)$
plus the ``non-forward" form factor $F_2(0)^N$. In analogy to
Eq.~(\ref{eq:JandT})  the relation between
the magnetic moment operator and the e.m. current:
$
\mu^i\propto\varepsilon^{ijk}\int d^3x\ x^j\ J_{\rm e.m.}^k\,
$ contains explicitly the coordinate operator $x^j$.
A detailed derivation and discussion of the angular momentum sum rule
(\ref{eq:original_ji_sr}) can be found in Refs.~\cite{Ji97b,Ji98b}.

Let us discuss here the role played by the D-term in the angular momentum
sum rule.
Unfortunately it is very hard to find an observable in which the GPDs
$H^q(x,\xi)$ and $E^q(x,\xi)$ enter as a sum. Also
we should keep in mind that in a hard exclusive
process, we have kinematically $\xi\neq 0$ and that the
GPDs $H(x,\xi)$, $E(x,\xi)$ enter observables with different
kinematical factors. Therefore, we shall discuss the angular momentum
sum rule (\ref{eq:original_ji_sr}) separately for the functions
$H$ and $E$.
We rewrite  Ji's
sum rule~(\ref{eq:original_ji_sr}) in the following
equivalent way \cite{Ji97a}

\be
\int_{-1}^1 dx\ x\ \sum_q H^q(x,\xi)&=&
M_2^Q +\frac 45\ d_1\ \xi^2\, ,
\nn
\int_{-1}^1 dx\ x\ \sum_q E^q(x,\xi)&=&\left( 2
J^Q- M_2^Q\right) -\frac 45\ d_1\ \xi^2\, .
\la{jisr}
\ee
Here $d_1$ is the first Gegenbauer coefficient in the expansion of
the D-term (\ref{dterm}), $J^Q=\sum_q J^q$ and $M_2^Q$ is the momentum
fraction carried by the quarks and antiquarks in the nucleon:
\be M_2^Q=
\sum_q \int_0^1dx\ x \ \left(q(x) + \bar q(x) \right)\, .
\ee

We see that the sum rule (\ref{jisr}) is sensitive only to the
combination $2J^Q- M_2^Q$. What can we say about this combination?
Asymptotically one has that
$\lim_{\mu\to\infty} 2J^Q- M_2^Q =0$ \cite{Ji96}.
At some finite scale, the value of $2 J^q-M_2^q$ which one would like to extract
from the data should be
compared with the contribution of the D-term on the {\em rhs} of the  sum rule
(\ref{jisr}). We shall see below that estimates in the chiral
quark-soliton model for the first Gegenbauer coefficient of
the D-term give the value $d_1\approx -4$, which leads at $\xi=0.15$
to a value of $0.07$ for the second term in Eq.~(\ref{jisr}).
Therefore at accessible
values of $\xi$ the extraction of the angular
momentum carried by the quarks from the observables requires
(among other things)
an accurate
knowledge of the D-term.
The size of the D-term can be determined from the observables
which are sensitive to the real part of the amplitude of the
corresponding process. An example of such an observable is the DVCS charge
asymmetry (accessed by reversing
the charge of the lepton beam), see a detailed discussion in Sec.~\ref{chap5_3_4}.

%

\subsection{Chiral properties of GPDs}
\label{chap3_4}
Here we shall argue that the spontaneously broken chiral symmetry
of QCD plays an important role in determining the properties of the
generalized parton distributions. We shall illustrate this on
examples of the D-term and the GPD $\widetilde E(x,\xi,t)$.

\subsubsection{Spontaneously broken chiral symmetry and the D-term}
\label{chap3_4_1}

First we will discuss how the physics of spontaneously broken
chiral symmetry  plays an important role in
determining the size and the sign of the D-term. In particular, this role is seen
clearly in
the case of the GPDs in the pion.
In this case the value of
the coefficient $d_1$ in the parametrization of
the pion
D-term (\ref{dterm}) can be computed in a model independent way and it is
strictly nonzero.
To compute the pion D-term we use the soft-pion theorem for the singlet GPD
in the pion derived in \cite{Pol99a}. This soft-pion  theorem
has been derived using the fact that the pion is a
(pseudo)Goldstone boson of the spontaneously broken chiral
symmetry. The theorem states
that the singlet GPD in the pion vanishes for $\xi = \pm 1$ and momentum transfer
squared $t=0$ (corresponding to to a pion with vanishing four momentum):
\be
\la{soft-pion}
\sum_q H_q^{(\pi)}(x,\xi= \pm 1)=0\, .
\ee
Hence (see also \cite{Pol99b}),
\be
\la{hpion}
\int_{-1}^1 dx \, x \, \left(\sum_q H_q^{(\pi)}(x,\xi) \right)
\, =\, \left(1 - \xi^2 \right) \, M_2^{Q\pi} \, .
\ee
Evaluating Eq.~(\ref{hpion}) at $\xi = 0$, determines~:
\be
\la{m2}
M_2^{Q\pi}= \int_0^1dx \, x \, \sum_q [q^{(\pi)}(x) + \bar q^{(\pi)}(x)]\, ,
\ee
being the fraction of the momentum carried by the quarks and antiquarks in the
pion.
As the highest power in $\xi$ in Eq.~(\ref{hpion}) (i.e. the term in
$\xi^2$) originates solely from the D-term, one easily obtains the
following expression for pion D-term~:
\be
\la{pionDterm}
D^{(\pi)}(z)=-\frac{5 M_2^Q} 4\ (1-z^2)
\biggl[\ C_1^{3/2}(z)
+ ...\biggr]\, .
\ee
We see therefore that the first Gegenbauer coefficient of the pion D-term
is {\em negative} and strictly nonzero. Its value is fixed by the chiral
relations in terms of the momentum fraction carried by the partons in the
pion.

\begin{figure}[h]
\epsfxsize=10cm
\centerline{\epsffile{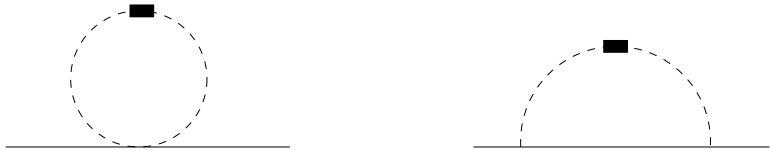}}
\caption{{\small Two-pion exchange
graphs giving the chiral contributions to the nucleon
form factors of the energy momentum tensor. The dashed
lines represent  the pion, and the solid lines represent the nucleon.
The black blob is the operator of the energy momentum tensor.}}
\label{iaa}
\end{figure}

The nucleon D-term is not fixed by general principles. However one
may expect that the contribution of the pion cloud of the nucleon
can be significant. Indeed in the chiral quark-soliton model,
which emphasizes the role of broken chiral symmetry in
the nucleon structure, the D-term is large and has the sign of the
pion D-term. Quantitatively
for the coefficients $d_1, d_3, d_5$ (see Eq.~\ref{dterm}),  the estimate
which is based on the calculation of GPDs
in the chiral quark soliton model \cite{Pet98}
at a low normalization point $\mu\approx 0.6$~GeV,
gives \cite{Kiv01b} (see also Sec.~\ref{chap4_4_1})~:
\begin{equation}
d_1 \approx -4.0, \hspace{1cm} d_3 \approx -1.2, \hspace{1cm} d_5 \approx -0.4,
\la{eq:dterm_exp}
\end{equation}
and higher moments (denoted by the ellipses in Eq.~(\ref{dterm}))
are small. Notice the negative sign of the Gegenbauer coefficients for
the D-term in Eq.~(\ref{eq:dterm_exp}), as obtained in the chiral
quark-soliton model. The coincidence of the signs in the nucleon and pion
D-terms hints that the D-term in the nucleon is intimately related
to the spontaneous breaking of the chiral symmetry.
The chiral contributions of the long range
two-pion exchange shown in Fig.~\ref{iaa} to the GPDs of the nucleon
are sizeable giving large and negative
contribution to the leading Gegenbauer coefficient of the nucleon D-term
$d_1$. The sign is related to the negative sign of the
pion D-term which is fixed by the soft pion theorem (\ref{pionDterm}).
\subsubsection{Pion pole contribution to the GPD $\widetilde E(x,\xi,t)$}
\label{chap3_4_2}

The role of the spontaneously broken chiral symmetry in the structure
of GPDs is seen particularly clearly in the case of the quark
helicity dependent GPD $\widetilde E(x,\xi,t)$.
We remind that this GPD satisfies
the sum rule of Eq.~(\ref{eq:ffsumruleet}), in terms of
the pseudoscalar nucleon nucleon form factor $h_A^q(t)$.
It is well known (see e.g. Ref.~\cite{Adl68}) that due to the
spontaneously broken chiral symmetry this form factor at small $t$
is dominated by the contribution of the pion pole of the form:

\be
\lim_{t\to m_\pi^2} h_A^q(t)=\frac 12\ \tau_{qq}^3\ \frac{4 g_A
m_N^2}{m_\pi^2-t}\, ,
\la{eq:ha_pipo}
\ee
where $g_A\approx 1.267$ is the nucleon isovector axial charge
and $\tau^3$ is the Pauli matrix in the flavor space.
The presence of the chiral singularity on the {\em rhs} of the
sum rule~(\ref{eq:ffsumruleet}) implies that one should also expect the
presence of the chiral singularity in the GPD
$\widetilde E^q(x,\xi,t)$ at small momentum transfer squared $t$ \cite{Fra98a}.
The form of this singularity has been specified in Refs.~\cite{Man99a,Fra99,Pen00a}
to be:

\be
\lim_{t\to m_\pi^2} \widetilde E^q(x,\xi,t)= \frac 12\
\tau_{qq}^3\ \frac{4 g_A
m_N^2}{m_\pi^2-t}\ \theta\left[\xi-|x|\right]\
\frac 1\xi\ \Phi_\pi\left(\frac x \xi\right)\, ,
\la{eq:etilde_chiral}
\ee
where $\Phi_\pi\left(\frac x \xi\right)$ is the pion distribution
amplitude entering e.g. description of the pion e.m. form factor at
large momentum transfer and the hard reaction $\gamma^*\gamma\to
\pi^0$. In Ref.~\cite{Pen00a} in the framework of the chiral
quark-soliton model
Eq.~(\ref{eq:etilde_chiral})
and the deviations from it
have been computed. The corresponding
results are discussed in Sec.~\ref{chap4_4}.

The presence of the pion pole singularity in the GPD $\widetilde E$
leads, in particular, to a strong dependence of the
differential cross section of, for example, hard $\pi^+$
production
on the $transverse$ polarization of the target
\cite{Fra99}.
The dependence of the {\em exclusive}
cross section on the transverse
polarization of the proton target has the
following dominant structure~\footnote{Detailed expressions
and quantitative estimates are shown in Sec.~\ref{chap_5_5_2}}:
\be
\nonumber
\sigma \propto && |S_\perp| \sin\Phi\
\frac{\sqrt{t_{\rm min}-t}}{-t+m_\pi^2}\\
&&\times
\left[ 2\left(\widetilde H^u(\xi,\xi, t)
-\widetilde H^d(\xi,\xi, t)\right)-\left(\widetilde H^u(-\xi,\xi, t)
-\widetilde H^d(-\xi,\xi, t)\right)\right]\, ,
\label{xsec}
\ee
where  $\sin\Phi$ is the azimuthal angle between
lepton plane and the plane spanned by 3-vectors of the
virtual photon and produced meson.
The specific $t$-dependence of the cross section is due to the contribution of
the chiral singularity (pion pole) to the
``truly non-forward" GPD $\widetilde E$ whose presence
is dictated by the chiral dynamics of QCD.

\subsection{Twist-3 GPDs}
\label{chap3_5}

Recently the DVCS amplitude  has been computed in
Refs.~\cite{Ani00,Pen00b,Bel00b,Rad00,Rad01b} including the terms
of the order $O(1/Q)$. The inclusion of such terms is
mandatory to ensure the electromagnetic gauge invariance of the
DVCS amplitude to the order $\sqrt{-t}/Q$. At the order $1/Q$ the
DVCS amplitude depends on a set of new generalized parton
distributions. Let us introduce generic generalized ``vector"
${\cal F}^\mu(x,\xi)$ and ``axial" $\widetilde {\cal F}^\mu(x,\xi)$
distributions\footnote{In this section we do not specify the
flavor index as all formulae here are valid for each flavor separately.}:

\be
{\cal F}^\mu(x,\xi,t)&=& \frac{\bar P^+}{2\pi} \int^\infty_{-\infty}d y^-\
e^{ix\bar P^+ y^-} \langle p^{\ \prime}| \pbar(-y/2)\gamma^\mu\psi (y/2)
 |p\rangle\Biggr|_{y^+=y_\perp=0} \, ,
\nonumber \\[4mm] \widetilde {\mathcal F}^\mu(x,\xi,t)&=& \frac{\bar P^+}{2\pi}
\int^\infty_{-\infty}d y^-\ e^{ix\bar P^+ y^-} \langle p^{\ \prime}|
\pbar(-y/2)\gamma^\mu\gamma_5\psi (y/2)
 |p\rangle \Biggr|_{y^+=y_\perp=0} \, .
 \la{Fdef}
\ee If in the above equations the index $\mu$ is projected onto
the ``plus" light-cone direction (i.e. $\mu=+$) we reproduce the
twist-2 GPDs, see definition~(\ref{eq:qsplitting}). The case of
$\mu=-$, according to the general twist counting rules of
Ref.~\cite{Jaf91} corresponds to the twist-4 GPDs which contribute
to the DVCS amplitude at the order of $O(1/Q^2)$. The twist-3
GPDs correspond to $\mu=\perp$. For a detailed discussion of the
DVCS observables to the twist-3 accuracy see Sec.~\ref{chap5_3_3}.
Here we first discuss the general properties of the twist-3 GPDs
and then will show that they can be related to the twist-2 GPDs in the
so-called Wandzura-Wilczek approximation.

\subsubsection{General properties of the twist-3 GPDs}
\label{chap3_5_1} Several properties of the twist-3 GPDs can be
derived without invoking any approximation.

\vspace{0.1cm}
\noindent \underline{ Forward limit}: In the forward limit the ``vector"
GPD ${\cal F}_\mu$ can be parametrized as follows:

\be
\lim_{\Delta\to 0} {\cal F}_\mu(x,\xi)= 2\bigl[ q(x) \widetilde p_\mu +
m_N^2 f_4(x) n_\mu \bigr]\, , \ee
where $q(x)$ is the unpolarized
forward quark (antiquark for $x\le 0$) distribution of twist-2,
$f_4(x)$ is the twist-4 quark distribution, see \cite{Jaf91}. Also we
introduced the light-cone vectors $\widetilde p^\mu$ and $n^\mu$,
the former has only ``plus" non-zero component $\widetilde
p^+=\bar P^+$ and the latter only ``minus" component $n^-=1/\bar
P^+$. As it could be expected, the twist-3 part of ${\cal F}_\mu$
disappears in the forward limit, because unpolarized parton
densities  of twist-3 are absent.

The forward limit for the twist-3 part of the ``axial" GPD
$\widetilde {\cal F}_{\mu_\perp}^{WW}$  is nontrivial:
\be
\nonumber \lim_{\Delta\to 0} \widetilde {\cal F}_{\mu_\perp}(x,\xi)= 2
S_\mu^\perp \Delta_T q(x)\, . \ee Here $\Delta_T q(x)$ is the
transverse spin quark distribution which can be measured in DIS on
a transversely polarized target.

\vspace{0.1cm}
\noindent \underline{Sum rules}: Performing the integral over $x$ of the
distributions ${\cal F}^\mu(x,\xi)$ and $\widetilde {\cal F}^\mu(x,\xi)$ given
by Eq.~(\ref{Fdef}) we obtain on the {\em rhs} the nucleon matrix
elements of the local vector and axial currents. This results in
the sum rules:

\be
\int_{-1}^1dx\ {\cal F}_\mu(x,\xi,t)&=&\spinn{\gamma_\mu}F_1(t)+
\spinn{\frac{i\sigma_{\mu\nu}\Delta^\nu}{2m_N}}F_2(t)\, ,\nn
\int_{-1}^1dx\ \widetilde {\cal F}_\mu(x,\xi,t)&=&
\spinn{\gamma_\mu\gamma_5}g_A(t)+
\frac{\Delta_\mu}{2m_N}\spinn{\gamma_5}h_A(t)\, .
\la{eq:sr0tw3} \ee
These sum rules can be viewed as a generalization
of the Burkhard-Cottingham sum rule for the transverse polarized
distribution $\Delta_T q(x)$ \cite{Bur70}. Indeed, in the forward
limit the sum rule (\ref{eq:sr0tw3}) for twist-3 GPD $\widetilde {\cal F}_{\mu_\perp}$
corresponds directly to the Burkhard-Cottingham sum rule
$$\int_0^1dx \Delta_T q(x)=\int_0^1dx \Delta q(x).$$

The second Mellin moment of GPDs ${\cal F}^\mu(x,\xi)$ and $\widetilde
 {\cal F}^\mu(x,\xi)$ can be also computed in terms of moments of twist-2
GPDs without invoking any approximation \cite{Pen00b,Kiv01c}.
Indeed taking the $x$-moment of Eq.~(\ref{Fdef}) we obtain on the
{\em rhs} the local operator of the type $\bar\psi \gamma_\mu (\gamma_5) \partialboth{}^+ \psi$
which with the help of QCD equations of motion can be reduced to
either the local operators of twist-2 or their
total derivatives. Let us illustrate how this works on the example
of the  twist-3 part of the ``vector" GPD ${\cal F}_{\mu_\perp}(x,\xi)$. From
the definition (\ref{Fdef}) we obtain:

\be
\int_{-1}^1 dx\ x\ {\cal F}^{\mu_\perp}(x,\xi)=\frac{i}{\bar P^+}\
\langle p^{\ \prime}|\bar \psi \gamma^\mu_\perp \partialboth{}^+\psi |p\rangle\, .
\la{m2step1}
\ee
Here $\nabla_\alpha$
is the gauge covariant derivative in the fundamental representation.
We rewrite
the operator in Eq.~(\ref{m2step1})  identically as:

\be
\nonumber
\bar \psi \gamma^\mu_\perp \partialboth{}^+\psi&=& \frac 12
\biggl[\bar \psi \gamma^\alpha \partialboth{}^\beta\psi+
\bar \psi \gamma^\beta \partialboth{}^{\alpha}\psi-
\frac 12 g^{\alpha \beta}
\bar \psi \gamma_\nu \partialboth{}^\nu\psi \biggr]\\
&+& \frac 12
\biggl[\bar \psi \gamma^\alpha \partialboth{}^\beta\psi-
\bar \psi \gamma^\beta \partialboth{}^{\alpha}\psi
\biggr]\biggr|_{\alpha=\mu_\perp,\beta=+}\, .
\ee
The first bracket is the operator of the twist-2 (it is symmetric
and traceless) and can be expressed in terms of the Mellin moments
of the twist-2 GPDs $H,E$. The operator in the second bracket
can be reduced to the total derivative of the axial current with
the help of QCD equation of motion for a massless quark:
 $\Dirac\nabla \psi=0$. Indeed, starting with the identities of
the type $\bar\psi \gamma_\alpha\gamma_\beta \Dirac\nabla\psi=0$
one derives the following identity:

\be
\bar \psi(x) \biggl[ \gamma_\alpha \partialboth_\beta- \gamma_\beta
\partialboth_\alpha\biggr] \psi(x)=
i\varepsilon_{\alpha\beta\rho\sigma}\
\partial^\sigma \biggl[\bar \psi(x)  \gamma^\rho
\gamma_5  \psi(x)\biggr]\, ,
\la{eq:opid} \ee
which can be used to reduce the second bracket in Eq.~(\ref{m2step1})
to the total derivative of the axial current. The latter is obviously
the operator of twist-2.

Puting everything together we obtain the following sum rules
\cite{Pen00b,Kiv01a}:

\be \nonumber \int_{-1}^1dx\ x\
 {\cal F}_{\mu_\perp}(x,\xi,t)&=&\spinn{\gamma^\perp_\mu}\ \\
\nonumber
&\times&\frac {1}2
\left[ \int_{-1}^1dx\ x
\left\{H(x,\xi,t)+E(x,\xi,t)\right\}+g_A(t)\right]\\
&+&\frac{\Delta^\perp_\mu}{4m_N} \bar N(p^{\ \prime})N(p)
 \frac{\partial}{\partial
\xi}\ \int_{-1}^1dx\ x \ E(x,\xi,t)\, .
\la{M1v} \ee
We see that the
GPD ${\cal F}_{\mu_\perp}$ is sensitive to the combination
$$\int_{-1}^1dx\ x \left(H(x,\xi,t)+E(x,\xi,t)\right)+g_A(t)$$
which is related to the spin structure of the nucleon in the
forward  limit:
\be \lim_{\Delta\to 0}\int_{-1}^1dx\ x
\left\{H^q(x,\xi,t)+E^q(x,\xi,t)\right\} + g_A(t)= 2 J^q + \Delta
q\, ,
\la{spin-tw3}
\ee
where $J^q$ is for a given flavor a fraction of the total angular momentum
of the nucleon carried by quarks and $\Delta q$ is the
corresponding fraction of the spin. To derive Eq.~(\ref{spin-tw3}) we
made use of angular momentum
 sum rule (\ref{eq:original_ji_sr}).
 We see that the twist-3 DVCS observables
can be used to probe
the spin structure of the
nucleon. In principle, this part of the amplitude can be
extracted from the data through the angular, spin and $Q$
dependence of the differential cross section
\cite{Die97,Bel00a,Bel01b}. As will be
discussed further in Sec.~\ref{chap5_3_4} and Sec.~\ref{chap5_3_5},
this may provide us with additional
possibility to probe $J^q$ in the nucleon.

In the case of the axial transverse GPD $\widetilde {\cal F}_{\mu_\perp}$ we
obtain in the same way the following result for the corresponding
second moment:
\be
\nonumber \int_{-1}^1dx\ x\ \widetilde
 {\cal F}_{\mu_\perp}(x,\xi,t)&=&\spinn{\gamma^\perp_\mu \gamma_5}\
\frac {1}2 \left[ \int_{-1}^1dx\ x\ \widetilde H(x,\xi)+ \xi^2
\{F_1(t)+ F_2(t)\}\right]\\&+& {\rm other\
structures} \, .
\la{M1a}
\ee
For simplicity we do not write all possible
spinor structures which appear on the {\em rhs}.
The sum rule (\ref{M1a}) can be viewed as the generalization
of the Efremov-Leader-Teryaev sum rule \cite{Efr97} for the polarized
structure function $\Delta_Tq(x)$. The reason for such sum rules is the
absence of the genuine twist-3 local gauge invariant operators of the form
$\bar\psi G \psi$ ($G$ is the gluon field strength) having dimension four.
The third and higher Mellin moments are not generically related to the moments of
twist-2 GPDs $H,E$ and $\widetilde H, \widetilde E$, they get
additional contributions from the genuine twist-3 local gauge invariant operators
of the form $\bar\psi G \psi$. For instance the third Mellin moment of
the GPDs ${\cal F}^\mu$ and $\widetilde {\cal F}^\mu$ has the following
structure:

\be
\nonumber
\int_{-1}^1 dx\ x^2\ {\cal F}^\mu (\widetilde {\cal F}^\mu)&=&
\frac{1}{3(\bar P^+)^2} \langle p^{\ \prime}|\bar\psi \gamma^+(\gamma_5) g\widetilde G^{+\mu }
\psi|p\rangle\\
&+&{\rm moments\ of\ } H,E,\widetilde H,\widetilde E \,.
\la{psiGpsi}
\ee
Here $\widetilde G^{\mu\nu}=\frac i2 \varepsilon^{\mu\nu\alpha\beta} G_{\alpha\beta}$
is a dual gluon field strength tensor.
Generically the value of the matrix element of the genuine twist-3
$\bar\psi G \psi$ operator in the above equation is comparable
with the size of matrix elements of twist-2 operators.
The genuine twist-3
matrix element in Eq.~(\ref{psiGpsi}) in the forward
kinematics has been estimated using QCD sum rules
\cite{Bra87,Ste95}, in the instanton model of the QCD vacuum
\cite{Bal98} and recently in lattice QCD \cite{Goc01}.

In the instanton model of the QCD vacuum the matrix elements of
genuine twist-3 operators are parametrically (and numerically) suppressed relative
to the corresponding matrix elements of twist-2 operators \cite{Bal98,Dre00}.
This suppression is due to the small packing fraction of
instantons in the QCD vacuum which is related to the ratio of the
average size of the instantons to their average relative
distance: $\frac{\bar \rho}{\bar R}\approx \frac 13$.
The dependence of the twist-3 to twist-2 parts of forward
matrix elements on this
small parameter  was obtained in Refs.~\cite{Bal98,Dre00} as:

\be
\frac{\rm twist-3}{\rm twist-2}\sim \frac{\bar\rho^4}{\bar R^4}
\log\left(\frac{\bar\rho^2}{\bar R^2}\right)\, .
\ee
Several indications for the smallness of the above ratio
have been provided recently
by measurements of the transverse spin structure
functions at SLAC \cite{Mit99,Bos00}.
We see that, at least for the forward matrix
elements, the genuine twist-3 operators can be neglected relative
to the twist-2 ones. If one does so, one can express the twist-3 GPDs
in terms of twist-2 ones. Such approximation is called
Wandzura-Wilczek (WW) approximation \cite{Wan77}.

The twist-3 GPDs in
the Wandzura-Wilczek approximation are discussed in the next
section.

\subsubsection{Wandzura-Wilczek approximation for the twist-3 GPDs}
\label{chap3_5_2}

The aim of this section is the calculation of the matrix elements
(\ref{Fdef}) in the
Wandzura-Wilczek (WW) approximation.
The operators entering the definition of the GPDs ${\cal F}^\mu$
and $\widetilde {\cal F}^\mu$ of Eq.~(\ref{Fdef}) can be decomposed
in a gauge invariant way
into two parts, one of which is fixed by the twist-2 GPDs $H,E,\widetilde H$
and $\widetilde E$, and another -- the ``genuine twist-3" part --
associated with operators of the type $\bar\psi G \psi$. The
latter operators are interaction dependent
and measure quark-gluon correlations in the target hadron.
The WW approximation consists in the {\em assumption} that the
nucleon matrix elements of the interaction dependent operators
are small and can be neglected.

To obtain the WW relation we make
use of the operator identities\footnote{Alternative derivation based on
generalized rotational invariance can be found in Ref.~\cite{Ani01}.}
derived
in~\cite{Bel00b,Kiv01c,Rad00,Kiv01a,Rad01a} on the basis of technique developed in
Ref.~\cite{Bra89}:
\be
&&\pbar ( x)\gamma_\mu \p (- x)=
\frac{\pert_\mu}{(\pert x)}\pbar ( x)\xslash \p (- x) \nonumber
\\
&+&\left[
\partial_\mu -\frac{\pert_\mu}{(x\pert)} (x\partial)
\right]\
 \frac12\int_0^1 d\alpha \left\{ e^{\bar \alpha
(x\pert)}+e^{-\bar \alpha (x\pert)} \right\}  \pbar (\alpha x)\xslash \p (-\alpha x)
\label{Oident1} \\
&+& \left[ i\varepsilon_{\mu
\alpha\beta\rho}x^\alpha\frac{\pert^\beta}{(\pert x)}
\partial^\rho\right]\
\frac12\int_0^1 d\alpha \left\{ e^{-\bar \alpha
(x\pert)} - e^{\bar \alpha (x\pert)} \right\}
\pbar (\alpha x)\xslash\gamma_5 \p (-\alpha x)+\dots
\nonumber
\ee
where $\bar \alpha=1-\alpha$ and ellipses stand for the
contributions of either twist-3 quark gluon operators or
twist-4 operators which both are neglected in the WW
approximation. An analogous expression for the operator $\pbar (
x)\gamma_\mu \gamma_5\p (- x)$ can be obtained from
Eq.~(\ref{Oident1}) by the replacement $\xslash\to\xslash\gamma_5$ on
{\em rhs} of this equation.
The symbol $\pert$ denotes the derivative with respect to the
total translation:
\be
\label{Dtot}
\pert_{\alpha}\left\{ \pbar(tx)
\Gamma [tx, -tx] \p(-tx) \right\} \equiv \left.
\frac{\partial}{\partial y^{\alpha}} \left\{ \pbar(tx + y) \Gamma
[tx + y, -tx + y] \p(-tx + y)\right\} \right|_{y \rightarrow 0},
\ee
with a generic Dirac matrix structure $\Gamma$ and $[x,y]
=\mbox{\rm Pexp}[ig\!\! \int_0^1\!\! dt\,(x-y)_\mu
A^\mu(tx+(1-t)y)] $. Note that in the matrix elements  the total
derivative can be easily converted into  the momentum transfer:
\be
\label{exmpl}
\langle p^{\ \prime}|\pert_\mu \pbar(tx)\Gamma [tx, -tx] \p(-tx)
|p\rangle=i(p^{\ \prime}-p)_\mu \langle p^{\ \prime}|\pbar(tx)\Gamma [tx, -tx] \p(-tx)
|p\rangle \ee and therefore in the matrix elements one can
associate $\pert$ with the momentum transfer $\Delta$.

The identity (\ref{Oident1}) allows to express the matrix elements
of the non-local ``vector" and ``axial" operators ({\em lhs}
of (\ref{Oident1})) in terms of
matrix elements of symmetric operators generated by non-local
operators
$\pbar ( x)\xslash (\gamma_5) \p (- x)$ standing on the {\em rhs}
of Eq.~(\ref{Oident1}). Since eventually (after performing differential
operations on the {\em rhs} of Eq.~(\ref{Oident1}))
the coordinate $x^\mu$ is put on the light-cone the symmetric
operators on the {\em rhs} are reduced to the twist-2 operators
and their total derivatives\footnote{Note that in the pioneering
approach of Ref.~\cite{Blu00} to the WW relations for GPDs
the total derivatives were neglected. In this paper the reader can find
a detailed discussion of the mathematical aspects of the twist decomposition
of light-cone operators.}.
This allows us to obtain the
WW-relations for twist-3 nucleon GPDs \cite{Bel00b,Kiv01a}:

\be
 {\cal F}^{WW}_\mu(x,\xi)&=&  \frac{\Delta_\mu}{2 m_N\xi}\spinn{}
E(x,\xi)- \frac{\Delta_\mu}{2\xi}\spinn{\Dirac n}(H+E)(x,\xi)
\nonumber\\
&+&\int_{-1}^{1}du\
G_\mu(u,\xi)W_{+}(x,u,\xi)+i\epsilon_{\perp \mu k}
\int_{-1}^{1}du\  \widetilde G^k (u,\xi)W_{-}(x,u,\xi)\, ,
\la{eq:F}
\\[4mm]
\widetilde {\cal F}^{WW}_\mu(x,\xi)&=&
\frac{\Delta_\mu}{2m_N} \spinn{\gamma_5}\widetilde E(x,\xi)-
\frac{\Delta_\mu}{2\xi}\spinn{\Dirac n\gamma_5}\widetilde H(x,\xi)
\nonumber\\
&+&\int_{-1}^{1}du\ \widetilde
G_\mu(u,\xi)W_{+}(x,u,\xi)+i \epsilon_{\perp \mu k}
\int_{-1}^{1}du\  G^k (u,\xi)W_{-}(x,u,\xi) \, .
\la{eq:Ft}
\ee
The following notations are used~:
\be
\la{eq:G}
G^\mu(u,\xi)&=& \spinn{\gamma^\mu_\perp}(H+E)(u,\xi)+
\frac{\Delta_\perp^\mu}{2\xi m_N} \spinn{}
\biggl[u\frac{\partial}{\partial u}+ \xi\frac{\partial}{\partial
\xi} \biggl] E(u,\xi) \nonumber \\[4mm]&&
-\frac{\Delta_\perp^\mu}{2\xi}
\spinn{\Dirac n}\biggl[u\frac{\partial}{\partial u}+
\xi\frac{\partial}{\partial \xi}\biggl] (H+E)(u,\xi) \, ,
\ee
\be
\la{eq:tG}
\widetilde G^\mu (u,\xi)& =&\spinn{\gamma^\mu_\perp\gamma_5} \widetilde H(u,\xi)
+\frac{\Delta_\perp^\mu}{2m_N} \spinn{\gamma_5}
\biggl[1+u\frac{\partial}{\partial u}+\xi\frac{\partial}{\partial
\xi}\biggl] \widetilde E(u,\xi) \nonumber\\[4mm]&&
-\frac{\Delta_\perp^\mu}{2\xi}\spinn{\Dirac n\gamma_5}
\biggl[u\frac{\partial}{\partial u}+\xi\frac{\partial}{\partial
\xi}\biggl] \widetilde H(u,\xi) \, .
\ee
The functions $W_{\pm}(x,u,\xi)$ can be called
Wandzura-Wilczek  kernels. They were
introduced in Ref.~\cite{Bel00b,Kiv01a} and are
defined as~:
\be
\la{eq:Wpm}
W_{\pm}(x,u,\xi)&=& \frac12\biggl\{
\theta(x>\xi)\frac{\theta(u>x)}{u-\xi}-
\theta(x<\xi)\frac{\theta(u<x)}{u-\xi} \biggl\} \nonumber
\\[4mm]&&\mskip-10mu \pm\frac12\biggl\{
\theta(x>-\xi)\frac{\theta(u>x)}{u+\xi}-
\theta(x<-\xi)\frac{\theta(u<x)}{u+\xi} \biggl\}.
\ee
We also
introduce the metric and totally antisymmetric tensors in the two
dimensional transverse plane ($\varepsilon_{0123}=+1$)~:
\be
\la{gt}
(-g^{\mu \nu})_\perp = -g^{\mu \nu}+ n^\mu \widetilde p^{ \nu}+n^\nu
\widetilde p^{ \mu},
\quad \epsilon^\perp_{\mu \nu}=
\epsilon_{\mu \nu \alpha\beta}n^\alpha \widetilde p^{\beta} \, .
\ee

In Sec.~\ref{chap5_3_4} and Sec.~\ref{chap5_3_5}
we shall use the WW-relations (\ref{eq:F},\ref{eq:Ft})
in order to estimate the contributions of the twist-3 effects to
various DVCS observables.

In the rest of this section, we  discuss properties of the
twist-3 GPDs peculiar for the WW approximation. Let us start with
a discussion of the properties of the WW-kernels (\ref{eq:Wpm}).
We introduce the
following notations for the action of the WW kernels on a function
$f(u,\xi)$:

\be
\label{conv}
W_\pm\otimes f [x,\xi]\equiv \int_{-1}^1du\
W_\pm(x,u,\xi) f(u,\xi)\, , \ee with $W_\pm$ given by
Eq.~(\ref{eq:Wpm}). We shall call ``WW transform'' the resulting
functions $W_\pm\otimes f [x,\xi]$ and we shall call the ``WW
transformation'' the action of the WW kernels.

\vspace{0.1cm}
\noindent\underline{ Limiting cases}.
 We consider two limiting cases of the WW transformation: the forward
limit $\xi\to 0$ and the `meson' limit $\xi\to 1$. In the forward
limit we easily obtain:

\be
\nonumber \lim_{\xi\to 0}
W_+\otimes f [x,\xi]&=& \theta(x\ge 0)\int_x^1 \frac{du}{u}\
f(u,\xi=0)- \theta(x\le 0)\int_{-1}^x \frac{du}{u}\ f(u,\xi=0)\,
,\\ \lim_{\xi\to 0} W_-\otimes f [x,\xi]&=&0\, .
\label{fwd} \ee
We can see that
the action of the $W_+$ in the forward limit reproduces the
Wandzura--Wilczek relation for the spin structure function $g_T$
\cite{Wan77}. The term with $\theta(x\ge 0)$ corresponds to the quark
distributions and the term with $\theta(x\le 0)$ to the antiquark
distributions. The $W_-$ kernel disappears in the forward limit, so that
this kernel is a `genuine non-forward' object.

In the limit $\xi\to 1$ the generalized parton distributions have
properties of meson distribution amplitudes. In this limit the WW
transforms have the form (we use the notation
$f(u,\xi=1)=\varphi(u)$):
\be \nonumber \lim_{\xi\to 1}
W_\pm\otimes f [x,\xi]&=&\frac 12 \biggl\{ \int_{-1}^x
\frac{du}{1-u}\ \varphi(u)\pm \int_{x}^1 \frac{du}{1+u}\
\varphi(u)\biggr\}\, ,
\label{meson}\ee
which corresponds to the WW relations
for the meson distribution amplitudes derived in \cite{Bal96,Ball98}.

We can see that WW transforms of generalized parton distributions
interpolate between WW relations for parton distributions and for the
meson distribution amplitude. The general form of the WW kernels
(\ref{eq:Wpm}) also allows us to derive WW relations for distribution amplitudes
of a meson of arbitrary spin.

\vspace{0.1cm}
\noindent \underline{ Mellin moments}: One can easily derive the Mellin
moments of the WW transform. The result has the form:

\be
\label{Mellin}
\int_{-1}^1dx \ x^N\ W_\pm\otimes f
[x,\xi]=\frac{1}{N+1} \int_{-1}^1 du \left[
\frac{u^{N+1}-\xi^{N+1}}{u-\xi}\pm
 \frac{u^{N+1}-(-\xi)^{N+1}}{u+\xi}\right] f(u,\xi)\, .
\ee
From this simple exercise we can see an important property of the WW
transformation, namely, if the function $f(u,\xi)$ satisfies the
polynomiality condition, {\it i.e.}:

\be
\label{poly}
\int_{-1}^1du\ u^N\ f(u,\xi)={\rm polynomial\ in\ } \xi\
{\rm of\ the\ order\ } N+1\, , \ee
its WW transform also satisfies
the same polynomiality condition.

\vspace{0.1cm}
\noindent\underline{Discontinuities}. In Refs.~\cite{Kiv01c,Rad00,Rad01a} it was
demonstrated that the twist-3 skewed parton distributions in the
WW approximation exhibit discontinuities at the points
$x=\pm\xi$. This feature is related to the properties of the WW
kernels. Let us compute the discontinuities of a WW transform at
the points $x=\pm \xi$:
\be \nonumber \lim_{\delta\to 0}
\left[ W_\pm\otimes f [\xi+\delta,\xi]-W_\pm\otimes f
[\xi-\delta,\xi] \right]&=&\frac 12 vp\int_{-1}^1 \frac{du}{u-\xi}
f(u,\xi)\, ,\\
\lim_{\delta\to 0} \left[ W_\pm\otimes f
[-\xi+\delta,\xi]-W_\pm\otimes f [-\xi-\delta,\xi]
\right]&=&\pm\frac 12 vp\int_{-1}^1 \frac{du}{u+\xi} f(u,\xi)\, .
\la{skachki}
\ee
Here $vp$ means an integral in the sense of {\it valeur
principal}. We see that for a very wide class of functions
$f(u,\xi)$, the discontinuity of the corresponding WW transforms
is nonzero. This feature of the WW transformation may lead to the
violation of the factorization for  the twist-3 DVCS amplitude.

Using general properties of the WW transformation (\ref{skachki}) and
Eqs.~(\ref{eq:F},\ref{eq:Ft}) one obtains that ${\cal F}_\mu^{WW}$ and
$\widetilde {\cal F}_\mu^{WW}$ have discontinuities at the points $x=\pm
\xi$ . But using a certain  symmetry of the Eqs.~(\ref{eq:F}) and
(\ref{eq:Ft}), one can find that some combinations of the distributions
${\cal F}_\mu^{WW}$ and $\widetilde {\cal F}_\mu^{WW}$  are free of
discontinuities. For example, using  Eq.~(\ref{skachki}) one can
see that the combination:

\be
\la{comb1} {\cal F}_\mu^{WW}(x,\xi)-i\varepsilon_{\perp\mu\rho}
\widetilde {\cal F}_\rho^{WW}(x,\xi)\, , \ee
has no discontinuity at
$x=\xi$. On the other hand, the  `dual' combination:
\be
\la{comb2}
 {\cal F}_\mu^{WW}(x,\xi)+ i\varepsilon_{\perp\mu\rho} \widetilde
 {\cal F}_\rho^{WW}(x,\xi)\, , \ee
is free of the discontinuity at
$x=-\xi$. The cancellation of
discontinuities in these particular combinations of the GPDs
ensures the factorization of the twist-3 DVCS amplitude on the
nucleon, as given in Sec.~\ref{chap5_3_3},

\vspace{0.1cm}
\noindent\underline{ Pion pole contribution:} The generalized parton
distribution $\widetilde E$ at small $t$ is dominated by chiral
contribution of the pion pole, see Eq.~(\ref{eq:etilde_chiral}).
In the WW relation (\ref{eq:F}) the twist-2 GPD
$\widetilde E$ enters only in the combination:
\be
\label{killpipo}
\left[ 1+ u\frac{\partial}{\partial
u}+\xi\frac{\partial}{\partial\xi} \right] \widetilde E(u,\xi)\, .
\ee
 One can easily see that the contribution of the pion pole
 (\ref{eq:etilde_chiral})
nullifies under the action of the differential operator in
Eq.~(\ref{killpipo}). The only place where the pion pole contribution
survives is the GPD $\widetilde {\cal F}_\mu$, see Eq.~(\ref{eq:Ft}). In this
way we obtain the following simple results for the contribution
of the pion pole to the twist-3 GPDs in the WW approximation:
\be
\label{pipo3}
\nonumber {\cal F}_\mu^{WW,{\rm pion\ pole}}(x,\xi)&=&0\, ,\\
\widetilde {\cal F}_\mu^{WW,{\rm pion\ pole}}(x,\xi)&=&
\frac{\Delta_\mu}{2m_N}\spinn{\gamma_5}\frac{4 g_A^2
m_N^2}{m_\pi^2-t}\ \frac 1\xi \Phi_\pi\left(\frac x\xi
\right)\theta(\xi-|x|)\, . \ee
We see an interesting result that
the pion pole contribution to the twist-3 GPDs is expressed in
terms of twist-2 pion distributions amplitude. This observation is
in nice agreement with the fact that in the WW approximation there
exists no twist-3 pion distribution amplitude associated with
vector- and axial-vector operators \cite{Bal96}.

\vspace{0.1cm}
\noindent \underline{ D-term contribution:} Now we discuss the
D-term
contribution to the twist-3 GPDs.
If we substitute the D-term contributions (\ref{addingDterm})
for the GPDs $H$ and $E$ into the WW
relations of Eqs.~(\ref{eq:F},\ref{eq:Ft}) we see that the result is:
\be
\label{Dterm3}
\nonumber {\cal F}_\mu^{\rm WW\
D-term}(x,\xi)&=&-\frac{\Delta_\mu}{2\xi\ m_N}\spinn{}
D\left(\frac{x}{\xi}\right)\ \theta(|x|\le\xi)\, ,\\ \widetilde
 {\cal F}_\mu^{\rm WW\ D-term}(x,\xi)&=&0\, . \ee
In some sense the D-term is an eigenfunction of the WW
transformation.

\subsection{GPDs for $N \to \Lambda, \Sigma$ transitions : SU(3) relations
and beyond}
\label{chap3_6}

The studies of hard exclusive processes with strangeness
production give an access to the flavor non-diagonal generalized
parton distributions \cite{Fra99b}.
These distributions correspond to the process where
a quark of one flavor is taken out of the
initial nucleon at the space-time point $y/2$, and then a quark
with another flavor (say a strange quark)
is put back exciting a hyperon  at the space-time
point $-y/2$. The strangeness changing distributions
for $N\to Y$ transitions ($Y=\Sigma, \Lambda$) can be defined as
(cf. Eq.~(\ref{eq:qsplitting})):
\begin{eqnarray}
&& {{\bar  P^{+}}\over {2\pi }}\, \int dy^{-}e^{ix\bar  P^{+}y^{-}}
\langle p^{\ \prime}, Y |\bar{s}_{\beta }(-y/2)  q_{\alpha}(y/2)
|p, N\rangle {\Bigg |}_{y^{+}=\vec{y}_{\perp }=0} \nonumber \\
&=& {1\over 4}\left\{ ({\gamma ^{-}})_{\alpha \beta }
\left[ H^{N\to Y}(x,\xi ,t)\; \bar{Y}(p^{'})\gamma ^{+}N(p)\,
\right. \right.\nonumber\\
&&\hspace{2cm}\left. +\, E^{N\to Y}(x,\xi ,t)\; \bar{Y}(p^{'})i\sigma ^{
+\kappa }{{\Delta _{\kappa }}\over {2m_{N}}}N(p)\right] \nonumber \\
&& \;+({\gamma _{5}\gamma ^{-}})_{\alpha \beta }
\left[ \tilde{H}^{N\to Y}(x,\xi ,t)\; \bar{Y}(p^{'})\gamma ^{+}\gamma_{5}N(p)\,
\right.\nonumber\\
&&\left. \left. \hspace{2cm}+\, \tilde{E}^{N\to Y}(x,\xi ,t)\;
\bar{Y}(p^{'})\gamma _{5}{{\Delta ^{+}}\over {2m_{N}}} N(p) \right]
\right\} , \;\;\;\;
\la{eq:strange}
\end{eqnarray}
where $s_\beta(-y/2)$ is the field operator of a strange quark, $q(y/2)$
is the field operator of a nonstrange quark (with $q=u$ or $d$ depending
on the charge of the hyperon $Y$).

The flavor non-diagonal GPDs provide a new tool to
study the non-perturbative structure of various nucleon-hyperon
transitions. Below we shall relate strangeness changing $N\to Y$ GPDs
to $N\to N$ GPDs with help of flavor $SU(3)$ relations. But
before doing this let us discuss some interesting features of the $N\to Y$
GPDs which are related to the $SU(3)$ symmetry breaking effects.

Let us start with a discussion of the form factor sum rules for
strangeness changing $N\to Y$ GPDs. Integrating the {\em lhs} of
Eq.~(\ref{eq:strange}) over $x$ one gets the local $\bar s q$ operator
sandwiched between a nucleon and a hyperon state. Such matrix
elements are parametrized in terms of vector and axial $N\to Y$
transition form factors entering the description of semi-leptonic
decays of hyperons (see e.g. \cite{Gar85})\footnote{We change the definitions
of the form factors $f_2, g_2$ and $g_3$ by a factor of two to have
correspondence with the standard definition of the nucleon form factors.}:

\[
\langle p^{\ \prime}, Y |\bar{s}\gamma^\mu q|p, N\rangle=\bar Y(p^{\prime})\Biggl[
f_1^{N\to Y} \gamma^\mu+
f_2^{N\to Y} \frac{i\sigma^{\mu\nu}\Delta_\nu}{2 m_N}+
f_3^{N\to Y} \frac{\Delta^\mu}{2 m_N}\Biggr] N(p)\, ,\]
\begin{equation}
\langle p^{\ \prime}, Y |\bar{s}\gamma^\mu\gamma_5 q|p, N\rangle=\bar Y(p^{\prime})\Biggl[
g_1^{N\to Y} \gamma^\mu+
g_2^{N\to Y} \frac{i\sigma^{\mu\nu}\Delta_\nu}{2 m_N}+
g_3^{N\to Y} \frac{\Delta^\mu}{2 m_N}\Biggr]\gamma_5 N(p).
\label{semilep-ff}
\end{equation}
We see that for the strangeness changing transitions two
additional form factors $f_3^{N\to Y}(t)$ and $g_2^{N\to Y}(t)$
related to the second class currents
($g_2$ is also often referred to as weak electricity ) appear.
Both these form-factors are proportional to flavor $SU(3)$
breaking ($\sim m_s$) effects and they are absent for transitions without
strangeness change.

The form factor sum rules for strangeness changing GPDs have the
following form:

\be
\int_{-1}^1dx\ H^{N\to Y}(x,\xi,t)&=&f_1^{N\to Y}(t)-\xi\frac{m_Y+m_N}{2
m_N}\ f_3^{N\to Y}(t)\, ,\nn
\int_{-1}^1dx\ E^{N\to Y}(x,\xi,t)&=&f_2^{N\to Y}(t)+\xi\ f_3^{N\to
Y}(t)\, ,\nn
\int_{-1}^1dx\
\widetilde H^{N\to Y}(x,\xi,t)&=&g_1^{N\to Y}(t)+\frac{m_Y-m_N}{2
m_N}\ g_2^{N\to Y}(t)\, ,\nn
\int_{-1}^1dx\
\widetilde E^{N\to Y}(x,\xi,t)&=&g_3^{N\to Y}(t)+\frac{1}{\xi}
\ g_2^{N\to Y}(t)\, .
\la{ff-sr-strange}
\ee
One sees that in contrast to the corresponding sum rules for the
GPDs without strangeness changing
(see Eqs.~(\ref{eq:ffsumruleh}-\ref{eq:ffsumruleet}))
the sum rules (\ref{ff-sr-strange}) get additional contributions
from the $SU(3)$ symmetry breaking form factors $f_3(t)$ and
$g_2(t)$, associated with the second class currents.
Note that these additional contributions also violate
the symmetry of GPDs under the transformation $\xi \to -\xi$
\cite{Man98a}. The reason is that for the derivation of this
symmetry the time inversion transformation has been used.
Under this transformation the initial and final baryon states
are exchanged. In the case when this transition occurs between baryons
belonging to the same symmetry multiplet the $T$-invariance gives the symmetry
property under $\xi\to -\xi$. However for the case of transitions
between baryons from different multiplets the $T$-invariance is not
restrictive (it gives only the realness of the GPDs).

Studies of semiletponic hyperon decays showed
(see e.g. \cite{Gar85}) that the assumption
of the $SU(3)$ flavor symmetry works very well for form factors
$g_1, f_1$, and $f_2$. Unfortunately the semileptonic decays of
hyperons are not sensitive to $f_3,g_2$ and $g_3$. On general
grounds we can state that the form factors $f_3$ and $g_2$ are
proportional to the explicit $SU(3)$ breaking, {\em i.e.} they are
proportional to the strange quark mass, therefore their effect can be
non-negligible.
For the form factor
$g_3(t)$ we can apply the generalized PCAC relations to obtain the
contribution of the kaon pole:

\be
g_3^{N\to Y}(t)\approx \frac{2 g_{KN Y} f_K
}{m_K^2-t}\, ,
\ee
where $g_{KNY}$ are the $KNY$ coupling constants, and $f_K\approx
159$~MeV is the kaon decay constant.
Comparing the $t-$dependence of this equation with the one of the
the pion pole contribution to
pseudoscalar nucleon form factor (\ref{eq:ha_pipo})
we notice a big difference
due to the large difference between the kaon and pion masses.
This shows that $SU(3)$ symmetry breaking effects are sizeable
for $g_3^{N\to Y}(t)$.

{}From the sum rules (\ref{ff-sr-strange}) we see that the $SU(3)$
symmetry breaking effects are suppressed for the GPDs $H,E$, and $\widetilde
H$, at least they are suppressed in the sum rules by
either $\xi$ (like for $H$ and $E$) or $m_Y-m_N$ (like for $\widetilde
H$). On contrary, the sum rule for GPD
$\widetilde E^{N\to Y}(x,\xi,t)$
shows that the flavor symmetry breaking effects can be very
strong for this GPD. Indeed the first term on the {\em rhs} of
the sum rule is proportional to $g_3^{N\to Y}(t)$ for which one
expects large symmetry breaking effects (see discussion above).
Additionally the weak electricity (symmetry breaking) form factor
enters the corresponding sum rule with the factor of $1/\xi$.
This, in particular, implies that in the limit $\xi\to 0$ the
function is divergent\footnote{One can introduce an
``ovecomplete" set of GPDs adding to the {\em rhs} of Eq.~(\ref{eq:strange})
one additional ``weak electricity"
GPD $\widetilde E^{N\to Y}_{WE}
\sigma^{+\nu}\Delta_\nu\gamma_5/2m_N$. In this way one avoids working
with GPDs non normalizable at $\xi=0$.},
and this divergency is dominated by
weak electricity. This observation opens an exciting possibility
to study the weak electricity of nucleon-hyperon transitions (otherwise hardly
accessible) in
hard exclusive strange meson production.
In this respect, the transverse spin azimuthal asymmetry in the production of
kaons on the nucleon is a very promising observable as it is sensitive to
the function $\widetilde E^{N\to Y}$, see Sec.~\ref{chap_5_5_4} and qualitative
discussion below.

Given the large expected $SU(3)$ symmetry breaking effects for the
GPD $\widetilde E^{N\to Y}$ we model this function by the
contribution of the kaon pole plus the weak electricity
contribution:

\be
\widetilde E^{N\to Y}(x,\xi,t)&=&
\frac{2 g_{KNY} m_N f_K}{m_K^2-t}\
\theta\left[\xi-|x|\right]\ \frac 1\xi\ \Phi_K\left(\frac x \xi\right)
\nn
&+&\frac 1\xi\ \widetilde E^{N\to Y}_{WE}(x,\xi,t)
\, ,
\la{hyperonEt}
\ee
where $\Phi_K(z)$ is the kaon distribution amplitude and $\widetilde E^{N\to Y}_{WE}(x,\xi,t)$
is the ``weak electricity" GPD normalized to the corresponding
weak electricity form factor $g_2^{N\to Y}(t)$:

\be
\int_{-1}^1dx\
\widetilde E^{N\to Y}_{WE}(x,\xi,t)&=&g_2^{N\to Y}(t)\, .
\ee
This GPD is proportional to explicit $SU(3)$ symmetry breaking
effects ($\sim m_s$). If for the weak electricity GPD we use the
following
rough (asymptotic) ansatz
$\widetilde E_{WE}^{N\to Y}=3/4 (1-x^2/\xi^2)/\xi\ g_2^{N\to Y}(t) $
the transverse spin asymmetry in production of the kaons  has the dominant contribution
($cf.$ Eq.~(\ref{xsec})):

\be
\nonumber
\sigma^{N\to Y K} \propto |S_\perp| \sin\Phi\ \sqrt{t_{\rm min}-t} \left[
\frac{ g_{KNY} m_N f_K}{-t+m_K^2}+
\frac{ g_2^{N\to Y}(t)}{2\xi }\right]
\widetilde H^{N\to Y}(\xi,\xi, t)\, .
\label{xsechyperon}
\ee
This expression illustrates that at small $\xi$ the measurement of
hard exclusive kaon production can give us (among other exciting things!) an access to
the symmetry breaking weak electricity form factor $g_2^{N\to
Y}(t)$. The contribution of the weak electricity is enhanced at
small $x_{B}$. For example for the $\gamma^* p\to \Lambda K^+$
processes at $-t=0.5$~GeV$^2$ and $x_{B}\approx 0.05$ the contribution of
the weak-electricity part in (\ref{xsechyperon}) is of the same
order as the contribution of the kaon pole. We take for this
estimate the value of $g_2^{p\to\Lambda}(0)\approx 0.1$ obtained
in Ref.~\cite{Kim98} in the framework of the chiral quark-soliton model.
This simple estimate shows that the studies of the second class
currents are feasible in hard exclusive reactions with production
of strangeness.

For the strangeness changing GPDs $H^{N\to Y},E^{N\to Y}$ and
$\widetilde H^{N\to Y}$ we can apply the flavor $SU(3)$ relations
 as it seems that for these GPDs the effects of the
symmetry breaking are small. The $SU(3)$ relations for these
functions have the following form \cite{Fra99}:

\be
H^{p\to \Lambda}(x,\xi,t)&=&-\frac{1}{\sqrt 6}\left(2 H^u-H^d-H^s
\right)\, ,\nn
H^{p\to \Sigma^0}(x,\xi,t)&=&-\frac{1}{\sqrt 2}\left(H^d-H^s
\right)\, ,\nn
H^{p\to \Sigma^+}(x,\xi,t)&=&-\left(H^d-H^s
\right)\, ,
\la{su3fromp}
\ee
and similar expressions for GPDs $E^{N\to Y}$ and
$\widetilde H^{N\to Y}$. For transitions from the neutron we can
analogously derive:
\be
H^{n\to \Lambda}(x,\xi,t)&=&-\frac{1}{\sqrt 6}\left(2 H^d-H^u-H^s
\right)\, ,\nn
H^{n\to \Sigma^0}(x,\xi,t)&=&-\frac{1}{\sqrt 2}\left(H^u-H^s
\right)\, ,\nn
H^{n\to \Sigma^-}(x,\xi,t)&=&-\left(H^u-H^s\right)\, .
\la{su3fromn}
\ee

The $SU(3)$ relations (\ref{su3fromp},\ref{su3fromn}) give us a
possibility to access strange quark distributions in the nucleon.
Indeed, using these relations we can derive the following obvious
relations [we show only the example of $\widetilde H$]:

\be
\lim_{\Delta\to 0}
\widetilde H^{p\to \Lambda}(x,\xi,t)&=&-\frac{1}{\sqrt 6}
\left(2 \Delta u(x)-\Delta d(x)-\Delta s(x)
\right)\, ,\nn
\lim_{\Delta\to 0}
\widetilde
H^{p\to \Sigma^0}(x,\xi,t)&=&-\frac{1}{\sqrt 2}\left(\Delta d(x)-\Delta s(x)
\right)\, ,\nn
\lim_{\Delta\to 0}
\widetilde
H^{p\to \Sigma^+}(x,\xi,t)&=&-\left(\Delta d(x)-\Delta s(x)
\right)\, .
\ee
This illustrates that measurements of strangeness changing GPDs
can give us important information about strange (anti-)quark
polarized distributions.

\subsection{GPDs for $N \to \Delta$ transition}
\label{chap3_7}

In the hard exclusive processes with an excitation of
the $\Delta$ one accesses the parton distributions for
$N\to\Delta$ transition \cite{Fra98a,Fra00}.
A knowledge of such distributions would
give us the detailed information about the physics of $N\to\Delta$
transition, in particular it would allow us to understand
the physics of this transition at the level of partons.

The $N\to\Delta$ quark distributions can be defined through
nondiagonal matrix elements of products of quark fields at
light--cone separation.
We give the definitions of $N\to \Delta$ generalized quark distributions
for $p\to\Delta^+$ transition.
The GPDs of other transitions, e.g.
$p\to\Delta^{++}$, $p\to\Delta^0$, etc. can be obtained
with the help of the following isospin factors:

\be
[p\to \Delta^{++}]=\sqrt{\frac 32}\ [p\to \Delta^+]\, , \qquad
{}[p\to \Delta^{0}]=\sqrt{\frac 12}\ [p\to \Delta^+]\, ,
\ee
and analogous relations for transitions from the neutron.
Keeping in mind this simple isospin relations,
 we introduce for the $p\to\Delta^+$ transition
 the following
three GPDs (cf. Eq.~\ref{eq:qsplitting}  with  $y^\mu=\lambda n^\mu$)):

\be
\label{HND-QCD-2} &&
\int \frac{d\lambda }{2\pi }e^{i\lambda x}\langle
\Delta, p^{\prime }|\bar \psi (-\lambda n/2){\Dirac n}\; \tau^3\
 \psi (\lambda n/2)|N, p\rangle =\\
 \nonumber
&&\sqrt{\frac{2}{3}} \; \bar \psi^\beta(p^{\ \prime}) \; \bigl[ H_M(x,\xi
,t) {\cal K}_{\beta\mu}^M n^\mu + H_E(x,\xi ,t)
{\cal K}_{\beta\mu}^E n^\mu + H_C(x,\xi ,t) {\cal
K}_{\beta\mu}^C n^\mu\bigr] \; N(p) \, . \ee Here
$n_\mu$ is a light-cone vector,
\be
n^2 &=& 0, \hspace{1cm} n\cdot \bar P \;\; = \;\; 1, \hspace{1cm}
\bar P^\mu=\frac 12 (p+p^{\ \prime})^\mu \label{n-normalization} \ee
$\Delta$ is the four--momentum transfer,
\begin{equation}
\Delta = p^{\prime }-p , \label{Delta-def}
\end{equation}
$\psi^\beta(p^{\ \prime})$ is the $\Delta$ Rarita--Schwinger spinor, $N(p)$
is the nucleon Dirac spinor. The $N\to\Delta$ quark distributions
$H_{M,E,C}(x,\xi ,\Delta ^2)$ are regarded as functions of the
variable $x$, the square of the four--momentum transfer,
$t$, and its longitudinal component
\begin{equation}
\xi =-\frac 12 (n\cdot \Delta ) . \label{xi-def}
\end{equation}
The covariants ${\cal K}_{\beta\mu}^{M,E,C}$ are magnetic dipole,
electric quadrupole and Coulomb quadrupole covariants \cite{Jon73}:
\be
\nonumber {\cal K}_{\beta\mu}^M&=& -i\frac{3(m_\Delta+m_N)
}{2m_N((m_\Delta+m_N)^2-t)}
\varepsilon_{\beta\mu\lambda\sigma}\bar P^\lambda
\Delta^\sigma\;,\\ {\cal K}_{\beta\mu}^E&=&-{\cal K}_{\beta\mu}^M-
\frac{6(m_\Delta+m_N)}{m_NZ(t)}
\varepsilon_{\beta\sigma\lambda\rho} \bar P^\lambda\Delta^\rho
\varepsilon^\sigma_{\mu \kappa\delta} \bar P^\kappa\Delta^\delta
\gamma^5\;, \label{K-def}
\\
\nonumber {\cal K}_{\beta\mu}^C&=&
-i\frac{3(m_\Delta+m_N)}{m_NZ(t)} \Delta_\beta(t
\bar P_\mu-\Delta\cdot\bar P\Delta_\mu)\gamma^5\;,
 \ee
where we introduced the notation $Z(t)=
[(m_\Delta+m_N)^2-t][(m_\Delta-m_N)^2-t]$. We choose
such definition of the $N\to\Delta$ quark distributions in order
to have the following sum rules:
\be
\int_{-1}^{1}dx\ H_{M,E,C}(x,\xi,t)=2\ G_{M,E,C}^*(t)\;,
\ee where $G_{M,E,C}^*(t)$ are the standard magnetic
dipole, electric quadrupole and Coloumb quadrupole
transition form factors respectively \cite{Jon73}.

Analogously to Eq.~(\ref{HND-QCD-2}) we introduce the following
quark helicity dependent $p\to\Delta^+$ generalized quark
distributions:

\be
\nonumber &&\int \frac{d\lambda }{2\pi }e^{i\lambda x}\langle
\Delta, p^{\prime }|\bar \psi (-\lambda n/2){\Dirac n}\gamma^5
\tau^3
 \psi (\lambda n/2)|N, p\rangle =\\
\nonumber && \bar \psi^\beta(p^{\ \prime}) \; \biggl[ C_1(x,\xi ,t)
n_\beta+C_2(x,\xi ,t)
\frac{\Delta_\beta(n\cdot\Delta)}{m_N^2} + C_3(x,\xi
,t)\frac 1{m_N}
 \{n_\beta {\Dirac \Delta}-\Delta_\beta {\Dirac n}\}\\
&& \hspace{1.cm} +\; C_4(x,\xi ,t) \frac{1}{m_N^2} \{\bar
P\cdot \Delta n_\beta-2\Delta_\beta\}\biggr] \; N(p) \;.
\label{HND-helicity-QCD-2} \ee Actually in this definition not all
the functions $C_i$ are independent, for example the function
$C_4$ can be absorbed in a redefinition of the functions $C_1$ and
$C_2$. We choose the ``overcomplete" definition
(\ref{HND-helicity-QCD-2}) to have simple ($\xi$-independent) sum
rules for the first Mellin
moment of generalized distributions $C_i$:

\be
\nonumber \int_{-1}^1 dx\ C_1(x,\xi,t)=2\ C_5^A(t)\, ,\quad
\int_{-1}^1 dx\ C_2(x,\xi,t)=2\ C_6^A(t) \, ,
\\ \int_{-1}^1
dx\
C_3(x,\xi,t)=2\ C_3^A(t)\, , \quad
\int_{-1}^1 dx\ C_4(x,\xi,t)=2\
C_4^A(t)\, . \label{axial-sumrules} \ee
Here $C_i^A(t)$ are axial
$N\to \Delta$ transition
form factors in the notations of Adler \cite{Adl75}:

\be
\nonumber &&\langle \Delta ^+, p^{\prime }|\bar \psi
(0)\gamma_\mu\gamma_5
 \psi (0)|p, p\rangle =\\
\nonumber && \bar \psi^\beta(p^{\ \prime}) \; \biggl[ C_5^A(t)
g_{\mu\beta}+C_6(t) \frac{\Delta_\beta\Delta_\mu}{m_N^2} +
C_3^A(t)\frac 1{m_N}
 \{g_{\beta\mu} {\Dirac \Delta}-\Delta_\beta \gamma_\mu\}\\
&& \hspace{1cm} +\; C_4^A(t) \frac{2}{m_N^2} \{\bar P\cdot
\Delta g_{\beta\mu}-\bar P_\mu \Delta_\beta\}\biggr] \; N(p) \;.
\label{Adlers-form-factors} \ee
In Sec.~\ref{chap4} we shall
derive large $N_c$ relations between the $N\to\Delta$ generalized quark
distributions and the usual $N\to N$ ones.

For the $N\to\Delta$ GPDs one can repeat almost word for word the
discussion of such issues as polynomiality, pion pole
contribution, etc. However in applications for hard exclusive
processes with excitation of $\Delta$ (see Sec.~\ref{chap5_4} and Sec.~\ref{chap_5_5_4})
we shall use only the large $N_c$ relations derived in Sec.~\ref{chap4_2}.
Therefore here we restrict ourselves to discussion of these relations. They have
the following form:

\be
H_M(x,\xi,t)&=&\frac{2}{\sqrt 3}\left[ E^u(x,\xi,t)
-E^d(x,\xi,t)\right]
\; , \nonumber\\
C_1(x,\xi,t)&=&
\sqrt 3\left[ \tilde H^{u}(x,\xi,t)- \tilde H^{d}(x,\xi,t)\right]\, , \nonumber\\
C_2(x,\xi,t)&=&
\frac{\sqrt 3}{4}\left[\tilde E^{u}(x,\xi,t)-\tilde E^{d}(x,\xi,t)\right]\, .
\la{ndNc}
\ee
All other $N\to\Delta$ GPDs are zero at leading order of $1/N_c$
expansion.
Let us also note that if we take the first
moment of the Eqs.~(\ref{ndNc}) and
take into account the sum rules for $N\to N$ GPDs
(\ref{eq:ffsumruleh}-\ref{eq:ffsumruleet})
we reproduce the well known result of
soliton models for the $N\to\Delta$ transition form factors.
Integrating $H_M$, given by the large $N_c$
relation of Eq.~\ref{ndNc},
over $x$ at $t=0$ gives the relation
between magnetic moments:
\be
\mu_{N\Delta}=\frac{1}{\sqrt 2}(\mu_p-\mu_n)\; ,
\ee
which is satisfied
rather well by the experimental data on magnetic moments.
As we discussed in Sec.~\ref{chap3_4_2}, the GPD $\widetilde E$
contains a contribution of the pion pole. From the large
$N_c$ relation for $C_2$
in Eq.~(\ref{ndNc}) we conclude that the $N\to \Delta$
GPD $C_2(x,\xi,t)$ also contain the pion pole contribution of the
form:
\be
\lim_{t\to m_\pi^2} C_2(x,\xi,t)= \sqrt 3\
\frac{g_A
m_N^2}{m_\pi^2-t}\ \theta\left[\xi-|x|\right]\ \frac 1\xi\ \Phi_\pi\left(\frac x
\xi\right)\, .
\la{eq:C2chiral}
\ee

The $N\to\Delta$ distribution $H_M(x,\xi,t)$ in the large
$N_c$ limit contains information about the isovector part
of the angular momentum of the nucleon carried by the quarks --
$J^u-J^d$. Indeed with help of angular momentum sum rule (\ref{jisr})
we easily obtain

\be
\lim_{t\to 0, N_c\to\infty}\int_{-1}^1dx\ x\ H_M(x,\xi,t)=
\frac{2}{\sqrt 3}\left[2\left(J^u-J^d\right)- M_2^u+M_2^d\right]
\,,
\ee
where, as usually:
$$M_2^q=\int_0^1 dx\ x \ \left[q(x)+\bar q(x)\right]\,.$$
Note that the contribution of the D-term does not appear
in the $N\to\Delta$ transition, because the D-term is
a flavor singlet quantity.

Obviously one can easily introduce GPDs for other $N\to N^*$ ($N^*$
is a nucleon resonance) transitions which enter description of
hard exclusive processes with excitation of baryonic resonances.
Such kind of reactions can be viewed as a new tool for the study of
baryon spectroscopy (spectroscopy by light-cone operators).

\subsection{Further generalizations: GPDs for $N \to \pi N$ transitions}
\label{chap3_8}

The  transitions of the nucleon to a $\pi N$
system with low invariant mass
can also be studied by light-cone probes encoded in the
$N\to\pi N$ GPDs\footnote{Previously $N\to\pi N$ GPDs were briefly
discussed in Ref.~\cite{Pol98}}
 which can be introduced as follows
(we only show the parametrization of the ``vector" bilocal quark operator):

\begin{eqnarray}
&& {{\bar  P^{+}}\over {2\pi }}\, \int dy^{-}e^{ix\bar  P^{+}y^{-}}
\langle  \pi(k) N(p^{\ \prime}) |\bar{\psi}(-y/2) \gamma^+  \psi(y/2)
|N(p)\rangle {\Bigg |}_{y^{+}=\vec{y}_{\perp }=0} \nonumber \\
&&=\frac{1}{f_\pi}
\left[ H_1^{\pi N}(x,\xi ,t,\alpha,t_\pi, m_{\pi N})\; \bar{N}(p^{'})\gamma ^{+}\gamma_5 N(p)\,
\right. \nonumber\\
&& +\, H_2^{\pi N}(x,\xi ,t,\alpha,t_\pi, m_{\pi N})\; \bar{N}(p^{'})
{{i\sigma ^{+\nu }
k _{\nu }}\over {2m_{N}}}\gamma_5 N(p) \nonumber \\
&& \;+
 {H_3}^{\pi N}(x,\xi ,t,\alpha,t_\pi, m_{\pi N})\; \bar{N}(p^{'})
 {{\Dirac k}\over {2 m_N}}\gamma_{5}N(p)\,
\nonumber\\
&& \left. +\;
{H_4}^{\pi N}(x,\xi ,t,\alpha,t_\pi, m_{\pi N})\;
\bar{N}(p^{'})\gamma_5 N(p) \right]. \;\;\;\;
\la{eq:piN}
\end{eqnarray}
Here $f_\pi\approx 92.4$~MeV is the pion decay constant and
we introduced the following kinematical variables:

\begin{equation}
\bar{P}\equiv \frac{p+p^{\ \prime}+k}{2}\, ,
\end{equation}
and the momentum transfer
\begin{equation}
\Delta \equiv p^{\ \prime}+k-p.
\end{equation}
\( n \) is a light-cone vector, i.e.~\( n^{2}=0 \), which is normalized
such that \( n\cdot \bar{P}=1 \).
The $N\to\pi N$ GPDs $H_i^{\pi N}$ are the functions of the
invariants on which the matrix elements
depend. These are the skewedness:
\begin{equation}
\xi \equiv -\frac{n\cdot \Delta }{2},
\end{equation}
the square of the momentum transfer \( t=\Delta^{2} \), the square of the
nucleon momentum difference \( t_{\pi }=(p^{\ \prime}-p)^{2} \), the invariant mass
of the final $\pi N$ system $m_{\pi N}$ and the
light-cone fraction \( \alpha  \), which
characterizes the longitudinal fraction of the final pion:
\begin{equation}
\alpha \equiv \frac{n\cdot k}{1-\xi }\, .
\end{equation}

These GPDs can be used to describe the hard
``semi-exclusive" processes of the type:

\be
\gamma^* + N\to \gamma + (\pi N) \, ,
\qquad \gamma^* + N\to M  +(\pi N) \, ,
\label{eq:reacsemiex}
\ee
in which the target nucleon dissociates into a low mass $\pi N$
system. The description of such reactions is important because in the
high energy experiments the resolution in the missing mass
of the recoil baryonic system can be rather low.
In such cases the data for e.g. $\gamma^* N\to \gamma N$
always include the $\gamma^* N\to \gamma (\pi N)$ `contamination'
due to a soft pion.

Another motivation for studies of $N\to\pi N$ GPDs is that in the
threshold region, where $m_{\pi N}\to m_\pi+ m_N$, $\alpha\to
m_\pi/(m_\pi+m_N)$, etc, the $\pi N$ GPDs can be expressed in
terms of nucleon and pion GPDs \cite{Mos01,Str01}.
Additionally one can develop
the systematic chiral perturbation theory for the $N\to \pi N$
GPDs. Indeed the first Mellin moment of the $N\to\pi N$ GPDs
(\ref{eq:piN}) corresponds to the amplitude of the pion
electroproduction off the nucleon. Near the threshold
this process has been described by chiral perturbation theory, for
the review see \cite{Ber94}. The second Mellin moment of
the $N\to\pi N$ GPDs corresponds to pion production by the
operator of the energy momentum tensor, this can be rephrased
as the pion production by a graviton. This simple example shows
that the studies of the hard reactions of the type (\ref{eq:reacsemiex})
open a new, large field for applications of chiral perturbation
theory. Measurements of these GPDs would give a new sensitive tool
for studies of chiral symmetry breaking mechanism in QCD.

The $N\to \pi N$ GPDs in the resonance region, i.e. for $m_{\pi N}\approx
m_R$, give us a new tool for studying of the resonance spectroscopy, i.e. the
spectroscopy with light-cone probes.

\newpage
\section{GPDs IN THE LARGE $N_c$ LIMIT}
\label{chap4}
The large $N_c$ limit in QCD
\cite{'tHo74,Wit79}
proved to be an useful tool to analyze
the properties of mesons and baryons.
The idea is that our world with the number of colors $N_c=3$ is
not qualitatively different from an imaginary world with large
number of colors. For the comprehensive
review of applications of the large $N_c$ expansion in hadronic physics see
Ref.~\cite{Mano98}. In this section we study the GPDs of the nucleon in the large
$N_c$ limit, first discussing general properties of the GPDs in
this limit and then presenting calculations of the GPDs in the specific
dynamical realization of the low-energy large $N_c$ QCD -- chiral
quark-soliton model.

\subsection{Generalized parton distributions at large $N_c$ : general properties}
\label{chap4_1}

In order to consider the generalized
parton distributions in the large $N_{c}$
limit we start from the quark correlator $\langle B_{2}|\psi
(x_{2})\bar{\psi} (x_{1})|B_{1}\rangle $. The Hartree picture of
the large $N_{c}$ nucleon described in \cite{Wit79} leads to the
following form of this correlation function \cite{Pob00}:
\[
\langle B_{1},{\bf p}_{1}|
\bar\psi _{s_{1}f_{1}}(x_{1}^{0}, {\bf
x}_{1})\psi _{s_{2}f_{2}}(x_{2}^{0},{\bf x}_{2})|B_{2}, {\bf
p}_{2}\rangle =2m_{B}N_{c}\int d^{3} {\bf X}e^{i({\bf p}_{2}-{\bf
p}_{1})\cdot {\bf X}}
\]
\begin{equation}
\times\int dR\phi_{B_{1}}^{\ast }(R)\phi_{B_{2}}(R)\left[
R_{f_{2}f_{2}^{\prime }}F_{s_{2}f_{2}^{\prime },s_{1}f_{1}^{\prime
}} \biggl(x_{1}^{0}-x_{2}^{0}, {\bf x}_{1}-{\bf X},{\bf
x}_{2}-{\bf X}\biggr) (R^{-1})_{ f_{1}^{\prime }f_{1}} \right]\, .
\label{psi-psi-bar-Hartree}
\end{equation}
Here $F_{s_{2}f_{2},s_{1}f_{1}}(x_{1}^{0}-x_{2}^{0},{\bf
x}_{1},{\bf x} _{2})$ is the correlation function $\langle \psi
(x_{2})\bar{\psi} (x_{1})\rangle $ corresponding to the static
solution of the Hartree equation, $s_{i}$ are Dirac spinor indices
and $f_i$ are $SU(2)$ isospin indices. The $x^{0}-y^{0}$ time
dependence expresses the fact that we deal with a static solution.
The solution $ F_{s_{2}f_{2},s_{1}f_{1}}(x_{2}^{0}-x_{1}^{0},{\bf
x}_{2},{\bf x}_{1})$ also violates the $SU(2)$ flavor and the
3-space translation invariance. Therefore in Eq.
(\ref{psi-psi-bar-Hartree}) we used plane waves $e^{i {\bf p}\cdot
{\bf X}}$ and the $SU(2)$ rotator wave functions $\phi_{B}(R) $
expressed in terms of Wigner finite rotation functions \cite{Adk83}:
\be
\phi _{S_3T_3}^{S=T}(R) &=&
\sqrt{2S+1}(-1)^{T+T_3}D_{-T_3,S_3}^{S=T}(R)
\label{wf1}\, , \ee
in
order to construct baryon states with given spin, isospin and
momentum.
The expression in the square brackets in Eq.~(\ref{psi-psi-bar-Hartree})
(where one puts $(x-y)^2=0$)
can be called a parton distribution of a soliton.

Since the exact form of the effective Hartree hamiltonian
corresponding to the large $N_{c}$ limit generically
is not known
we do not
know the function $
F_{s_{2}f_{2},s_{1}f_{1}}(x_{2}^{0}-x_{1}^{0},{\bf x}_{2},{\bf
x}_{1})$ either\footnote{In the next section
we shall use the chiral quark-soliton model to compute this function}.
However, following \cite{Wit79} we assume that
this solution has the spin-flavor symmetry:
\begin{equation}
S_{s_{2}s_{2}^{\prime }}(R)R_{f_{2}f_{2}^{\prime
}}F_{s_{2}^{\prime }f_{2}^{\prime },s_{1}^{\prime }f_{1}^{\prime
}} \biggl(x^{0}-y^{0},O_{space}(R) {\bf x},O_{space}(R){\bf
y}\biggr) (R^{-1})_{f_{1}^{\prime }f_{1}}S_{s_{1}^{\prime
}s_{1}}(R^{-1})
\end{equation}
\begin{equation}
=F_{s_{2}f_{2},s_{1}f_{1}}(x^{0}-y^{0},{\bf x},{\bf y})
\label{F-hedgehog}
\end{equation}
Here we use the notation $S_{s_{2}s_{2}^{\prime }}(R)$ for spin
rotations and $O_{space}(R)$ for space rotations, which are parametrized
by a $SU(2)$ matrix $R$.

Using the expression (\ref{psi-psi-bar-Hartree}), one can obtain any
quark parton distribution function
in the large $ N_{c}$ limit. Using the spin-flavor symmetry (\ref{F-hedgehog}) of
the soliton distribution (the expression in square brackets of
Eq.~(\ref{psi-psi-bar-Hartree})) we can write explicitly the
dependence on the orientation angles $R$ of particular $\bar \psi \psi$
operators entering the definition of GPDs, see
Eq.~(\ref{eq:qsplitting}).
In the leading order of $1/N_c$ we have:
\be
\nonumber
\frac{1}{2\pi}
\int d y^-  e^{i x \bar P^+ y^-}\langle {\rm
sol}\ P^{\prime }|\bar \psi_{f'} (-y/2) \gamma^+
 \psi_f (y /2)|{\rm sol}\  P\rangle\biggr|_{y^+=y_\perp=0} \\ {}
=H_0(x,\xi,t) \delta_{ff'} -i \frac{H_1(x,\xi,t)}{m_N} \frac12
\Tr(R^\dagger M^{(ff')} R\tau_j)\cdot \varepsilon_{3jk}
\Delta_k \, , \label{spdsol1} \ee
where $R$ is $SU(2)$ matrix which
defines the orientation of the soliton
in the spin-flavor space, $M^{(ff')}$ is the matrix, of which
all elements  are
zero except the element in row $f$ and column $f'$ which is unity.
Furthermore in Eq.~\ref{spdsol1}, $H_{0,1}(x,\xi,t)$ are the
universal GPDs of the soliton. Analogously
we can write for the quark helicity dependent GPDs of the soliton:
\be
\nonumber
\frac{1}{2\pi}
\int d y^-  e^{i x \bar P^+ y^-}\langle {\rm
sol}\ P^{\prime }|\bar \psi_{f'} (-y/2)\gamma^+ \gamma_5
 \psi_f (y/2)|{\rm sol}\ P\rangle\biggr|_{y^+=y_\perp=0} \\
=\Bigl[ \widetilde H_0(x,\xi,t)\delta^{i3} -
\widetilde E_1(x,\xi ,t)
 \frac{\Delta^3\Delta^i}{m_N^2}\Bigr]
\frac12 \Tr(R^\dagger M^{(ff')} R\tau_i) \, . \label{spdsol2}
\ee
Now to get the GPDs  for a baryon with particular quantum
numbers (or for transition GPDs) we have to sandwich equations
(\ref{spdsol1},\ref{spdsol2}) between rotational states (\ref{wf1}) of the
soliton which corresponds to baryons from the octet and/or decuplet, i.e.
\be
\langle S'=T',S'_3,T'_3|\ldots| S=T,S_3,T_3\rangle &=& \int
dR\;\phi^{\ast\;S'=T'}_{S'_3T'_3}(R) \; \ldots \; \phi^{S=T}_{S_3T_3}(R)\,.
\label{spisosp}
\ee

We note here that in the derivation of
Eqs.~(\ref{spdsol1},\ref{spdsol2}) we used that
in the large $N_c$ limit
the baryon  masses rise linearly with $N_c$,
with mass splitting $m_{\Delta}-m_N\sim 1/N_c$. Therefore in the
first approximation we can
neglect nucleon-$\Delta$ mass difference and consider them as
nonrelativistic particles if $-t\ll m_N^2$. It is useful to work in
the Breit frame, where:
\be
\nonumber
\vec p^{\ \prime}&=&-\vec p=\vec\Delta/2\sim N_c^0\, ,\\
\bar P^\mu&=&\frac 12 (P'^0+P^0,0,0,0)\approx (m_N,0,0,0)\sim (N_c,0,0,0)\, ,
\ee
we introduce a light-like vector $n$ which is
normalized by $n\cdot \bar P=1$. In the large
$N_c$ limit, we can choose it in the form
\be
n^\mu=\frac1{m_N}(1,0,0,-1)\sim \frac 1{N_c}(1,0,0,-1)\;,
\ee
which leads to
\be t\sim N_c^0 \quad {\rm and}\quad
\xi=-\frac 12 (n\cdot\Delta)\approx \frac{\Delta^3}{2m_N}\sim 1/N_c\;.
\ee
These relations fix the large
$N_c$ counting rules for baryon masses and kinematical variables
in the problem.

Let us first  apply the expressions for GPDs of the soliton
 (\ref{spdsol1},\ref{spdsol2})
to the case of
$N\to N$ GPDs. The projection of the {\em rhs} of Eqs.~(\ref{spdsol1},\ref{spdsol2})
onto particular nucleon states
can be done with help of the standard integral:
\be
\int dR\ \phi^{\ast\;S'=T'}_{S'_3T'_3}(R)  \phi^{S=T}_{S_3T_3}(R)\
\frac 12\ \Tr\left(R^\dagger\tau^a R\tau^b\right)=
\frac{\sqrt{2S+1}}{\sqrt{2S'+1}}
C_{T -T_3; 1 a}^{T' -T'_3}\ C_{S S_3; 1 b}^{S' S'_3}\, ,
\ee
where $C_{j_1 m_2; j_2 m_2}^{j m}$ is the Clebsch-Gordan
coefficient.
Performing the projection we observe that in the leading order of the $1/N_c$
expansion only the following flavor combinations
of nucleon GPDs are nonzero:
\be
H^u(x,\xi,t)+H^d(x,\xi,t), \quad
\widetilde H^u(x,\xi,t)-\widetilde H^d(x,\xi,t)\, ,
\label{largeH}
\ee
and
\be
E^u(x,\xi,t)-E^d(x,\xi,t), \quad \widetilde E^u(x,\xi,t)-\widetilde
E^d(x,\xi,t)\, .
\label{largeE}
\ee
Complementary combinations are zero in the leading order of the
$1/N_c$ expansion.
Further, taking into account the $N_c$ counting rules for kinematical
variables and for the nucleon form factors
(see e.g. Ref.~\cite{Das95,Mano98}) we obtain
the following counting rules for the nucleon GPDs:
\be
\nonumber
\left.
\begin{array}{cr}
H^u(x,\xi,t)+H^d(x,\xi,t) \\
\widetilde H^u(x,\xi,t)-\widetilde H^d(x,\xi,t)
\end{array}
\right\} \sim N_c^2 f\left(N_c x,N_c\xi,t\right), \\
\left.
\begin{array}{cr}
H^u(x,\xi,t)-H^d(x,\xi,t) \\
\widetilde H^u(x,\xi,t)+\widetilde H^d(x,\xi,t)
\end{array}
\right\} \sim N_c f\left(N_c x,N_c\xi,t\right),
\label{NccountingH}
\ee
and
\be
\nonumber
E^u(x,\xi,t)-E^d(x,\xi,t)\sim N_c^3 f\left(N_c x,N_c\xi,t\right), \\
\widetilde E^u(x,\xi,t)-\widetilde E^d(x,\xi,t) \sim N_c^4 f\left(N_c
x,N_c\xi,t\right),\nn
E^u(x,\xi,t)+E^d(x,\xi,t)\sim N_c^2 f\left(N_c x,N_c\xi,t\right),
\nn
\widetilde E^u(x,\xi,t)+\widetilde E^d(x,\xi,t)
\sim N_c^3 f\left(N_c x,N_c\xi,t\right)\, .
\label{NccountingE}
\ee
In the above equations $f(x_1,x_2,t)$ is a generic function stable
in the large $N_c$ limit. Also note that for the ``truly
non-forward" functions $E,\widetilde E$ the overall powers of $N_c$
are higher than for the GPDs $H,\widetilde H$ because of the
kinematical factor $1/m_N\sim 1/N_c$ and $\xi/m_N\sim 1/N_c^2$
in the definition of the GPDs $E$ and $\widetilde E$ respectively
(see Eq.~(\ref{eq:qsplitting})).

Taking the forward limit $\xi\to 0, t\to 0$ of the large-$N_c$
counting rules (\ref{NccountingH}) we reproduce the corresponding
counting rules for polarized and unpolarized nucleon parton
distributions derived in Refs.~\cite{Dia96b,Dia97}.

Let us analyse the large $N_c$ behaviour of particular
contributions to GPDs. We start with the D-term contribution,
see Sec.~\ref{chap3_2}. The large-$N_c$ counting for the D-term
can be extracted from either (\ref{NccountingH}) or from
(\ref{NccountingE}), because the D-term contributes to both functions $H$
and $E$ in the same way (up to the sign). For the isosinglet
component of the D-term we obtain:

\be
D^u(z)+D^d(z)\sim N_c^2 f(z)\, .
\label{NccountingDisosc}
\ee
When we turn to the isvector D-term we see that
that the counting rules for $H^u-H^d$ and $E^u-E^d$ are different
the latter being the dominant one in the large $N_c$ limit. It implies
that for the counting of the large $N_c$ behaviour of the isovector D-term
$D^u(z)-D^d(z)$ one has to use Eq.~(\ref{NccountingH}). One obtains:

\be
D^u(z)-D^d(z)\sim N_c f(z)\, .
\label{NccountingDisov}
\ee
Comparing Eqs.~(\ref{NccountingDisosc}) and
(\ref{NccountingDisov}),
one sees that the isovector D-term is suppressed relative to the
isoscalar one in the large $N_c$ limit.

Next let us check that the pion pole contribution of
Eq.~(\ref{eq:etilde_chiral}) to $\widetilde E^u-\widetilde E^d$
corresponds to the general counting rule (\ref{NccountingE}).
Indeed, taking into account that $g_A\sim N_c$ and $m_N\sim N_c$
we  easily reproduce (\ref{NccountingE}) from
Eq.~(\ref{eq:etilde_chiral}).

In section~\ref{chap4_4} we shall discuss the results obtained
in the chiral quark-soliton
model for the ``large" GPDs (\ref{largeH},\ref{largeE}).
Before doing this we derive the large $N_c$ relations between
$N\to N$ and $N\to \Delta$ GPDs introduced in Sec.~\ref{chap3_7}.

\subsubsection{GPDs for octet to decuplet transition in the large
$N_c$ limit} \label{chap4_2}

In order to derive the large $N_c$ relations between the $N\to \Delta$
GPDs and usual ones we shall use the fact that in the large $N_c$
limit the nucleon and $\Delta$ states are simply different rotational
excitations of the same classical object -- soliton \cite{Fra98a}.

To obtain the relations we sandwich
Eqs.~(\ref{spdsol1},\ref{spdsol2}) between rotational states of the
soliton which corresponds to baryons from octet and decuplet.
Doing this we obtain the expressions for $N\to N$ and $N\to\Delta$
GPDs in terms of the universal functions $H_{0,1}$ and $\widetilde
H_{0,1}$, {\em i.e.} the GPDs of the soliton introduced by
Eqs.~(\ref{spdsol1},\ref{spdsol2}). Since both $N\to N$ and $N\to\Delta$
GPDs are expressed in terms of the {\em same} GPDs of a soliton
$H_{0,1}$ and $\widetilde H_{0,1}$
we obtain the following relations [The $N\to\Delta$ GPDs are defined
by Eqs.~(\ref{HND-QCD-2},\ref{HND-helicity-QCD-2})]:

\be
H_M(x,\xi,t)&=&\frac{2}{\sqrt 3}\left[ E^u(x,\xi,t)
-E^d(x,\xi,t)\right] \; , \nonumber\\ C_1(x,\xi,t)&=& \sqrt
3\left[ \tilde H^{u}(x,\xi,t)- \tilde H^{d}(x,\xi,t)\right]\, ,
\nonumber\\ C_2(x,\xi,t)&=&\frac{\sqrt 3}{4}\left[\tilde
E^{u}(x,\xi,t)-\tilde E^{d}(x,\xi,t)\right]\, . \la{ndNc2} \ee
All
other $N\to\Delta$ GPDs are zero at leading order of $1/N_c$
expansion. The properties of these relations were discussed in
Sec.~\ref{chap3_7}.

\subsection{GPDs in the chiral-quark soliton model}
\label{chap4_4}
Model calculations of the nucleon GPDs are important firstly to
understand how
non-perturbative mechanisms  generate various
structures in GPDs. In this respect the chiral models should be
very useful because as we saw in Sec.~\ref{chap3_4}  the
spontaneously broken chiral symmetry plays a prominent role
in the physics of GPDs.
Secondly, model calculations allow us to develop an intuition about
the structure of the GPDs which is important for the construction of
phenomenological parameterizations for GPDs.

Model calculations of nucleon GPDs were first performed in
Ref.~\cite{Ji97a} where GPDs were computed in the MIT bag model.
Unfortunately in the bag model the chiral symmetry is not
respected and also the non field theoretic nature of this model
leads to problems with positivity of antiquark distributions.
Calculations of the nucleon GPDs in the chiral quark-soliton model
\cite{Dia88} do not show these deficiencies and
 have been
performed in Refs.~\cite{Pet98,Pen00a}. The chiral quark-soliton
model is field theoretic model of the nucleon
which stresses the dominant role of spontaneous chiral symmetry
breaking in the dynamics of the nucleon bound state.
The review of foundations of this model and its applications to
calculations of the parton distributions can be found in
\cite{Dia98,Dia00}. The review of applications to calculations of
nucleon form factors and many static observables can be found in
Refs.~\cite{Chr96,Alk96}. We refer the reader to above reviews for details.
Here we remind only some basics of this model.

The chiral quark-soliton model  is essentially based on the
$1/N_c$ expansion.  Although in reality the number of colors $N_c=3$,
the academic limit of large $N_c$ is known to be a useful guideline. At
large $N_c$ the nucleon is heavy and can be viewed as a classical
soliton of the pion field \cite{Wit79,Adk83}. In the previous
section we derived from this large-$N_c$ semiclassical picture of
the nucleon general large $N_c$ counting rules for various nucleon
GPDs (see Eqs.~(\ref{NccountingH},\ref{NccountingE})) and
relations between $N\to N$ and $N\to\Delta$ GPDs (see
Eq.~(\ref{ndNc2})). In order to compute the nucleon GPDs we have
to specify the dynamics of the low energy QCD.
Here we work
with the effective chiral action given by the functional integral over
quarks in the background pion field \cite{Dia83,Dia86}:

\begin{eqnarray}
\exp \left( iS_{{\rm eff}}[\pi (x)]\right) =\int D\psi D\bar \psi \;\exp
\left( i\int d^4x\,\bar \psi (i\dd -MU^{\gamma _5})\psi \right) ,
\nonumber
\end{eqnarray}
\begin{equation}
U\;=\;\exp \left[ i\pi ^a(x)\tau ^a\right] ,\hspace{1cm}U^{\gamma
_5}\;=\;\exp \left[ i\pi ^a(x)\tau ^a\gamma _5\right] \;=\;\frac{1+\gamma _5}
2U+\frac{1-\gamma _5}2U^{\dagger }.  \label{FI}
\end{equation}
Here $\psi $ is the quark field, $M$ the effective quark mass, which is
due to the spontaneous breakdown of chiral symmetry (generally
speaking, it is momentum dependent), and $U$ is the $SU(2)$ chiral pion
field. The effective chiral action given by Eq.~(\ref{FI}) is known
to contain automatically the Wess--Zumino term and the four-derivative
Gasser--Leutwyler terms, with correct coefficients.
Equation (\ref{FI}) has been
derived from the instanton model of the QCD vacuum \cite{Dia86}, which
provides a natural mechanism of chiral symmetry breaking and enables one to
express the dynamical mass $M$ and the ultraviolet cutoff intrinsic in
(\ref{FI}) through the $\Lambda _{QCD}$ parameter. The ultraviolet
regularization of the effective theory is provided by the specific momentum
dependence of the mass, $M(p^2)$, which drops to zero for momenta of
order of the inverse instanton size in the instanton vacuum,
$1/\rho \sim 600\, {\rm MeV}$.
We also note the the inverse instanton size sets the
renormalization scale for computed parton distributions at
$\mu=0.6$~GeV.

An immediate application of the effective chiral theory (\ref{FI}) is the
chiral
quark-soliton model of baryons of Ref.~\cite{Dia88}.
According to this model nucleons can
be viewed as $N_c$ ``valence" quarks bound by a self-consistent pion
field (the ``soliton") whose energy coincides with the aggregate
energy of the quarks of the negative-energy Dirac continuum. Similarly to
the Skyrme model the large $N_c$ limit is needed as a parameter to justify the use of
the mean-field approximation; however, the $1/N_c$--corrections can be
--- and, in some cases, have been --- computed \cite{Chr96}.

Let us remind the reader how the nucleon is described in the effective
low-energy theory Eq.~(\ref{FI}). Integrating out the quarks in (\ref{FI}) one
finds the effective chiral action,
\begin{equation}
S_{{\rm eff}}[\pi ^a(x)]=-N_c\,\mbox{Sp}\log D(U)\,,\hspace{1cm}
D(U)\;\;=\;\;i\partial _0-H(U),  \label{SeffU}
\end{equation}
where $H(U)$ is the one-particle Dirac hamiltonian,
\begin{equation}
H(U)=-i\gamma ^0\gamma ^k\partial _k+M\gamma ^0U^{\gamma _5}\,,  \label{hU}
\end{equation}
and $\mbox{Sp}\ldots $ denotes the functional trace.
For a given time-independent pion field $U=\exp(i\pi^a({\bf x})\tau^a)$ one
can determine the spectrum of the Dirac hamiltonian,
\begin{equation}
H\Phi_n = E_n \Phi_n.  \label{Dirac-equation}
\end{equation}
It contains the upper and lower Dirac continua (distorted by the presence of
the external pion field), and, in principle, also discrete bound-state
level(s), if the pion field is strong enough. If the pion field has unity
winding number, there is exactly one bound-state level which travels all the
way from the upper to the lower Dirac continuum as one increases the spatial
size of the pion field from zero to infinity \cite{Dia88}. We denote the
energy of the discrete level as $E_{{\rm lev}},\;\;-M\leq E_{{\rm lev}}\leq
M $. One has to occupy this level to get a non-zero baryon number state.
Since the pion field is color blind, one can put $N_c$ quarks on that level
in the antisymmetric state in color.

The limit of large $N_c$ allows us to use the mean-field approximation to
find the nucleon mass. To get the nucleon mass one has to add
$N_cE_{{\rm lev}}$ and the energy of the pion field.
Since the effective chiral lagrangian
is given by the determinant (\ref{SeffU}) the energy of the pion field
coincides exactly with the aggregate energy of the lower Dirac continuum,
the free continuum subtracted. The self-consistent pion field is thus found
from the minimization of the functional \cite{Dia88}

\begin{equation}
m_N = \min_U \; N_c\left\{E_{{\rm lev}}[U] \; + \;
\sum_{E_n<0}(E_n[U]-E_n^{(0)})\right\}.  \label{nm}
\end{equation}
{From} symmetry considerations one looks for the minimum in a hedgehog ansatz:
\begin{equation}
U_c({\bf x}) \; = \; \exp\left[i\pi^a({\bf x})\tau^a\right] \; = \;
\exp\left[i n^a \tau^a P(r)\right], \hspace{1cm} r \; = \; |{\bf x}|,
\hspace{1cm} {\bf n} \; = \; \frac{{\bf x}}{r} ,  \label{hedge}
\end{equation}
where $P(r)$ is called the profile of the soliton.

The minimum of the energy (\ref{nm}) is degenerate with respect to
translations of the soliton in space and to rotations of the soliton field
in ordinary and isospin space. For the hedgehog field (\ref{hedge}) the two
rotations are equivalent.
The projection on a nucleon state with
given spin ($S_3$) and isospin ($T_3$) components
is obtained by integrating over all spin-isospin rotations,
$R$, see Eq.~(\ref{spisosp}).
Analogously, the projection on a nucleon state with given momentum
${\bf P}$ is obtained by integrating over all shifts, ${\bf X}$, of the
soliton,
\be
\langle {\bf P^\prime}|\ldots|\ {\bf P}\rangle
&=& \int d^3{\bf X}\;e^{i({\bf P^\prime-P})\cdot{\bf X}}\; \ldots
\label{totmom}
\ee

Recently it was shown that the chiral quark--soliton model of the nucleon
provides a framework
for a successful calculation not only of nucleon form factors
\cite{Chr96} but also
of the nucleon
parton distributions, both unpolarized and polarized, at a low normalization
point \cite{Dia96b,Dia97,Pob96,Wei97,Gam98,Wak98a,Wak98b,Pob99}.
Because both cases are limiting cases of the GPDs, the chiral
quark-soliton model is a promising tool to calculate them.
Since the chiral quark-soliton model is
 a quantum field--theoretical
description of the nucleon, with explicit quark degrees of
freedom, it
allows an unambiguous identification of the quark as well as antiquark
distributions in the nucleon. We were able to demonstrate that all
general properties of the quark and antiquark distributions (positivity,
sum rules, Soffer inequalities \cite{Pob96,Pob00}, {\em etc.}).

All GPDs can be expressed in terms of
a matrix element of a non-local quark bilinear operator in
the nucleon states with definite 4-momenta $P,P'$ and spin and
isospin components, see general large $N_c$
expression Eq.~(\ref{psi-psi-bar-Hartree}).
Since
in the chiral quark-soliton model the dynamics of the mean-field is
specified, one can write the explicit expression for  correlator of
the type of
Eq.~(\ref{psi-psi-bar-Hartree}).
[We write
explicitly all the flavor ($f,g=1,2$) and the Dirac ($i,j=1,...,4$)
indices for clarity]:
\[
\bra{{\bf P'},S'_3,T'_3}\, \bar\psi_{fi}(x^0,{\bf x}) \,
\psi^{gj}(y^0,{\bf y}) \,
\ket{{\bf P},S_3,T_3}  =
\]
\[
2m_N N_c \int d^3{\bf X}
e^{i({\bf P'}-{\bf
P'})\cdot {\bf X}}
\int dR \; \phi_{S'_3T'_3}^\dagger(R)  \phi_{S_3T_3}(R)
\]
\be
\times
\biggl[R_{g^\prime}^g
\sum\limits_{\scriptstyle n\atop \scriptstyle{\rm
occup.}}\exp[iE_n(x^0-y^0)] \; \Phi_{n,f^\prime i}^\dagger({\bf
x-X}) \,
\, \Phi_n^{g^\prime j}({\bf y-X}) (R^\dagger )^{f^\prime}_f
\biggr]  \,.
\label{annihilate}
\ee
The functions $\Phi_n$ are eigenstates of energy $E_n$ of the Dirac
hamiltonian (\ref{hU}) in the external (self-consistent) pion field $U_c$,
they can be found numerically by diagonalization of the hamiltonian.
In this way we compute the function $F$ in general large-$N_c$
Eq.~(\ref{psi-psi-bar-Hartree}).

\subsubsection{Results for $H$ and $E$}
\label{chap4_4_1}

Having the expression (\ref{annihilate}) for nucleon matrix
elements of quark bilocal operators in the chiral quark-soliton
model
we can now express
various nucleon GPDs as the sum over occupied quark orbitals in
the external pion field $U_c$. We start with GPDs $H$ and
$E$. One immediately observes that, in
accordance with general large-$N_c$ counting rules (\ref{NccountingH}
\ref{NccountingE}), in the leading $N_c$ order
only the flavor singlet part of
$H(x,\xi ,t)$ and the flavor--nonsinglet
part of $E(x,\xi ,t)$ are non-zero. They are given, respectively,
by
\be
\lefteqn{
\sum\limits_q H^q(x,\xi ,\Delta ^2)  \;\; = \;\;
\frac{N_cm_N}{2\pi }\int dz^0\int
d^3{\bf X} \; \exp [i\mbox{{\boldmath $\Delta$}}\cdot {\bf X}] }
\nonumber \\
&&\times \sum\limits_{{\scriptstyle}\atop {\scriptstyle {\rm occup.}}}\exp
\{iz^0[(x+\xi)m_N-E_n]\} \; \Phi _n^{\dagger }({\bf X}
)(1+\gamma ^0\gamma ^3)\Phi _n({\bf X}-z^0{\bf e}_3)\, ,
\label{H-singlet-general} \\
\lefteqn{E^{u}(x,\xi ,t)-E^{d}(x,\xi ,t)
 =  -\frac{iN_cm_N^2}{3\pi} \int dz^0\int d^3{\bf X} \; \exp [i
\mbox{{\boldmath $\Delta$}}\cdot {\bf X}]\  \frac{\epsilon ^{3jk}\Delta
^j}{\Delta^2}}
&&
\nonumber \\
&&\times \sum\limits_{{\scriptstyle}\atop {\scriptstyle {\rm occup.}}}\exp
\{iz^0[(x+\xi)m_N-E_n]\} \; \Phi _n^{\dagger }({\bf X})\tau
^k(1+\gamma ^0\gamma ^3)\Phi _n({\bf X}-z^0{\bf e}_3)\, .
\label{E-nonsinglet-general}
\ee
The isovector part of $H(x,\xi ,\Delta ^2)$ and the isosinglet part
of $E(x,\xi ,\Delta ^2)$ appear only in the next--to--leading order
of the $1/N_c$--expansion, {\em i.e.}, after taking into account
the finite angular velocity of the soliton rotation.
\par
First
we would like to demonstrate that the two limiting cases
of the GPDs
--- usual parton distributions and elastic form factors --- are
correctly reproduced within the chiral quark--soliton model.
Taking in Eq.(\ref{H-singlet-general}) the forward limit,
$\Delta\rightarrow 0$, one recovers the formula for the usual
singlet (anti--) quark distributions in the chiral quark-soliton
 model which was
obtained in Ref.~\cite{Dia96b}. Thus the forward limit,
Eq.~(\ref{eq:dislimit}), is reproduced. On the other hand, integrating
Eqs.~(\ref{H-singlet-general},\ref{E-nonsinglet-general}) over $-1\le
x\le 1$ one obtains (up to corrections parametrically small in $1/N_c$)
the expressions for the electromagnetic formfactors of the nucleon
derived in Ref.~\cite{Dia88}:
\begin{equation}
\nonumber
\int_{-1}^1 dx\ \sum\limits_q H^q(x,\xi ,t) =
N_c \int d^3{\bf X}\exp[i\mbox{{\boldmath $\Delta$}}\cdot {\bf X}]
\sum\limits_{{\scriptstyle}\atop {\scriptstyle {\rm occup.}}}
\Phi _n^{\dagger }({\bf X})\Phi _n({\bf X})
= F_1^{\rm (T=0)}(t)\, ,
\label{H-singlet-sum-rule}
\end{equation}
\[
\int_{-1}^1 dx\ \left[ E^{u}(x,\xi ,t)-E^{d}(x,\xi ,t)\right] =
\frac{i N_cm_N}{\Delta^2} \varepsilon_{3ki}\Delta^i
\int d^3{\bf X}\exp[i\mbox{{\boldmath $\Delta$}}\cdot {\bf X}]
\]
\begin{equation}
\times\sum\limits_{{\scriptstyle}\atop {\scriptstyle {\rm occup.}}}
\Phi _n^{\dagger }({\bf X})\gamma^0\gamma^3\tau^k
\Phi _n({\bf X})=
F_2^{(\rm T=1)}(t)\, .
\label{E-nonsinglet-sum-rule}
\end{equation}
The electromagnetic formfactors computed in the chiral quark
soliton model on the basis of these formulae compare rather well
with the experimentally measured ones up to momenta of order
$-t\sim 1$~GeV$^2$ \cite{Chr96}.
One can also check that \cite{Pet98,Pen00a} the GPDs given by
Eqs.(\ref{H-singlet-general},\ref{E-nonsinglet-general}) satisfy
the polynomiality condition (\ref{pc}).
\begin{figure}[h]
\epsfxsize=7.5 cm
\centerline{\epsffile{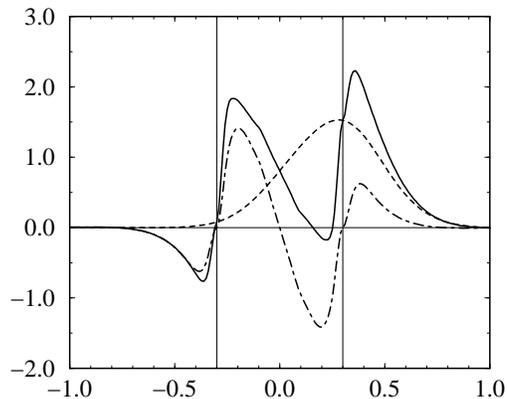}}
\caption{\small The GPD  $H^u (x, \xi, t)+H^d (x, \xi, t)$
as a function of $x$ at $\xi = 0.3$
and $t=t_{\rm min}=-0.35$~GeV$^2$. Dashed curve: contribution from valence
level. Dashed-dotted curve:contribution of the Dirac continuum.
Solid curve: the total distribution (sum of the dashed and dashed-dotted
curves). The vertical lines mark the crossover points $x=\pm\xi$.}
\label{fig:modelH025}
\end{figure}

Eqs.~(\ref{H-singlet-general},\ref{E-nonsinglet-general}) express
the GPDs as a sum over quark single--particle levels in the
soliton field. This sum runs over {\it all} occupied levels, including
both the discrete bound--state level and the negative
polarized Dirac continuum.
We remind the reader that in the case of usual parton distributions
it was demonstrated that in order to ensure the positivity of the antiquark
distributions it is essential to take into account the contributions
of {\it all} occupied levels of the Dirac Hamiltonian \cite{Dia96b}.
The so-called ``valence level approximation'' advocated
in Refs.~\cite{Wei97,Gam98} for structure
functions in the chiral quark-soliton model
leads to unacceptable {\it negative}
antiquark distributions. We shall see below that also in the
off-forward case the contribution of the Dirac continuum drastically
changes the shape of the distribution function, leading to
characteristic crossovers of $H(x,\xi,\Delta)$ at $|x|=\xi$,
in addition
this contribution generates the large and negative D-term.
\par
Now we present results for calculations of the contributions
of the discrete bound--state level and the negative Dirac continuum
to Eqs.~(\ref{H-singlet-general},\ref{E-nonsinglet-general}).
Technical details and methods of calculations can be found in
Refs.~\cite{Pet98,Pen00a}.
\begin{figure}[h]
\epsfxsize=7.5 cm
\centerline{\epsffile{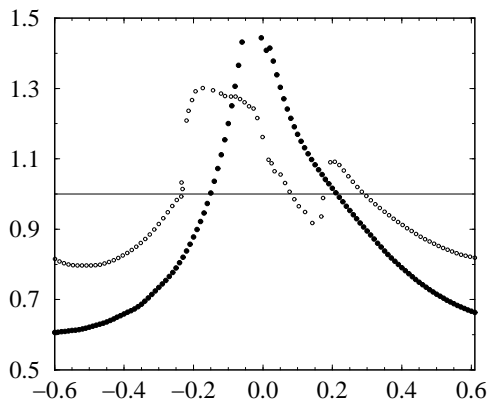}}
\caption[]{\small
Ratio Eq.~(\protect{\ref{R-ratio}}) as a function of $x$ showing the deviation from
the factorization ansatz for the $t$-dependence of the GPD $H$ for the values of $\xi=0$
(filled circles) and $\xi=0.2$ (open circles) with
$t_0=-4 m_N^2\xi^2/(1-\xi^2)$ and $t_1=t_0-0.5$~GeV$^2$.}
\label{fig:RH}
\end{figure}

In Fig.~\ref{fig:modelH025} a typical shape of the GPD
$H^u(x,\xi,t)+H^d(x,\xi,t)$ as a function of the variable $x$
is shown at $\xi=0.3$
and $t=t_{\rm min}$, with $$t_{\rm min}=-4 m_N^2\xi^2/(1-\xi^2).$$
For $\xi=0.3$ one has $t_{\rm min}=-0.35$~GeV$^2$.
On this figure we
show separately the contributions of the valence level and of the
Dirac continuum as well as their sum.
We see that the Dirac continuum contribution is essential
especially in the region $-\xi\le x\le \xi$, also in the region $x\le
-\xi$,
corresponding to minus antiquark distributions, the Dirac continuum
contribution ensures the positivity of antiquarks.
We also computed several Mellin moments at various values of $\xi$ and $t$
(in the model calculations the values of $\xi$ is limited by
$\xi^2\le -t/(4 m_N^2-t)$) and with the help of Eq.~(\ref{D-term-generation})
we determined the value of the first few coefficients in a
Gegenbauer expansion of the D-term (\ref{dterm}). The result
extrapolated to the value of $t=0$ is:

\be
\sum_{q=u,d,s...}
D^q(z)=(1-z^2)\biggl[ -4.0 \ C_1^{3/2}(z)
-1.2 \ C_3^{3/2}(z) -0.4\ C_5^{3/2}(z) + ...\biggr]\, .
\la{dterm-num}
\ee
We shall use this result for the D-term in Sec.~\ref{chap5} for
estimates of various observables.

Calculations of GPDs in the chiral quark-soliton model
give a possibility to study
their $t-$dependence.
In modeling the $t$-dependence of GPDs usually the
factorization ansatz:
$H(x,\xi,t)= H(x,\xi ) F_1(t)$
is adopted.
\begin{figure}[h]
\epsfxsize=7.5 cm
\centerline{\epsffile{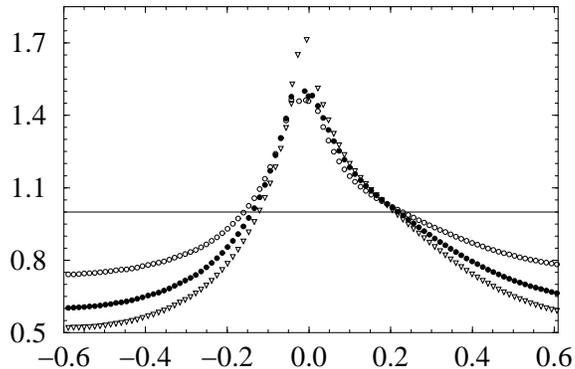}}
\caption[]{\small
Ratio Eq.~(\protect{\ref{R-ratio}}) as a function of $x$
for a fixed value of  $\xi=0$ and for $t_0=0$ and for
$t_1=-0.25$ (open circles), $-0.5$ (filled circles), $-0.75$ (triangles)~GeV$^2$. }
\label{fig:RHxi0}
\end{figure}
In order to illustrate numerically the non-factorazibility
of the
$t$-dependence of $ H(x,\xi,t)$ we show in Fig.~\ref{fig:RH}
the following ratio :

\be
R =
\frac{H(x,\xi,t_0)}{
 H(x,\xi,t_1)}\
\frac{\int_{-1}^{1}dx\  H(x,\xi,t_1)}{\int_{-1}^{1}
dx\ H(x,\xi,t_0)}\, ,
\label{R-ratio}
\ee
[$H=H^u+H^d$]
as a function of $x$ taking
$t_0=t_{\rm min}$ and $t_1=t_{\rm min}-0.5$~GeV$^2$,
and $\xi=0$ (filled circles), $\xi=0.2$ (open circles).
If the $t$-dependence of the GPDs factors out then $R=1$ independently of
$t_0, t_1$ and $x, \xi$.
{From} Fig.~\ref{fig:RH} we see that the ratio (\ref{R-ratio})
can deviate considerably  from unity and it depends in a
non-trivial way on both variables $x$ and $\xi$.

To illustrate the dependence of the ratio (\ref{R-ratio})
on the value of $t_1$ we plot in Fig~\ref{fig:RHxi0} this ratio
as the function of $x$ with $\xi=0$ and $t_0=0$ for
several values of $t_1=-0.25,0.5, 0.75$~GeV$^2$.
The results of the chiral quark-soliton
model on the $t-$dependence of GPDs shown in
Fig.~\ref{fig:RH} and Fig.~\ref{fig:RHxi0}
illustrate that the assumption of factorization of
the $t-$dependence for GPDs should be taken with caution and that
more work on $t-$dependence of phenomenological
parametrizations of GPDs should be done.

It is interesting to note  that the features of the $t$-dependence of
GPDs shown on Fig.~\ref{fig:RHxi0} can be qualitatively reproduced
by the following ansatz for $H^q(x,\xi=0,t)$:

\be
H^q(x,\xi=0,t)=\frac{1}{x^{\alpha' t}}\ q(x)\, ,
\label{Regge-parametrization}
\ee
where $\alpha'$ can be interpreted as
the slope of a Regge trajectory.
This Regge theory motivated ansatz reproduces all qualitative
feature of Fig.~\ref{fig:RHxi0} obtained in the chiral
quark-soliton model. As a curiosity we note that a rough fit of the curves
on Fig.~\ref{fig:RHxi0} by the ansatz
(\ref{Regge-parametrization}) gives
$\alpha'\approx 0.8$~GeV$^{-2}$, amazingly close to the slopes of meson
Regge trajectories.

\begin{figure}[h]
\epsfxsize=7.5 cm
\centerline{\epsffile{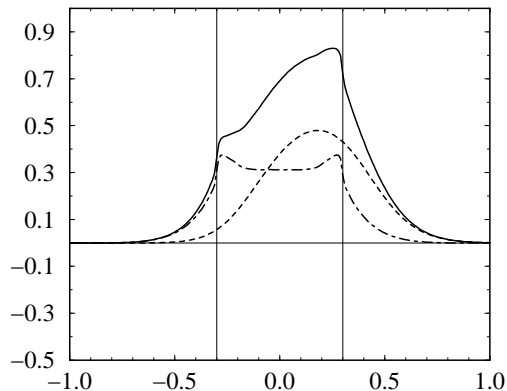}}
\caption{\small The GPD  $E^u (x, \xi, t)-E^d (x, \xi, t)$
as a function of $x$ at $\xi = 0.3$
and $t=t_{\rm min}=0.35$~GeV$^2$.
Dashed curve: contribution from valence
level. Dashed-dotted curve:contribution of the Dirac continuum.
Solid curve: the total distribution (sum of the dashed and dashed-dotted
curves). The vertical lines mark the crossover points $x=\pm\xi$.}
\label{fig:modelE025}
\end{figure}

Let us now discuss results for the GPD $E^u-E^d$.
In Fig.~\ref{fig:modelE025} we plot the GPD $E^u (x, \xi, t)-E^d (x, \xi, t)$
at $\xi=0.3$ and $t=t_{\rm min}$. We see that in this case the
Dirac continuum contribution is also essential and it generates a
symmetric sea contribution to the GPD $E^u-E^d$. Computing several Mellin moments of
the function $E^u (x, \xi, t)-E^d (x, \xi, t)$ we are able to
extract the value of the isovector D-term $D^u(z)-D^d(z)$. The
leading-$N_c$
result for this flavor combination is zero,
i.e. $$D^u(z)-D^d(z)= 0.$$
This is
in accordance with general large $N_c$ counting rules for the
D-term (\ref{NccountingDisov}).
Therefore we conclude that the D-term is
independent of quark flavor.

\begin{figure}[h]
\epsfxsize=7.5 cm
\centerline{\epsffile{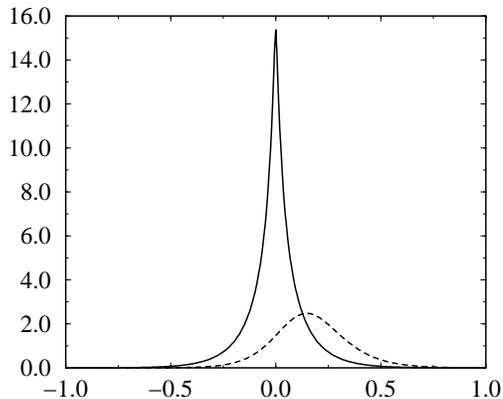}}
\caption{\small The GPD  $E^u (x, \xi=0, t=0)-E^d (x, \xi=0, t=0)=e^u(x)-e^d(x)$
as a function of $x$.
Dashed curve: contribution from valence
level. Solid curve: contribution of the Dirac continuum.}
\label{fig:modelE0}
\end{figure}
For the construction of the phenomenological model for the GPD
$E^q$ we shall need the forward limit of this functions which we
denote as $e^q(x)$:
\be
e^q(x)=\lim_{\Delta\to 0} E^q(x,\xi,t)\, .
\la{e-small-def}
\ee
In Fig.~\ref{fig:modelE0} we present the result of the model
calculation of $e^u(x)-e^d(x)$. We see that it consists of the
valence contribution and of the sea part peaked around $x=0$.
With help of sum rules for the function $e^u(x)-e^d(x)$:
\be
\int_{-1}^1dx\ \left[e^u(x)-e^d(x)\right]&=&\kappa^p-\kappa^n\,
,\nn
\int_{-1}^1dx\ x\ \left[e^u(x)-e^d(x)\right]&=&2 \left[J^u-J^d\right]-
M_2^u+M_2^d\,
\ee
where $\kappa^N$ is an anomalous magnetic moment of a nucleon $N=p,n$
and $J^q$ and $M_2^q$ are the parts of
angular and of total momentum carried by a quark,
we obtain in the model:
\[
\kappa^p-\kappa^n=1.0_{val}+2.22_{sea}=3.22\ \
vs.\ \ 3.7\ ({\rm exp.}),
\]
\be
2 \left[J^u-J^d\right]-
M_2^u+M_2^d=0.18_{val}+0_{sea}=0.18\, .
\ee
Here we show also the individual contributions from the valence
and sea parts.
We see that the value of the anomalous magnetic
moment is dominated by the sea contribution and its value is close to the
experimental number. The model prediction for $2 J^{u-d}-2 M_2^{u-d}$ is
dominated by the valence contribution.

\begin{figure}[h]
\epsfxsize=7.5 cm
\centerline{\epsffile{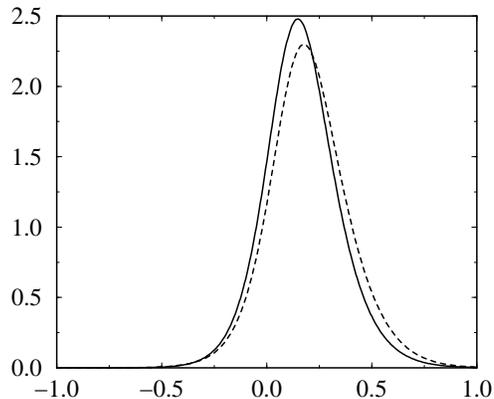}}
\caption{\small Comparison of shapes of the valence level
contributions to $u(x)+d(x)$ (solid curve) and to $e^u(x)-e^d(x)$ (dashed curve).
Both contributions are normalized to unity.}
\label{fig:valEvsH}
\end{figure}

Since the  combination $J^u-J^d$ of angular momenta is dominated
by the valence part of the function $e^u(x)-e^d(x)$\footnote{As
shown in Ref.~\cite{Pob99} the $M_2^u-M_2^d$ is also dominated by
valence level contributions.}
it is
instructive to compare the shapes of the valence contributions to
$e^q(x)$ and to $q(x)$ with each other. The corresponding comparison is shown in
Fig.~\ref{fig:valEvsH} where we plot
$$\frac{e^u_{val}(x)-e^d_{val}(x)}{\int_{-1}^1dx\ (e^u_{val}(x)-e^d_{val}(x))}
\ \ {\rm versus}\ \
\frac{q^u_{val}(x)+q^d_{val}(x)}{\int_{-1}^1dx\ (q^u_{val}(x)+q^d_{val}(x))}\, ,$$
computed in the chiral quark-soliton model.
We see that the shapes of both these valence distributions in the model
are very close. This observation will motivate us in
Sec.~\ref{chap5_2c} for the ansatz:
\be
e^q_{val}(x)=A^q q_{val}(x)\, ,
\ee
where $A^q$ is a $x$-independent constant.

The qualitative features of the GPDs $H$ and $E$ obtained in
the chiral quark-soliton model will be used in Sec.~\ref{chap5_2}
for the construction of phenomenological parametrization of these
distributions.

\subsubsection{Results for $\widetilde H$ and $\widetilde  E$}
\label{chap4_4_2}
Now we turn to the results for the quark helicity dependent
GPDs $\widetilde H$ and $\widetilde  E$. Expressions for
these distributions in terms of a summation over
occupied quark orbitals in the external pion mean field can be
found in Ref.~\cite{Pen00a}. Here we discuss only the main results.

According to the general large $N_c$ counting rules derived in
Sec.~\ref{chap4_1} at the leading order only isovector combinations
of GPDs $\widetilde H^u-\widetilde H^d$ and $\widetilde E^u-\widetilde E^d$
are non-zero. This general counting is fulfilled in the chiral
quark-soliton model.

Before presenting the numerical results for the GPDs
$\widetilde H$ and $\widetilde E$ let us discuss specific
contribution to these GPDs originating from the long range pion tail
of the pion mean-field.
The behaviour of the mean pion field at large distances is governed
by linearized equations of motion and PCAC:

\be
\lim_{|\vec x|\to \infty }U(x) \, =\, 1+ \frac{3 g_A}{8\pi f_\pi^2 |\vec x|^2}
\Big( 1+m_\pi |\vec x| \Big) \,
\frac{i x^a\tau^a}{|\vec x|}
\exp(-m_\pi |\vec x|)\; ,
\label{pi-tail}
\ee
where $g_A\approx 1.267$ is the axial charge of the nucleon,
$f_\pi\approx 92.4$~MeV is the pion decay constant.
Contribution of this long-range tail to $\widetilde E$
can be computed analytically (see details in Ref.~\cite{Pen00a})
with the result:

\be
\label{E-1-sym-res-mp-pi}
&& \widetilde E^{u-d}_{\pi}(x,\xi ,t)
\, =\, \frac{m_N F(t)}{f_\pi^2}
 \; \int \frac{d^4p}{(2\pi )^4} \;
\delta \left[ \Bigl(x- \xi \Bigr) m_N- p^0+p^3 \right] \\
\nonumber
&& \, \times \,
\frac{M^{1/2} \Big[ ( p+\Delta)^2 \Big]}{ \Big[ (p+\Delta)^2-M^2+i0 \Big]} \,
 \,
\frac{M^{3/2} \Big[ ( p-\Delta)^2 \Big]}{\Big[ (p-\Delta)^2-
M^2+i0 \Big] }
\, +\, \Biggl( \xi \to -\xi, \;\; \mbox{{\boldmath $\Delta$}}
\to -\mbox{{\boldmath $\Delta$}}
\Biggr)\, .
\ee
Here $M[p^2]$ is the momentum dependent constituent quark
mass as obtained in the theory of the instanton vacuum \cite{Dia86}.
It provides the UV regularization of the effective theory
(\ref{FI}).
In the Eq.~(\ref{E-1-sym-res-mp-pi}) we introduced the following form factor:
\be
F(-\vec k^2) \, =\, \frac{4 m_N^2 f_\pi^2}{3k^3}\int d^3 x\ \exp(i\vec k
\cdot \vec x) \, \, {\rm Tr} \Big[ \Big( U(\vec{x})-1 \Big) \tau^3 \Big]\, .
\label{Fff}
\ee
The small momentum asymptotic of this form factor can be obtained
by substituting into Eq.~(\ref{Fff}) the long-range asymptotic of the
mean-field (\ref{pi-tail}). The result is:
\be
\lim_{t\to m_\pi^2} F(t) \, =\,
\frac{4 g_A m_N^2}{m_\pi^2-t}\, .
\ee

Now the crucial observation \cite{Pen00a} is that the integral
over $p$ in Eq.~(\ref{E-1-sym-res-mp-pi}) coincides
$exactly$ (up to trivial renaming of variable) with the expression
for the light-cone pion distribution amplitude in the instanton model
of the QCD vacuum \cite{Pet99}.
Therefore the expression
(\ref{E-1-sym-res-mp-pi}) for the $\widetilde E_\pi$ at $t\to m_\pi^2$
can be written in the compact form:

\be
\widetilde E^{u-d}(x,\xi,t)=  \frac{4 g_A
m_N^2}{m_\pi^2-t}\
\theta\left[\xi-|x|\right]\ \frac 1\xi\ \Phi_\pi\left(\frac x \xi\right)\,.
\ee
This is exactly the general result obtained in Sec.~\ref{chap3_4_2},
see Eq.~(\ref{eq:etilde_chiral}).

\begin{figure}[h]
\epsfxsize=11. cm
\vspace{-0.5cm}
\centerline{\epsffile{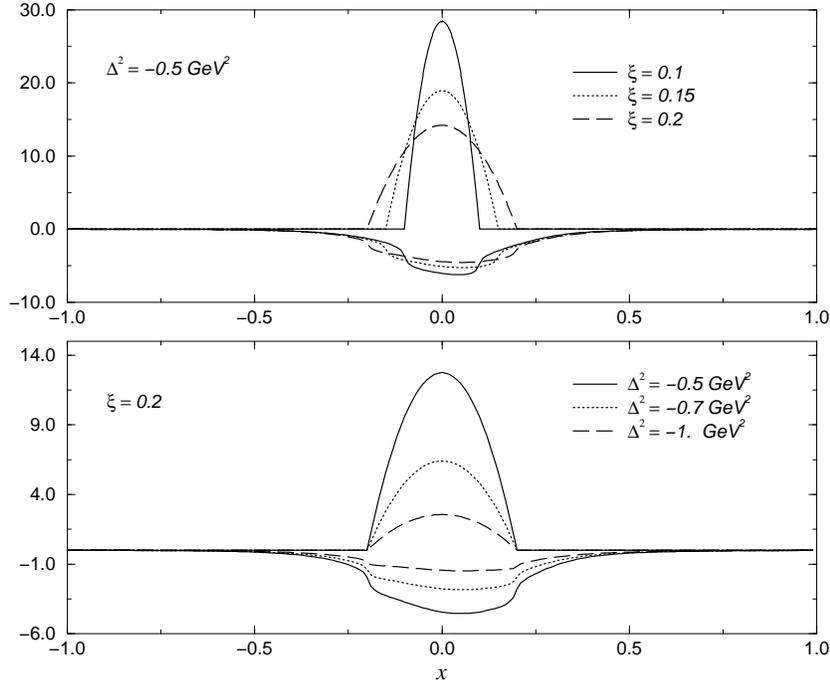}}
\vspace{-.25cm}
\caption{\small
Comparison of pion pole contribution and non-pole part
of the isovector GPD $\widetilde E$ at various values of $t$ and $\xi$.
The positive curves correspond to pion pole contributions.}
\label{fig:ea}
\end{figure}
In Ref.~\cite{Pen00a} the non-pole
contributions to $\widetilde E^u-\widetilde E^d$
have been computed in the chiral quark-soliton model. The
corresponding results are presented in Fig.~\ref{fig:ea}.
We see that for wide range of $t$ and $\xi$ the pion pole
contribution dominates, the non-pole part being negative.
\begin{figure}[h]
\epsfxsize=8. cm
\vspace{-.25cm}
\centerline{\epsffile{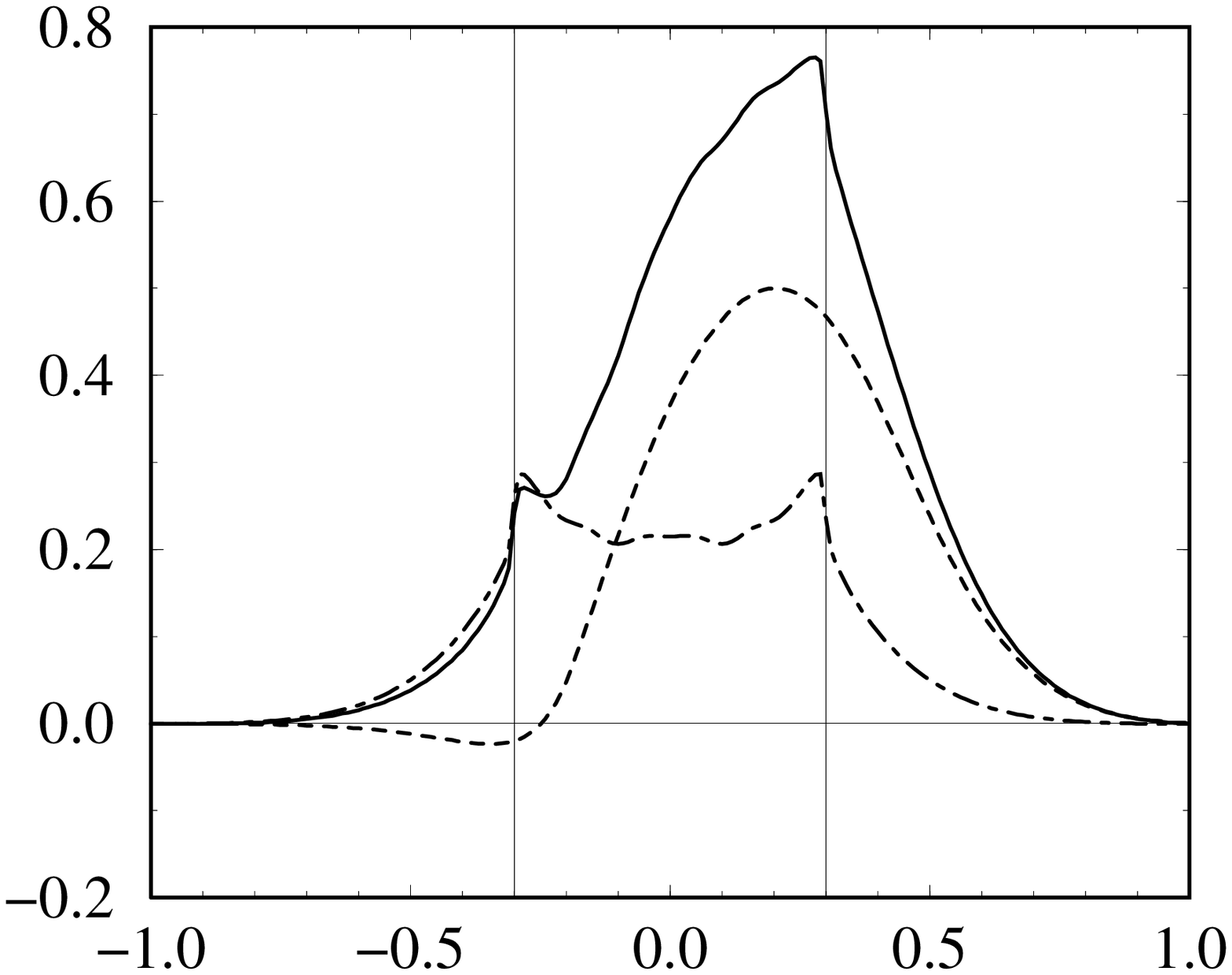}}
\vspace{-.5cm}
\caption{\small The GPD
$\widetilde H^u (x, \xi, t)-\widetilde H^d (x, \xi, t)$
as a function of $x$ at $\xi = 0.3$
and $t=t_{\rm min}=-0.35$~GeV$^2$. Dashed curve: contribution from valence
level. Dashed-dotted curve:contribution of the Dirac continuum.
Solid curve: the total distribution (sum of the dashed and dashed-dotted
curves). The vertical lines mark the crossover points $x=\pm\xi$.}
\label{fig:modelHt03}
\end{figure}
The result of the chiral quark-soliton model for
$\widetilde H^u (x, \xi, t)-\widetilde H^d (x, \xi, t)$ at
$\xi=0.3$ and $t=t_{\rm min}$ is presented in
Fig.~\ref{fig:modelHt03}.
We observe the sizeable contribution of the Dirac continuum.

\begin{figure}[h]
\epsfxsize=8. cm
\vspace{-.2cm}
\centerline{\epsffile{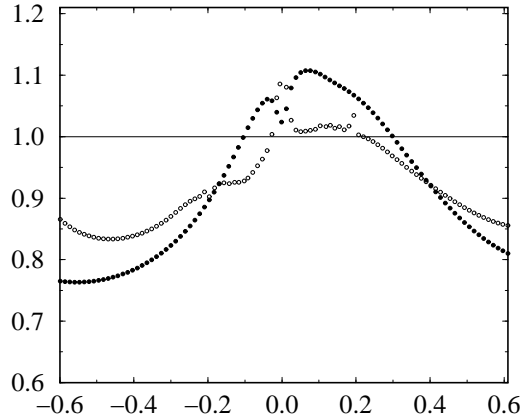}}
\vspace{-.5cm}
\caption[]{\small
The same as Fig.~\protect{\ref{fig:RH}} but for the GPD
$\widetilde H^u (x, \xi, t)-\widetilde H^d (x, \xi, t)$.}
\label{fig:RHt}
\end{figure}

To illustrate the non-factorizability of the $t-$dependence
also for the GPD $\widetilde H$ we plot in Fig.~\ref{fig:RHt}
the ratio
analogous to Eq.~(\ref{R-ratio}) with the same values of all parameters
replacing GPD
$H^u (x, \xi, t)+H^d (x, \xi, t)$ by
$\widetilde H^u (x, \xi, t)-\widetilde H^d (x, \xi, t)$.
We observe again a noticeable  deviation from the factorization ansatz
for the $t-$dependence, though in this case the deviation
is smaller.

With these lessons from chiral quark-soliton
model calculations we next turn to a discussion
of the phenomenological applications of the GPDs to various hard
exclusive reactions.

\newpage
\section{GPDs FROM HARD ELECTROPRODUCTION REACTIONS}
\label{chap5}


\subsection{Introduction}
\label{chap5_1}

In this section, we show how one can access the GPDs through the 
measurement of hard exclusive reactions. 
As discussed in Sec.~\ref{chap3_1}, in leading twist there are
four GPDs for the nucleon, 
i.e. $H$, $E$, $\tilde H$ and $\tilde E$, which 
are defined for each quark flavor ($u$, $d$, $s$). 
These GPDs depend upon the different longitudinal momentum fractions 
$x + \xi$ ($x - \xi$) of the initial (final) quark and upon the
overall momentum transfer $t$ to the nucleon. Therefore,  
these functions contain a wealth of new nucleon structure information. 
\newline
\indent
In the following, it will be shown how the different GPDs 
enter in the observables for different hard exclusive processes. 
We will discuss in detail the 
deeply virtual Compton scattering (DVCS) process on the nucleon and hard
electroproduction of vector and pseudoscalar mesons (HMP) on the nucleon. 
Besides the hard processes involving nucleon GPDs, it will also be 
discussed how the $N \to \Delta$ DVCS process and the hard
electroproduction of $\pi \Delta$ final states depend upon GPDs for the $N
\to \Delta$ transition, providing new information on the quark structure of
the $\Delta$ resonance.  
Furthermore, it will be discussed how 
the electroproduction of kaon-hyperon ($K Y$) final states contains 
GPDs for the $N \to Y$ transitions, yielding information on the
quark structure of hyperons.  
\newline
\indent
It is clear that to extract all those different GPDs 
from hard electroproduction reactions and to map out their dependencies 
on the three variables ($x, \xi, t$) is a quite challenging program. 
As a first step towards this goal, one strategy is to use  
phenomenological parametrizations for the GPDs which include all 
general constraints and which are flexible enough 
to allow for a fit to different hard electroproduction observables.   
The aim of this section is to present such phenomenological
and physically motivated parametrizations of the GPDs 
and to test then their sensitivity on observables.

\subsection{Phenomenological parametrization of GPDs}
\label{chap5_2}

In the following, we discuss successively the parametrizations for 
the GPDs $H^q$, $E^q$, $\tilde H^q$ and $\tilde E^q$
for each quark flavor $q$ (= $u, d, s$). 
As we address here mainly hard exclusive reactions in the valence
region at relatively small values of $-t$ ( $< 1$ GeV$^2$ ),
we start by parametrizing the GPDs at $t = 0$ and discuss firstly their $x$
and $\xi$ dependencies. For each GPD, we then come back to their 
$t$-dependence.

\subsubsection{Parametrization of the GPD $H^q$} 
\label{chap5_2a}

\begin{figure}[hp]
\epsfxsize=7.5 cm
\centerline{\epsffile{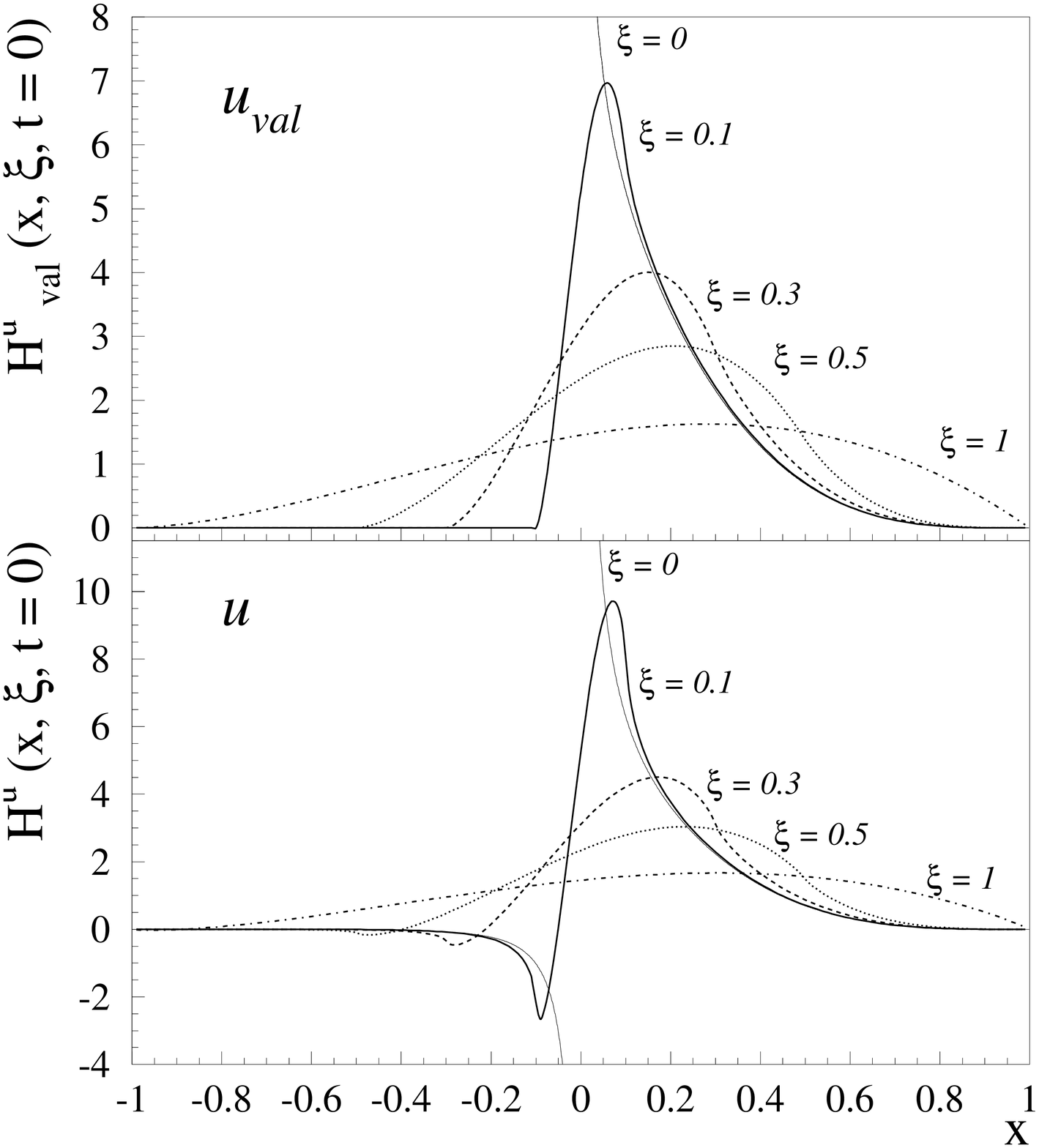}}
\caption{\small Double distribution part to the 
$u$-quark GPD $H^u$ at $t = 0$ for different $\xi$, 
using the ansatz (\ref{eq:dd2}) as desribed in the text
for $b_{val} = b_{sea} = 1$. 
Upper panel~: valence $u$-quark GPD, lower panel~: total (valence and sea) 
$u$-quark GPD. The thin lines ($\xi = 0$) show the ordinary forward  
$u$-quark distributions (MRST98 parametrization at $\mu^2$ = 2 GeV$^2$).}
\label{fig:up_H}
\epsfxsize=7.5 cm
\centerline{\epsffile{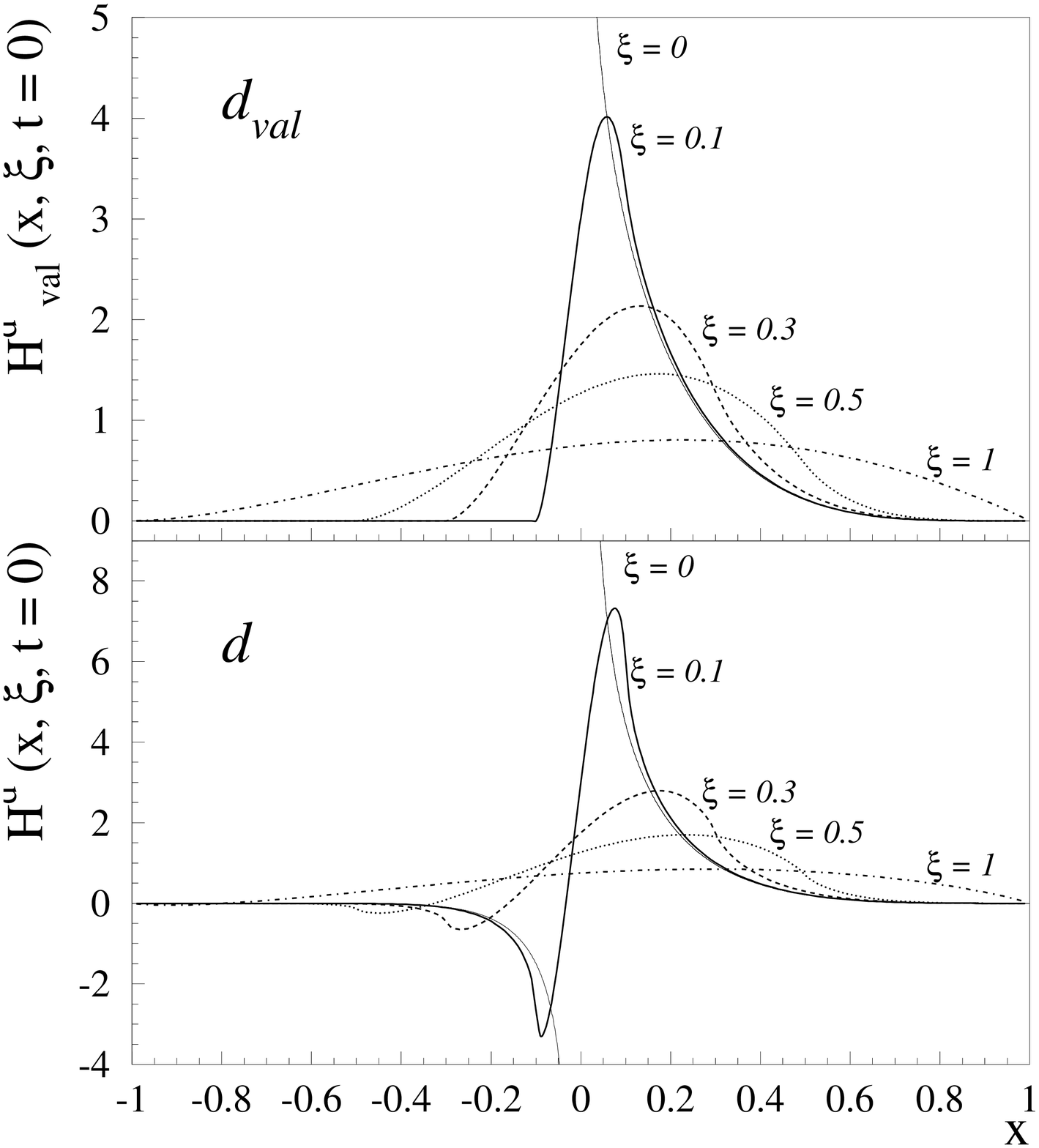}}
\caption{\small The same as Fig.~\ref{fig:up_H} but for the 
double distribution part to the $d$-quark GPD $H^d$ at $t = 0$.}
\label{fig:down_H}
\end{figure}

\begin{figure}[h]
\epsfxsize=9 cm
\centerline{\epsffile{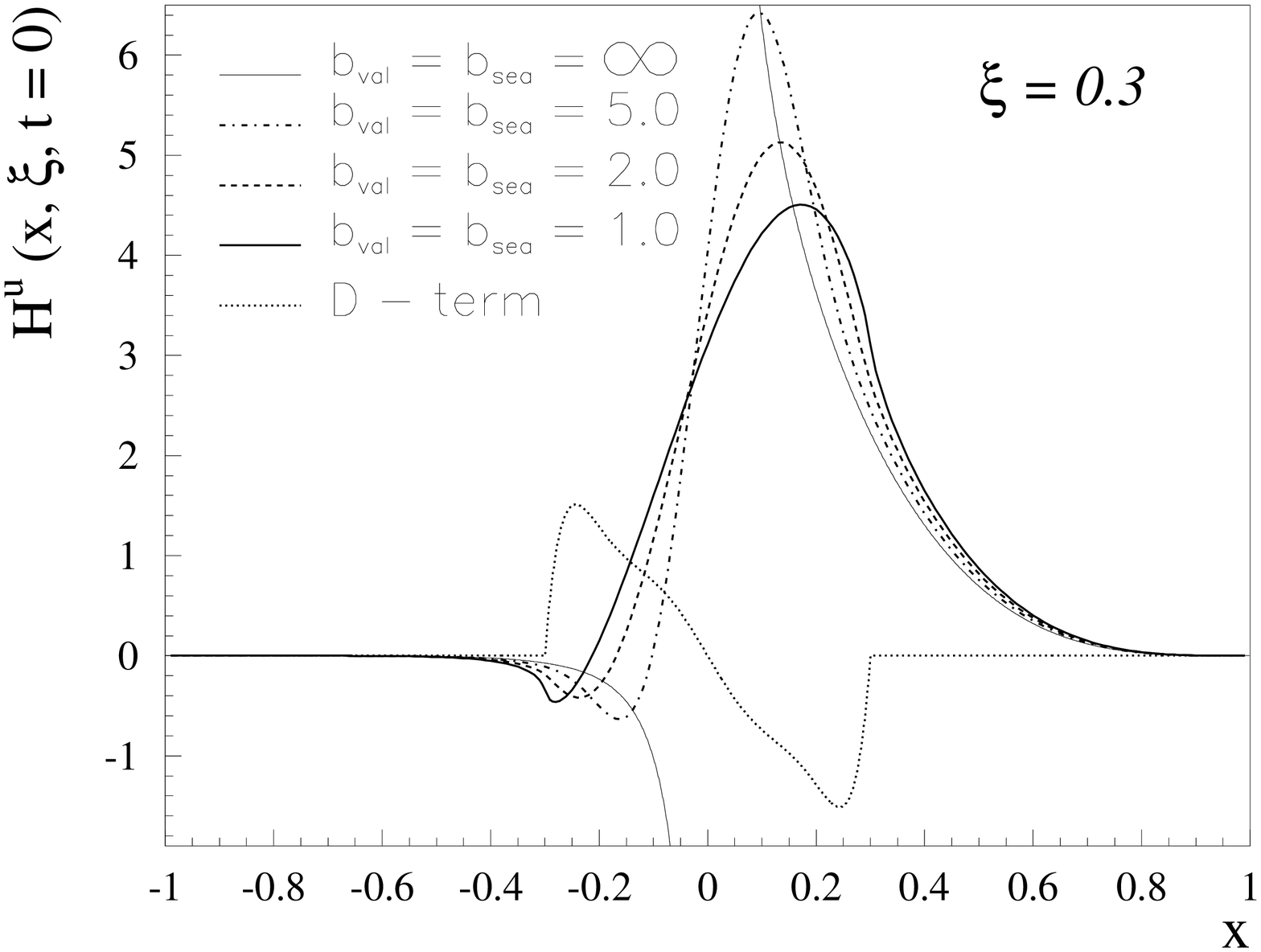}}
\caption{\small Double distribution part to the 
GPD  $H^u (x, \xi, t = 0)$ at $\xi = 0.3$ 
for different values of the parameters $b_{val}$, $b_{sea}$. 
The limit  $b_{val} = b_{sea} = \infty$
corresponds to the forward $u$-quark distribution. The D-term 
contribution to $H^u$, as it comes out from the chiral quark soliton model
(see Eq.~(\ref{dterm-num})) is shown by the dotted curve, and should be 
added to the double distribution part to obtain the total GPD $H^u$.}
\label{fig:upb_H}
\end{figure}

\begin{figure}[h]
\epsfxsize=8 cm
\centerline{\epsffile{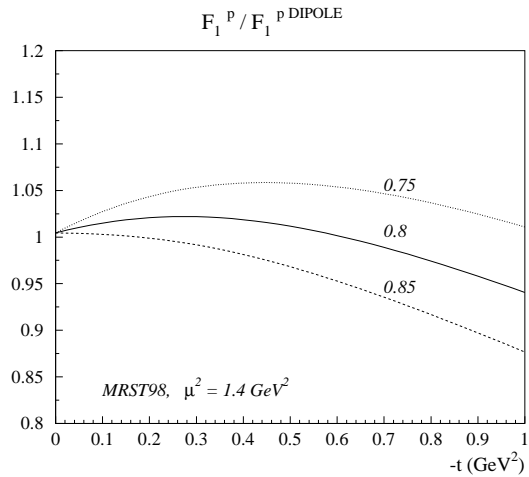}}
\caption{\small The ratio of the proton Dirac form factor
obtained from Eq.~(\ref{Regge-F1}) to the dipole form of the Dirac
form factor (\ref{dipolef1}) for different values of the slope 
$\alpha'$ (in GeV$^{-2}$) as indicated in the figure.}
\label{fig:factf1p}
\end{figure}

\begin{figure}[h]
\epsfxsize=8 cm
\centerline{\epsffile{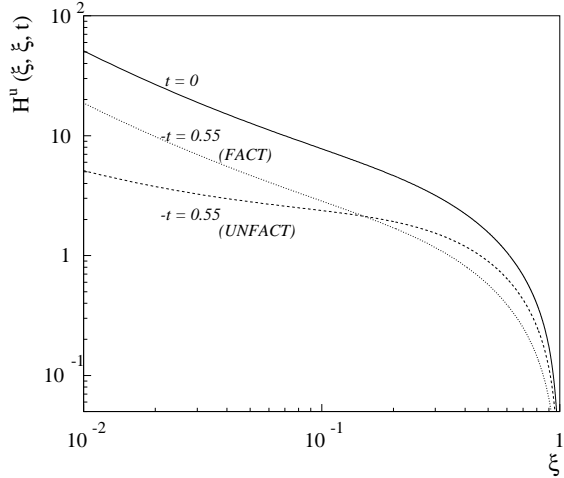}}
\caption{\small Comparison of the $\xi$-dependence of the `envelope' function 
$H^u(\xi, \xi, t)$ at two values of $-t$ (in GeV$^2$).
The function $H^u(\xi, \xi, 0)$ (full curve) 
is calculated according to the double 
distribution ansatz of Eq.~(\ref{eq:dd2}) with $b_{val} = b_{sea} = 1$, using 
the MRST98 parametrization at a scale $\mu^2$ = 1.4 GeV$^2$ for the forward 
$u$-quark distribution as input. For the function $H^u(\xi, \xi, t)$ at 
$- t$ = 0.55 GeV$^2$, the factorized model of Eq.~(\ref{eq:factt}) 
(dotted curve), is compared with the unfactorized parametrization 
of Eq.~(\ref{eq:ddunpolregge}) (dashed curve).}
\label{fig:upt_xiscan}
\end{figure}

For the function $H^q$ (for each flavor $q$),
the $t$-independent part $H^q(x, \xi) \equiv H^q(x, \xi, t = 0)$ 
is parametrized by a two-component form~:
\be
H^q(x, \xi) \,=\, H^q_{DD}(x, \xi) \,+\,
\theta(\xi-|x|)\, \frac{1}{N_f} \, D(\frac{x}{\xi})\, ,
\la{eq:dd}
\ee
where $H^q_{DD}$ is the part of the GPD which is obtained as a
one-dimensional section of a two-variable double distribution (DD) $F^q$
\cite{Rad99}, imposing a particular dependence on the 
skewedness $\xi$, as given by Eq.~(\ref{dd2})
\be
H^q_{DD}(x,\xi)=
\int_{-1}^{1}d\beta\
\int_{-1+|\beta|}^{1-|\beta|} d\alpha\
\delta(x-\beta-\alpha\xi)\  F^q(\beta,\alpha)\ \, .
\la{eq:dd2}
\ee
The D-term contribution $D$ in Eq.~(\ref{eq:dd})
completes the parametrization of GPDs, restoring the correct polynomiality
properties of GPDs \cite{Ji98b,Pol99b} as discussed in 
Sec.~\ref{chap3_2}.
It has a support only for $|x| \leq |\xi|$, 
so that it is `invisible' in the forward limit.
Because the D-term is an isoscalar contribution, it adds the same
function for each flavor. In Eq.~(\ref{eq:dd}), $N_f$ = 3 is the number of
 active flavors $(u, d, s)$.
\newline
\indent
For the double distributions, entering Eq.~(\ref{eq:dd2}),
we follow Radyushkin's suggestion \cite{Rad99} to use the following model 
\be
F^q(\beta, \alpha) = h(\beta, \alpha) \ q(\beta) ,
\la{eq:ddunpol}
\ee
where $q(\beta)$ is the forward quark distribution (for the flavor
$q$) and where $h(\beta, \alpha)$ denotes a profile function. 
In the following estimates, we parametrize the profile
function through a one-parameter ansatz, following Refs.~\cite{Rad99,Rad01b}~:
\begin{eqnarray}
h(\beta , \alpha) = 
 \frac{\Gamma(2b+2)}{2^{2b+1}\Gamma^2(b+1)}\
\frac{\bigl[(1-|\beta|)^2-\alpha^2\bigr]^{b}}{(1-|\beta|)^{2b+1}}\; .
\la{eq:profile}
\end{eqnarray}
In Eq.~(\ref{eq:profile}),
the parameter $b$ characterizes the strength of the $\xi$
dependence of the GPD $H(x,\xi)$. The limiting case $b\to\infty$
corresponds to the $\xi$ independent ansatz for the GPD,
i.e. $H(x,\xi)=q(x)$, as used in Refs.~\cite{Gui98,Vdh98}.
The power $b$ in Eq.~(\ref{eq:profile}) is a free parameter
for the valence contribution ($b_{val}$)
and for the sea/antiquark contribution ($b_{sea}$) to the GPD,
which can be used in such an approach as fit parameters in the 
extraction of GPDs from hard electroproduction observables.
As an example, the twist-2 DVCS predictions of 
Ref.~\cite{Vdh99} and the twist-3 DVCS predictions of Ref.~\cite{Kiv01b} 
correspond to the choice $b_{val} = b_{sea} = 1.0$.
In the following, we discuss the dependence of DVCS
observables on the shape of the profile function, i.e. on the choice
of the parameters $b_{val}$ and $b_{sea}$.
\newline
\indent
In Eq.~(\ref{eq:ddunpol}), we use the phenomenological forward quark
distributions $q(\beta)$ as measured from DIS as input, 
ensuring the correct forward limit for the GPDs, i.e. $H^u(x,0)=u(x)$,
$H^d(x,0)=d(x)$, $H^s(x,0)=s(x)$. 
Specifically, the calculations shown in the following were performed using the
NLO quark distributions of \cite{Mar98} (more precisely, their
so-called ``central gluon parametrization'').
We checked explicitely however that the results in the valence region
practically do not change when using another current parametrization
for the NLO quark distributions, as those parametrizations are very
close to each other in the valence region which we consider here. 
As an illustration, we show in Figs.~\ref{fig:up_H}, \ref{fig:down_H} the 
parametrization of Eqs.~(\ref{eq:dd}-\ref{eq:profile}) for the double
distribution part to the $u$-quark GPD $H^u(x, \xi)$ and $d$-quark GPD
$H^d(x, \xi)$ respectively, for the parameter value $b_{val} = b_{sea}= 1$. 
One sees how the GPDs approach the forward $u$- and $d$-quark
distributions as $\xi \to 0$. Note that the negative $x$-range at $\xi
= 0$ corresponds to the anti-quark distribution, i.e. 
$q(-x) = - \bar q(x)$.
\newline
\indent
Fig.~\ref{fig:upb_H} shows, at a fixed value of $\xi = 0.3$, 
the dependence of the DD part to the GPD $H^u$ on the
parameters $b_{val}, b_{sea}$. In the limit 
$b_{val} = b_{sea} = \infty$, one finds back the foward $u$-quark
distribution. 
\indent
\newline
\indent
We next discuss the parametrization of the $t$-dependence of the 
double distribution part to the GPD $H^q(x, \xi, t)$. 
As for the hard exclusive reactions which we
discuss $t$ should be a small scale compared to the hard scale $Q^2$ of the
reaction, we will only be concerned here with the 
small $-t < 1$~GeV$^2$ region. 
The main constraint which one has on the $t$-dependence of the GPDs
comes through the sum rules of 
Eqs.~(\ref{eq:ffsumruleh}-\ref{eq:ffsumruleht}), through which the first
moment of the GPDs is given by the elastic nucleon form factors.
For the GPD $H^q$ and for a quark of flavor $q$, 
the first moment is determined through Eq.(\ref{eq:ffsumruleh}).
in terms of the elastic Dirac form factors $F_1^q(t)$ 
for a quark of flavor $q$ in the nucleon. 
In Eq.~(\ref{eq:vecff}), the strange form factor \( F_{1}^{s} \), 
which is not so well known and expected to be small,
is set equal to zero in the following numerical evaluations.
\newline
\indent
The simplest parametrization of the $t$ dependence of the GPD $H^q$ 
in the small $-t$ region which fulfills the sum rule 
of Eq.~(\ref{eq:ffsumruleh}), consists of the factorized ansatz 
for the $t$-dependence~:
\begin{eqnarray}
H^u(x, \xi, t) \,&=&\, H^u(x, \xi) \; F_1^u(t) \, / \, 2\;, \nonumber\\
H^d(x, \xi, t) \,&=&\, H^d(x, \xi) \; F_1^d(t) \;, \nonumber\\
H^s(x, \xi, t) \,&=&\, 0 \;, 
\la{eq:factt}
\end{eqnarray}
where $F_1^u(t)$ and $F_1^d(t)$ are determined through
Eq.~(\ref{eq:vecff}) using the empirical parametrizations for the
proton and neutron Dirac form factors. 
Besides satisfying the sum rule constraint, the ansatz of Eq.~(\ref{eq:factt})
also yields the correct forward limit for $u$- and $d$-quark
distributions, i.e. 
$H^u(x, \xi = 0, t = 0) = H^u(x, \xi = 0) = u(x)$ and 
$H^d(x, \xi = 0, t = 0) = H^d(x, \xi = 0) = d(x)$. 
\newline
\indent
Note that the sum rule of Eq.~(\ref{eq:ffsumruleh}) only gives a
 constraint for the valence quark part to the GPD, as the sea-quark
 contribution drops out of this sum rule. Indeed at $\xi = 0$, where
 the GPDs reduce to the forward quark distributions, one sees from  
$q(x) = q_{val}(x) + \bar q(x)$ and $q(-x) = -\bar q(x)$ that the
sea-quark and the anti-quark contributions cancel each other in the
sum rule of Eq.~(\ref{eq:ffsumruleh}).  
One could therefore use a different $t$-dependence for the sea-quark
part to the GPD $H^q$. However, in lack of more information, we adopt here
the same $t$-dependence for both valence and sea-quark parts. 
\newline
\indent
Such factorized forms for the $t$-dependence 
as in Eq.~(\ref{eq:factt}) have been used in practically all 
theoretical estimates for hard exclusive reactions at small $-t$ up to now. 
It remains to be investigated how realistic such parametrizations are
if one goes away from the $t = 0$ region. In particular, the 
$t$-dependence of the differential cross section for 
$\rho^0$ electroproduction on the proton, in which the
GPDs $H^u$ and $H^d$ enter (see further), 
has been measured at HERMES at values of 
$x_B$ around 0.1 - 0.15 \cite{Tyt01}. These cross sections clearly display 
an exponential fall-off with $-t$ in the range $t_{min} < -t < 0.4$~GeV$^2$. 
A form factor type ansatz as in Eq.~(\ref{eq:factt}), even if it gives
the correct normalization at $t_{min}$, deviates over this 
$t$ range already noticeably from an exponential fall-off. 
\newline
\indent
We also saw in Sec.~\ref{chap4_4_1} that the calculations of GPDs
in the chiral quark-soliton model indicate a deviation
from the factorization ansatz (\ref{eq:factt})
for the $t-$dependence of GPDs, as seen from 
Figs.~\ref{fig:RH},\ref{fig:RHxi0}. The chiral quark-soliton model 
results for the $t-$dependence of GPDs in the small $-t$ region and 
at a low normalization point can be described
by a simple Regge theory motivated ansatz
(\ref{Regge-parametrization}). If one uses such a simple ansatz
for $H^q(x,\xi=0,t)$ then the nucleon form factors follow from 
the sum rule Eq.~(\ref{eq:ffsumruleh}):

\be
F_1^q(t)=\int_0^1 dx\ \frac{1}{x^{\alpha' t}}\ q_{val}(x)\, .
\la{Regge-F1}
\ee
Here $q_{val}(x)$ is the valence quark distribution at a low
normalization point. The ansatz of Eq.~(\ref{Regge-F1}) is 
only valid for the small $-t$ region, because at larger values of
$-t$ the integral in Eq.~(\ref{Regge-F1}) is dominated by the large
$x$ region, for which a Regge ansatz does not hold. 
Furthermore, note that the model of Eq.~(\ref{Regge-F1}) for 
the nucleon Dirac form factor 
as well as the ansatz of Eq.~(\ref{Regge-parametrization})
is valid only at a low normalization point. Strictly speaking
the form of the ansatz of Eq.~(\ref{Regge-parametrization}) should
be changed when one goes to a higher normalization point. In
particular, when increasing the normalization
point, the slope $\alpha'$ should decrease \cite{Bro94,Fra96}. This
should lead for instance to the disappearence of the shrinkage of
the diffractive cone in hard exclusive processes when increasing
$Q^2$ \cite{Fra96}.
\newline
\indent
In Fig.~\ref{fig:factf1p} we show the ratio of the proton Dirac form
factor $F_1^p(t)$
obtained from Eq.~(\ref{Regge-F1}), using the 
MRST98 parametrization for the valence quark distributions at the scale
$\mu^2=1.4$~GeV$^2$ (which will be relevant in the discussion of 
observables further on) to the dipole form of this form factor, given by:

\be
F_1^{\rm dipole}(t)=\frac{1- (1 + \kappa^p) \ t/4m_N^2}{1-t/4m_N^2}\
\frac{1}{(1-t/0.71)^2}\, .
\la{dipolef1}
\ee
The result is shown for several values of the slope $\alpha'=0.75, 0.8$ and
$0.85$~GeV$^{-2}$, which are close to the phenomenological slope
of Regge trajectories. It is seen from Fig.~\ref{fig:factf1p} 
that the simple model
of Eq.~(\ref{Regge-F1}) with e.g. $\alpha'$ = 0.8~GeV$^{-2}$ 
gives a rather satisfactory description of the proton Dirac form factor 
over the range $0 < -t < 1$ GeV$^2$.
This finding therefore suggests that the Regge theory motivated 
ansatz of Eq.~(\ref{Regge-parametrization})
can be used as a guide for a more realistic parametrizations 
of the $t-$dependence of the GPDs. 
From the theoretical point of view it can
give a clue for understanding the interplay between the Regge
and partonic pictures of hard processes.
\newline
\indent
The $\xi$-dependence of the Regge type ansatz of 
Eq.~(\ref{Regge-parametrization}) for $H(x, \xi = 0, t)$ can be restored 
from Eq.~(\ref{dd2}) by using the following model for the double distribution~:
\begin{eqnarray}
F^q(\beta, \alpha, t) = h(\beta, \alpha) \ q(\beta) \, 
{1 \over {|\beta|^{\alpha' \, t}}} \, .
\la{eq:ddunpolregge}
\end{eqnarray}
where $h(\beta, \alpha)$ is a profile function as in Eq.~(\ref{eq:profile}).
\newline
\indent
In Fig.~\ref{fig:upt_xiscan}, we compare the factorized model 
 of Eq.~(\ref{eq:factt}) for the 
`envelope' function $H(\xi, \xi, t)$ (i.e. along the line $x = \xi$) 
with the unfactorized model of Eq.~(\ref{eq:ddunpolregge}). 
One sees that the unfactorized model leads to an increasingly 
larger reduction when 
going to smaller $\xi$ (typical for a Regge type ansatz) 
and to an enhancement at the larger $\xi$.

Since the theoretical status of the ans\"atze of the type 
(\ref{Regge-parametrization},\ref{eq:ddunpolregge})
still remains to be clarified, we restrict ourselves 
in the discussion of most observables, 
to the simple factorization ansatz (\ref{eq:factt}) 
unless explicitely indicated. 
In some observables (like spin asymmetries for hard meson production)
the $t-$dependence of the GPDs tends to cancel 
in the ratio of cross sections and
therefore these observables are not very sensitive to the 
explicit form for the $t-$dependence of the 
GPDs. However, for absolute cross sections 
and also for the DVCS single spin asymmetry,
the ansatz for the $t-$dependence of the GPDs can influence the
prediction, as will be shown in Sec.~\ref{chap5_3_5}.

\subsubsection{Parametrization of the D-term} 
\label{chap5_2b}

\begin{figure}[h]
\vspace{.5cm}
\epsfxsize=12cm
\centerline{\epsffile{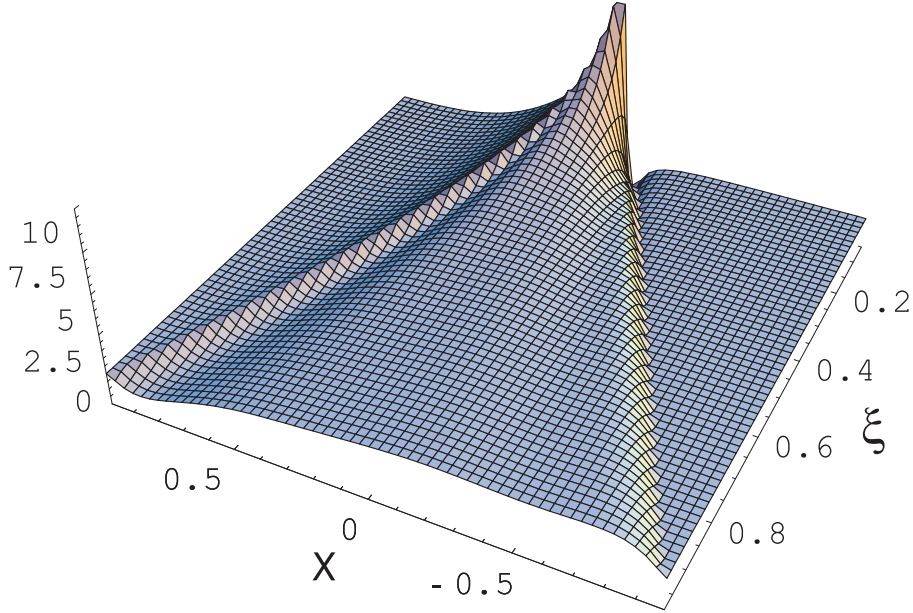}}
\caption{\small The $x$ and $\xi$ dependence of the 
GPD $H^u(x, \xi, t = 0)$ for the $u$-quark distribution, 
including the double distribution part 
(for the value $b_{val} = b_{sea} = 1$) and the D-term.}
\label{fig:gpddt}
\end{figure}

To complete the parametrization of the GPD $H$, we next specify the
D-term part in Eq.~(\ref{eq:dd}).  
The D-term contributes only to the singlet GPDs $H(x,\xi)$,
i.e. yields the same contribution for each quark flavor. 
\newline
\indent
As discussed in Sec.~\ref{chap3_2}, the function $D(z) \equiv D(z, t = 0)$ in
Eq.~(\ref{eq:dd}) is an odd function of its argument, i.e. $D(-z) = -
D(z)$, and can be expanded in odd Gegenbauer polynomials 
as given by Eq.~(\ref{dterm}). 
The moments $d_1, d_3, d_5$ in such an expansion have been estimated 
in Eq.~(\ref{dterm-num}) using the calculation 
of GPDs in the chiral quark soliton model \cite{Pet98}
at a low normalization point $\mu\approx 0.6$~GeV.
In principle, the D-term also has a scale dependence but it will be
neglected as for the moment the uncertainties 
in the modeling of the D-term are larger than the
logarithmic scale dependence, for the scales relevant for present and
planned experiments. 
\newline
\indent
The general expansion of the D-term in Eq.~(\ref{dterm}) can be
used in phenomenological fits to data, 
and the moments $d_1, d_3, d_5, ...$ can
be used as fit parameters to be extracted from hard electroproduction 
observables which are sensitive to the D-term. 
For the following estimates, we will use the values of 
Eq.~(\ref{dterm-num}) obtained from the chiral quark soliton model
calculation. The resulting D-term is represented in
Fig.~(\ref{fig:upb_H}) for one quark flavor (i.e. by dividing
Eq.~(\ref{dterm-num}) by $N_f = 3$. 
Note the negative sign of the Gegenbauer coefficients for
the D-term in Eq.~(\ref{dterm-num}), as obtained in the chiral
quark-soliton model, which fixes the sign of the D-term relative to
the double distribution part to the GPD $H^q$.
\newline
\indent 
In Fig.~\ref{fig:gpddt}, we display  
the full $x$ and $\xi$ dependence of the parametrization of Eq.~(\ref{eq:dd}) 
for the GPD $H^u(x, \xi, t = 0)$, i.e.  
including both the DD part and the D-term. By moving along lines of constant 
$\xi$, one nicely sees how the sensitivity 
of the GPDs to the D-term part increases relative to the 
double distribution part when increasing the skewedness
parameter $\xi$. In the end, at $\xi$ = 1, the D-term contribution dominates 
the GPD $H$.
\newline
\indent
The $t$-dependence of the D-term is again not constrained through
the sum rule of Eq.~(\ref{eq:ffsumruleh}), because the D-term drops
out of this sum rule. In absence of more information, we will assume 
the same $t$-dependence for the D-term as for the DD contribution to $H^q$. 
The study of the $t$-dependence of the $D$-term is a subject for future work.

\subsubsection{Parametrization of the GPD $E^q$ and quark contribution
  to the proton spin}
\label{chap5_2c}

The parametrization of \( E^q \), which corresponds 
to a nucleon helicity flip amplitude, 
is more difficult as we don't have the DIS constraint
for the \( x \)-dependence in the forward limit. 
In the amplitudes for hard electroproduction processes at small $-t$, 
\( E^q \) is multiplied by a momentum transfer and
therefore its contribution is suppressed at small \( t \) in the observables.
\newline
\indent
One contribution to the GPD $E^q$ is however determined through the 
polynomiality condition as explained in Sec.~\ref{chap3_2}, which requires
that the D-term contributes with opposite sign to $H$ and $E$. This
leads to the D-term contribution for $E^q$~:  
\begin{eqnarray}
E^q_{\rm D-term}(x,\xi,t)\;=\;-\,\theta(\xi-|x|)\, {1 \over {N_f}} \, 
D\left(\frac{x}{\xi}, t\right) \, , 
\la{eq:dtermcontrib2} 
\end{eqnarray}
which guarantees that the D-term contribution is cancelled in the combination
$H+E$. Therefore, the D-term part does not enter 
in the angular momentum sum rule of Eq.~(\ref{eq:original_ji_sr}). 
\newline
\indent
Similarly to Eq.~(\ref{eq:dd}) for $H^q$, one can parametrize $E^q$ 
by adding a double distribution part to the D-term of 
Eq.~(\ref{eq:dtermcontrib2}). This leads to the parametrization~:
\begin{eqnarray}
E^q(x,\xi,t)&=&E^q_{DD}(x, \xi, t) \,-\, \theta(\xi-|x|)\, {1 \over {N_f}} \, 
D\left(\frac{x}{\xi}, t \right) \, , 
\la{eq:parame} 
\end{eqnarray}
where $E^q_{DD}$ is the double distribution part. 
\newline
\indent
In the forward limit, the DD part reduces to the function 
$e^q(x) \equiv E^q_{DD}(x, \xi = 0, t = 0)$ which is a priori unknown.
However one constraint exists on the normalization of $E^q_{DD}$ due
to the form factor sum rule Eq.~(\ref{eq:ffsumrulee}) 
in terms of the Pauli form factor $F_2^q(t)$ for the quark of flavor $q$ in
the proton. The quark form factors $F_2^u$ and $F_2^d$ 
are normalized at $t = 0$ through the anomalous
magnetic moments of proton and neutron, i.e. $F_2^p(0) = \kappa^p = 1.793$ and 
$F_2^n(0) = \kappa^n = -1.913$. Defining $\kappa^u \equiv F_2^u(0)$
and $\kappa^d \equiv F_2^d(0)$, one finds~:
\begin{eqnarray}
\kappa^u \,&=&\, 2 \, \kappa^p \,+\, \kappa^n \,=\, 1.673 \, ,\nonumber\\
\kappa^d \,&=&\, \kappa^p \,+\, 2 \, \kappa^n \,=\, -2.033 \, .
\la{eq:f2ud}
\end{eqnarray}  
Taking Eq.~(\ref{eq:ffsumrulee}) in the forward limit ($\xi = 0$, $t = 0$)
leads to the normalization condition 
\begin{eqnarray}
&& \int _{-1}^{+1}dx\; e^{q}(x)\, =\, \kappa^{q}\; .
\la{eq:vece2sumrule} 
\end{eqnarray}
As a first guess for the function $e^q(x)$, which is consistent with
this normalization condition, one can assume that it has the same $x$
dependence as a valence quark distribution, i.e.  
\begin{eqnarray}
e^u(x) \,&=&\, {1 \over 2} \, u_{val}(x) \; \kappa^u \;, \nonumber\\
e^d(x) \,&=&\, d_{val}(x) \; \kappa^d \;, \nonumber\\
e^s(x) \,&=&\, 0 \;.  
\la{eq:factte}
\end{eqnarray}
Indeed, it is seen from Fig.~\ref{fig:valEvsH} 
that the chiral quark-soliton model calculation for the 
isovector function $e^u(x) - e^d(x)$ 
(which is leading in the large $N_c$ limit) 
contains such a valence quark component.
\newline
\indent
Using the ansatz of Eq.~(\ref{eq:factte}), one obtains a prediction for 
the total angular momentum of the quarks in the nucleon, 
evaluating Ji's sum rule at $\xi = 0$~: 
\be
J^q \;=\; {1 \over 2} \int_{-1}^1dx\ x\, \left[\, q(x) \,+\, e^q(x) 
\, \right]\;. 
\la{eq:jipol}
\ee
By defining the total fraction of the proton momentum carried by the quarks
and antiquarks of flavor $q$ as
\be
M_2^q \,=\, \int_0^1dx \, x \, \left[ \, q(x) + \bar q(x) \, \right]\, =\, 
\int_0^1dx \, x \, \left[ \, q_{val}(x) + 2 \, \bar q(x) \, \right] \, ,
\la{eq:meproton}
\ee
and the momentum fraction carried by the valence quarks as 
\be
M_2^{q_{val}} \,=\, \int_0^1dx \, x \, q_{val}(x) \, ,
\la{mevalproton}
\ee
the ansatz of Eq.~(\ref{eq:factte}) for $e^q(x)$ leads to an estimate  
for the total angular momentum carried by the $u$, $d$ and $s$-quarks in the
proton, $J^u$, $J^d$, and $J^s$ respectively, as~:
\be
J^u \,&=&\, {1 \over 2} \left( M_2^u \,+\, {1 \over 2} \kappa^u \;
  M_2^{u_{val}} \right) \, , 
\la{eq:ju} \\
J^d \,&=&\, {1 \over 2} \left( M_2^d \,+\, \kappa^d \;
  M_2^{d_{val}} \right) \, ,
\la{eq:jd} \\
J^s \,&=&\, {1 \over 2} \, M_2^s \, .
\la{eq:js}
\ee
 
\begin{table}[h]
{\centering \begin{tabular}{|c|c|c|c|}
\hline
&&& \\ 
&$M_2^{q_{val}}$ ($\mu^2$ = 1 GeV$^2$)  & $M_2^{q}$ ($\mu^2$ = 1 GeV$^2$) & 
$2 \, J^q$ ($\mu^2$ = 1 GeV$^2$)\\
&&& \\
\hline 
\hline
$u$ & 0.34 & 0.40 & 0.69 \\
\hline 
$d$ & 0.14 & 0.22 & -0.07 \\
\hline 
$s$ & 0 & 0.03 & 0.03 \\
\hline 
$u + d + s$ & 0.49 & 0.65 & 0.65 \\
\hline 
\hline
\end{tabular}\par}
\caption{\small Estimate of $2 \, J^q$ at the scale $\mu^2$ = 1 GeV$^2$ 
according to Eqs.~(\ref{eq:ju}-\ref{eq:js})
using the valence model of Eq.~(\ref{eq:factte}).
For the forward parton distributions, 
the MRST98 parametrization (so-called `central gluon
parametrization' of Ref.~\cite{Mar98}) has been used.
\label{table_5_1}}
\end{table}

In Table~\ref{table_5_1}, we show the 
values of the momentum fractions $M_2^{q_{val}}$ and $M_2^q$ at the
scale $\mu^2$ = 1 GeV$^2$, using the MRST98 parametrization \cite{Mar98} 
for the forward parton distributions.
We also show the estimate for $J^u$, $J^d$, and $J^s$ of 
Eqs.~(\ref{eq:ju}-\ref{eq:js})  at the same scale. 
These estimates lead to a large fraction (69 \%) of the total angular
momentum of the proton carried by the $u$-quarks and a relatively
small contribution carried by the $d$ and $s$-quarks. 
Using this estimate, the sum of $u$, $d$, and $s$-quark contributions is 
around 65 \% at this low scale. 
This estimate is in good agreement with the recent lattice QCD estimate 
of Ref.~\cite{Mat00} for $J^q$, which found the total value $ 2 \, J^q = 0.60
\pm 0.14$ at a comparable low scale, and the QCD sum rule estimate 
of Ref.~\cite{Bal97}.
\newline
\indent
One furthermore sees from Table~\ref{table_5_1}
that $2 J^u + 2 J^d \approx M_2^u + M_2^d$ to a rather high accuracy. 
For the separate
$u$ and $d$ contributions however, one sees from Table~\ref{table_5_1} that 
in the ansatz of Eq.~(\ref{eq:factte}), 
$2 J^u$ is quite different from $M_2^u$, and similarly $2 J^d$ is
quite different from $M_2^d$. 
This leads us to the observation that the following 
remarkable relation between nucleon anomalous magnetic moments 
and valence quark distributions in the proton seems to hold in general 
\footnote{ At this point one could  say: `` .....cuius rei
demonstrationem mirabilem sane detexi. Hanc marginis exiguitas non
caperet"}~:
\begin{eqnarray}
\kappa^u \, \int_0^1dx\ x \ \left[\, u(x)-\bar u(x) \,\right] \approx
-2 \, \kappa^d \, \int_0^1dx\ x \ \left[ \, d(x)-\bar d(x) \, \right]\, .
\la{eq:remarkrel}
\end{eqnarray}
Such a relation allows us to express the ratio of 
anomalous magnetic moments of proton and neutron in terms of the 
proton momentum fractions $M_2^{q_{val}}$ carried by 
the valence quarks (see definition of Eq.~(\ref{mevalproton})) as follows:
\begin{eqnarray}
\frac{\kappa^p}{\kappa^n} \,=\, -\,{1 \over 2} \, 
\frac{4 \, M_2^{d_{val}}+ M_2^{u_{val}}}{ M_2^{d_{val}}+ M_2^{u_{val}}}\, .
\la{eq:kpkn}
\end{eqnarray}
\begin{figure}[h]
\epsfxsize=9 cm
\centerline{\epsffile{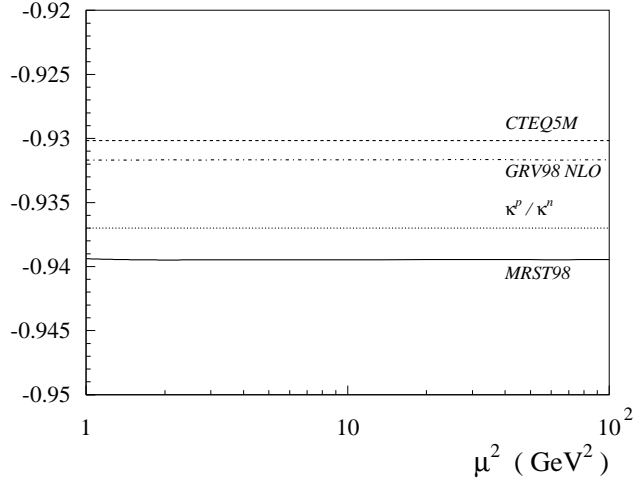}}
\caption{\small Scale dependence of the {\it rhs} of Eq.~(\ref{eq:kpkn}) for  
three different NLO forward parton distributions : 
MRST98 \cite{Mar98} (full curve), CTEQ5M \cite{Lai00} (dashed curve), and 
GRV98 NLO(${\overline{\mathrm{MS}}}$) \cite{Glu98} (dashed-dotted
curve). Also shown is the {\it lhs} of Eq.~(\ref{eq:kpkn}), i.e. the 
value $\kappa^p / \kappa^n$ (dotted curve). 
}
\label{fig:remarkrel}
\end{figure}
To check the validity of the above relation, 
we show in Fig.~\ref{fig:remarkrel} 
the scale dependence of the {\it rhs} of Eq.~(\ref{eq:kpkn}) 
for three current parametrizations of the NLO parton distributions.
Fig.~\ref{fig:remarkrel} shows 
that the scale dependence drops out of the {\it rhs} of Eq.~(\ref{eq:kpkn}),
although the numerator and denominator separately clearly have a scale
dependence. In addition, it is seen from Fig.~\ref{fig:remarkrel}, 
for the three parametrizations of
the NLO parton distributions, that the relation of Eq.~(\ref{eq:kpkn}) is
numerically verified to an accuracy at the one percent level! Both of
these observations suggest that Eq.~(\ref{eq:kpkn}) holds and that 
the unpolarized valence $u$ and $d$-quark forward 
distributions contain a non-trivial information 
about the anomalous magnetic moments of the proton and neutron. 
The relation of Eq.~(\ref{eq:remarkrel}) can for example be checked 
in a theory with weak coupling as well as in chiral perturbation theory.
Note that in a simple quark model with $M_2^{u_{val}} = 2 \,
M_2^{d_{val}}$, one recovers from Eq.~(\ref{eq:kpkn}) the result from $SU(6)$
symmetry, i.e. $\kappa^p$ = - $\kappa^n$. In quark model language, 
the relation of Eq.~(\ref{eq:kpkn}) implies that the breaking of the 
$SU(6)$ symmetry follows some rule which is encoded in the valence quark
distributions. 
\newline 
\indent
Because Eq.~(\ref{eq:remarkrel}) holds to such good
accuracy, when applying it to the estimates given in 
Eqs.~(\ref{eq:ju},\ref{eq:jd}), it leads to the result
that $2 J^u + 2 J^d \approx M_2^u + M_2^d$. For the separate
$u$ and $d$ contributions however, it was already noted 
from Table~\ref{table_5_1} that in the ansatz of Eq.~(\ref{eq:factte}), 
$2 J^u$ is quite different from $M_2^u$, and similarly $2 J^d$ is
quite different from $M_2^d$. 
One therefore concludes that the valence ansatz of
Eq.~(\ref{eq:factte}) for $e^q(x)$ leads to the result that the {\it sum}
over all quarks yields the same 
contribution to the proton spin as to the proton momentum, 
at a scale as low as $\mu^2 \sim 1$ GeV$^2$.   
\newline
\indent
In Table~\ref{table_5_2}, we show the estimate 
of Eqs.~(\ref{eq:ju}-\ref{eq:js}) for $2 J^u + 2 J^d + 2 J^s$ at different
scales, using the three different NLO forward parton distributions
discussed in Fig.~\ref{fig:remarkrel}. 
One firstly sees that the three estimates are very
close. One furthermore sees that $J^u + J^d + J^s$
decreases (very slowly) with increasing scale. 
The scale dependence of the total quark ($J^q$) and gluon 
($J^g$) contributions to the
nucleon spin has been calculated in Ref.~\cite{Ji96}. In particular, it
has been shown that as $\mu \to \infty$, there exists a fixed point solution~:
\begin{eqnarray}
J^q(\infty) \,=\, {1 \over 2} \, {{3 \, N_f} \over {16 \,+\, 3 N_f}}\,
,
\la{eq:fixedpoint}
\end{eqnarray}
and $J^g$ follows from $J^q + J^g = 1/2$.
Applying Eq.~(\ref{eq:fixedpoint}) for $N_f = 3$, 
one obtains that the total quark contribution to the nucleon spin
in the limit $\mu \to \infty$ amounts to 36 \%.  
\begin{table}[h]
{\centering \begin{tabular}{|c|c|c|c|}
\hline
&&& \\ 
& $2 \, J^q$ & $2 \, J^q$ & $2 \, J^q$ \\
& ($\mu^2$ = 2.5 GeV$^2$)  
& ($\mu^2$ = 10 GeV$^2$)  
& ($\mu^2$ = 100 GeV$^2$)  \\
&&& \\
\hline 
\hline
$u + d + s$ & 0.60 & 0.55 & 0.51 \\
(MRST98) &&& \\
\hline 
$u + d + s$ & 0.59 & 0.55 & 0.52 \\
(CTEQ5M) &&& \\
\hline 
$u + d + s$ & 0.59 & 0.55 & 0.51 \\
(GRV98 NLO) &&& \\
\hline 
\hline
\end{tabular}\par}
\caption{\small Dependence of the estimate for $2 J^u + 2 J^d + 2 J^s$ 
on the scale $\mu^2$, using the MRST98 \cite{Mar98}, CTEQ5M
\cite{Lai00}, and GRV98 NLO(${\overline{\mathrm{MS}}}$) \cite{Glu98}
forward parton distributions.
\label{table_5_2}}
\end{table}

One can further interpret the spin structure of the proton by 
decomposing the quark contribution to the 
proton spin in an intrinsic part ($\Delta q$) and an orbital part
($L^q$) \cite{Ji98b}, for each quark flavor, as~:
\begin{eqnarray}
J^q(\mu) \;=\; {1 \over 2} \, \Delta q(\mu) \;+\; L^q(\mu) \, .
\la{eq:jqdecomp}
\end{eqnarray}
The flavor decomposed polarized quark contributions $\Delta u$,
$\Delta d$ and $\Delta \bar q$ to the proton spin have been
determined by SMC \cite{Ade98} at a scale $\mu^2$ = 10 GeV$^2$, 
and by HERMES \cite{Ack99} at a scale $\mu^2$ = 2.5 GeV$^2$ from 
the measurement of semi-inclusive spin asymmetries. 
Both experiments are in good agreement
when extrapolated to the same scale $\mu^2$ = 2.5 GeV$^2$, 
as has been done in Ref.~\cite{Ack99}. 
The sum of the quark intrinsic spin contributions to the proton spin 
yields about 30~\% at this scale.
By using our estimate for $J^q$, and the measured values 
$\Delta q$ for the different quark flavors, one can extract from 
Eq.~(\ref{eq:jqdecomp}) the corresponding orbital angular momentum
contributions $L^q$ for the $u$, $d$, and $s$-quarks to the proton
spin, as is shown in Table~\ref{table_5_3}. 
One sees from Table~\ref{table_5_3} that at the scale 
$\mu^2 = 2.5$~GeV$^2$, our estimate for the total quark contribution
to the proton spin yields about 60\%, which is nearly entirely due to
the $u$-quarks. The measured intrinsic spin contributions show that both
$\Delta u$ and $\Delta d$ are large, and have opposite sign.  
Our estimate leads then to the picture where the
quark orbital angular momenta contribute about 30~\% to the proton
spin, of comparable size as the intrinsic spin contribution. For the
separate flavors, Table~\ref{table_5_3} shows that for the $u$-quarks,
the dominant contribution to the proton spin comes from the intrinsic
contribution $\Delta u$, with only a small amount due to orbital
angular momentum. For the $d$-quarks however, 
the intrinsic contribution $\Delta d$ is nearly entirely 
cancelled by a large orbital contribution, yielding 
a small net value for $J^d$.

\begin{table}[h]
{\centering \begin{tabular}{|c|c|c|c|}
\hline
&&& \\ 
& $2 \, J^q$ ($\mu^2$ = 2.5 GeV$^2$) 
& $\Delta q$ ($\mu^2$ = 2.5 GeV$^2$) 
& $2 \, L^q$ ($\mu^2$ = 2.5 GeV$^2$) \\
& & (HERMES) & \\
&&& \\
\hline 
\hline
$u$         &  0.61 &  0.57 $\,\pm\,$ 0.04 & 0.04 $\,\mp\,$ 0.04 \\
\hline 
$d$         & -0.05 & -0.25 $\,\pm\,$ 0.08 & 0.20 $\,\mp\,$ 0.08 \\
\hline 
$s$         &  0.04 & -0.01 $\,\pm\,$ 0.05 & 0.05 $\,\mp\,$ 0.05 \\
\hline 
$u + d + s$ &  0.60 &  0.30 $\,\pm\,$ 0.10 & 0.30 $\,\mp\,$ 0.10 \\
\hline 
\hline
\end{tabular}\par}
\caption{\small Estimate of $2 \, L^q$ at the scale  $\mu^2$ =
  2.5~GeV$^2$ from Eq.~(\ref{eq:jqdecomp}).   
The value of $2 \, J^q$ is obtained 
according to the model estimate of Eqs.~(\ref{eq:ju}-\ref{eq:js})
using the MRST98~\cite{Mar98} parton distributions at the same scale.  
The values for $\Delta q$ (including both quark and anti-quark contributions) 
are taken from the HERMES experiment 
\cite{Ack99} at this same scale. We added the errors from Ref.~\cite{Ack99}
in quadrature.
\label{table_5_3}}
\end{table}

Having discussed a valence parametrization for the function $e^q(x)$ 
up to now, we next investigate a possible sea quark contribution to $e^q(x)$. 
It is seen from Fig.~\ref{fig:modelE0} that the chiral quark soliton model 
calculation for the leading (at large $N_c$) isovector GPD $E^u - E^d$ 
contains such a sea quark component, implying that the valence model of 
Eq.~(\ref{eq:factte}) for $e^q(x)$ is incomplete. 
This sea quark component to $e^q(x)$ is
symmetric in $x$, in contrast to the sea quark contribution
to $q(x)$, which is antisymmetric in $x$. As the chiral quark soliton
model calculation indicates that this sea quark component of $e(x)$ is a very
narrowly peaked function around $x = 0$, we propose to
parametrize it by a $\delta$ function in $x$. Adding this piece to
the valence parametrization of Eqs.~(\ref{eq:factte}) for $u$ and 
$d$-quark flavors, leads to the following parametrization for $e^q(x)$~: 
\begin{eqnarray}
e^u(x) \,&=&\,  A^u \; u_{val}(x)  \;+\; B^u \; \delta(x) \, ,
\nonumber \\
e^d(x) \,&=&\,  A^d \; d_{val}(x)  \;+\; B^d \; \delta(x) \, , 
\nonumber \\
e^s(x) \,&=&\, 0 \;, 
\la{eq:etotparam}
\end{eqnarray}
where the parameters $A^u, A^d$ are related to $J^u, J^d$ 
through the total angular momentum sum rule Eq.~(\ref{eq:jipol}) as~:
\begin{eqnarray}
A^q \,=\, {{2 \, J^q \,-\, M_2^q} \over {M_2^{q_{val}}}} \, ,
\la{eq:Aq}
\end{eqnarray}
and where the parameters $B^u, B^d$ follow from the first sum rule 
Eq.~(\ref{eq:vece2sumrule}) as~:
\begin{eqnarray}
B^u \,&=&\, 2 \, \left[ \, {1 \over 2} \, \kappa^u \,-\, 
{{2 \, J^u \,-\, M_2^u} \over {M_2^{u_{val}}}} \right]\, , \\
B^d \,&=&\, \kappa^d \,-\, 
{{2 \, J^d \,-\, M_2^d} \over {M_2^{d_{val}}}}\, .
\la{eq:Bq}
\end{eqnarray}
One sees that the total angular momenta carried by 
$u$- and $d$-quarks, $J^u$ and
$J^d$, enter now directly as fit parameters in the parametrization of
Eq.~(\ref{eq:etotparam}), in contrast to the valence model 
for $e^q(x)$ of Eq.~(\ref{eq:factte}), where the values $J^u$ and $J^d$ are
fixed as in Eqs.~(\ref{eq:ju},\ref{eq:jd}).  
Therefore, such a parametrization as in Eq.~(\ref{eq:etotparam}) 
can be used to see the sensitivity of hard electroproduction 
observables on $J^u$ and $J^d$, as will be shown further on. 
\newline
\indent 
The physical interpretation of the sea quark part of $e^q(x)$ 
in Eqs.~(\ref{eq:etotparam}) can be understood as being due to
``vector meson exchange'' because $e^q(x)$ is normalized to $\kappa^q$. 
In a vector meson dominance picture, 
the isovector (isoscalar) anomalous magnetic
moments can be mainly understood as due to exchange of a $\rho$ ($\omega$)
meson, through its tensor coupling to the nucleon.  
\newline
\indent
Having specified the model for the forward distribution $e^q(x)$, 
i.e. Eq.~(\ref{eq:etotparam}), we
now turn to the $\xi$-dependence of the GPD $E^q_{DD}(x, \xi) \equiv
E^q_{DD}(x, \xi, t = 0)$ in Eq.~(\ref{eq:parame}). 
We generate the $\xi$-dependence of $E^q_{DD}(x, \xi)$ through a double
distribution $K^q(\beta, \alpha)$ as given by Eq.~(\ref{dd2e}) 
similar as in Eq.~(\ref{eq:dd2})~:
\begin{eqnarray}
E^q_{DD}(x,\xi) \,=\,
\int_{-1}^{1}d\beta\
\int_{-1+|\beta|}^{1-|\beta|} d\alpha\ \,
\delta(x-\beta-\alpha\xi)\,  K^q(\beta,\alpha)\ \, .
\la{eq:dd3}
\end{eqnarray}
The double distribution  $K^q(\beta, \alpha)$ is then obtained,
in an analogous way as in Eq.~(\ref{eq:ddunpol}), 
by multiplying the forward distribution
$e^q(\beta)$ with a profile function as~:
\be
K^q(\beta, \alpha) = h(\beta, \alpha) \, e^q(\beta) \, ,
\la{eq:ddk}
\ee
and where the profile function $h(\beta, \alpha)$ 
is taken as in Eq.~(\ref{eq:profile}).
Using the parametrization of Eq.~(\ref{eq:etotparam}), 
one obtains then for the GPD $E^q_{DD}(x,\xi)$ the expression~:
\begin{eqnarray}
E^q_{DD}(x,\xi) \,=\,E^{q_{val}}_{DD}(x,\xi) \,+\,
B^q \, \frac{\Gamma(2b+2)}{2^{2b+1}\Gamma^2(b+1)}\,
{1 \over \xi} \, \theta(\xi  - |x|) \,
\left( 1 - {{x^2} \over {\xi^2}} \right)^b \, ,
\la{eq:edd}
\end{eqnarray}
where the first (second) term in Eq.~(\ref{eq:edd}) 
is the DD part originating from the valence (sea) contribution to $e^q$ 
respectively in Eq.~(\ref{eq:etotparam}). 
In the following, we take the power $b$ in
Eq.~(\ref{eq:edd}), which enters in the profile
function in Eq.~(\ref{eq:ddk}), equal to the value $b = 1$. 
This is motivated because this corresponds to ``vector meson exchanges''
with an asymptotic distribution amplitude.  

\begin{figure}[hp]
\epsfxsize=8.25 cm
\centerline{\epsffile{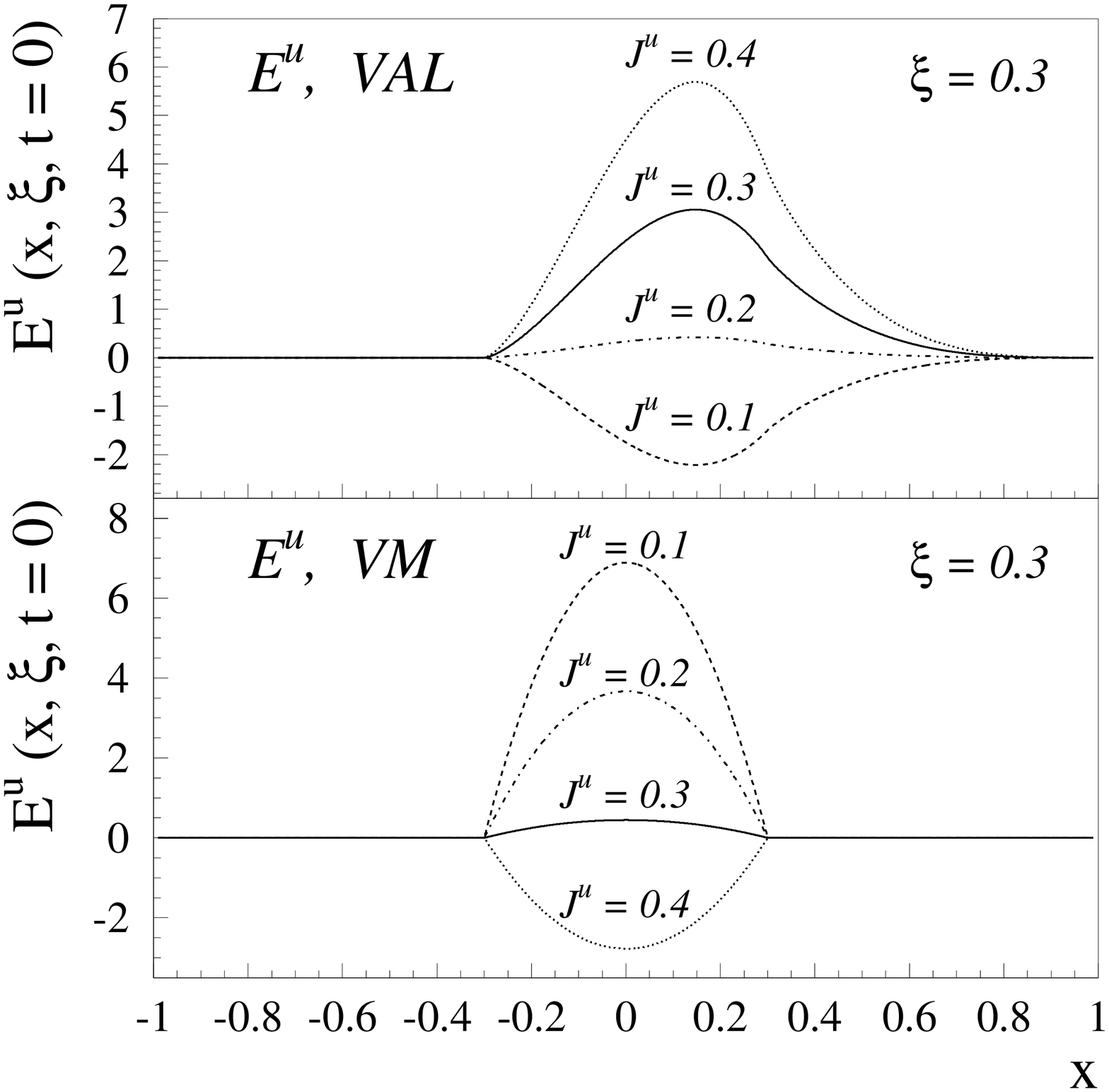}}
\caption{\small Double distribution part to the 
$u$-quark GPD $E^u$ at $t = 0$ using the ansatz of
Eq.~(\ref{eq:etotparam}), for different values of $J^u$ as indicated
on the curves. 
Upper panel~: valence part (VAL), lower panel~: sea
contribution due to vector meson exchange (VM).}
\label{fig:up_E}
\epsfxsize=8.25 cm
\centerline{\epsffile{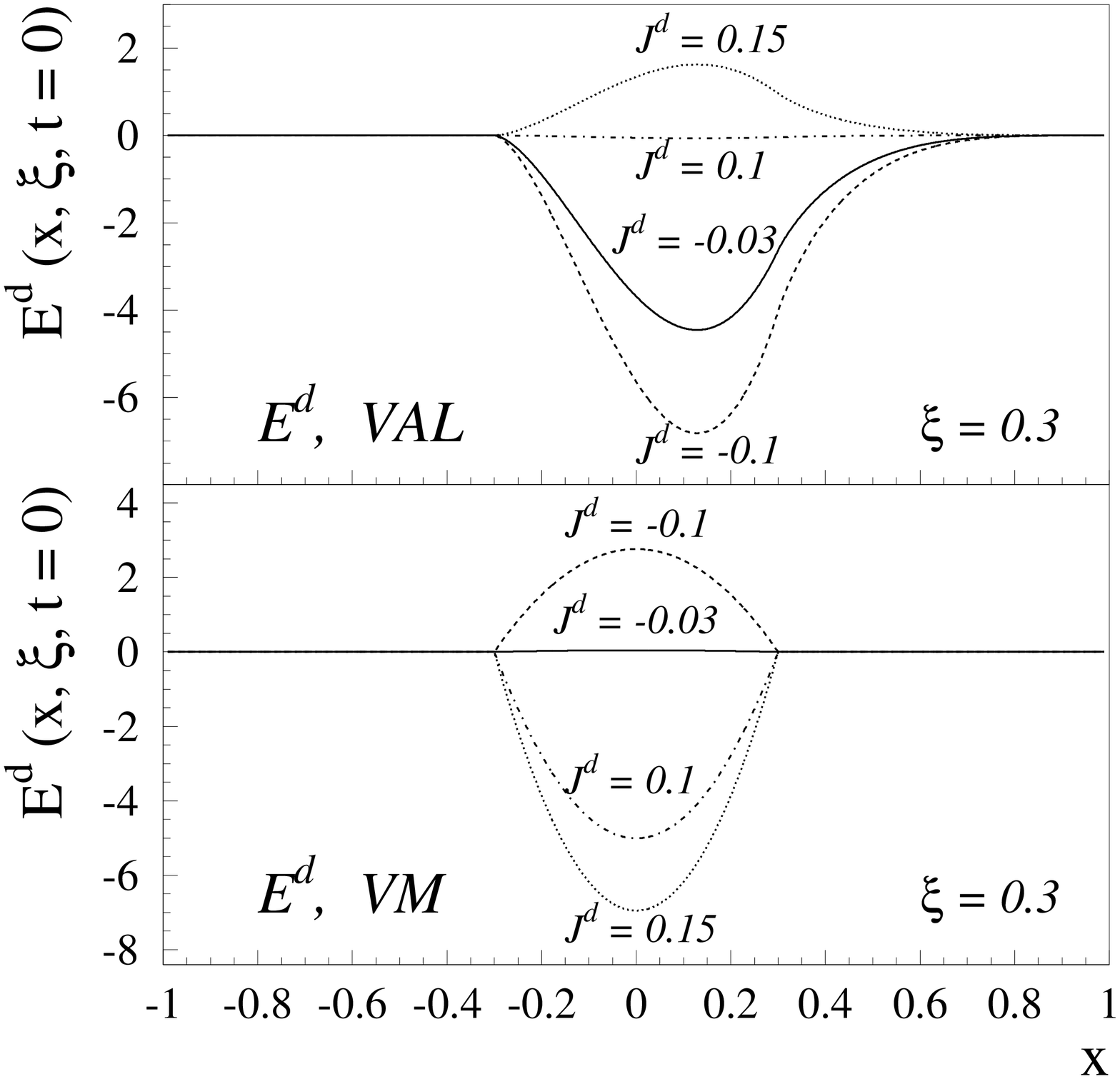}}
\caption{\small The same as Fig.~\ref{fig:up_E} but for the 
double distribution part to the $d$-quark GPD $E^d$ at $t = 0$, for
different values of $J^d$.}
\label{fig:down_E}
\end{figure}

In Figs.~\ref{fig:up_E},\ref{fig:down_E}, we show the parametrization
of Eq.~(\ref{eq:edd}) for $E^u_{DD}$ and $E^d_{DD}$ respectively at
a fixed value $\xi = 0.3$. The two components in Eq.~(\ref{eq:edd})
are displayed for different values of $J^u$ ($J^d$), which are the
free parameters in $E^u_{DD}$ ($E^d_{DD}$) respectively. 
The valence contribution is calculated using the forward
MRST98 valence quark distributions at a scale $\mu^2$ = 2 GeV$^2$. 
It is seen that the sea contribution due to ``vector meson exchange''
displays the profile of an asymptotic distribution amplitude. One also
sees from Figs.~\ref{fig:up_E},\ref{fig:down_E}, that the two
components are correlated as the integral of their sum is fixed to
$\kappa^q$. In particular, for the value $J^u \approx 0.32$, the sea
contribution to $E^u$ vanishes and the ansatz of
Eq.~(\ref{eq:etotparam}) reduces to the valence ansatz of
Eq.~(\ref{eq:factte}). Similarly, for the value  $J^d \approx -0.03$,
the sea contribution to $E^d$ vanishes. For other values of $J^u$
($J^d$) however, the relative contribution of both components changes.
As both components have a different shape in $x$, one gets a sensitivity on
$J^u$ and $J^d$ in observables where the GPDs $E^u$ and $E^d$ enter, as
will be shown further on. 

\begin{figure}[h]
\epsfxsize=9 cm
\centerline{\epsffile{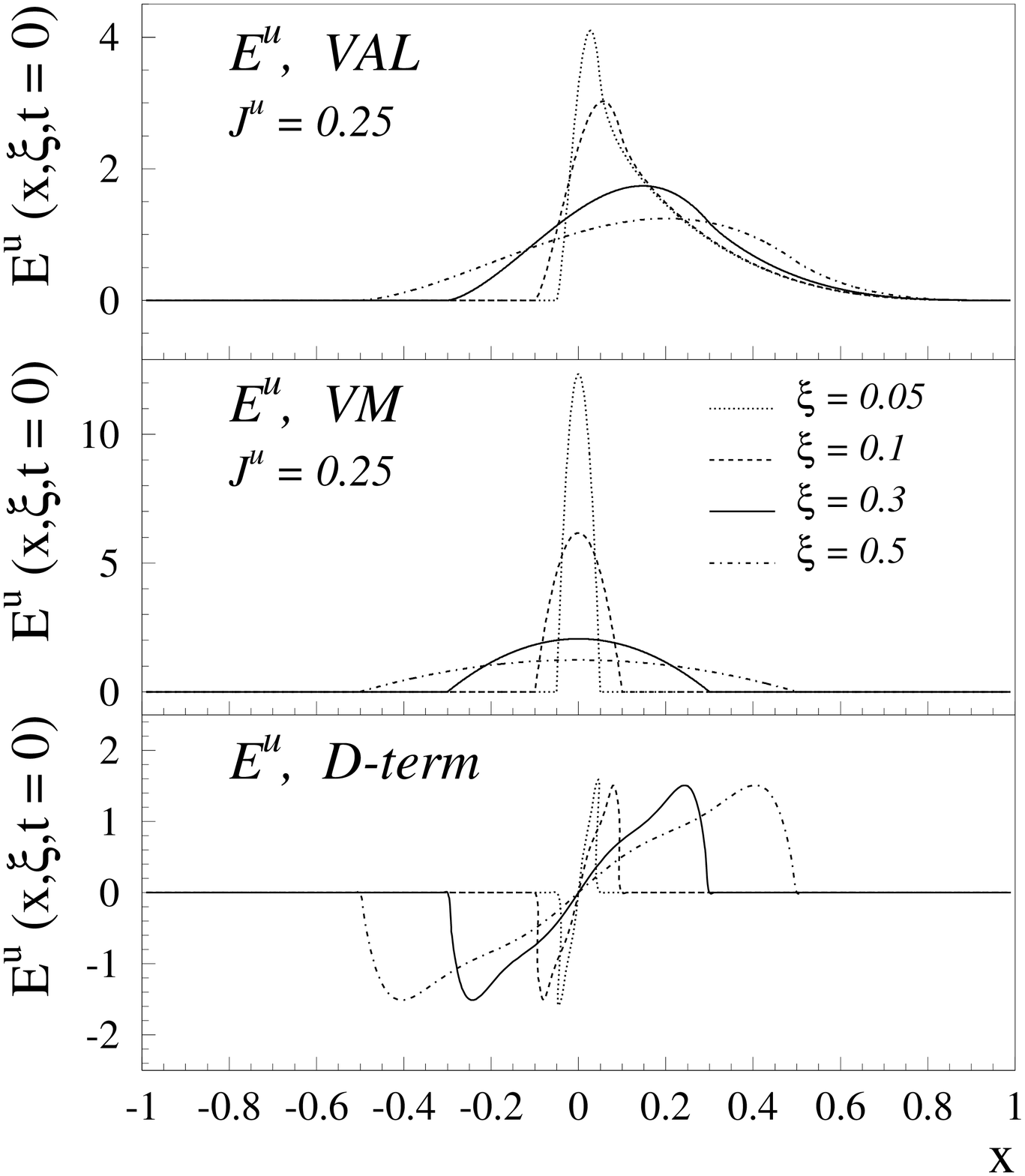}}
\caption{\small The three contributions of the model for the 
$u$-quark GPD $E^u$ at $t = 0$, for different values of $\xi$ as
indicated on the figure. 
Upper panel~: valence part (VAL) in the ansatz of
Eq.~(\ref{eq:etotparam}) for $J^u = 0.25$;
middle panel~: sea contribution (second term in Eq.~(\ref{eq:etotparam})) 
due to vector meson exchange (VM) for $J^u = 0.25$; 
lower panel~: D-term contribution of Eq.~(\ref{eq:dtermcontrib2}) to $E^u$.}
\label{fig:E_up_xi}
\end{figure}

In Fig.~\ref{fig:E_up_xi}, we show the $\xi$ dependence of the
three contributions of the model for $E^u$. Besides the valence and
sea contributions in the double distribution part of
Eq.~(\ref{eq:dd3}) which are shown at a fixed value of $J^u$ = 0.25, 
the third contribution originates from the D-term of 
Eq.~(\ref{eq:dtermcontrib2}), which contributes with opposite sign to
$E^u$ as compared to $H^u$, and therefore drops out of the spin sum rule. 
Notice also from Fig.~\ref{fig:E_up_xi} that for small $\xi$, the sea
contribution approaches a $\delta$-function in our parametrization. 
\newline
\indent
Having specified the parametrization of $E^q(x, \xi, t = 0)$, we next discuss
the $t$-dependence of the GPD $E^q(x, \xi , t)$. 
This $t$-dependence of $E^q$ is constrained through the first sum rule
of Eq.~(\ref{eq:ffsumrulee}), which is given by the Pauli form factor
$F_2(t)$. As stated before, we are only concerned here with hard
exclusive reactions in the small $-t$ region, with $-t$ small 
compared to the hard scale $Q^2$. 
In this small $-t$ region, the simplest parametrization 
which is consistent with this sum rule constraint consists of a
factorized ansatz 
\begin{eqnarray}
E^q(x, \xi, t) \;=\; E^q(x, \xi) \; 
{1 \over {(1 - t / 0.71 )^2}} \, , 
\la{eq:dipoleff}
\end{eqnarray}
where the $t$-dependence has been expressed through a dipole form
factor, which is known to be a
good parametrization for the $t$-dependence of the nucleon magnetic form
factors. In all estimates at small $-t$ which contain $E^q$, the
factorized form (in $t$) of Eq.~(\ref{eq:dipoleff}) is used. 
As discussed for the GPD $H^q$, 
a more realistic estimate of the $t$-dependence of the GPD $E^q$ might 
also be given by a Regge ansatz analogous to 
Eqs.~(\ref{Regge-parametrization},\ref{eq:ddunpolregge}), which is  
postponed to a future work. 
\newline
\indent
To summarize this section, we constructed a three component model 
for the GPDs $E^q$ as illustrated in Fig.~\ref{fig:E_up_xi}. 
The valence component was fixed in terms of valence forward quark 
distributions. This assumption is supported by the calculation
in the chiral quark-soliton model (see Sec.~\ref{chap4_4}).   
For the sea part, we included the two physically important contributions
due to the D-term and the component associated with the 
``vector meson exchange''. 
The D-term contribution is required by the polynomiality condition 
(see Sec.~\ref{chap3_2}), whereas the ``vector meson exchange'' component
is motivated by the vector meson dominance model for the nucleon 
anomalous magnetic moments.

\subsubsection{Parametrization of the GPD $\widetilde H^q$} 
\label{chap5_2d}

To model the GPD \( \widetilde{H}^{q} \), 
we need the corresponding polarized quark distributions.
For this we follow the work of Ref.~\cite{Lea98}, where a next to
leading order QCD analysis of inclusive polarized deep-inelastic lepton-nucleon
scattering was performed and which yields an excellent fit of the data.
In this analysis, the input polarized densities 
(at a scale \( \mu^{2} \) = 1 GeV\( ^{2} \)) are given by : 
\begin{eqnarray}
&& \Delta u_{V}(x,\mu^{2})=\eta _{u}\, A_{u}\; x^{0.250}\; 
u_{V}(x,\mu^{2})\; ,\nonumber \\
&& \Delta d_{V}(x,\mu^{2})=\eta _{d}\, A_{d}\; x^{0.231}\; 
d_{V}(x,\mu^{2})\; ,\nonumber \\
&& \Delta \bar{q}(x,\mu^{2})=\eta _{\bar{q}}\, A_{S}\; x^{0.576}\;
S(x,\mu^{2})\; ,
\la{eq:poldistr} 
\end{eqnarray}
where a $SU(3)$ symmetric sea has been assumed, i.e. 
\( \Delta \bar{q}\equiv \Delta \bar{u} \)
= \( \Delta \bar{d} \) = \( \Delta \bar{s} \) and \( S \) represents the
total sea. On the {\it rhs} 
of Eqs.~(\ref{eq:poldistr}), the normalization factors
\( A_{u},A_{d},A_{S} \) are determined so that the first moments of
the  polarized densities are given by 
\( \eta _{u},\eta _{d},\eta _{\bar{q}} \) respectively.
For the valence quark densities, \( \eta _{u} \) and \( \eta _{d} \) were
fixed in Ref.~\cite{Lea98} by the octet hyperon \( \beta  \) decay constants
which yield : 
\begin{equation}
\eta _{u}\; =\; \int _{0}^{+1}dx\; \Delta u_{V}(x)\; \approx 0.918\; ,
\hspace {1.cm}
\eta _{d}\; =\; \int _{0}^{+1}dx\; \Delta d_{V}(x)\; \approx -0.339\; .
\end{equation}
The first moment of the polarized sea quark density was determined by the fit
of Ref.~\cite{Lea98} which gives \( \eta _{\bar{q}}\approx -0.054 \). 
\newline
\indent 
Our ansatz for \( \widetilde{H}^{q} \) is to generate it from a double 
distribution $\widetilde F^q(\beta, \alpha, t)$, analogous to Eq.~(\ref{dd2}), 
for which we take 
\begin{eqnarray}
\widetilde{F}^{u}(\beta,\alpha ,t) &\,=\,& h(\beta,\alpha) \; 
\Delta u_{V}(\beta)
\;{g_{A}^{u}(t)}/{g_{A}^{u}(0)}\;, \nonumber \\
\widetilde{F}^{d}(\beta,\alpha ,t) &\,=\,& h(\beta,\alpha) \; 
\Delta d_{V}(\beta)
\;{g_{A}^{d}(t)}/{g_{A}^{d}(0)} \;,
\la{eq:factpol}
\end{eqnarray}
where $h(\beta, \alpha)$ is the profile function of Eq.~(\ref{eq:profile}). 
The parameter $b$ in Eq.~(\ref{eq:profile}) can again be chosen as a free 
parameter in such a parametrization, but we will show all estimates 
with the value $b = 1$ for the GPD $\widetilde H$.
\newline
\indent
One can check that Eq.~(\ref{eq:dislimitp}) is verified by construction and
that, using Eq.~(\ref{eq:axff}) to evaluate the
axial form factors \( g_{A}^{u} \) and \( g_{A}^{d} \), the sum rule of
Eq.~(\ref{eq:ffsumruleht}) is satisfied within 10~\%.
\newline
\indent
The parametrization of the antiquark contribution to $\widetilde H^q$ 
remains to be investigated in future work.
This might be interesting to study since a remarkable prediction 
of the chiral quark soliton model, noted first in Ref.~\cite{Dia96b,Dia97}, is
the strong flavour asymmetry of polarized antiquarks. Such a feature
is missing in other models like, for instance, pion cloud models, see 
discussion in Ref.~\cite{Dres00b}.
\newline
\indent
We stress that the factorization of the $t$-dependence
of GPDs is not expected to hold when increasing the value of $-t$. 
For example, the calculations of the GPD $\widetilde H$ in the 
chiral quark soliton model \cite{Pen00a} also show 
a deviations from the factorization ansatz for the $t$-dependence of GPDs
as is shown in Fig.~\ref{fig:RHt}.
In the calculations presented here, we limit ourselves to
relatively small values of $-t$ and assume the
factorization ansatz.

\subsubsection{Parametrization of the GPD $\widetilde E^q$} 
\label{chap5_2e}

As for the GPD $E^q$, the GPD $\widetilde E^q$, 
corresponding with the induced pseudoscalar transition at the nucleon
side in the matrix element of Eq.~(\ref{eq:qsplitting}), also vanishes 
in the forward direction and does not contribute to DIS. 
However, it was shown in Ref.~\cite{Man99b,Fra99,Pen00a} 
that the pion exchange, which contributes 
to the region \( -\xi \leq x\leq \xi  \) of \( \widetilde E \), may
be non negligible at small \( t \) due to the proximity of the pion pole at
\( t=m_{\pi }^{2} \). 
As discussed in Sec.~\ref{chap4_4_2}, it was found \cite{Pen00a} 
in the chiral quark-soliton model calculation 
that in the limit $t \to m_\pi^2$ the function $\widetilde E$ exactly
reduces to this pion pole contribution. 
Furthermore it is seen from Fig.~\ref{fig:ea}, that in the model
calculation for $\widetilde E$ the pion pole part to $\widetilde E$ dominates
over a wide range of $t$ and $\xi$ values. 
\newline
\indent
According to these findings, the main contribution to $\widetilde E$ is
obtained by evaluating \( \widetilde E \) 
assuming it is entirely due to the pion pole.
According to this hypothesis, since
the pion exchange is isovector, one has
\begin{equation}
\widetilde E^{u}=-\widetilde E^{d}={1\over 2}\; \widetilde E_{\pi -pole}\; .
\la{eq:etildeisov}
\end{equation}
 The \( t \) -dependence of \( \widetilde E_{\pi -pole}(x,\xi ,t) \) is fixed
by the sum rule of Eq.~(\ref{eq:ffsumruleht}) in terms of the  
pseudoscalar form factor \( h_{A}(t) \). 
\newline
\indent
In the region \( -\xi \leq x\leq \xi  \), the quark and antiquark couple to
the pion field of the nucleon. Therefore, this coupling should be proportional
to the pion distribution amplitude. For the latter we 
adopt the asymptotic form.
Expressing the quark's longitudinal momentum fraction \( z \) in the pion in
the symmetric range \( -1\leq z\leq 1 \), the asymptotic distribution amplitude
\( \Phi _{as} \) is given by \( \Phi _{as}(z)=3/4 \) \( (1-z^{2}) \), and
is normalized as \( \int _{-1}^{+1}dz\; \Phi _{as}(z)=1 \). The light-cone
momentum fractions of the quark and antiquark in the pion 
are respectively given by \( (1+x/\xi )/2 \) and \( (1-x/\xi )/2 \). 
Therefore, \( \widetilde E_{\pi -pole} \) is finally modelled in the 
estimates as~: 
\begin{equation}
\widetilde E_{\pi -pole}\; =\; \theta \left( -\xi \leq x\leq \xi \right) 
\; h_{A}(t)\; {1\over \xi }\; \Phi _{as}\left( {x\over \xi }\right) \; ,
\la{eq:etildepipole}
\end{equation}
which satisfies the sum rule of Eq.~(\ref{eq:ffsumruleet}).


\newpage
\subsection{Deeply virtual Compton scattering (DVCS)}
\label{chap5_3}

In order to access the GPDs experimentally, the deeply virtual Compton
scattering (DVCS) process
\be
\gamma^*(q)+N(p)\to \gamma(q') +N(p') \, ,
\la{eq:proc}
\ee
was proposed as a practical tool in Ref.~\cite{Ji97b}.
In the Bjorken regime, where the photon virtuality $Q^2$ is large,
with $x_B \equiv Q^2/2 (p \cdot q)$ fixed,
the leading order DVCS amplitude in the forward direction
is given by the handbag diagrams of
Fig.~\ref{fig:handbags} \cite{Mul94,Rad96a,Ji97b}.
These handbag diagrams express the factorization of the DVCS amplitude
in a hard scattering part (which is exactly calculable in PQCD) and
a soft, non-perturbative nucleon structure part, represented by the
lower blobs in Fig.~\ref{fig:handbags}. This soft, non-perturbative object
is parametrized in terms of the GPDs as given by Eq.~(\ref{eq:qsplitting}).
The factorization of the DVCS amplitude in a hard and soft part
has been proven in Ref.~\cite{Ji98a,Rad98,Col99}
for the leading power in $Q$ and all logarithms.
\newline
\indent
To calculate the DVCS amplitude, one starts from its definition as a
nucleon matrix element of the $T$-product of two electromagnetic currents~:
\begin{eqnarray}
H^{\mu\nu}=- i\int d^4x\ e^{-i (q\cdot x)}\langle p'|T\left[J_{\rm
e.m.}^\mu (x) J_{\rm e.m.}^\nu(0)\right]|p\rangle\, ,
\la{eq:Tdef}
\end{eqnarray}
where the four-vector index $\mu$ ($\nu$) refers to the virtual (real)
photon. The DVCS amplitude is obtained from the DVCS tensor of
Eq.~(\ref{eq:Tdef}) by contracting with the photon polarization vectors as~:
\begin{eqnarray}
T \; = \; \varepsilon_\mu (q) \; \varepsilon_\nu^* (q') \; H^{\mu \nu} \, ,
\end{eqnarray}
where $\varepsilon_\mu (q)$ ( $\varepsilon_\nu^* (q')$ )
are the polarization vectors of the virtual (real) photons respectively.
\newline
\indent
In the following, we firstly introduce the kinematical variables
entering the DVCS process. We then discuss the DVCS amplitude of
Eq.~(\ref{eq:Tdef}) in the leading power in $Q$ (twist-2 accuracy), and
show its dependence upon the GPDs. In order to be able to extract the
twist-2 GPDs from DVCS observables at accessible values of the hard
scale $Q$, it is necessary to have an estimate of the effect of power
suppressed (higher twist) contributions to those observables. The
first power correction to the DVCS amplitude is of order $O(1/Q)$,
hence it is called twist-3.
We therefore discuss subsequently, the twist-3 corrections to the DVCS
amplitude, which have been derived and calculated recently by several groups
\cite{Ani00,Pen00b,Bel00b,Rad00,Kiv01a,Kiv01b,Kiv01c,Rad01a,Bel00c,Bel01b}
using different approaches.
We then explore the experimental opportunities and discuss the
different DVCS observables.

\subsubsection{Kinematical variables of the DVCS process}
\label{chap5_3_1}

As explained in Sec.~\ref{chap3_1}, the calculation of the DVCS
amplitude in the Bjorken regime and the parametrization of the non-perturbative
matrix elements, representing the lower blobs in the handbag diagrams,
is performed in a frame where the virtual photon momentum
\( q^{\mu } \) and the average nucleon momentum
\(\bar P^{\mu } \) are collinear along the $z$-axis
and in opposite direction.
To calculate the DVCS amplitude,
it is therefore convenient to introduce lightlike vectors
along the positive and negative \( z \)-directions
as \( \tilde{p}^{\mu }= \bar P^{+}/\sqrt{2}(1,0,0,1) \)
and \( n^{\mu }=1/ \bar P^{+}\cdot 1/\sqrt{2}(1,0,0,-1) \) respectively,
satisfying $\tilde p \cdot n = 1$.
\newline
\indent
In this frame, the physical momenta entering the DVCS process
(\ref{eq:proc}), where $p (p')$ are
the momenta of the initial (final) nucleon and $q (q')$ are the momenta of
the initial (final) photon respectively,
have the following decomposition \cite{Gui98}~:
\begin{eqnarray}
 &  & \bar P^{\mu } \,=\, {1\over 2}\left( p^{\mu }+p'^{\mu }\right)
\,=\, \tilde{p}^{\mu }+{{\bar{m}^{2}}\over 2}\, n^{\mu }\; ,
\la{eq:dvcsp} \\
&  & q^{\mu }=-\left( 2\xi ^{'}\right) \,
\tilde{p}^{\mu }+\left( {{Q^{2}}\over {4\xi ^{'}}}\right) \, n^{\mu}\; ,
\la{eq:dvcsq} \\
&  & \Delta ^{\mu }\equiv p^{'\mu }-p^{\mu }
\,=\,-\left( 2\xi \right) \, \tilde{p}^{\mu }
+\left( \xi \, \bar{m}^{2}\right) \, n^{\mu }+\Delta^{\mu }_{\perp }\; ,
\la{eq:dvcsxi} \\
&  & q^{'\mu }\equiv q^{\mu }-\Delta ^{\mu } \,=\,-2\left( \xi ^{'}-\xi
 \right) \, \tilde{p}^{\mu }+\left( {{Q^{2}}\over {4\xi ^{'}}}-\xi \,
 \bar{m}^{2}\right) \, n^{\mu }-\Delta ^{\mu }_{\perp }\; ,
\la{eq:dvcsqp}
\end{eqnarray}
where $\Delta_\perp$ is the perpendicular component of the
momentum transfer $\Delta$
( i.e. $\tilde p \cdot \Delta_\perp = n \cdot \Delta_\perp = 0$ ),
and where the variables \( \bar{m}^{2} \), \( \xi ^{'} \) and \( \xi  \) are
given by
\begin{eqnarray}
&  & \bar{m}^{2}\,=\,{m_{N}}^{2}-{{\Delta ^{2}}\over 4}\; ,
\la{eq:dvcskin} \\
&  & 2\xi ^{'}\,=\,{{\bar P\cdot q}\over {\bar{m}^{2}}}\,
\left[ -1+\sqrt{1+{{Q^{2}\,\bar{m}^{2}}\over {(\bar P\cdot q)^{2}}}}\right] \;
\stackrel{Bj}{\longrightarrow }\; \frac{x_{B}}{1-\frac{x_{B}}{2}}\; ,\\
&  & 2\xi \,=\, 2\xi ^{'}\, {{Q^{2}-\Delta ^{2}}\over {Q^{2}
+\bar{m}^{2}(2\xi ^{'})^{2}}}\;
\stackrel{Bj}{\longrightarrow }\; \frac{x_{B}}{1-\frac{x_{B}}{2}}\; .
\la{eq:dvcskin3}
\end{eqnarray}
\newline
\indent
To twist-3 accuracy, Eqs.~(\ref{eq:dvcsp}-\ref{eq:dvcsqp}) reduce to
\be
\bar P &\,=\,& \tilde p, \hspace{4.2cm}
\Delta \,=\, -2 \,\xi \, \bar P \,+\, \Delta_\perp, \, \nonumber \\
q &\,=\, &- 2 \, \xi \, \bar P \,+\, \frac{Q^2}{4\xi} \,n,
\hspace{2cm}
q' \,=\, \frac{Q^2}{4\xi} \,n \,-\, \Delta_\perp \, .
\la{eq:kintw3}
\ee

\subsubsection{Twist-2 DVCS tensor}
\label{chap5_3_2}

Using the parametrization of Eq.~(\ref{eq:qsplitting})
for the bilocal quark operator,
the DVCS tensor in leading order in $Q$, $H^{\mu \nu}_{\mathrm{DVCS-LO}}$,
follows from the handbag diagrams of Fig.~\ref{fig:handbags} as
\cite{Mul94,Ji97b,Rad96a}~:
\begin{eqnarray}
H^{\mu \nu }_{\mathrm{DVCS-LO}}
&\,=\,& {1\over 2}\, (-g^{\mu \nu})_\perp \;
\int _{-1}^{+1}dx \; C^+(x, \xi) \,
\left[ \, H^{p}_{DVCS}(x,\xi ,t)\; \bar{N}(p^{'}) \, {\Dirac n} \, N(p)\right.
\nonumber\\
&&\left. \hspace{4.25cm} +\, E^{p}_{DVCS}(x,\xi ,t)\;
\bar{N}(p^{'})i\sigma ^{\kappa \lambda }
{{n_{\kappa }\Delta _{\lambda }}\over {2m_{N}}}N(p)\right] \nonumber \\
&+& {i\over 2}\, (\epsilon^{\nu \mu})_\perp
\; \int _{-1}^{+1}dx \; C^-(x,\xi) \,
\left[\, \tilde{H}^{p}_{DVCS}(x,\xi ,t) \;
\bar{N}(p^{'})\, {\Dirac n} \gamma_{5} \, N(p) \right.\nonumber\\
&&\left. \hspace{4.cm}+\,\tilde{E}^{p}_{DVCS}(x,\xi ,t) \;
\bar{N}(p^{'})\gamma_{5}{{\Delta \cdot n}\over {2m_{N}}}N(p) \,\right] ,\;
\la{eq:dvcsampl}
\end{eqnarray}
where the symmetrical and antisymmetrical twist-2 tensors were
introduced in Eq.~(\ref{gt}) as~:
\begin{eqnarray}
(-g^{\mu \nu})_\perp \,=\, -g^{\mu \nu}+ n^\mu \tilde{p}^{\nu} +
n^\nu  \tilde{p}^{\mu},
\quad \quad (\epsilon_{\mu \nu})_\perp \,=\,
\epsilon_{\mu \nu \alpha\beta}n^\alpha  \tilde{p}^{\beta} \, .
\la{eq:gt}
\end{eqnarray}
In Eq.~(\ref{eq:dvcsampl}), we also introduced the coefficient functions
$C^{\pm}(x, \xi)$, which are defined as~:
\be
C^\pm(x,\xi)=\frac{1}{x-\xi+i\varepsilon}\pm
\frac{1}{x+\xi-i\varepsilon}.
\la{eq:alf}
\ee
\newline
\indent
In the DVCS process on the proton, the GPDs for each quark flavor
enter in the combination
\begin{eqnarray}
H^{p}_{DVCS} \, =\,
{4\over 9}H^{u}\, +\, {1\over 9}H^{d}\, +\, {1\over 9}H^{s}\; ,
\la{eq:hdvcsp}
\end{eqnarray}
and similarly for \( \tilde{H} \), \( E \) and \( \tilde{E} \).
The tensor for the DVCS process on the neutron is given by
Eq.~(\ref{eq:dvcsampl}) by replacing $H^p_{DVCS}$ by
\begin{eqnarray}
H^{n}_{DVCS} \, &=&\,
{1\over 9}H^{u}\, +\, {4\over 9}H^{d}\, +\, {1\over 9}H^{s}\; ,
\la{eq:hdvcsn}
\end{eqnarray}
and similarly for \( \tilde{H} \), \( E \) and \( \tilde{E} \).
Note that in Eqs.~(\ref{eq:hdvcsp},\ref{eq:hdvcsn})
the flavor dependent GPDs $H^u$, $H^d$ and $H^s$
in our notation always refer to the corresponding
quark flavor in the proton, e.g. $H^u \equiv H^{u/p}$ as explained in
Sec.~\ref{chap5_2}.
\newline
\indent
One sees from the {\it rhs} of the DVCS amplitude
of Eq.~(\ref{eq:dvcsampl}) that the GPDs $H, \tilde H, E, \tilde E$ enter
in a convolution integral over the quark momentum fraction $x$.
This is a qualitative difference compared to the case of DIS,
where one is only sensitive (through the optical theorem)
to the imaginary part of the forward double virtual Compton
amplitude. In the case of DIS, the convolution integral
collapses and the quark momentum fraction $x$ is fixed to the
kinematical quantity $x_B$.
In contrast, the non-forward DVCS amplitude of Eq.~(\ref{eq:dvcsampl})
displays a richer structure and has both real and imaginary parts.
For the imaginary part, the convolution integral again collapses, and
the value $x$ is fixed to the kinematically determined skewedness
variable $\xi$. Therefore, the imaginary part of the DVCS amplitudes
is directly proportional to the GPDs evaluated along the line $x =
\xi$ (e.g. in Fig.~\ref{fig:gpddt}), and
measures in this way the `envelope functions'
$H(\xi, \xi, t)$, $E(\xi, \xi, t)$,
$\tilde H(\xi, \xi, t)$, and $\tilde E(\xi, \xi, t)$.
\newline
\indent
The $x$-dependence of the GPDs away from the line $x = \xi$ is
contained in the principal value integral of the real part of the DVCS
amplitude Eq.~(\ref{eq:dvcsampl}). As the variable $x$ is not a
kinematically accessible quantity, the extraction of GPDs away from
the line $x = \xi$ is a non-trivial task. The strategy which we
propose here is to start from physically motivated parametrizations
for those GPDs as outlined in Sec.~\ref{chap5_2}. One can then extract
the parameters entering those parametrizations from different
observables, which have a different sensitivity to the real and
imaginary part of the hard electroproduction amplitudes, as shown
further on.

\subsubsection{Twist-3 DVCS tensor}
\label{chap5_3_3}

The DVCS amplitude of Eq.~(\ref{eq:dvcsampl}) displays a scaling
behavior as it is independent of $Q$.
One easily sees however that this amplitude is
not complete when going beyond the leading order in $Q$.
Although the twist-2 DVCS amplitude of Eq.~(\ref{eq:dvcsampl}), is exactly
gauge invariant with respect to the virtual photon, i.e.
\( q_{\mu }\, H^{\mu \nu }_{\mathrm{DVCS-LO}}=0 \),
electromagnetic gauge invariance is violated however by the
real photon except in the forward direction.
This violation of gauge invariance is a higher twist effect
compared to the leading order term $ H^{\mu \nu }_{\mathrm{DVCS-LO}} $.
Since $q^{'}_{\nu }\, H^{\mu \nu }_{\mathrm{DVCS-LO}}
= - (\Delta_{\perp })_\nu \,  H^{\mu \nu }_{\mathrm{DVCS-LO}}$,
an improved DVCS amplitude linear in $\Delta _{\perp }$ has been
proposed in Refs.~\cite{Gui98,Vdh99} to restore gauge invariance (in the
nonforward direction) in a heuristic way~:
\begin{equation}
H^{\mu \nu }_{\mathrm{DVCS}}\, =\, H^{\mu \nu }_{\mathrm{DVCS-LO}}\, +\,
{{\bar P^{\nu }}\over {\left( \bar P \cdot q^{'}\right) }}\;
\left( \Delta _{\perp }\right) _{\lambda }\,
H^{\mu \lambda}_{\mathrm{DVCS-LO}}\; ,
\la{eq:dvcsgaugeinv}
\end{equation}
leading to a correction term of higher order in $Q$ to the twist-2
DVCS amplitude, because $( \bar P \cdot q^{'} ) = O(Q^2)$.
\newline
\indent
Recently, the DVCS amplitude on the nucleon \footnote{The DVCS
amplitude to twist-3 accuracy has also been derived for the analogous
case of a  pion target in Refs.~\cite{Ani00,Rad00,Rad01a}.}
to the order $O(1/Q)$
has been derived explicitely in a parton model approach \cite{Pen00b} and
in a light-cone expansion framework \cite{Bel00b}, yielding the result~:
\be
&&H^{ \mu \nu} = \frac12 \int_{-1}^1 dx\quad \biggl\{ \left[(-g^{\mu
\nu})_\perp-\frac{\bar P^\nu\Delta_\perp^\mu}{(\bar P \cdot q')}
\right] \; n^\beta {\cal F}_\beta (x,\xi) \; C^+(x,\xi)
\la{eq:Tdvcstw3} \\
&&- \left[(-g^{\nu k})_\perp
-\frac{\bar P^\nu\Delta_\perp^k}{(\bar P \cdot q')}  \right]
i ({\epsilon_{k }}^{\mu})_\perp \; n^\beta {\cal \widetilde F}_\beta (x,\xi)
\; C^-(x,\xi) \nonumber \\
&&- \frac{(q+4\xi \bar P)^\mu}{(\bar P \cdot q)}
\left[(-g^{\nu k})_\perp
-\frac{\bar P^\nu\Delta_\perp^k}{(\bar P \cdot q')}  \right]
\left\{ {\cal F}_k(x,\xi) \; C^+(x,\xi)- i (\epsilon_{k \rho})_\perp
{\cal \widetilde F}^\rho (x,\xi) \; C^-(x,\xi)\right\} \biggr\} \, , \nonumber
\ee
where the functions ${\cal F}_\mu$ and ${\cal \widetilde F}_\mu$ are given by
Eqs.~(\ref{eq:F},\ref{eq:Ft}).
\newline
\indent
In the expression Eq.~(\ref{eq:Tdvcstw3})
for the DVCS amplitude to the twist-3 accuracy,
the first two terms correspond to the scattering of transversely
polarized virtual photons. This part of the amplitude,
containing $n_\beta \, {\cal F}^\beta$ and $n_\beta \,
{\cal \widetilde F}^\beta$, depends
only on the twist-2 GPDs $H,E$ and $\widetilde H, \widetilde E$ and was
anticipated in Refs.~\cite{Gui98,Vdh99}.
The third term in Eq.~(\ref{eq:Tdvcstw3}) corresponds to
the contribution of the longitudinal polarization of the virtual
photon. Defining the polarization vector of the virtual photon as
\be
\varepsilon_L^\mu(q)=\frac{1}{Q}\biggl(
2\xi \bar P^\mu+\frac{Q^2}{4\xi} n^\mu \biggr)\, ,
\la{eq:epsL}
\ee
we can easily calculate the DVCS amplitude for longitudinal
polarization of the virtual photon ($L\to T$ transition),
which is purely of twist-3~:
\be
(\varepsilon_L)_\mu \; H^{\mu\nu}=\frac{2\xi}{Q}\int_{-1}^1dx
\ \Biggl( {\cal F}_\perp^\nu\  C^+(x,\xi)- i\varepsilon_\perp^{\nu k}
{\cal \widetilde F}_{\perp k}\  C^-(x,\xi)
\Biggr)\, .
\la{eq:LtoT}
\ee
It is therefore seen that this term depends only
on new `transverse' GPDs ${\cal F}_\perp^\mu$ and
${\cal \widetilde F}_\perp^\mu$,
which can be related to the twist-2 GPDs
$H,E,\widetilde H$ and $\widetilde E$ with help of
Wandzura-Wilczek relations as discussed in Sec.~\ref{chap3_5}, and
which are given by Eqs.~(\ref{eq:F}-\ref{eq:tG}).
\newline
\indent
The DVCS amplitude of Eq.~(\ref{eq:Tdvcstw3})
is electromagnetically gauge invariant, i.e.
\be
q_\mu H^{\mu\nu}=(q-\Delta)_\nu H^{\mu\nu}= 0\, .
\la{eq:Ginv}
\ee
formally to the accuracy $1/Q^2$. In order to have `absolute' transversality
of the amplitude (i.e. such that Eq.~(\ref{eq:Ginv}) is satisfied exactly)
we keep in the expression (\ref{eq:Tdvcstw3})
terms of the order $\Delta^2/Q^2$, by applying the prescription of
Eq.~(\ref{eq:dvcsgaugeinv}),  i.e.
\be
-g^{\mu\nu}_\perp \, \to \, -g^{\mu\nu}_\perp \,-\,
\frac{\bar P^\nu\Delta_\perp^\mu}{(\bar P \cdot q')} \, ,
\la{eq:transvers}
\ee
for the twist-3 terms in the amplitude. Formally such terms are beyond
the twist-3 accuracy and they do not form a complete set of $1/Q^2$
contributions, but we prefer to work with the DVCS amplitude,
satisfying Eq.~(\ref{eq:Ginv}) exactly.
\newline
\indent
A more systematic treatment of corrections of the order $t / Q^2$ for DVCS
still remains to be done. Recently, the target mass corrections,
which induce correction terms in powers of $M^2$/$Q^2$,
have been addressed in Ref.~\cite{Bel01b}.
\newline
\indent
To calculate the twist-3 DVCS amplitude,
the convolution of the leading order Wilson coefficients (\ref{eq:alf}) with
the WW kernels in Eq.~(\ref{eq:Tdvcstw3}) can be performed
analytically, as shown in Refs.~\cite{Kiv01b,Bel00c}.
In particular, it has been shown that the
strongest (integrable) singularity of the resulting integrand
is logarithmic only (see \cite{Kiv01b} for technical details).
\newline
\indent
Finally, let us note that in Eq.~(\ref{eq:Tdvcstw3}),
the quark flavor dependence in the
GPDs ${\cal F}_\mu$ ($\widetilde {\cal F}_\mu$) is implicitely understood, i.e.
\be
{\cal F}_\mu \, (\widetilde {\cal F}_\mu) \to
\sum_{q=u,d,s}e_q^2\ {\cal F}_\mu^q \, (\widetilde {\cal F}_\mu^q) \, .
\la{eq:flav}
\ee

\subsubsection{DVCS cross section and charge asymmetry}
\label{chap5_3_4}

\begin{figure}[h]
\epsfxsize=8.5cm
\centerline{\epsffile{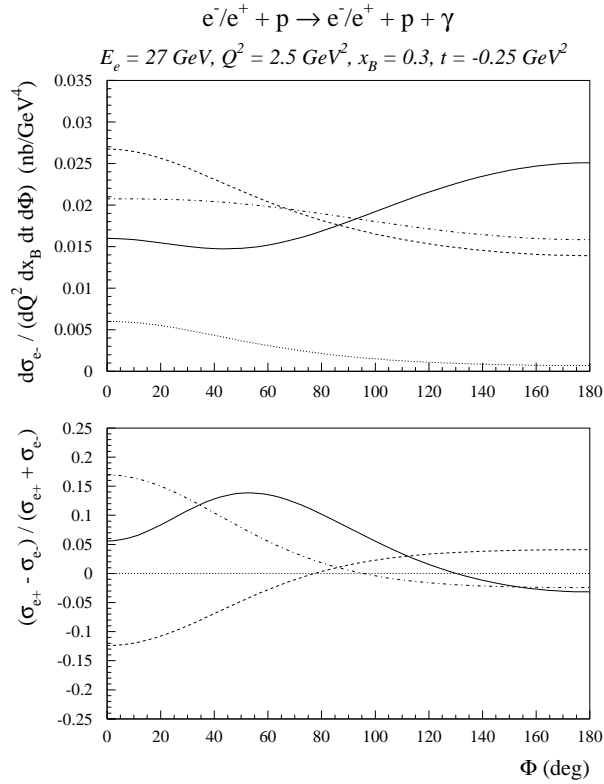}}
\caption{\small Invariant cross section for the
$e^- p \to e^- p \gamma$ reaction (upper panel) and
DVCS charge asymmetry (lower panel) at $E_e$ = 27 GeV, for the DVCS
kinematics as indicated in the figure.
Dotted curves : BH
contribution; dashed curves : BH + twist-2 DVCS (without
D-term); dashed-dotted curves : BH + twist-2 DVCS (with D-term);
full curves : BH + twist-2 + twist-3 DVCS (with D-term).
The calculations are performed
using the values $b_{val} = b_{sea} = 1$.}
\label{fig:cross1_hermes}
\end{figure}

In this section, we present results for the DVCS cross sections.
We give all results for the invariant cross section of the
$e p \to e p \gamma$ reaction, which is differential with respect to
$Q^2$, $x_B$, $t$, and out-of-plane angle $\Phi$ ($\Phi =
0^o$ corresponds to the situation where the real photon is emitted in
the same half plane as the leptons). The invariant $e p \to e p
\gamma$ cross section is given by~:
\begin{eqnarray}
{{d \sigma} \over {d Q^2 \, d x_B \, d t \, d \Phi}} \, = \,
{1 \over {(2 \pi)^4 \, 32}} \, \cdot \, {{x_B \, y^2} \over {Q^4}}
\, \cdot \, \left( 1 + {{4 m_N^2 x_B^2} \over {Q^2}}\right)^{-1/2}
\, \cdot \, \biggl| T_{BH} + T_{FVCS} \biggr|^2 \, ,
\label{eq:invcs}
\end{eqnarray}
where $m_N$ is the nucleon mass, $y \equiv (p \cdot q) / (p \cdot k)$,
and $k$ is the initial lepton four-momentum.
In the $e p \to e p \gamma$ reaction, the final photon can be emitted
either by the proton or by the lepton.
The former process is referred to as the fully VCS process
(amplitude $T_{FVCS}$ in Eq.~(\ref{eq:invcs})), which includes
the leptonic current. The process where the photon is emitted from the
initial or final lepton is referred to as the
Bethe-Heitler (BH) process (amplitude $T_{BH}$ in
Eq.~(\ref{eq:invcs})), and can be calculated exactly.
For further technical
details of the $e p \to e p \gamma$ reaction and observables, we refer
to Refs.~\cite{Gui98,Bel01a,Vdh00b}
\footnote{see Ref.~\cite{Vdh00a} for a calculation of the QED radiative
corrections to the VCS process.}.
\newline
\indent
When calculating DVCS observables, we show all
results with the exact expression for the BH amplitude, i.e. we do
{\it not} expand the BH amplitude in powers $1/Q$.
Also in the actual calculations, we use the exact kinematics for the
four-momenta of the participating particles as given by
Eqs.~(\ref{eq:dvcsp}-\ref{eq:dvcsqp}).
In this way we take (partially) kinematical higher twists into account.
Furthermore, when refering to the twist-2 DVCS results, we include
those higher twist effects which restore exact transversality of the
amplitude as expressed through Eq.~(\ref{eq:transvers}). Similarly, when
referring to the twist-3 DVCS results (provided by
a longitudinally polarized virtual photon), we include the gauge
restoring higher twist terms.
\newline
\indent
Because the BH amplitude contains two lepton electromagnetic couplings
in contrast to the FVCS process, the interference between BH and
FVCS process changes sign when comparing the $e^+ p \to e^+ p \gamma$ and
 $e^- p \to e^- p \gamma$ processes. Therefore, in the difference of
cross sections $\sigma_{e^+} - \sigma_{e^-}$, the BH (whose amplitude
is purely real) drops out. This difference measures the real part
of the BH-FVCS interference \cite{Bro72}
\begin{equation}
\sigma_{e^+} - \sigma_{e^-} \sim  \Re  \left[ T^{BH} {T^{FVCS}}^* \right] \;,
\end{equation}
and therefore is proportional
to the {\it real} (principle value integral) part of the
DVCS amplitude. In this way, the difference $\sigma_{e^+} - \sigma_{e^-}$
is sensitive to the GPDs away from the line $x = \xi$
(as e.g. in Fig.~\ref{fig:gpddt}).
\newline
\indent
In the following, we firstly show results for DVCS cross sections and
DVCS charge asymmetries
$(\sigma_{e^+} - \sigma_{e^-}) / (\sigma_{e^+} + \sigma_{e^-})$.
\newline
\indent
In Fig.~\ref{fig:cross1_hermes}, the $\Phi$-dependence of the $e p
\to e p \gamma$ cross section is shown for a lepton (either electron or
positron) of $E_e$ = 27 GeV (accessible at HERMES).
The kinematics corresponds to the valence
region ($x_B = 0.3$) and to a ratio $-t / Q^2$ = 0.1 (Remark that
increasing the ratio $-t/Q^2$, increases the higher twist effects).

\begin{figure}[h]
\epsfxsize=8.5cm
\centerline{\epsffile{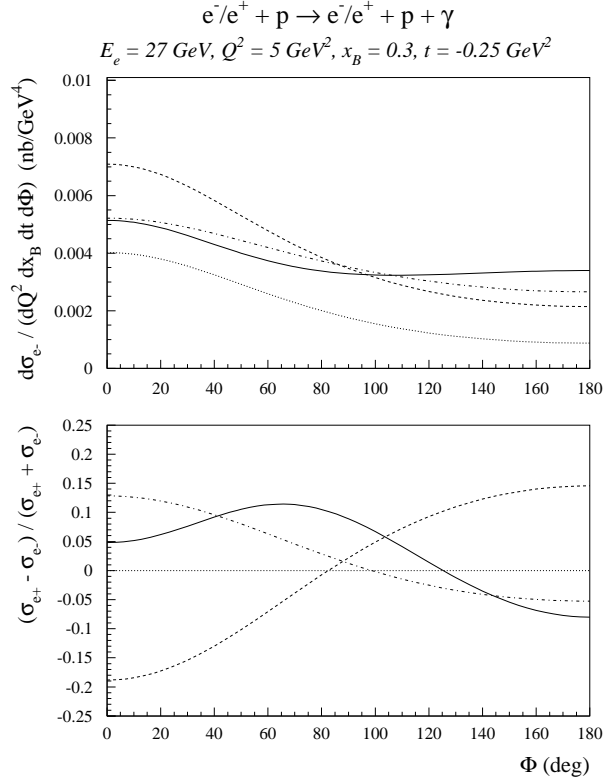}}
\caption{\small Same as Fig.~\ref{fig:cross1_hermes},
but for $Q^2$ = 5 GeV$^2$.}
\label{fig:cross2_hermes}
\end{figure}

It is firstly seen from Fig.~\ref{fig:cross1_hermes}, that in these
kinematics, the pure twist-2 DVCS process without D-term contribution
(dashed curves) dominates the $e p \to e p \gamma$ cross section
compared to the BH process (which is sizeable only around $\Phi$ =
0$^o$, where it reaches its maximal value).
In absence of the BH, the twist-2 DVCS cross section would give a
$\Phi$-independent cross section which practically
saturates the $e p \to e p \gamma$ cross section at $\Phi$ = 180$^o$,
where the BH is vanishingly small.
The $\Phi$-dependence of the dashed curve in
Fig.~\ref{fig:cross1_hermes} can
be understood as the sum of this constant twist-2 DVCS cross section,
the cross section for the BH process,
and the interference of the BH with the relatively
small real part of the DVCS amplitude
(in the valence kinematics, $x_B \simeq 0.3$, shown in
Fig.~\ref{fig:cross1_hermes},
the ratio of real to imaginary part of the DVCS amplitude without
D-term, is around 15 \%).
\newline
\indent
When adding to the twist-2 DVCS amplitude the purely real D-term
contribution, the resulting cross section for the full twist-2 DVCS process
(dashed-dotted curves in Fig.~\ref{fig:cross1_hermes}) shows a noticeable
sensitivity to the D-term.
At $\Phi$ = 180$^o$ (where the BH `contamination' is very small),
the predominantly imaginary DVCS amplitude
(in absence of the D-term contribution), and the
purely real D-term amplitude have only a small interference. In this
region, the DVCS cross section is enhanced by about 10 \%, when using the
chiral quark soliton model estimate of
Eq.~(\ref{eq:dterm_exp}) for the D-term Eq.~(\ref{dterm}).
However, around $\Phi$ = 0$^o$, the purely real D-term amplitude
interferes maximally with the BH. This interference is destructive
for the electron reaction and constructive for the positron reaction.
Therefore, the effect of the D-term can be very clearly seen in the
$\Phi$-dependence of the charge asymmetry as shown on the lower panel
of Fig.~\ref{fig:cross1_hermes}.
Including the D-term contribution, the full twist-2 DVCS charge
asymmetry changes sign and obtains a rather large value ($\approx 0.15$)
at $\Phi$ = 0$^o$. On the other hand, at  $\Phi$ = 180$^o$,
where the interference with the BH (and hence the difference between
$e^-$ and $e^+$) is small, the charge asymmetry is correspondingly small.
The pronounced $\Phi$-dependence of the charge asymmetry and its
value at $\Phi$ = 0$^o$, provides therefore a useful observable
to study the D-term contribution to the GPDs,
and to check the chiral quark soliton model estimate of
Eqs.~(\ref{dterm-num}).
\newline
\indent
We next study the twist-3 effects on the DVCS cross section
in the WW approximation.
It is seen from Fig.~\ref{fig:cross1_hermes} (full curves)
that they induce an additional (approximate) $\cos \Phi$ structure
in the cross section.
The twist-3 effects would induce an exact $\cos \Phi$ structure only
when the BH amplitude is approximated by its leading term in an
expansion in $1/Q$, neglecting its additional $\Phi$-dependence.
In all calculations, we keep however the full
$\Phi$-dependence of the BH, due to the lepton propagators, which
complicates the interference at the lower $Q$. One sees from
Fig.~\ref{fig:cross1_hermes} that the interference of the twist-3 amplitude
with the BH + twist-2 DVCS amplitude is destructive for
$\Phi \lesssim 90^o$, and constructive for $\Phi \gtrsim 90^o$.
Around $\Phi = 0^o$, the twist-3 effects reduce the full twist-2 DVCS
cross section (including the D-term) by about 25 \%, whereas around
$\Phi = 180^o$, they largely enhance the twist-2 DVCS
cross section (by about 55 \%).
\newline
\indent
To see how the size of the twist-3 effects on the $e p \to e p \gamma$ cross
section decreases when increasing the value of $Q^2$ at fixed $x_B$ and $t$,
the cross section and charge asymmetry are shown
in Fig.~\ref{fig:cross2_hermes}
at the same $E_e$, $x_B$, and $t$, as in Fig.~\ref{fig:cross1_hermes},
but at a value $Q^2$ = 5 GeV$^2$.
One firstly sees that going to higher $Q^2$, at the same $E_e$, $x_B$,
and $t$, enhances the relative contribution of the BH process compared
to the DVCS process. Consequently, the cross section follows
much more the $\Phi$-behavior of the BH process. For the twist-2
cross section, one sees again clearly the effect of the D-term which
leads, through its interference with the BH amplitude, to a charge
asymmetry of opposite sign as compared to the one for the twist-2 DVCS
process without D-term.
\newline
\indent
The twist-3 effects in the kinematics of Fig.~\ref{fig:cross2_hermes}, where
to ratio $t/Q^2$ is only half the value of Fig.~\ref{fig:cross1_hermes},
are correspondingly smaller. It is furthermore seen that the twist-3
effects induce an (approximate) structure $\sim \, A \, \cos (2 \Phi)$
in the charge asymmetry ($A \approx$ - 0.08 in the kinematics of
Fig.~\ref{fig:cross2_hermes}).
Only for out-of-plane angles $\Phi \gtrsim 130^o$
one sees a deviation from this simple structure induced
by the twist-3 amplitude, due to the more complicated $\Phi$-dependence
when calculating the BH amplitude exactly.

Besides exploiting the interference between the DVCS and BH contributions to
the $e p \to e p \gamma$ reaction, as e.g. through the charge asymmetry,
one can also study the DVCS amplitude directly in a kinematical
regime where it dominates over  the BH amplitude. Indeed, it was found
in Refs.~\cite{Gui98,Vdh98} that at some fixed $Q^2$, $x_B$ and $t$,
one favors the DVCS contribution over the BH by going to higher beam energy.
In particular, at intermediate values of $x_B \sim 0.15 - 0.3$, one is in such a regime at 200 GeV beam energy at COMPASS,
where a DVCS experiment has been proposed \cite{d'Ho99}.

\subsubsection{Electron single spin asymmetry (SSA)}
\label{chap5_3_5}

\begin{figure}[h]
\epsfxsize=8.75cm
\centerline{\epsffile{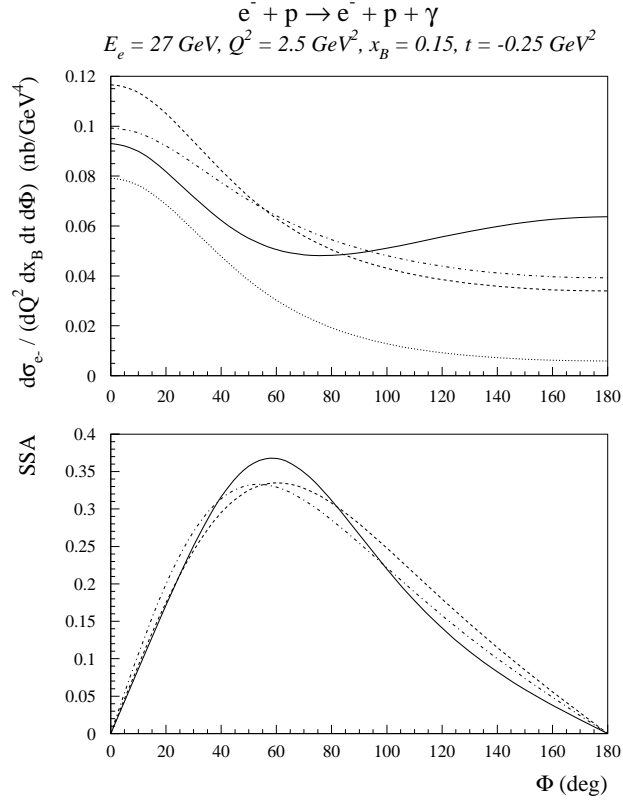}}
\caption{{\small Invariant cross section for the
$e^- p \to e^- p \gamma$ reaction (upper panel) and
DVCS single spin asymmetry (lower panel) at $E_e$ = 27 GeV, for the
kinematics as indicated in the figure.
Curve conventions as in Fig.~\ref{fig:cross1_hermes}. }}
\label{fig:asymm1}
\end{figure}

At intermediate lepton beam energies, one can also extract the imaginary part
of the interference between the FVCS and BH amplitudes through the
$\vec e p \to e p \gamma$ reaction with a polarized lepton beam,
by measuring the out-of-plane angular dependence of the produced
photon \cite{Kro96}.
It was found in Refs.~\cite{Gui98,Vdh98} that the resulting electron
single spin asymmetry (SSA)
\be
{\cal A}^{\rm{SSA}} \;=\; {{\sigma_{e, h = +1/2} - \sigma_{e, h = -1/2}} \over
{\sigma_{e, h = +1/2} + \sigma_{e, h = -1/2}}} \, ,
\la{eq:ssa}
\ee
with $\sigma_{e, h}$ the cross section for an electron of helicity $h$,
can be sizeable for HERMES ($E_e$ = 27 GeV) and JLab ($E_e$ = 4 - 11 GeV)
beam energies.
The helicity difference $(\sigma_{e, h = +1/2} - \sigma_{e, h = -1/2})$
vanishes for the BH process and depends linearly on the
imaginary part of the DVCS amplitude. This helicity difference is therefore
directly proportional to the GPDs along the line $x = \xi$, and maps out
the 'envelope' function $H(\xi, \xi, t)$,
and analogously for $E$, $\tilde H$ and $\tilde E$.
\newline
\indent
In the following, we discuss the SSA to twist-3 accuracy and show the
sensitivity of the SSA to different parametrizations of the GPDs in kinematics
at HERMES ( $E_e$ = 27 GeV ) and JLab ( $E_e$ = 4.23 GeV ),
where this SSA has been measured recently.
\newline
\indent
In Fig.~\ref{fig:asymm1}, we show the $\Phi$-dependence
of the DVCS cross section and of the single spin asymmetry (SSA)
for the same values of $Q^2$ and $t$ as in Figs.~\ref{fig:cross1_hermes} and
\ref{fig:cross2_hermes}, and for a value of $x_B = 0.15$, accessible at
HERMES. Note that, due to parity invariance, the SSA is odd in $\Phi$.
We therefore display only half of the $\Phi$ range,
i.e. $\Phi$ between 0$^o$ and 180$^o$.

\begin{figure}[hp]
\epsfxsize=9cm
\centerline{\epsffile{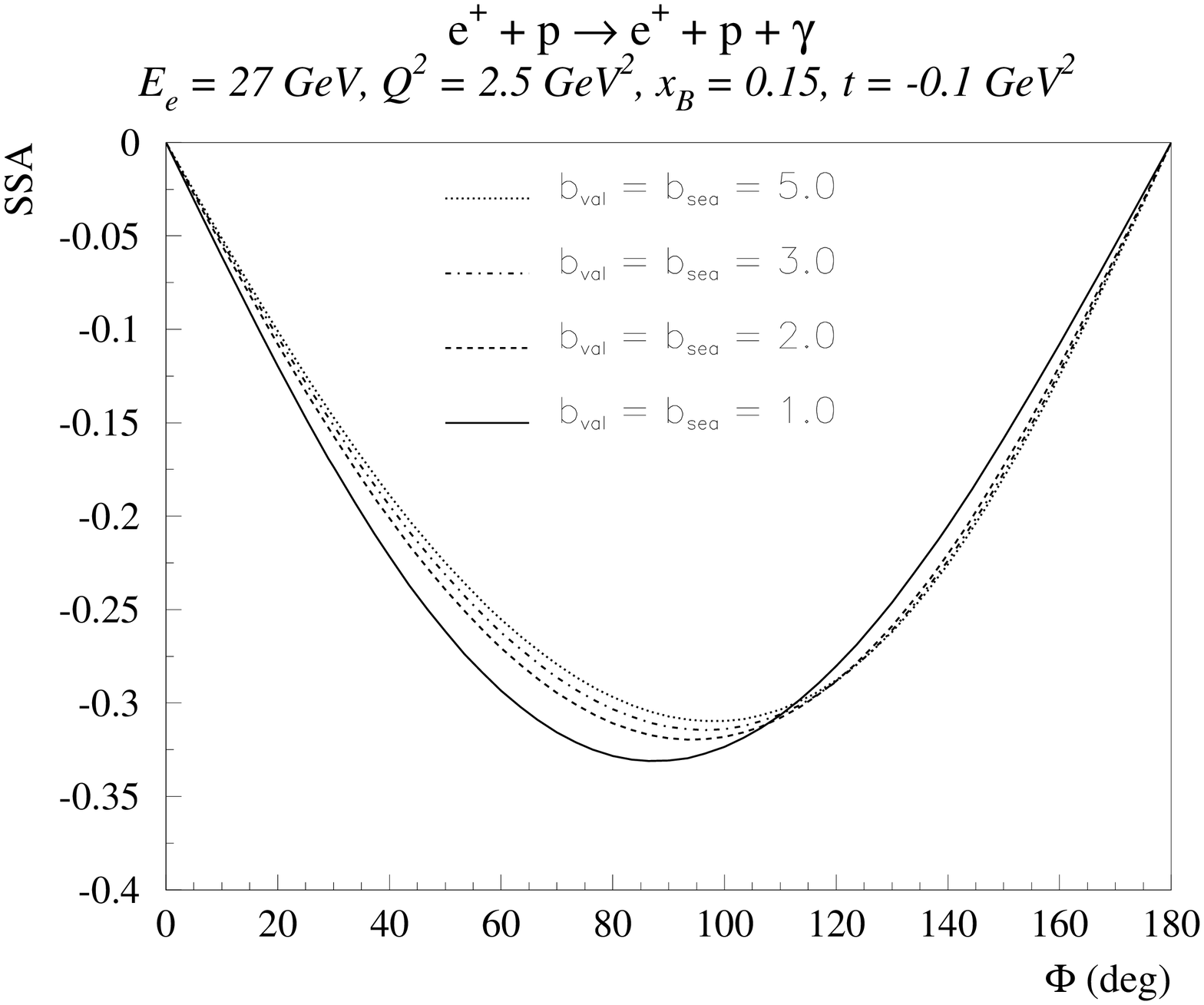}}
\caption{\small Sensitivity of the single spin asymmetry for the
$e^+ p \to e^+ p \gamma$ reaction, in the HERMES kinematics \cite{Air01},
to different values of the parameters  $b_{val}$ and $b_{sea}$ as
indicated, which enter in the parameterization of the GPDs $H^u$ and $H^d$
( see Sec.~\ref{chap5_2a} ).
All calculations include the twist-3 effects to the DVCS amplitude.}
\label{fig:ssa_b_hermes}
\epsfxsize=9cm
\centerline{\epsffile{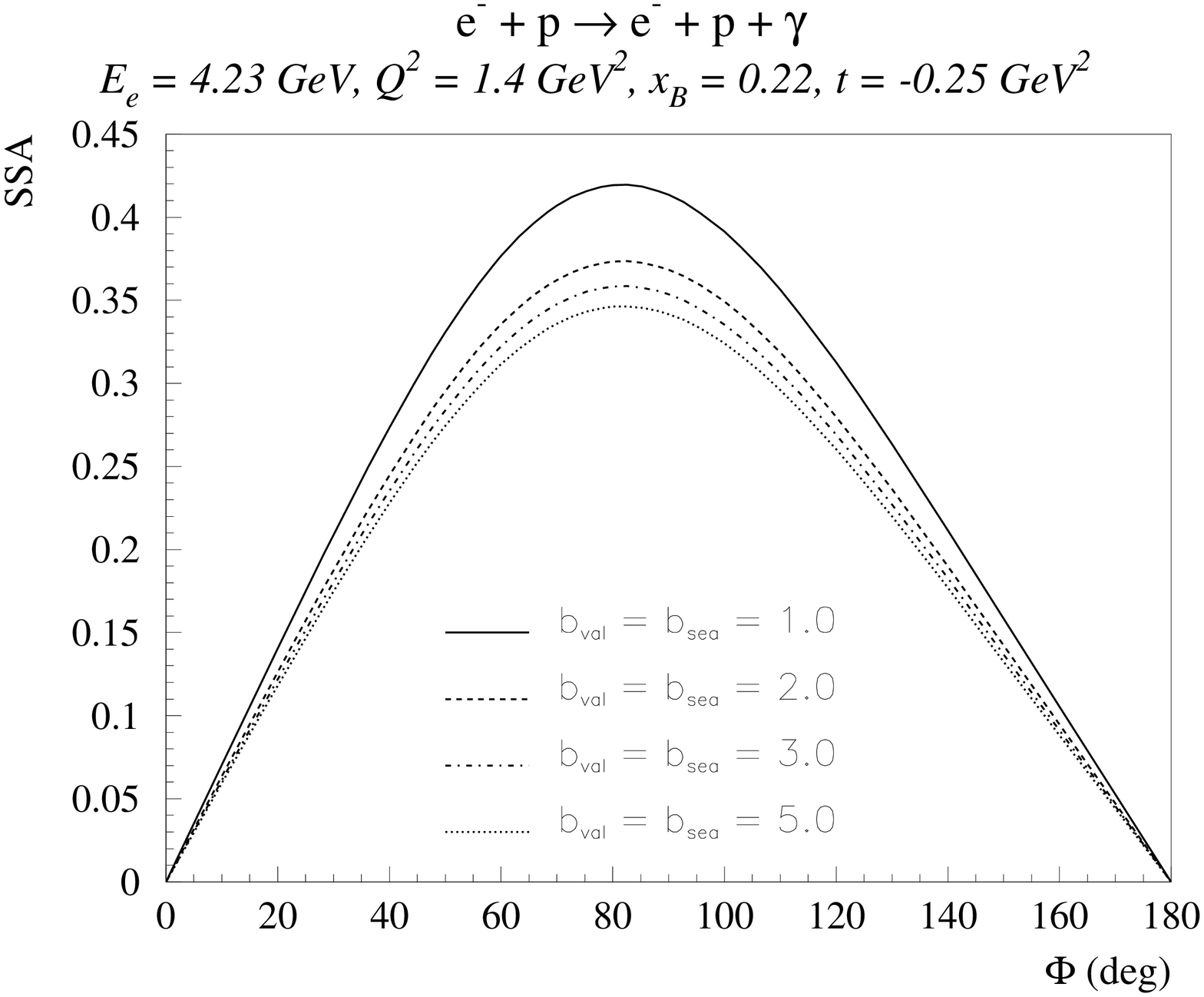}}
\caption{\small Sensitivity of the single spin asymmetry for the
$e^- p \to e^- p \gamma$ reaction, in the CLAS kinematics \cite{Step01},
to different values of the parameters  $b_{val}$ and $b_{sea}$ as
indicated, which enter in the parameterization of the GPDs $H^u$ and $H^d$
( see Sec.~\ref{chap5_2a} ).
All calculations include the twist-3 effects to the DVCS amplitude.}
\label{fig:ssa_b_clas}
\end{figure}

When comparing Figs.~\ref{fig:cross1_hermes}, \ref{fig:cross2_hermes} with
Fig.~\ref{fig:asymm1}, it is firstly seen that the cross
sections for the $e p \to e p \gamma$ process increases strongly when
decreasing $x_B$ at fixed $Q^2$ and fixed $t$, mainly due to the
growth of the BH amplitude.
Due to this large BH amplitude, the relative twist-3 effect
on the cross section in Fig.~\ref{fig:asymm1} is smaller than the ones in
Figs.~\ref{fig:cross1_hermes}, \ref{fig:cross2_hermes}. However, the large BH
amplitude leads to a large value for the SSA through its interference
with the DVCS process. At (pure) twist-2 level, the SSA originates
from the interference of the imaginary part of the
DVCS amplitude and the (real) BH amplitude.
In case the BH is approximated by its leading term in an expansion in 1/$Q$,
the twist-2 SSA displays a pure $\sin \Phi$ structure. Due to the more
complicated $\Phi$-dependence of the BH at the lower values of $Q^2$,
this form gets distorted and its maximum displaced.
Note that, for practical considerations,
our ``twist-2'' DVCS calculations include kinematical higher
twist terms as well as the gauge restoring terms according to
Eq.~(\ref{eq:transvers}). Their effect can be seen in the slight change
in the SSA (of the percent level), due to the DVCS process by itself
(i.e. when increasing the real part of the DVCS amplitude by adding
the D-term contribution, the curves for the SSAs are slightly displaced).
\newline
\indent
We next discuss the twist-3 effects, calculated in WW approximation,
on the SSA for the DVCS process. One sees from Fig.~\ref{fig:asymm1} that the
twist-3 corrections induce an (approximate)
$\sin (2 \Phi)$ structure in the SSA.
The amplitude of the $\sin (2 \Phi)$ term is however rather
small and the twist-3 effects change the SSA by less than 5 \% in
the kinematics corresponding to $t / Q^2 = 0.1$. It was checked that
at $x_B = 0.3$ and for a value $t / Q^2 = 0.1$,
the twist-3 effects on the SSA are of similar size.
One therefore observes that although the twist-3 effects
in WW approximation can provide a sizeable contribution
to the real part of the amplitude (see Fig.~\ref{fig:cross1_hermes}),
they modify the imaginary part, and hence the SSA, to a much lesser extent.
\newline
\indent
In Figs.~\ref{fig:ssa_b_hermes} and \ref{fig:ssa_b_clas} we investigate the
sensitivity of the SSA to different parametrizations of the GPD $H$
in kinematics
of a HERMES experiment \cite{Air01}
and of a JLab experiment by the CLAS Collaboration
\cite{Step01,Bur01b,Guid01}.
\newline
\indent
The calculations in the previous figures
( Figs.~\ref{fig:cross1_hermes} - \ref{fig:asymm1} ) have been performed
using the values $b_{val} = b_{sea} = 1$ for the parameters which enter
in the profile function of Eq.~(\ref{eq:profile}) to reconstruct the GPD $H$.
In Figs.~\ref{fig:ssa_b_hermes} and \ref{fig:ssa_b_clas},  the SSA to twist-3
accuracy is shown for different values of $b_{val}$ and $b_{sea}$.
One sees that increasing the value of $b_{val}$ and $b_{sea}$ decreases the
size of the SSA. It is seen from Fig.~\ref{fig:upb_H} that by increasing
$b_{val}$ and $b_{sea}$, one approaches the forward quark distribution
( which corresponds to the limit $b_{val} = b_{sea} = \infty$ ).
One may use this sensitivity to extract the parameters
$b_{val}$ and $b_{sea}$ from accurate measurements of the SSA.
For example, one sees from Fig.~\ref{fig:ssa_b_clas} that the maximum of the
SSA is reduced to about 80 \% of its value when increasing the values of
$b_{val}$ and $b_{sea}$ over the indicated range.

\newpage

\begin{figure}[h]
\epsfxsize=9cm
\centerline{\epsffile{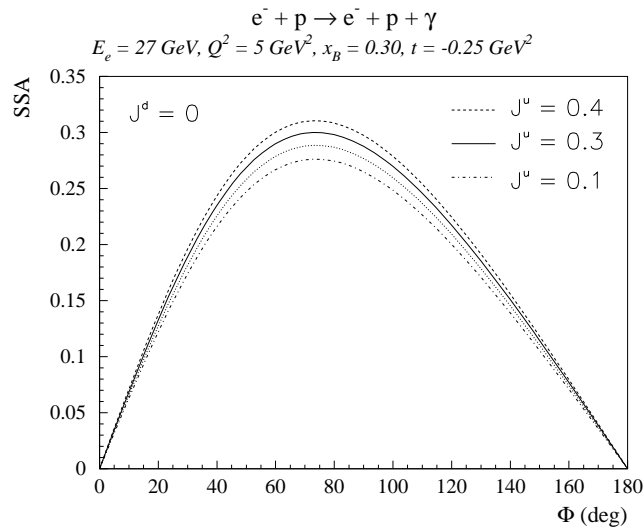}}
\caption{\small Sensitivity of the SSA for the
$e^- p \to e^- p \gamma$ reaction, in HERMES kinematics,
to different values of the quark contributions to the proton spin,
$J^u$ and $J^d$, which enter
in the parametrization of the GPDs $E^u$ and $E^d$
( see Sec.~\ref{chap5_2c} ).
The dotted curve corresponds to the calculation when neglecting the
contribution of the GPDs $E^u$ and $E^d$.
The calculations are performed for the BH + twist-2 DVCS amplitude.}
\label{fig:ssa_e_27}
\end{figure}

The previous figures for DVCS observables have all been obtained by
neglecting the GPD $E$ (except for its D-term contribution of
Eq.~(\ref{eq:dtermcontrib2})). In Fig.~\ref{fig:ssa_e_27}, we investigate
the effect of the GPD $E$ on the DVCS SSA, using the three-component
parametrization ( D-term, valence contribution,
and ``vector-meson'' (VM) contribution ) of
Eqs.~(\ref{eq:parame}, \ref{eq:edd}). This parametrization allows to see
directly the sensitivity of observables to the total angular momentum
contribution of the $u$-quark $J^u$) and $d$-quark ($J^d$) to the
proton spin.
\newline
\indent
In Fig.~\ref{fig:ssa_e_27}, we show the SSA in kinematics accessible
at HERMES for different values of $J^u$ (for a value $J^d = 0$)
corresponding to the values shown in Fig.~\ref{fig:up_E}.
Because the SSA is sensitive to the imaginary part of the DVCS amplitude
(apart from a small contribution to the SSA due to the DVCS process alone,
i.e. in absence of the BH process),
it is mainly sensitive to the valence component of the GPD $E^u$
(upper plot in Fig.~\ref{fig:up_E}). Indeed, the D-term and VM contributions
to the GPD $E$ give rise to a real amplitude. By comparing
Fig.~\ref{fig:up_E} (upper plot) and Fig.~\ref{fig:ssa_e_27},  one sees
that neglecting the contribution of the GPD $E^u$, corresponds approximately
to the value $J^u \sim 0.2$. Furthermore, one sees that varying the value
of $J^u$ from $J^u = 0.1$ to $J^u = 0.4$ changes the SSA by about
10\% of its value. Since the sensitivity of the DVCS SSA to
$J^u$ (and $J^d$) is modest, it calls for precise measurements
as well as detailed studies of higher-twist effects and NLO corrections.
However, the physical importance of $J^u$ and $J^d$
warrants such dedicated efforts.

\begin{figure}[h]
\epsfxsize=9cm
\centerline{\epsffile{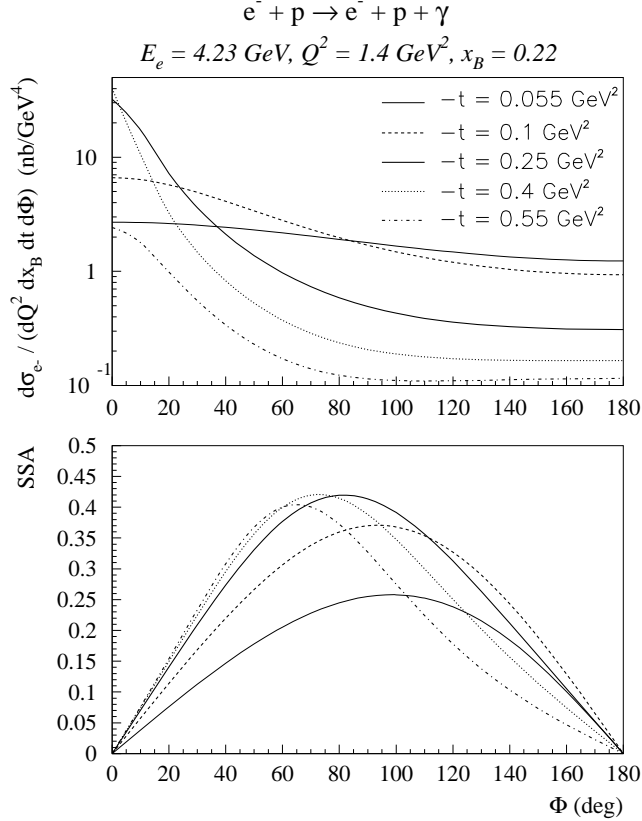}}
\caption{\small Effect of the twist-3 corrections on the
$\Phi$-dependence of the $e^- p \to e^- p \gamma$ invariant
cross section (upper panel)
and SSA (lower panel) in CLAS kinematics,
for different values of the momentum transfer $-t$ as indicated in the figure.
All calculations represent BH + twist-2 + twist-3 DVCS cross
sections for $b_{val} = b_{sea} = 1.0$.}
\label{fig:ssa_clas_tdep}
\end{figure}

In Fig.~\ref{fig:ssa_clas_tdep}, we investigate the $t$-dependence of
the DVCS cross section and SSA to twist-3 accuracy in CLAS kinematics.
One sees from Fig.~\ref{fig:ssa_clas_tdep} that both cross section and
SSA display a qualitative change with increasing values of $-t$
(note that the SSA vanishes at $t = t_{min}$). Furthermore, it is seen that
when increasing the value of $-t$ at fixed $Q^2$ and fixed $x_B$, the
shape of the SSA changes from the ``$\sin \Phi$'' dependence and adopts
an increasingly larger ``$\sin (2 \Phi)$'' twist-3 component.
This qualitative change in the $t$-dependence is important to
investigate experimentally to have confidence in the extraction of GPDs
from a fit to SSA data.

\begin{figure}[ht]
\epsfxsize=9cm
\centerline{\epsffile{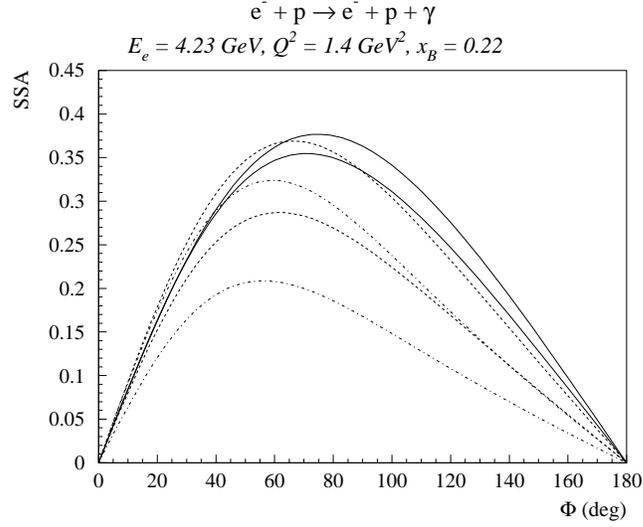}}
\caption{\small Single spin asymmetry for the
$e^- p \to e^- p \gamma$ reaction in CLAS kinematics, for
two models for the $t$-dependence of the GPD $H$.
The factorized model of Eq.~(\ref{eq:factt}) (thick upper curves)
is compared to the unfactorized (Regge inspired)
model of Eqs.~(\ref{Regge-parametrization},\ref{eq:ddunpolregge})
(thin lower curves) for different values of $-t$~:
0.25 GeV$^2$ (full curves),
0.4 GeV$^2$ (dashed curves), 0.55 GeV$^2$ (dashed-dotted curves).
All calculations represent BH + twist-2 DVCS cross
sections for $b_{val} = b_{sea} = 1.0$.}
\label{fig:ssa_clas_regge}
\end{figure}

In Fig.~\ref{fig:ssa_clas_regge}, we investigate the sensitivity of
the DVCS SSA to different models for the $t$-dependence of the GPD $H$
(we perform this comparison for the twist-2 DVCS amplitude).
We compare the factorized model of Eq.~(\ref{eq:factt}) which has been used
in all calculations before with an unfactorized, Regge inspired model of
Eqs.~(\ref{Regge-parametrization},\ref{eq:ddunpolregge})
(with $\alpha'$ = 0.8 GeV$^{-2}$),
which has an exponential
$t$-dependence and exhibits shrinkage at small values of $\xi$,
as was shown in Fig.~\ref{fig:upt_xiscan}.
One sees from Fig.~\ref{fig:ssa_clas_regge} that at $-t$ = 0.25 GeV$^2$,
the unfactorized form leads to a slightly reduced SSA compared to the
factorized one. However, for a value of $-t$ = 0.55 GeV$^2$, one already
sees a noticeable reduction of the unfactorized model over the factorized one,
and the SSA gets reduced to about 65 \% of its value when using an
unfactorized ansatz. Such important sensitivity clearly points out the
need to map out experimentally the $t$-dependence of the SSA
carefully in order to extract GPDs.
It also stresses the need from the theoretical side to study the
parametrization of the $t$-dependence of GPDs in more detail in the future.

\newpage
\subsection{DVCS with $N \to \Delta$ transition}
\label{chap5_4}

To derive the $N \rightarrow \Delta$ DVCS amplitudes, we use
analogously as for $N \rightarrow N$ DVCS 
a frame where the virtual photon momentum $q^\mu$ and the 
average $N \Delta$ momentum $\bar P^\mu$, defined in 
Eq.~(\ref{n-normalization}), are collinear along the
$z$-axis and in opposite directions. By denoting the lightlike vector
along the positive and negative $z$-directions as $\tilde p^\mu$ and
$n^\mu$ respectively, we can decompose the physical momenta of the 
$N \rightarrow \Delta$ DVCS process as follows :
\begin{eqnarray}
&&\hspace{-1cm} \bar P^{\mu }={1\over 2}\left( p^{\mu }+p'^{\mu }\right) \; 
=\tilde{p}^{\mu }+{{\bar m_{N \Delta}^{2}}\over 2}\, n^{\mu }\; ,\\
&&\hspace{-1cm} q^{\mu }=-\left( 2\xi ^{'}\right) \, 
\tilde{p}^{\mu }+\left( {{Q^{2}}\over {4\xi ^{'}}}\right) \, n^{\mu }\; ,\\
&&\hspace{-1cm} \Delta ^{\mu }\equiv p^{'\mu }-p^{\mu }=
-\left( 2\xi \right) \, \tilde{p}^{\mu } 
\,+\, \left[ \xi \, \bar m_{N \Delta}^{2} 
+ {1 \over 2} \left( m_\Delta^2 - m_N^2 \right) \right] \, n^{\mu } 
\,+\,\Delta ^{\mu }_{\perp } \;,
\la{eq:nddvcsxi} \\
&&\hspace{-1cm} q^{'\mu }\equiv q^{\mu }-\Delta ^{\mu } = 
-2\left( \xi ^{'}-\xi \right) \, \tilde{p}^{\mu }
\,+\, \left[ {{Q^{2}}\over {4\xi ^{'}}}-\xi \, \bar m_{N \Delta}^{2} 
- {1 \over 2} \left( m_\Delta^2 - m_N^2 \right) \right] \, n^{\mu } 
\,-\, \Delta ^{\mu }_{\perp },
\la{eq:nddvcsqp} 
\end{eqnarray}
 where the variables \( \bar m_{N \Delta}^{2} \), 
\( \xi ^{'} \) and \( \xi  \) are given by 
\begin{eqnarray}
&& \bar m_{N \Delta}^{2}= {1 \over 2} \left({m_{N}}^2 + {m_{\Delta}}^2 \right)
-{{\Delta ^{2}}\over 4}\; ,
\la{eq:nddvcskin} \\
&& 2\xi ^{'}={{\bar P \cdot q}\over {\bar m_{N \Delta}^{2}}}\, \left[
  -1+\sqrt{1+{{Q^{2}\, \bar m_{N \Delta}^{2}} \over 
{(\bar P \cdot q)^{2}}}}\right] \;, \\
&& 2\xi =2\xi ^{'}\,\cdot\, {{Q^{2}-\Delta ^{2} - 
\left( 2 \xi^{'}\right) \left( m_{\Delta}^2 - m_{N}^2 \right) }\over 
{Q^{2}+\bar m_{N \Delta}^{2}(2\xi ^{'})^{2}}}\;. 
\la{eq:nddvcskin3} 
\end{eqnarray}

\begin{figure}[h]
\vspace{-1.25cm}
\epsfxsize=10.4 cm
\epsfysize=13 cm
\centerline{\epsffile{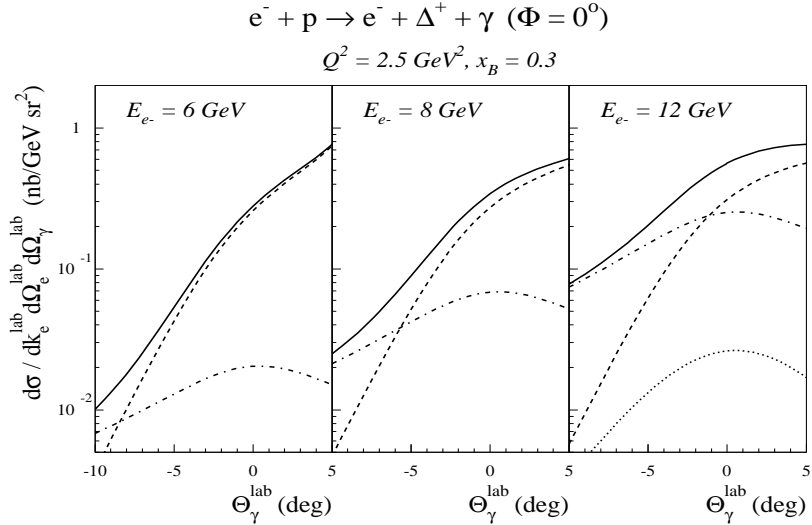}}
\vspace{-3.5cm}
\caption[]{\small Angular distribution for the
$N \rightarrow \Delta$ DVCS  cross section at $Q^2$ = 2.5 GeV$^2$,
$x_B$ = 0.3, and for different beam energies accessible at JLab.
A comparison is shown between the BH (dashed curves),
the $N \rightarrow \Delta$ DVCS (dashed-dotted curves)
and their coherent sum (full curves). We also show the
contribution of the $C_1$ GPD (dotted curve at 12 GeV).}
\label{fig:ndeltadvcs_jlab}
\end{figure}

The leading order $N \rightarrow \Delta$ DVCS tensor follows from the
handbag diagrams (Fig.~\ref{fig:handbags} where the outgoing nucleon 
is replaced by a $\Delta^+$). 
By parametrizing the $N \rightarrow \Delta$ 
matrix elements of the vector and axial vector bilocal quark operators as in 
Eqs.~(\ref{HND-QCD-2}) and (\ref{HND-helicity-QCD-2}) respectively,
the leading order $N \rightarrow \Delta$ DVCS tensor 
$H^{\mu \nu}$, defined in an analogous way as in Eq.~(\ref{eq:Tdef}),  
is given by~: 
\begin{eqnarray}
H^{\mu \nu }\left( \gamma^* p \rightarrow \gamma \Delta^+ \right)
&=& {1\over 2}\, \left[ \tilde{p}^{\mu }n^{\nu }+\tilde{p}^{\nu
 }n^{\mu }-g^{\mu \nu }\right] \;
\int _{-1}^{+1}dx \; C^+(x,\xi) \nonumber \\
&\times& \, {\sqrt{2 \over 3}} \; {1 \over 6}\;
\bar \psi^\beta(p^{'}) \,
\left\{\,H_{M}(x,\xi ,t)\; \left( - {\cal K}_{\beta\kappa}^M \right)
\, n^\kappa \right. \nonumber\\
&&\hspace{2.3cm}\;+\;H_{E}(x,\xi,t)\;
\left(- {\cal K}_{\beta\kappa}^E \right)
\, n^\kappa \nonumber\\
&&\hspace{2.3cm} \left.\;+\;H_{C}(x,\xi,t)\;
\left(- {\cal K}_{\beta\kappa}^C \right) \, n^\kappa
\right\} \,N(p) \nonumber\\
&+& {1\over 2}\, \left[ i \, \varepsilon^{\nu \mu \kappa \lambda }
\, n_{\kappa} \tilde{p}_{\lambda }\right] \;
\int _{-1}^{+1}dx \; C^-(x, \xi) \nonumber \\
&\times& {1 \over 6}\;
\bar \psi^\beta(p^{'}) \,
\left\{ C_1(x,\xi ,t)\, n_\beta \,+\,
C_2(x,\xi ,t) \, \Delta_\beta \,
{{\Delta \cdot n}\over {m_N^2}} \right.\nonumber\\
&&\hspace{1.5cm}\,+\; C_3(x,\xi ,t) \, {1\over {m_N}}
 \left( {\Dirac \Delta \, n_\beta - {\Dirac n} \, \Delta_\beta}\right)
 \nonumber\\
&&\left. \hspace{1.5cm}\,+\; C_4(x,\xi ,t) \, {2\over {m_N^2}}
 \left( {\Delta \cdot \bar P}\, n_\beta - \Delta_\beta\right)
\right\} N(p)\;,
\la{eq:ndeldvcsampl}
\end{eqnarray}
where the factor 1/6 in Eq.~(\ref{eq:ndeldvcsampl}) results from the
quadratic quark charge combination ($e_u^2 - e_d^2$)/2. 
\newline
\indent
To give estimates for the $N \rightarrow \Delta$ DVCS amplitudes, 
we need a model for the $N \rightarrow \Delta$ GPDs which appear in 
Eq.~(\ref{eq:ndeldvcsampl}). 
Here we will be guided by the large $N_c$ relations discussed in 
Sec.~\ref{chap3_7}. These relations connect the $N \rightarrow \Delta$
GPDs $H_M, C_1$ and $C_2$ to
the $N \rightarrow N$ isovector GPDs 
$E^u - E^d$, $\tilde H^u - \tilde H^d$ and $\tilde E^u - \tilde E^d$ 
respectively, as expressed by Eq.~(\ref{ndNc}). 
All other (sub-dominant) GPDs, which 
vanish at leading order in the $1/N_c$ expansion,
are set equal to zero in our estimates for the processes involving 
the $N \to \Delta$ GPDs. 
For the $N \rightarrow N$ GPDs, we use 
the phenomenological $\xi$-dependent ansatz as discussed in 
Sec.~\ref{chap5_2}.
\newline
\indent
For the $N \rightarrow \Delta$ DVCS in the near forward direction, unlike the 
$N \rightarrow N$ DVCS case, the axial transitions 
(distributions $C_1$ and $C_2$) are numerically much more
important than the vector transition (distribution $H_M$). 
This is because the $H_M$ comes
with a momentum transfer ($\Delta$) in the tensor ${\cal
  K}_{\beta\kappa}^M$ as seen from Eq.~(\ref{K-def}). 
In contrast, the GPD 
$C_1$, which is proportional to the polarized quark distribution
in the forward limit ($\Delta \rightarrow 0$), enters with no momentum
transfer in Eq.~(\ref{eq:dvcsampl}).
Besides $C_1$, the distribution $C_2$ is numerically
most important as it contains a pion-pole contribution. 
The $\pi^0$ pole contribution to the $N \rightarrow \Delta$ DVCS amplitude
can be evaluated analytically by using :
\begin{equation}
\int _{-1}^{+1}dx\left[ {1\over {x-\xi +i\epsilon }}
-{1\over {x+\xi -i\epsilon }
}\right] \; {1 \over 6} \, C_{2, \pi-pole}(x,\xi ,t)\; =\;
-\, {1\over {2\xi }}\; {\sqrt{3} \over 4} \; h_{A}(t)\;,
\end{equation}
with the pion-pole part of the pseudoscalar form factor $h_{A}(t)$ given by
Eq.~(\ref{eq:ha_pipo}). 
This leads then for the $\pi^0$ pole contribution to the $N \rightarrow
\Delta$ DVCS tensor 
\begin{equation}
H_{\pi - pole}^{\mu \nu }\left( \gamma^* p \rightarrow \gamma \Delta^+\right) 
\;=\; {1\over 2} \, i \, \varepsilon ^{\nu \mu \kappa \lambda } 
\, n_{\kappa} \tilde{p}_{\lambda} \; 
{{g_A \, \sqrt{3}} \over {-t + m_\pi^2}} \; 
\bar \psi^\beta(p^{'}) \, \Delta_\beta \, N(p) \;.
\end{equation}
\newline
\indent
The $N \rightarrow \Delta$ DVCS process can be accessed experimentally
through the $e p \rightarrow e \Delta^+ \gamma$ reaction.  
In Fig.~\ref{fig:ndeltadvcs_jlab}, we show our predictions for the
fivefold differential $e p \rightarrow e \Delta^+ \gamma$ cross
sections in the JLab energy range. 
As for the $e p \rightarrow e p \gamma$ reaction, there is also for the 
$e p \rightarrow e \Delta^+ \gamma$ reaction an analog contribution 
when the photon is emitted from the electron line, which is the
Bethe-Heitler (BH) process. 
In comparison with the BH for the $e p \rightarrow e p \gamma$
reaction, the BH process in the 
$e p \rightarrow e \Delta^+ \gamma$ reaction is 
reduced by an order of magnitude in the near forward direction. 
Firstly, this is due to the photon propagator in the BH process,
which goes like $1/t$, which is smaller for
the $N \rightarrow \Delta$ case due to the larger value of
$t_{min}$ in the unequal mass case compared to the $N \rightarrow N$
transition. Secondly, the electromagnetic $N \rightarrow \Delta$
vertex in the BH process contains an additional
momentum transfer (magnetic $N \to \Delta$ transition), 
which results in an additional suppression at the small angles.
\newline
\indent
Fig.~\ref{fig:ndeltadvcs_jlab} shows that if one stays away from the BH
peaks by going in the opposite halfplane (negative angles on 
Fig.~\ref{fig:ndeltadvcs_jlab}) as the electron lines, the $N
\rightarrow \Delta$ DVCS process shows a fast increase with energy and
starts to dominate over the BH already around 10 GeV.  
\newline
\indent
As mentioned above, the $N \rightarrow \Delta$ DVCS in the near
forward direction is dominated by the axial transitions  
$C_1$ and $C_2$. The vector transitions, which dominate
the $N \rightarrow N$ DVCS are suppressed here by a momentum transfer
in the corresponding tensors. Among the axial transitions, the pion
pole contribution in $C_2$ dominates in the valence region. 
This is also seen in Fig.~\ref{fig:ndeltadvcs_jlab}, where the $C_1$ 
contribution to the $N \rightarrow \Delta$ DVCS amplitude is shown
separately (at 12 GeV). 

Due to dominance of the pion pole in the valence region, our predictions
for $N \rightarrow \Delta$ DVCS are relatively model
independent. Therefore, they can be used to estimate the background for
the DVCS on the proton as it may be difficult in an actual DVCS experiment
to pin down the final state unambiguously. If one does not have the
resolution to distinguish the
$e p \rightarrow e \gamma \Delta^+ \rightarrow e \gamma \pi^0 p$ reaction
from the $e p \rightarrow e \gamma p$ reaction, an estimate of the 
former process may be helpful to quantify the background from the 
$e p \rightarrow e \gamma \Delta^+$ contribution.

\newpage
\subsection{Hard meson electroproduction (HMP)}
\label{chap5_5}

\begin{figure}[ht]
\vspace{-1.2cm}
\epsfxsize=11 cm
\epsfysize=15. cm
\vspace{-1.5cm}
\centerline{\epsffile{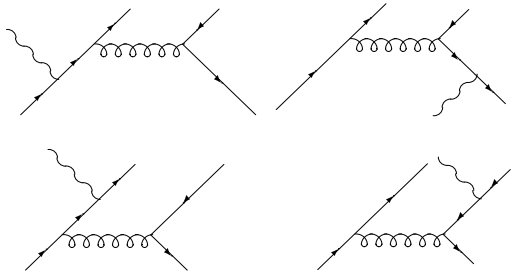}}
\vspace{-9.cm}
\caption[]{\small Leading order diagrams for the quark exchange contribution 
to hard meson electroproduction.}
\label{fig:4diaghard}
\end{figure}

In the previous sections, we discussed how the GPDs enter into
different DVCS observables. 
As discussed in Sec.~\ref{chap1}, 
the GPDs reflect the structure of the nucleon 
independently of the reaction
which probes the nucleon. In this sense, they are universal quantities and can
also be accessed, in different flavor combinations, through the hard exclusive
electroproduction of mesons 
- \( \pi ^{0,\pm },\eta ,...,\rho ^{0,\pm },\omega ,\phi ,... \) - 
(see Fig.~\ref{fig:factmeson}) for which a QCD factorization proof was given
in Refs.~\cite{Col97}. According to Ref.~\cite{Col97}, the
factorization applies when the virtual photon is longitudinally 
polarized, which corresponds to a small size configuration 
compared to a transversely polarized photon. 
More technically, for a longitudinally polarized photon 
the end-point contributions in the meson wave function are power
suppressed. Furthermore, it was shown that the cross section for a transversely
polarized photon is suppressed by 1/\( Q^{2} \) compared to a longitudinally
polarized photon.
\newline
\indent
In the hard scattering regime, the leading order diagrams for meson
production are shown in Fig.~\ref{fig:4diaghard} 
(\( T_{H} \) part in Fig.~\ref{fig:factmeson})
\footnote{For neutral isoscalar mesons with positive C-parity 
(e.g. $f_0$, $f_2$, ...), besides the 
quark diagrams shown in Fig.~\ref{fig:4diaghard}, an additional 
diagram contributes with 2 collinear gluons projecting onto the meson 
wavefunction \cite{Leh00}.}. 
Because the quark helicity is conserved in the hard scattering process, 
the meson acts as a helicity filter. 
In particular, the leading order perturbative QCD predicts \cite{Col97}
that the longitudinally polarized vector 
meson channels (\( \rho ^{0,\pm }_{L} \),
\( \omega _{L} \), \( \phi _{L} \)) are sensitive only to the unpolarized
GPDs (\( H \) and \( E \)) whereas the pseudo-scalar channels (\( \pi ^{0,\pm },\eta ,... \))
are sensitive only to the polarized GPDs (\( \tilde{H} \) and \( \tilde{E} \)).
In comparison to meson electroproduction reactions,
we recall that DVCS depends at the same time on \textit{both} the unpolarized
(\( H \) and \( E \)) and polarized (\( \tilde{H} \) and \( \tilde{E} \))
GPDs. This property makes the hard meson electroproduction reactions 
complementary to the DVCS process, as it provides an additional tool 
to disentangle the different GPDs.

It was shown in Ref.~\cite{Die99} that the leading 
twist contribution to exclusive electroproduction of 
transversely polarized vectors mesons vanishes at all orders
in perturbation theory. 

According to the above discussion, we give predictions only for the
meson electroproduction observables which involve 
a longitudinal virtual photon. The
longitudinal \( \gamma ^{*}_{L}+p\rightarrow M+p \) two-body cross section
\( d\sigma _{L}/dt \) is given by~:
\begin{eqnarray}
{{d\sigma _{L}}\over {dt}}={1\over {16\pi \left( s-m_{N}^{2}\right) \Lambda (s,-Q^{2},m_{N}^{2})}}\; {1\over 2}\sum _{h_{N}}\sum _{h^{'}_{N}}|{\mathcal{M}}^{L}(\lambda _{M}=0,h^{'}_{N};h_{N})|^{2}\; ,
\la{eq:mescross}
\end{eqnarray}
where \( {\mathcal{M}}^{L} \) is the amplitude
for the production of a meson with helicity \( \lambda _{M}=0 \) by a 
longitudinal photon, and where \( h_{N} \), \( h^{'}_{N} \) 
are the initial and final nucleon helicities.
Furthermore in Eq.~(\ref{eq:mescross}), the 
standard kinematic function \( \Lambda (x,y,z) \) is defined by 
\begin{equation}
\Lambda (x,y,z)=\sqrt{x^{2}+y^{2}+z^{2}-2xy-2xz-2yz}\; .
\end{equation}
 which gives \( \Lambda (s,-Q^{2},m_{N}^{2})=2m_{N}|\vec{q}_{L}| \), where
\( |\vec{q}_{L}| \) is the virtual photon momentum in the {\it lab} system. 
As the way to extract \( d\sigma _{L}/dt \) from the 
measured fivefold electroproduction cross
section is a matter of convention, all results for \( d\sigma _{L}/dt \) 
are given with the choice of the flux factor of Eq.~(\ref{eq:mescross}).

\subsubsection{Hard electroproduction of vector mesons}
\label{chap_5_5_1}

\begin{figure}[h]
\vspace{.5cm}
\epsfxsize=6.5 cm
\epsfysize=9.2 cm
\centerline{\epsffile{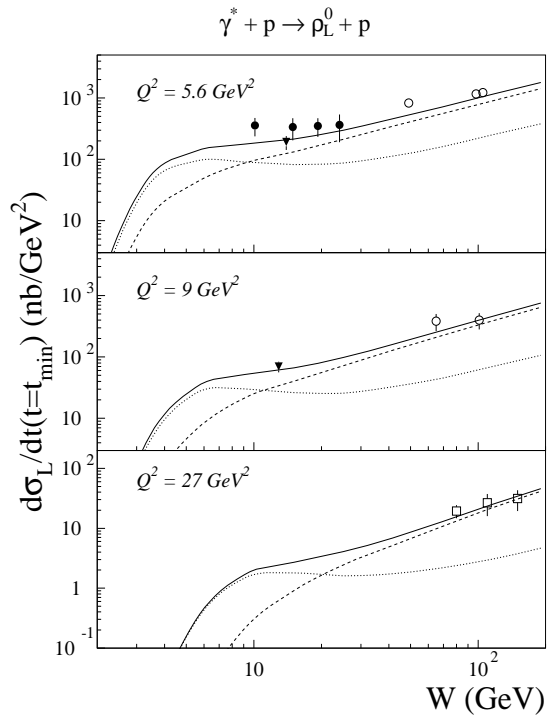}}
\vspace{-.25cm}
\caption{\small Longitudinal forward differential cross section for 
$\rho^0_L$ electroproduction. Calculations compare the quark exchange
mechanism (dotted curves) with the two-gluon
exchange mechanism (dashed curves) and the sum of both (full
curves). The calculations include the corrections due to the intrinsic
transverse momentum dependence for the quark exchange \cite{Vdh99} as
well as for the two-gluon exchange \cite{Fra96}. 
The data are from NMC (triangles) \cite{Arn94}, 
E665 (solid circles) \cite{Ada97}, 
ZEUS 93 (open circles) \cite{Der95} 
and ZEUS 95 (open squares) \cite{Bre98}. 
Figure from Ref.~\cite{Vdh99}.}
\label{fig:rhotot2}
\end{figure}

In this section, we consider the vector meson electroproduction processes 
$\gamma ^{*}_{L} + N \rightarrow V_L + N$ at large $Q^2$ on the nucleon $N$, 
where $V_L$ ($\rho^0_L$, $\rho^+_L$, $\omega_L$, ...) 
denotes the produced vector meson with longitudinal polarization. 
The longitudinal polarization component of the vector meson is
obtained experimentally through its decay angular distribution. 
$\rho^0$ electroproduction data both at high $W$ 
( $ 30 < W < 140$ GeV ) \cite {Adl00} as well as at intermediate 
values of $W$ ( $ 3.8 < W < 6.5$ GeV ) \cite{Ack99,Tyt01} 
support the fact that s-channel helcity conservation (SCHC) holds to
good accuracy. This implies that a longitudinally polarized vector
meson originates from a longitudinally polarized photon. Therefore,
when SCHC holds, one can access the leading order longitudinal
photon cross section for vector meson production solely by measuring the decay
angular distribution of the vector meson, i.e. without performing a
Rosenbluth separation of longitudinal and transverse photon cross
sections.  
\newline
\indent
In the valence region, the vector meson amplitude in leading order in $Q$ 
is obtained from the hard scattering diagrams of Fig.~\ref{fig:4diaghard}. 
At small $x_B$, $\rho^0$ and $\omega$ electroprodution can also proceed 
through a two gluon exchange mechanism, 
as discussed in \cite{Bro94,Fra96,Rad96b} to which we refer for details.  
We restrict ourselves in the following discussion 
to the quark exchange contribution, and will mostly show results 
in the valence region. 
\newline
\indent
The resulting amplitudes \( {\mathcal{M}}^{L} \) for electroproduction 
of longitudinally polarized vector mesons 
were calculated in Refs.~\cite{Man98a,Vdh98},  
yielding the following expression~: 

\begin{eqnarray}
{\mathcal{M}}^{L}_{V_{L}}&\,=\,& -ie\, {4\over 9}\, {1\over {Q}}\; 
\left[ \, \int _{0}^{1}dz{{\Phi_{V_L }(z)}\over z}\right] 
\; {1 \over 2} \; (4\pi \alpha _{s})\; 
\nonumber\\
&\times&
\left\{ A_{V_L  N} \;\bar{N}(p^{'})\, {\Dirac n} \,N(p)
\,+\, B_{V_L  N} \; \bar{N}(p^{'}) \, i\sigma^{\kappa\lambda} 
{{n_{\kappa }\Delta _{\lambda }}\over {2 \, m_{N}}} \,N(p)\right\} \, ,
\la{eq:rhoampl} 
\end{eqnarray}
where $\Phi_{V_L }(z)$ is the distribution amplitude (DA) for a 
longitudinally polarized vector meson, and where $\alpha_s$ is the 
strong coupling constant.  
\newline
\indent
For $\rho^0_L \, p$ electroproduction on the proton, 
the amplitudes $A$ and $B$ 
in Eq.~(\ref{eq:rhoampl}) are given by \cite{Man98a,Vdh98}~:
\be
A_{\rho^0_L \, p} \,&=&\, \int_{-1}^1 dx \; 
{1 \over {\sqrt 2}} \, {\left(e_u \ H^u \,-\, e_d \ H^d\right)} 
\; \left\{ {{1} \over {x - \xi + i \epsilon}}
+ {{1} \over {x + \xi - i \epsilon}} \right\} , 
\la{eq:arhoo} \\
B_{\rho^0_L \, p} \,&=&\, \int_{-1}^1 dx \;
{1 \over {\sqrt 2}} \, {\left(e_u \ E^u \,-\, e_d \ E^d\right)} 
\;\left\{ {{1} \over {x - \xi + i \epsilon}}
+ {{1} \over {x + \xi - i \epsilon}} \right\} ,
\la{eq:brhoo}
\ee
where $e_u = +2/3$  ($e_d = -1/3$) are the $u$ ($d$) quark 
charges respectively.  
\newline
\indent
For $\rho^+_L \, n$ electroproduction on the proton, 
the amplitudes $A$ and $B$ are given by 
\cite{Man99a,Vdh99}~:
\be
A_{\rho^+_L \, n} \,&=&\, - \, \int_{-1}^1 dx \; 
{\left(H^u \,-\, H^d\right)}
\; \left\{ {{e_u} \over {x - \xi + i \epsilon}}
+ {{e_d} \over {x + \xi - i \epsilon}} \right\} , 
\la{eq:arhop} \\
B_{\rho^+_L \, n} \,&=&\, - \, \int_{-1}^1 dx \;
{\left(E^u \,-\, E^d\right)}
\;\left\{ {{e_u} \over {x - \xi + i \epsilon}}
+ {{e_d} \over {x + \xi - i \epsilon}} \right\} .
\la{eq:brhop}
\ee
For $\omega_L \, p$ electroproduction on the proton, 
the amplitudes $A$ and $B$ are given by 
\cite{Man98a,Vdh98}~:
\be
A_{\omega_L \, p} \,&=&\, \int_{-1}^1 dx \; 
{1 \over {\sqrt 2}} \, {\left(e_u \ H^u \,+\, e_d \ H^d\right)}
\; \left\{ {{1} \over {x - \xi + i \epsilon}}
+ {{1} \over {x + \xi - i \epsilon}} \right\} , 
\la{eq:aome} \\
B_{\omega_L \, p} \,&=&\, \int_{-1}^1 dx \;
{1 \over {\sqrt 2}} \, {\left(e_u \ E^u \,+\, e_d \ E^d\right)}
\;\left\{ {{1} \over {x - \xi + i \epsilon}}
+ {{1} \over {x + \xi - i \epsilon}} \right\} .
\la{eq:bome}
\ee
\newline
\indent
One sees from Eqs.~(\ref{eq:arhoo}-\ref{eq:bome}) that the amplitudes for 
the hard electroproduction of longitudinally polarized 
$\rho^0$, $\rho^+$ and $\omega$ mesons involve 
different flavor combinations of the GPDs $H^u(x, \xi, t)$, $H^d(x, \xi, t)$, 
and analogously for $E^u(x, \xi, t)$, $E^d(x, \xi, t)$. 
Therefore, measuring those different channels allows to make a 
flavor decomposition of the GPDs $H^q$ and $E^q$.
\newline
\indent
In order to extract the GPDs from cross sections of 
hard meson electroproduction, one needs to specify the meson distribution 
amplitude $\Phi_{V_L }(z)$ entering in Eq.~(\ref{eq:rhoampl}). 
For the longitudinally polarized $\rho^0$ meson, an 
experimental estimate of the second Gegenbauer coefficient of its 
distribution amplitude has been given for the first time 
in Ref.~\cite{Cle00}, and was found to 
be $a_2^{(\rho)}=-0.1\pm 0.2$ at a scale $\langle \mu^2\rangle=21.2$~GeV$^2$.
Unfortunately the precision in the determination of $a_2^{(\rho)}$
is still too low to discriminate between different model predictions
for this quantity  ( QCD sum rules:
$0.18\pm 0.1$ \cite{Bal96}, $0.08\pm 0.02$ \cite{Bak98} and
instanton model: $-0.14$ \cite{Pol99a} all at a scale around 
$\mu \approx 1$ GeV ).
However all these analyses point to relatively small values for the second 
Gegenbauer coefficient and favor therefore a DA for the longitudinally 
polarized \( \rho  \) meson that is rather close to its asymptotic form. 
\newline
\indent
In view of these findings, we use in all calculations 
shown below, the asymptotic DA for longitudinally polarized vector mesons~: 
\begin{equation}
\Phi_{V_L}(z) \,=\, f_{V}\; 6 \, z \, (1-z) \; ,
\la{eq:vectormda}
\end{equation}
 with \( f_{\rho }\approx  \) 0.216 GeV and 
\( f_{\omega }\approx  \) 0.195 GeV, determined from the electromagnetic decay
\( V\rightarrow e^{+}e^{-} \). 
\newline
\indent 
One sees from Eq.~(\ref{eq:rhoampl}) that the 
leading order electroproduction amplitude for longitudinally polarized 
vector mesons is of order $1/Q$, 
in contrast to the leading order DVCS amplitude which is constant in $Q$.  
This difference is due to the additional gluon propagator 
in the hard scattering amplitude for the leading order 
meson electroproduction amplitude (see Fig.~\ref{fig:4diaghard}).  
This also leads to the dependence of the leading order amplitude on 
the strong coupling constant $\alpha_s(\sim Q^2)$. At large scales $Q^2$, 
the running coupling constant is given by its 
expression from perturbative analyses.   
However, the average virtuality of the exchanged gluon in the leading order
meson electroproduction amplitudes can be considerably less than the external
$Q^2$, which is therefore not the ``optimal'' choice for the 
renormalization scale. This puts some caveats on the applicability 
of the leading order meson electroproduction amplitude 
to analyse absolute cross sections at accessible scales. 
The corrections to the leading order vector meson electroproduction amplitude 
(both in powers of $1/Q$ and in $\alpha_s$) 
is still an open question to be addressed in  future work. 
\newline
\indent
We next compare the results for $\rho^0_L$ electroproduction 
cross sections to the available data. Because the leading order amplitude 
of Eq.~(\ref{eq:rhoampl}) is of order $1/Q$, it predicts a $1/Q^6$ scaling 
behavior for the cross section $d \sigma_L / dt$. By measuring the 
$Q^2$ behavior of the meson electroproduction cross section,  
one may study how fast one approaches the scaling regime predicted by 
perturbative QCD. 
In particular, the measurement of hard electroproduction 
reactions in the region 
\( Q^{2}\approx 1-20 \) GeV\( ^{2} \), which is accessible experimentally, 
may provide information on 
the importance of power corrections to the leading order amplitudes. 
One source of power corrections is evident from the structure of the
matrix element of Eq.~(\ref{eq:qsplitting}) which defines the GPDs,
where the quarks are taken at zero transverse separation. This amounts
to neglect, at leading order, the quark's transverse
momentum compared to its large longitudinal (+ component) momentum. 
A first study of these corrections due to the quark's intrinsic transverse
momentum, assuming a gaussian dependence,  
has been performed in Ref.~\cite{Vdh99}. 
At virtualities $Q^2$ of a few GeV$^2$, these corrections were 
found to lead to a sizeable reduction of the cross section
\cite{Vdh99}, e.g. a factor 4 reduction at $x_B \approx 0.3$ 
and $Q^2 \approx 4$ GeV$^2$ ( see Ref.~\cite{Vdh99} for details ).  
A comparison of these calculations to existing 
$\rho^0_L$ electroproduction data in the few GeV$^2$ range is shown in 
Fig.~\ref{fig:rhotot2}.  

\begin{figure}[h]
\vspace{.5cm}
\epsfxsize=9.cm
\centerline{\epsffile{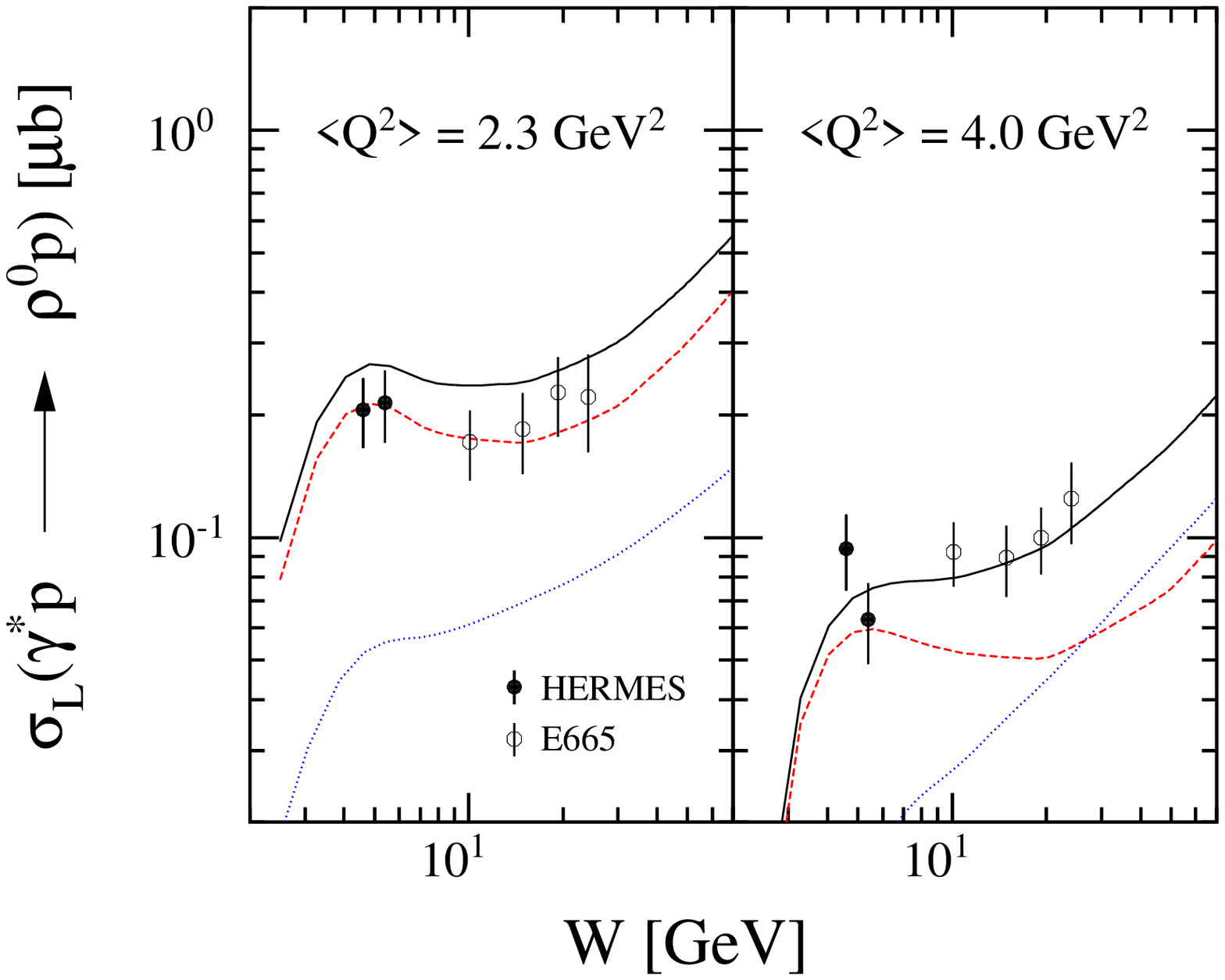}}
\vspace{-.1cm}
\caption{\small Longitudinal cross section for 
$\rho^0_L$ electroproduction. Calculations compare the quark exchange
mechanism (dashed curves) with the two-gluon
exchange mechanism (dotted curves) and the sum of both (full
curves). The calculations include the corrections due to intrinsic
transverse momentum dependence as calculated in Ref.~\cite{Vdh99}. 
The data are from E665 (open circles) \cite{Ada97} 
and HERMES (solid circles) \cite{Air00}. Figure from Ref.~\cite{Air00}.}
\label{fig:rhohermes}
\end{figure}

As the  \( \gamma^{*}_L\, p\longrightarrow \rho ^{0}_{L}\, p \) 
reaction has mostly been measured at small values of $x_B$ 
\cite{Der95,Bre98,Adl00} ( or equivalently large values of the 
c.m. energy $W$ of the $\gamma^* p$ system 
\footnote{$W$ is expressed in terms of 
$x_B$ as~: $W^2 = m_N^2 \,+\, Q^2 \, (1 - x_B)/x_B$. } ), 
the calculations 
in Fig.~\ref{fig:rhotot2} for $\rho^0_L$ electroproduction can reveal 
how the valence region is approached, with decreasing value of $W$, 
where one is sensitive to the quark GPDs. 
For the purpose of this discussion, we call the mechanism
proceeding through the quark GPDs of Fig.~\ref{fig:factmeson}, 
the quark exchange mechanism (QEM). 
Besides the QEM, $\rho^0$ electroproduction at large $Q^2$ and small
$x_B$ proceeds predominantly through a perturbative two-gluon exchange
mechanism (PTGEM) as studied in Ref.~\cite{Bro94,Fra96}. 
To compare to the data at intermediate $Q^2$, the
power corrections due to the parton's intrinsic transverse momentum
dependence were implemented in both mechanisms 
using the estimate of Ref.~\cite{Vdh99}. The comparison of the calculations 
with the data in Fig.~\ref{fig:rhotot2} show that the PTGEM
explains well the fast increase of the cross section 
at high c.m. energy ($W$), 
but substantially underestimates the data at the lower energies. This is
where the QEM is expected to contribute since $x_B$ is then in the
valence region. The results including the QEM describe 
the change of behavior of the data at lower $W$ quite nicely. 
\newline
\indent
Recently, $\rho^0_L$ data have also been obtained by the HERMES
Collaboration for $Q^2$ up to 5 GeV$^2$ and around 
$W \approx$ 5 GeV \cite{Air00}. 
The comparison of the calculations for $\rho^0_L$  
with those data is shown in Fig.~\ref{fig:rhohermes}. 
One sees from Fig.~\ref{fig:rhohermes} a clear dominance of the QEM 
in the intermediate $W$ range as predicted in Refs.~\cite{Vdh98,Vdh99}. 
When including the model estimate for the power corrections, 
a fairly good agreement with these longitudinal
$\rho^0$ electroproduction data \cite{Air00} is obtained, 
as seen from Fig.~\ref{fig:rhohermes}.

A dedicated experiment to investigate the onset of the scaling behavior of 
$\rho^0_L$ electroproduction in the valence region 
($Q^2 \approx 3.5$ GeV$^2$ and $x_B \approx$ 0.3) is also planned in the very 
near future \cite{Guid98} using the CLAS detector at JLab. 
\newline
\indent
Besides the $\rho^0_L$ electroproduction cross section, the 
$\omega_L$ electroproduction cross section has also been measured 
recently by the HERMES Collaboration \cite{Tyt01} at intermediate values of 
$W \approx 5$~GeV. 
The measurement of the ratio  of $\omega_L / \rho^0_L$ 
electroproduction cross sections at these lower values of $W$ allows 
one to see the departure from the diffractive regime, 
at small $x_B$, where this cross section ratio 
is given by $\omega_L : \rho^0_L = 1 : 9$. 
In the valence region, around $x_B \approx 0.3$, the 
quark exchange mechanism of Fig.~\ref{fig:factmeson} leads to a 
ratio $\omega_L : \rho^0_L \approx 1 : 5$ \cite{Col97,Vdh99}, 
which is nearly twice as large as in the diffractive region. 
This larger ratio of around 0.2 for the 
$\omega_L / \rho^0_L$ cross sections at larger values of 
$x_B$ is in good agreement with the recent HERMES measurements \cite{Tyt01}, 
and also supports the dominance of the quark exchange process of 
Fig.~\ref{fig:factmeson}.
\newline
\indent
It is also instructive to compare the $t$-dependences 
for vector meson electroproduction cross sections. 
In particular we found that a factorized ansatz in $t$ for the GPD 
as in Eq.~(\ref{eq:factt}) gives a much less steep $t$-dependence 
around $x_B \simeq 0.1$ as the unfactorized (Regge motivated) 
form of Eq.~(\ref{eq:ddunpolregge}). 
In particular, the unfactorized form of Eq.~(\ref{eq:ddunpolregge}) 
leads, in the region $x_B \simeq 0.1$, to an exponential $t$ dependence 
$d \sigma_L \sim e^{-B \, |t|}$ in the small $-t$ region.
When using the value $\alpha '$ = 0.8 GeV$^{-2}$ for the Regge 
slope parameter in Eq.~(\ref{eq:ddunpolregge}), we found for the 
$t$-slope of the (twist-2) cross section $d \sigma_L$ 
for $\rho^0_L$ electroproduction the value~:  
$B_{\rho^0} \simeq$ 6.5 GeV$^{-2}$. 
This value is remarkably close to the experimentally extracted 
slope parameters for $\rho^0$ electroproduction on the proton which has been 
measured at HERMES \cite{Tyt01} in the kinematic range 
$0.7 < Q^2 < 5.0$ GeV$^2$ and $4.0 < W < 6.0$ GeV, and   
which was found to be \cite{Tyt01}~: 
\begin{eqnarray}
B_{\rho^0} \,=\, 7.08 \,\pm\, 0.14 \, (\mathrm{stat.}) \, 
\mbig{$^{+0.58}_{-0.08}$} \, (\mathrm{syst.}) \; \mathrm{GeV}^{-2} \, .
\end{eqnarray}
This hints that a Regge type (unfactorized) 
form as in Eq.~(\ref{eq:ddunpolregge}) might be a reasonable starting point 
for a more realistic parametrization of the $t$-dependence of GPDs. 
The $t$-slope of the vector meson electroproduction cross sections was
also addressed in Ref.~\cite{Fra96} where it was shown that
higher-twist contributions are able to explain the change of the $t$-slope
with $Q^2$. In particular the value of $B_{\rho^0}$ decreases with
increasing $Q^2$. 
\newline
\indent
Although the data for the vector meson electroproduction channels 
at intermediate values of $x_B$ discussed above 
point towards the dominance of the quark 
exchange mechanism, it would be too premature at the present stage 
to try to extract quark GPDs from these 
vector meson electroproduction cross sections. 
To reach this goal, one first 
needs to get a better theoretical control over 
the power (higher-twist) corrections, which is an important topic for  
future work. 
\newline
\indent
Besides the cross section $\sigma_L$, the second
observable which involves only longitudinal amplitudes and which is
therefore a leading order observable for hard exclusive
meson electroproduction, is the transverse spin asymmetry, ${\cal A}_{V_L N}$ 
(TSA) for a proton target polarized perpendicular to the reaction
plane (or the equivalent recoil polarization observable). 
For definiteness let us consider the following (azimuthal) asymmetry~:
\be
{\cal A}=\frac{1}{|S_\perp|}
\frac{
\int_0^{\pi}d\beta \, \sigma (\beta)-
\int_{\pi}^{2\pi}d\beta \, \sigma(\beta)
}{\int_{0}^{2\pi}d\beta \, \sigma(\beta)} \, ,
\la{eq:tsa}
\ee
where $\beta$ is the angle between the transverse proton spin 
$S_\perp$ and the reaction plane 
(spanned by the virtual photon and the produced meson). 
\newline
\indent
For the electroproduction of longitudinally polarized vector mesons, 
induced by a longitudinal virtual photon, we 
found this transverse spin asymmetry to be given by 
(~compare with the corresponding Eq.~(\ref{eq:piasymm}) for the pion )~:
\be                                         
{\cal A}_{V_L N} = - \, {{2 \, |\Delta_\perp|} \over {\pi}} \,
\frac{{\rm{Im}} (A B^*) / m_N}
{|A|^2 \, (1-\xi^2) - |B|^2 \, \left(\xi^2 + t / (4 m_N^2) \right) 
- {\rm{Re}}(AB^*) \, 2 \, \xi^2} \, ,
\la{eq:rhopasy}
\ee
which is proportional to the modulus 
$|\Delta_\perp|$ of the perpendicular component of the momentum transfer 
$\Delta$ of Eq.~(\ref{eq:dvcsxi}), and 
where $A$ and $B$ are given by Eqs.~(\ref{eq:arhoo}-\ref{eq:bome}) 
for the different vector meson channels. 
One sees that the transverse spin asymmetry is proportional to the 
imaginary part of the {\it interference} of the amplitudes $A$ and $B$, 
which contain the GPDs $H$ and 
$E$ respectively. Therefore, it depends {\it linearly} on the GPD $E$. 
Note that in contrast, both in the DVCS cross sections and SSA as well as 
in the longitudinal cross sections for 
(longitudinally polarized) vector mesons, 
the GPD $E$ only enters besides a large contribution of the GPD $H$. 
In order to increase the sensitivity to the GPD $E$ 
in those observables, one needs to increase 
the value of the momentum transfer $\Delta$, because the function $E$ is 
kinematically suppressed in the amplitudes when $\Delta \to 0$. 
Since the value of 
$\Delta$ should remain small however in comparison with the hard scale $Q$ 
in order not to be totally dominated by higher twist effects, 
the DVCS cross sections, SSA and vector meson cross sections give only a 
limited handle in the extraction of the GPD $E$. 
In this context, 
the transverse spin asymmetry of Eq.~(\ref{eq:rhopasy}) provides   
a unique observable to extract the GPD $E$.  
\newline
\indent
Besides, also the theoretical uncertainties and open questions 
which we discussed for the 
meson electroproduction cross sections largely disappear for the 
transverse spin asymmetry. 
Indeed, because the transverse spin asymmetry 
involves a ratio of cross sections, 
the dependence on the distribution amplitude and the strong coupling constant 
in Eq.~(\ref{eq:rhoampl}) drops out in this ratio. This also suggest that the 
transverse spin asymmetry is less sensitive to pre-asymptotic effects and 
that the leading order expression of Eq.~(\ref{eq:rhopasy})
is already accurate  
at accessible values of $Q^2$ (in the range of a few GeV$^2$).

\begin{figure}[hp]
\epsfxsize=8.3cm
\centerline{\epsffile{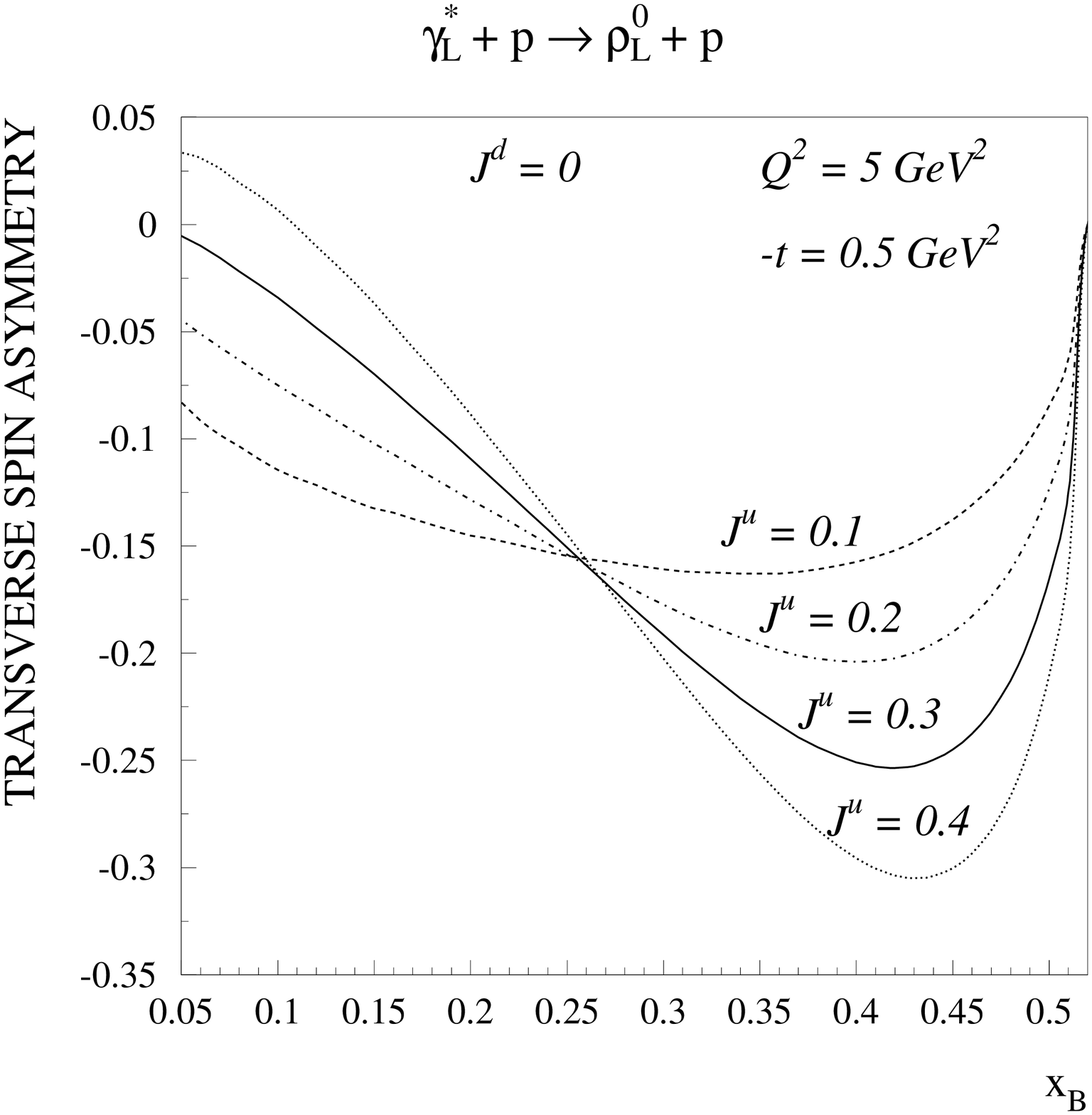}}
\caption{\small Transverse spin asymmetry for the 
$\gamma^*_L \vec p \to \rho_L^0 p$ reaction at $t$ = -0.5 GeV$^2$ and
$Q^2$ = 5 GeV$^2$ . 
The estimates are given using the three-component model for $E^u$ and
$E^d$ as described in Sec.~\ref{chap5_2c}, 
where $J^d$ was fixed to the value $J^d = 0$. 
The curves show the sensitivity to the value of  
$J^u$ as indicated on the curves, 
which enters in the model for the GPD $E^u$, shown in Fig.~\ref{fig:up_E}.
For the GPDs $H^u$, $H^d$ and for the valence part of $E^u$, $E^d$, 
the values $b_{val} = b_{sea} = 1$ have been used.} 
\label{fig:rhoo_asy_0p5}
\epsfxsize=8.3cm
\centerline{\epsffile{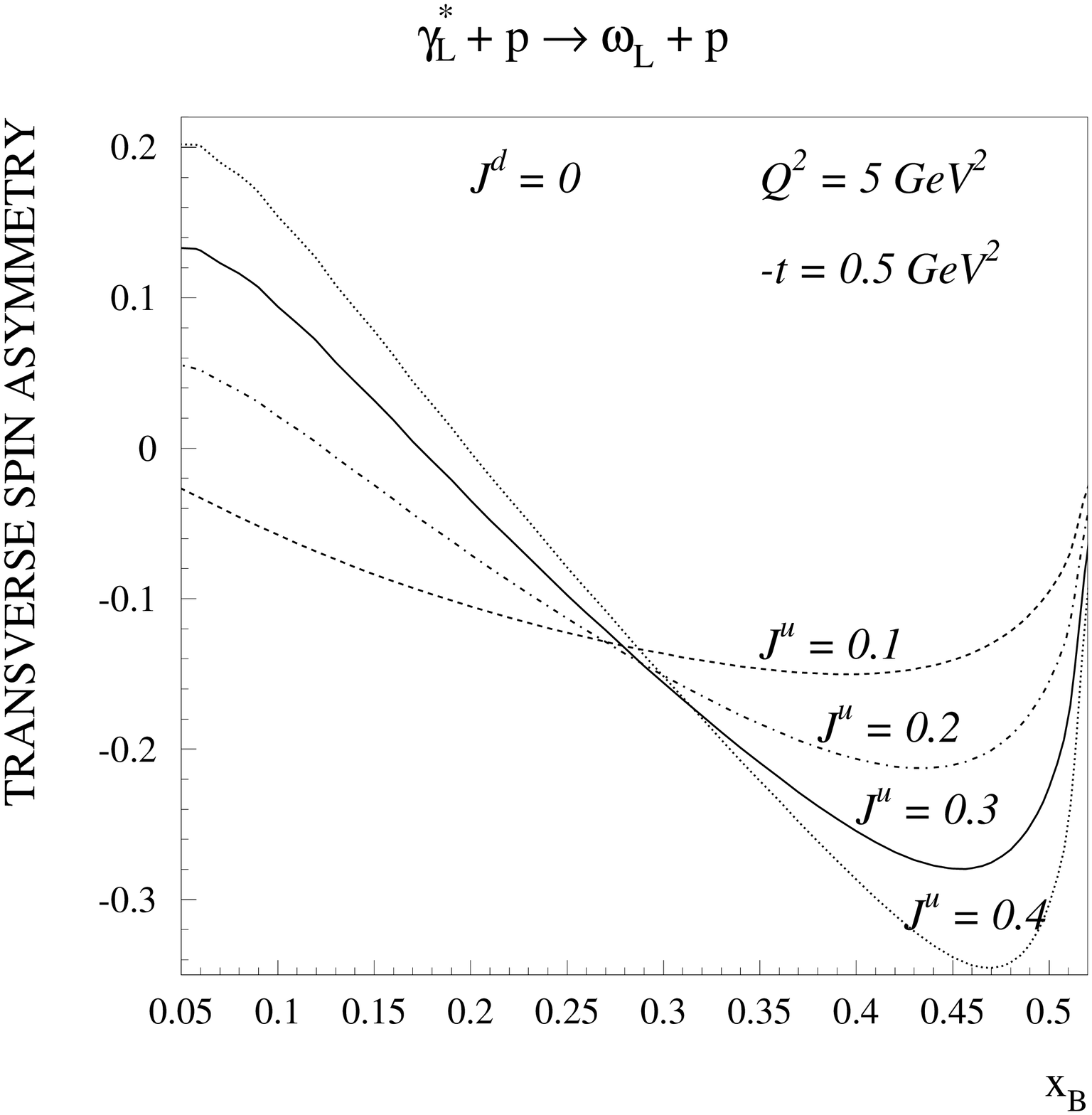}}
\caption{\small Same as Fig.~\ref{fig:rhoo_asy_0p5} but for the  
$\gamma^*_L \vec p \to \omega_L p$ reaction.} 
\label{fig:ome_asy_0p5}
\end{figure}

\begin{figure}[h]
\vspace{-.75cm}
\epsfxsize=7cm
\centerline{\epsffile{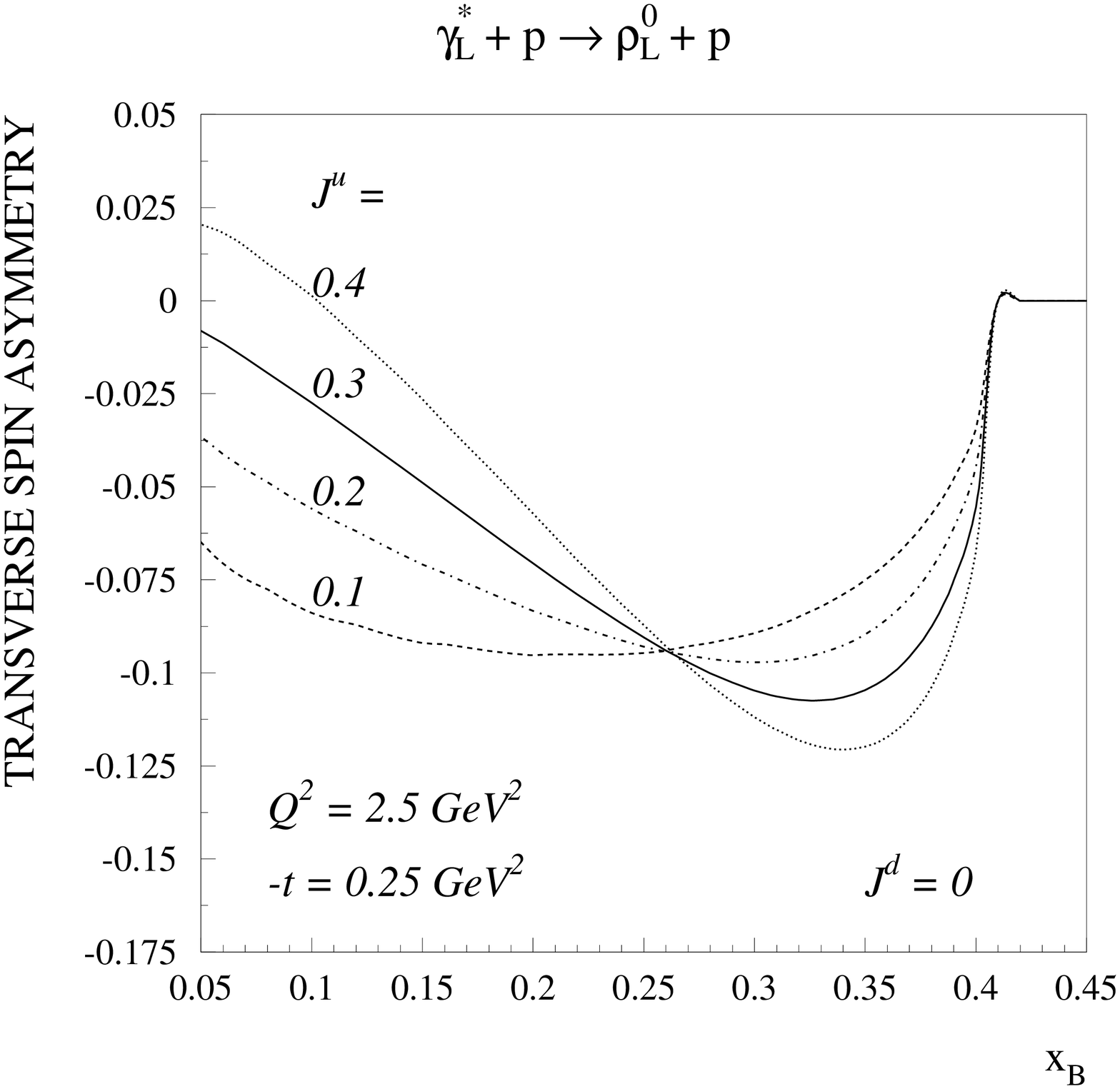}}
\vspace{-.25cm}
\caption{\small Same as Fig.~\ref{fig:rhoo_asy_0p5}, but for 
$t$ = -0.25 GeV$^2$ and $Q^2$ = 2.5 GeV$^2$.} 
\label{fig:rhoo_asy_0p25}
\end{figure}

Due to its linear dependence on the GPD $E$, the transverse spin asymmetry 
for longitudinally polarized vector mesons opens up the perspective to 
extract from it the total angular momentum contributions $J^u$ and 
$J^d$ of the $u-$ and $d$-quarks to the proton spin. 
Indeed, in the three-component parametrization for 
$E^u$ and $E^d$ presented in 
Eqs.~(\ref{eq:parame}, \ref{eq:edd}), the values of $J^u$ and $J^d$ 
enter as free parameters (see Figs.~\ref{fig:up_E}, \ref{fig:down_E}).
We therefore investigate in the following the dependence of the 
transverse spin asymmetry for $\rho^0_L$, $\omega_L$ and $\rho^+_L$ 
electroproduction on $J^u$ and $J^d$. 
Due to the different $u$- and $d$-quark content of the vector mesons, 
the asymmetries for the $\rho^0_L$, $\omega_L$ and $\rho^+_L$ channels 
are sensitive to different combinations of $J^u$ and $J^d$. 
From Eqs.~(\ref{eq:arhoo}-\ref{eq:bome}), one finds that the  
$\rho^0_L$ production is mainly sensitive to the combination $2 J^u + J^d$, 
$\omega_L$ to the combination $2 J^u - J^d$, and 
$\rho^+_L$ to the isovector combination $J^u - J^d$.
\newline
\indent
In Figs.~\ref{fig:rhoo_asy_0p5} and \ref{fig:ome_asy_0p5} we firstly show 
the $x_B$ dependence of the transverse spin asymmetry for 
$\rho^0_L$ and $\omega_L$ production at $Q^2$ = 5 GeV$^2$ and 
$-t$ = 0.5 GeV$^2$. In these figures, we fixed the value of $J^d = 0$, 
which is close to the valence estimate of Eq.~(\ref{eq:jd}), for which 
the VM part of $E^d$ vanishes (see Fig.~\ref{fig:down_E}).
For this value of $J^d$, we then demonstrate 
in Figs.~\ref{fig:rhoo_asy_0p5},\ref{fig:ome_asy_0p5} 
the sensitivity of the $\rho^0_L$ and $\omega_L$ transverse spin asymmetries 
to different values of $J^u$. Note that the value $J^u \approx 0.3$ would 
correspond to the valence estimate of Eq.~(\ref{eq:ju}), as is also 
seen from the vanishing of the VM part of $E^u$ in Fig.~\ref{fig:up_E} 
for this value. 
\newline
\indent
From Figs.~\ref{fig:rhoo_asy_0p5},\ref{fig:ome_asy_0p5}, one sees that 
the transverse spin asymmetries for $\rho^0_L$ and $\omega_L$ 
electroproduction display a pronounced sensitivity to $J^u$ around 
$x_B \approx 0.4$, where asymmetries are predicted in the -15 \% to -30 \% 
range according to the value of $J^u$. 
This transverse spin asymmetry for $\rho^0_L$ and $\omega_L$ 
at large $x_B$ gets enhanced by the (isoscalar) D-term,  which contributes to 
both processes. In particular,  it gives a large real part to the amplitude 
$B$ in Eq.~(\ref{eq:rhopasy}) which is then multiplied by the imaginary part 
of the amplitude $A$, proportional to $H(\xi, \xi, t)$. 
At these larger values of $x_B$ a precise extraction of the values of 
$J_u$ and $J_d$ from the $\rho^0_L$ and $\omega_L$ asymmetries requires a 
good knowledge of the D-term which can be measured e.g. through the charge 
asymmetry for DVCS as discussed in Sec.~\ref{chap5_3_4}. 
At smaller values of $x_B$, where the D-term contribution shrinks away, 
the sensitivity to $J^u$ and $J^d$ arises through both the real and imaginary 
parts of $B$, which depend on $J^u$ through the VM contribution to $E^u$ 
(giving a purely real part to $B$) 
and the valence contribution to $E^u$ (giving predominantly an imaginary 
contribution to $B$). 
\newline
\indent
For comparison, we show in Fig.~\ref{fig:rhoo_asy_0p25} 
the transverse spin asymmetry for $\rho^0_L$ 
at the smaller value of $-t$ = 0.25 GeV$^2$ 
(the asymmetry of Eq.~(\ref{eq:rhopasy}) is proportional to $|\Delta_\perp|$) 
and $Q^2$ = 2.5 GeV$^2$. Those values are at present already  
accessible at HERMES. It will therefore be very 
interesting to provide a first measurement of this asymmetry 
in the near future for a transversely polarized target. 

Although we showed the transverse spin asymmetries in 
Figs.~\ref{fig:rhoo_asy_0p5},\ref{fig:ome_asy_0p5}, for the value 
$J^d = 0$, one can vary this value and exploit the different dependence of the 
$\rho^0_L$ (sensitive to $2 J^u + J^d$) and 
$\omega_L$ (sensitive to $2 J^u - J^d$) transverse spin asymmetries to  
extract information on the different flavor contributions. 

\begin{figure}[hp]
\epsfxsize=8.3cm
\centerline{\epsffile{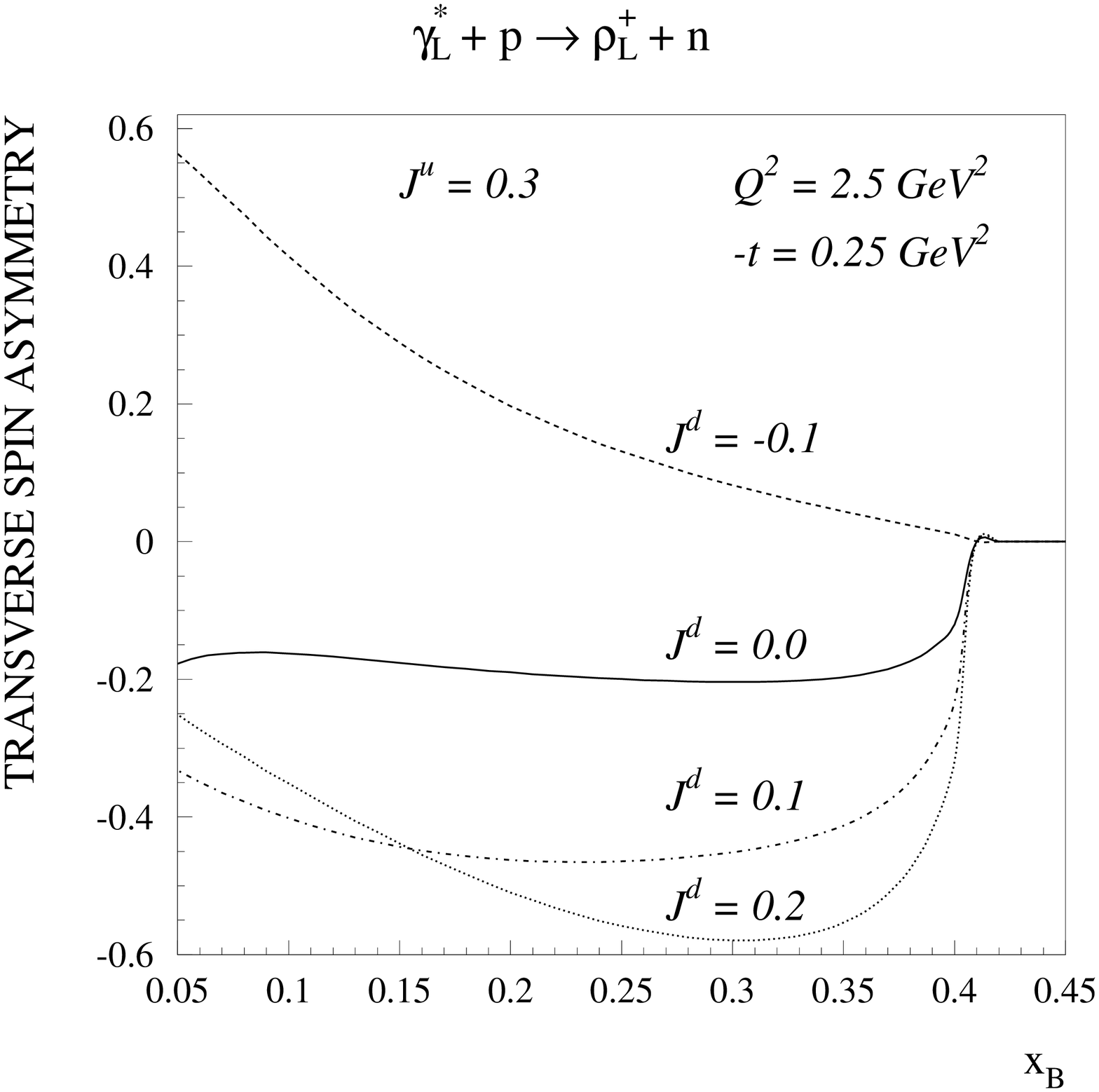}}
\caption{\small Transverse spin asymmetry for the 
$\gamma^*_L \vec p \to \rho_L^+ n$ reaction at $t$ = -0.25 GeV$^2$ and
$Q^2$ = 2.5 GeV$^2$ . 
The estimates are given using the three-component model for $E^u$ and
$E^d$, where $J^u$ was fixed to the value $J^u = 0.3$. 
The curves show the sensitivity to the value of  
$J^d$ as indicated on the curves, 
which enters in the model for the GPD $E^d$, shown in 
Fig.~\ref{fig:down_E}.} 
\label{fig:rhop_asy_0p25}
\epsfxsize=8.3cm
\centerline{\epsffile{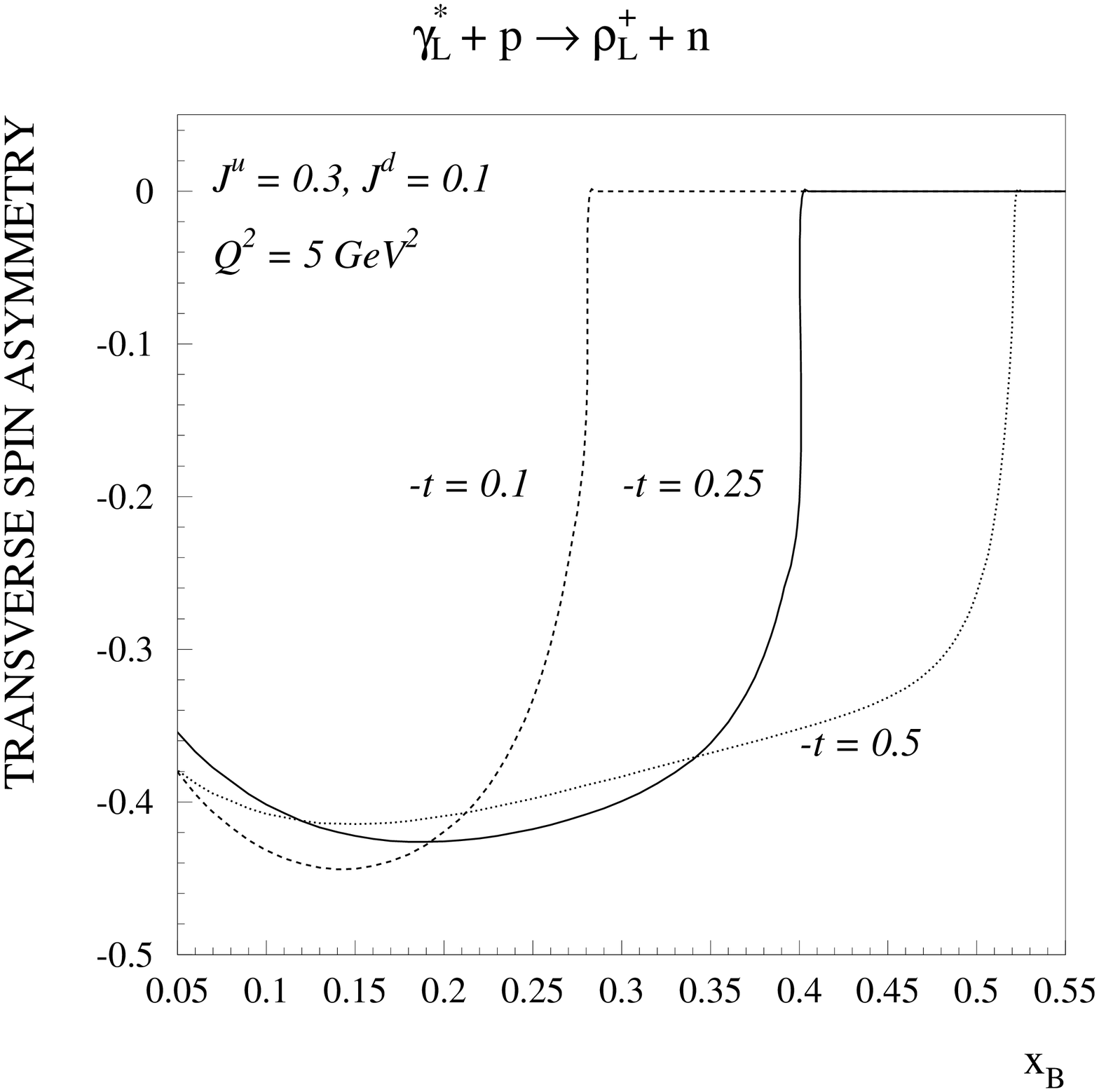}}
\caption{\small Transverse spin asymmetry for the 
$\gamma^*_L \vec p \to \rho_L^+ n$ reaction at  
$Q^2$ = 5 GeV$^2$ for different values of $t$, as indicated on the
curves (in GeV$^2$). 
The estimates are given using the three-component model for $E^u$ and
$E^d$, where $J^u$ and $J^d$ were fixed as indicated on the figure.} 
\label{fig:rhop_asy_tdep}
\end{figure}
\newpage

One is directly sensitive to the difference $J^u - J^d$ 
by measuring the transverse spin asymmetry for $\rho^+_L$ production. 
The corresponding asymmetry is shown in Fig.~\ref{fig:rhop_asy_0p25}
for a value of $J^u$ = 0.3 and for different values of $J^d$. 
One observes that this asymmetry can be very large over a wide range 
in $x_B$. Because for $\rho^+_L$, the isoscalar D-term contribution 
is absent, this large value for the asymmetry mainly originates from the 
product of the imaginary part of the amplitude 
$A$ ( proportional to $H(\xi,\xi,t)$ ) 
with the real part of the amplitude $B$ ( mainly proportional to the VM part 
of $E^u - E^d$ which contains the sensitivity to $J^u - J^d$ ). 
\newline
\indent
In Fig.~\ref{fig:rhop_asy_tdep}, we also show the 
$t$-dependence of the $\rho^+_L$ 
transverse spin asymmetry. One observes how by increasing the value of $-t$, 
the kinematical accessible region over which one can access this asymmetry 
gets larger. 
\newline
\indent
Summarizing this section, we have identified 
the transverse spin asymmetry for longitudinally polarized vector mesons 
as a unique observable which depends {\it linearly} on the GPD $E$. 
These asymmetries for $\rho^0_L$, $\omega_L$ and $\rho^+_L$ 
display a large sensitivity to the total angular momentum contributions 
$J^u$ and $J^d$ of the $u$- and $d$-quarks to the proton spin. As these 
asymmetries represent a ratio of cross sections, several theoretical 
uncertainties drop out. Also from the experimental point of view, 
such ratios are much easier to measure than absolute cross sections. 
A first measurement of these asymmetries will therefore be of high interest 
in the quest of our understanding of the quark contributions 
to the proton spin.

\subsubsection{Hard electroproduction of pions}
\label{chap_5_5_2}

In this section, we discuss the hard pion electroproduction processes 
$\gamma ^{*}_{L} + p \rightarrow \pi^+ + n$ and  
$\gamma ^{*}_{L} + p \rightarrow \pi^0 + p$.
\newline
\indent
The pion electroproduction amplitude \( {\mathcal{M}}^{L}_\pi \) 
for a longitudinal virtual photon is also obtained 
in leading order in $Q$ from the hard scattering 
diagrams of Fig.~\ref{fig:4diaghard}. This leads to the amplitude~: 

\begin{eqnarray}
{\mathcal{M}}^{L}_\pi &\,=\,& -ie\, {4\over 9}\, {1\over {Q}}\; 
\left[ \, \int _{0}^{1}dz{{\Phi_{\pi}(z)}\over z}\right] 
\; {1 \over 2} \; (4\pi \alpha _{s})\; 
\nonumber\\
&\times&
\left\{ A_{\pi N} 
\; \bar{N}(p^{'})\, {\Dirac n} \gamma_{5} \, N(p)
\,+\, B_{\pi N} 
\; \bar{N}(p^{'}) \, \gamma _{5} {{\Delta \cdot n}\over {2m_{N}}} \,N(p) 
\right\} \, ,
\la{eq:piampl} 
\end{eqnarray}
where $\Phi_{\pi}(z)$ is the pion distribution amplitude, 
and where $\alpha_s$ is the strong coupling constant.  
\newline
\indent
For $\pi^0 \, p$, the amplitudes $A$ and $B$ 
in Eq.~(\ref{eq:piampl}) are given by \cite{Man98a,Vdh98}~:
\be
A_{\pi^0 \, p} \,&=&\, \int_{-1}^1 dx \; 
{1 \over {\sqrt 2}} \, 
{\left(e_u \ \widetilde H^u \,-\, e_d \ \widetilde H^d\right)} 
\; \left\{ {{1} \over {x - \xi + i \epsilon}}
+ {{1} \over {x + \xi - i \epsilon}} \right\} , 
\la{eq:apio} \\
B_{\pi^0 \, p} \,&=&\, \int_{-1}^1 dx \;
{1 \over {\sqrt 2}} \, 
{\left(e_u \ \widetilde E^u \,-\, e_d \ \widetilde E^d\right)} 
\;\left\{ {{1} \over {x - \xi + i \epsilon}}
+ {{1} \over {x + \xi - i \epsilon}} \right\} .
\la{eq:bpio}
\ee
\newline
\indent
For $\pi^+ \, n$ electroproduction on the proton, 
the amplitudes $A$ and $B$ are given by 
\cite{Man99b,Fra99}~:
\be
A_{\pi^+ \, n} \,&=&\, - \, \int_{-1}^1 dx \; 
{\left(\widetilde H^u \,-\, \widetilde H^d\right)}
\; \left\{ {{e_u} \over {x - \xi + i \epsilon}}
+ {{e_d} \over {x + \xi - i \epsilon}} \right\} , 
\la{eq:apip} \\
B_{\pi^+ \, n} \,&=&\, - \, \int_{-1}^1 dx \;
{\left(\widetilde E^u \,-\, \widetilde E^d\right)}
\;\left\{ {{e_u} \over {x - \xi + i \epsilon}}
+ {{e_d} \over {x + \xi - i \epsilon}} \right\} .
\la{eq:bpip}
\ee

In the amplitudes for pion electroproduction of Eq.~(\ref{eq:piampl}),
the pion distribution amplitude \( \Phi_\pi  \) enters. 
Recent data \cite{Gro98} for the \( \pi ^{0}\gamma ^{*}\gamma  \)
transition form factor up to \( Q^{2} \) = 9 GeV\( ^{2} \) are in 
agreement \footnote{See however Ref.~\cite{Bak01} for a recent 
discussion of the deviations from the asymptotic distribution amplitude 
for the pion.} 
with the asymptotic form of the distribution amplitude, given by~: 
\begin{equation}
\label{eq:piondistr}
\Phi _{\pi }(z)=\sqrt{2} \, f_{\pi }\; 6z(1-z)\; ,
\end{equation}
 with \( f_{\pi } \) = 0.0924 GeV from the pion weak decay.
\newline
\indent
As discussed in Sec.~\ref{chap3_4_2}, the function 
$\widetilde E^u - \widetilde E^d$ contains a strong pion pole singularity 
in the limit $t \to m_\pi^2$. The pion pole gives a dominant contribution 
to the hard $\pi^+$ electroproduction amplitude 
\( {\mathcal{M}}^{L}_{\pi^+} \) in  the valence region, 
which can be worked out explicitely. 
Using the pion pole formula of Eq.~(\ref{eq:ha_pipo}) 
for the induced pseudoscalar form factor \( h_{A}(t) \), 
and by using the PCAC relation \( g_{A}/f_{\pi }=g_{\pi NN}/m_{N} \),
where \( g_{\pi NN} \) is the \( \pi NN \) coupling constant, 
one obtains for the pion pole part of the 
amplitude \( {\mathcal{M}}^{L}_{\pi ^{\pm }} \)~:
\begin{equation}
{\mathcal{M}}^{L}_{\pi ^{+}}\left( \pi ^{+}-{\mathrm{p}ole}\right) \;
=\; ie\, \sqrt{2}\, Q\, F_{\pi }\left( Q^{2}\right) \; {{g_{\pi
      NN}}\over {-t+m_{\pi }^{2}}}\; \bar{N}(p^{'})\gamma _{5}N(p)\;,
\label{eq:lopipole}
\end{equation}
where \( F_{\pi} \) represents the pion electromagnetic form factor. 
When using an asymptotic distribution amplitude for the pion, 
the leading order pion pole amplitude is obtained by using in 
Eq.~(\ref{eq:lopipole}) the asymptotic pion form factor \( F_{\pi}^{as} \), 
which is given by~: 
\begin{equation}
F_{\pi }^{as}\left( Q^{2}\right) 
\; =\; {{16\pi \alpha _{s}\, f_{\pi }^{2}}\over {Q^{2}}}\; .
\la{eq:piasff}
\end{equation}
\newline
\indent
Besides the $\pi$ hard electroproduction, one can also study the
$\eta$ and $\eta'$ hard electroproduction. 
For the $\eta$ electroproduction, when neglecting the effects of the
QCD axial anomaly, the corresponding amplitudes $A$ and $B$ as in 
Eq.~(\ref{eq:piampl}) are given by \cite{Man98a}~:
\be
\hspace{-.25cm} A_{\eta \, p} \, &=& \int_{-1}^1 dx \; 
{1 \over {\sqrt 6}} \, 
{\left(e_u \ \widetilde H^u \,+\, e_d \ \widetilde  H^d 
\,-\,2 \, e_s \ \widetilde H^s \right)} 
\; \left\{ {{1} \over {x - \xi + i \epsilon}}
+ {{1} \over {x + \xi - i \epsilon}} \right\} , \; 
\la{eq:aeta} \\
\hspace{-.25cm} B_{\eta \, p} \, &=& \int_{-1}^1 dx \;
{1 \over {\sqrt 6}} \, 
{\left(e_u \ \widetilde E^u \,+\, e_d \ \widetilde E^d 
\,-\,2 \, e_s \ \widetilde E^s \right)} 
\;\left\{ {{1} \over {x - \xi + i \epsilon}}
+ {{1} \over {x + \xi - i \epsilon}} \right\} . \;
\la{eq:beta}
\ee
The effects of the QCD axial anomaly have been studied in
Ref.~\cite{Eid99}. 

\begin{figure}[h]
\epsfxsize=7.5 cm
\epsfysize=9. cm
\centerline{\epsffile{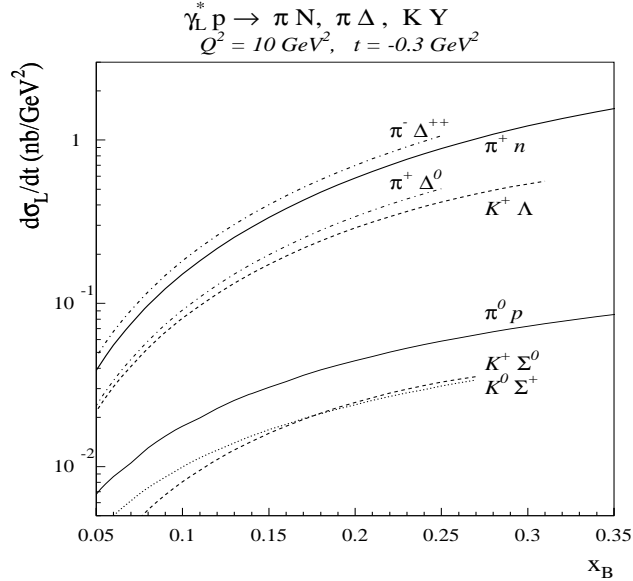}}
\vspace{-1.2cm}
\caption{\small Leading order predictions for the 
$\pi N$, $\pi \Delta$ and $K Y$ 
longitudinal electroproduction cross sections 
at $t$ = -0.3~GeV$^2$, as function of $x_B$
(plotted up to the $x_B$ value for which $t = t_{min}$). 
Results for pion and charged kaon channels are given using an
asymptotic distribution amplitude. For $K^0 \Sigma^+$, 
predictions are shown using the Chernyak-Zhitnitsky \cite{Che84} 
distribution amplitude with antisymmetric
part : $\eta^a_K = 0.25$. Figure from Ref.~\cite{Fra00}.}
\label{fig:pionkhyp}
\end{figure}

We show in Fig.~\ref{fig:pionkhyp} the leading order cross sections for 
various pseudoscalar meson production channels.
As discussed before for the absolute cross sections for 
vector meson electroproduction, 
also for pseudoscalar meson production the (higher-twist) power corrections 
to the cross section have not yet been worked out systematically. 
Therefore, these leading order results could receive sizeable corrections 
(see e.g. Ref.~\cite{Vdh99} for a first estimate of such corrections). 
To be on the safe side, we therefore show in Fig.~\ref{fig:pionkhyp} 
the leading order predictions at a relatively large scale 
($Q^2$ = 10 GeV$^2$), where we use for the running coupling constant 
$\alpha_s$ its expression from perturbative analyis at a scale 
$Q^2$ = 10 GeV$^2$.  
\newline
\indent
By comparing in Fig.~\ref{fig:pionkhyp}, the $\pi^+ n$ and $\pi^0 p$ channels, 
one immediately sees the prominent contribution of the 
charged pion pole to the $\pi^+$ cross section. The cross section for the 
$\pi^0$ channel, where this pion pole is absent, is more than a decade lower 
for $x_B \geq 0.15$. Going to smaller values of $x_B$, on the other hand, 
one finds an increasing ratio $\pi^0 : \pi^+$, due to the amplitudes $A$ of 
Eqs.~(\ref{eq:apio},\ref{eq:apip}). This was also noticed in 
Ref.~\cite{Man99b}. One can also understand the ratio $\pi^0 : \pi^+$ at 
smaller $x_B$ by comparing the amplitude of 
Eqs.~(\ref{eq:apio}) and (\ref{eq:apip}). 
One sees that in the $\pi^0 p$ channel, the GPDs $\widetilde H^u$ and 
$\widetilde H^d$ enter with the same sign, whereas in the $\pi^+ n$ channel, 
they enter with opposite signs. These GPDs $\widetilde H^u$ and 
$\widetilde H^d$ are constructed starting 
from the polarized quark distributions $\Delta u$ and $\Delta d$ as discussed 
in Sec.~\ref{chap5_2d}. Because $\Delta u$ and $\Delta d$ have opposite 
signs, this leads to a partial cancellation for the $\pi^0 p$ channel 
compared to the $\pi^+ n$ channel, yielding a ratio 
$\pi^0 : \pi^+ \approx 1 : 5$ for the non-pole part of the $\pi^+$ channel 
\cite{Vdh99}. 
\newline
\indent
For the $\eta p$ channel, using the ansatz for the GPDs $\widetilde H^q$ 
discussed in Sec.~\ref{chap5_2d} in Eqs.~(\ref{eq:aeta},\ref{eq:beta}), 
a slightly smaller cross section was found as for $\pi^0$, 
i.e. $\eta : \pi^0 \approx$ 4 : 5 \cite{Vdh99}, comparable with the ratio 
$\eta : \pi^0 \approx$ 2 : 3 found in Ref.~\cite{Man98a}.
This ratio is also very sensitive to the $x$-dependence of the 
polarized $u$- and $d$-quark distributions, giving a possibility to 
cross-check various parametrizations for those polarized quark
distributions. 
Besides, when including the effects of the axial anomaly 
and $SU(3)$ breaking, markedly
different results were obtained in Ref.~\cite{Eid99}, 
i.e. $\pi^0 : \eta : \eta'$ = 27 : 1.6 : 1.4, which will be
interesting to check experimentally.  
\vspace{0cm}

As we discussed already for the hard electroproduction of 
longitudinally polarized vector mesons there is, 
besides the cross section $\sigma_L$, 
a second relevant observable, which involves only 
longitudinal amplitudes and which is of leading order. 
This is the transverse spin asymmetry for a proton target 
polarized perpendicular to the reaction plane, see Eq.~(\ref{eq:tsa}).
\newline
\indent 
For the hard electroproduction of $\pi N$ final states, 
the transverse spin asymmetry ${\cal A}_{\pi N}$ is given by \cite{Fra99}~: 
\begin{eqnarray}
{\cal A}_{\pi N} = {{2 \, |\Delta_\perp|} \over {\pi}} \,
\frac{{\rm{Im}}(A B^*) \, 4 \, \xi \, m_N}{|A|^2  \, 4 \, m_N^2 \, (1-\xi^2) 
\,-\, |B|^2 \, \xi^2 \, t \,- \, {\rm{Re}}(AB^*) \, 8 \, \xi^2 \, m_N^2 } \, ,
\la{eq:piasymm}
\end{eqnarray}
where $A$ and $B$ are defined as in Eqs.~(\ref{eq:apio}-\ref{eq:bpip}). 

\begin{figure}[h]
\vspace{-.25cm}
\epsfxsize=7.5 cm
\epsfysize=9. cm
\centerline{\epsffile{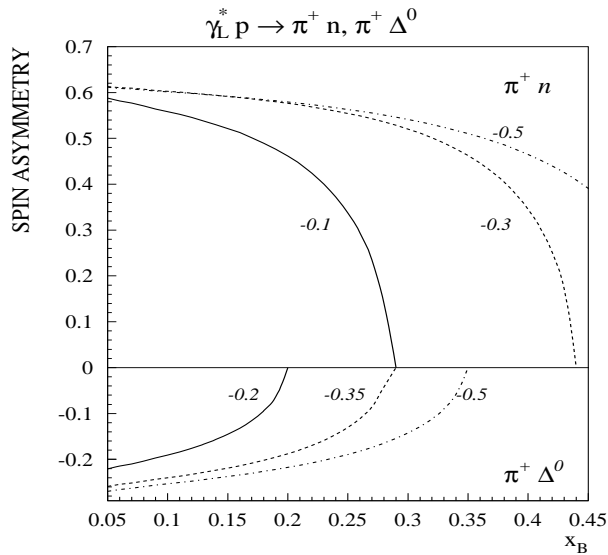}}
\vspace{-1.5cm}
\caption{\small Transverse spin asymmetry ${\cal A}_{\pi N}$
  for the longitudinal electroproduction of 
$\pi^+ n$ and $\pi^+ \Delta^0$, 
at different values of $t$ (indicated on the curves in GeV$^2$). 
Figure from Ref.~\cite{Fra00}.}
\label{fig:pidelasymm}
\end{figure}

In Fig.~\ref{fig:pidelasymm} (upper plot), we display the transverse 
spin asymmetry ${\cal A}_{\pi^+n}$ for $\pi^+ n$ electroproduction 
at several values of $-t$, as a function of $x_B$.
For the GPDs $\widetilde H^u$ and $\widetilde H^d$, we use the 
phenomenological parametrization as discussed in 
Sec.~\ref{chap5_2d}. 
It is obvious from Fig.~\ref{fig:pidelasymm} that 
the predicted ${\cal A}_{\pi^+n}$ is large and comparable with
the results of \cite{Fra99} in which the GPDs computed in the chiral
quark soliton model \cite{Pen00a} were used. 
As a consequence, the investigation of these processes 
can provide unique tests of the soliton type approach to baryon structure. 
\newline
\indent
The transverse spin asymmetries are likely to be less sensitive 
to higher twist effects 
and to NLO (in $\alpha_s$) corrections. The latter corrections have been 
calculated recently in Ref.~\cite{Bel01d} and were found to be sizeable as well as strongly scheme dependent for absolute cross sections, but nearly 
drop out of the transverse spin asymmetries. 
Hence these asymmetries can be explored 
already at presently accessible scales 
at HERMES and JLab, using a transversely polarized target.
\newline
\indent
The HERMES collaboration reported already preliminary 
a rather large asymmetry for $\pi^+$ electroproduction in case of a
longitudinally polarized target along the beam direction \cite{Tho01}. 
As was noted in Ref.~\cite{Pol00}, 
such a measurement includes a small target polarization 
component transverse to the virtual photon direction
$S_\perp$, leading to the following form for the polarized cross section 
$\sigma_P$~:
\begin{equation}
\sigma_P \sim \sin\Phi \biggl[S_\perp \sigma'_L + S_L
\sigma'_{LT}\biggr]
\, ,
\la{eq:lsa}
\end{equation}
where $\sigma'_L$ receives only contributions from the longitudinal
amplitude which dominates in the hard regime for the exclusive
channel. The cross section $\sigma'_{LT}$ contains the interference of
longitudinal and transverse amplitudes, which is suppressed by
a power of the hard scale ($1/Q$) relative to $\sigma'_L$. Note, however
that the transverse spin component $S_\perp$ 
drops as $1/Q$ relative to the longitudinal spin component $S_L$ for a
target polarized along the beam direction, so that in this case both terms in 
Eq.~(\ref{eq:lsa}) are of the same order in $1/Q$.
Therefore, to extract the transverse spin asymmetry unambiguously, 
an experiment using a transversely polarized target is really needed.  
\newline
\indent
A measurement of ${\cal A}_{\pi N}$
can also provide an important help in the
extraction of the pion electromagnetic form factor. 
For $x_B \leq 0.15$ where one reaches values 
$t_{min} \sim 2 m_\pi^2$, we find that the  
pion pole constitutes about 70 \% to the longitudinal cross section.
Measurement of the transverse spin asymmetry, which is an interference between 
the amplitudes $A$ and $B$ in Eq.~\ref{eq:piampl}, 
would help to constrain the non-pole term, and 
in this way help to get a more reliable extraction of the pion form factor.

\subsubsection{Hard electroproduction of strangeness}
\label{chap_5_5_3}

In this section, we turn to strangeness hard electroproduction reactions 
$\gamma ^{*}_{L} + p \rightarrow K + Y$.  
In leading order, the kaon electroproduction amplitude is given by an 
expression analogous as Eq.~(\ref{eq:piampl}).
\newline
\indent
For the charged kaon electroproduction channels, the 
amplitudes $A$ and $B$ are given by \cite{Fra99,Fra00}~:
\begin{eqnarray}
\label{eq:akpy}
A_{K^+ Y} &=& \int_{-1}^1 dx \; \widetilde H^{p \rightarrow Y}
\left\{ {{e_u} \over {x - \xi + i \epsilon}}
+ {{e_s} \over {x + \xi - i \epsilon}} \right\} , \\
B_{K^+ Y} &=& \int_{-1}^1 dx \; \widetilde E^{p \rightarrow Y}
\left\{ {{e_u} \over {x - \xi + i \epsilon}}
+ {{e_s} \over {x + \xi - i \epsilon}} \right\} ,
\label{eq:bkpy}
\end{eqnarray}
where $e_s$ is the $s$-quark charge, and $Y$ is the produced 
hyperon ($Y = \Lambda, \Sigma^0$).
To estimate the GPD $\widetilde H^{p \to Y}$, 
we use the $SU(3)$ relations as discussed in Sec.~\ref{chap3_6}.
The SPD $\widetilde E^{p \to Y}$  
contains a charged kaon pole contribution, given by~:
\begin{equation}
B^{pole}_{K^+ Y} = - 3 / (2 \xi) \, \eta^s_K
\, f_K \, g_{KNY}\, (2 m_N) / (-t + m_K^2) \, ,
\label{eq:kpole}
\end{equation}
where $f_K \simeq$ 159 MeV is the kaon decay constant, 
and where $g_{KNY}$ are the $KNY$ coupling
constants. In line with our use of $SU(3)$ relations to estimate
$\tilde H$, we use also $SU(3)$ predictions for the coupling
constants~: $g_{K N \Lambda}/\sqrt{4 \pi} \approx -3.75$
and $g_{K N \Sigma}/\sqrt{4 \pi} \approx 1.09$, which 
are compatible with those obtained from a Regge fit to high energy kaon
photoproduction data \cite{Guid97}. 
In Eq.~(\ref{eq:kpole}), $\eta^s_K$ is defined as~:
\begin{equation}
\left\{ \begin{array}{c} \eta^s_K \\ 
\eta^a_K  \\ \end{array} \right\} 
\,\equiv\, {2 \over 3} \, \int _{-1}^{+1}d \zeta \,
\left\{ \begin{array}{c} \Phi^s_K\left(\zeta\right) \\ 
\zeta \; \Phi^a_K\left(\zeta\right) \\ \end{array} \right\} 
\, {1 \over {1 - \zeta^2}} \, ,
\label{eq:czkaon}
\end{equation}
where $\Phi^s_K(\zeta)$ is the symmetric (in $\zeta$) part of the 
distribution amplitude (DA) of the kaon, 
which runs over the range -1 to +1 in the notation 
of Eq.~(\ref{eq:czkaon}). One has $\eta^s_K$ = 1 for an asymptotic DA
( $\Phi^s(\zeta) = 3/4 (1 - \zeta^2)$ ), and
$\eta^s_K$ = 7/5 for the Chernyak-Zhitnitsky (CZ) kaon DA \cite{Che84}.
Furthermore, 
in Eq.~(\ref{eq:czkaon}) we have also defined - for further use -
$\eta^a_K $ for the antisymmetric part $\Phi^a_K(\zeta)$ of the kaon DA. 
The latter is due to $SU(3)_f$ symmetry breaking effects. 
\newline
\indent
For the $K^0 \Sigma^+$ electroproduction, the expressions for
$A$ and $B$, allowing for both a symmetric and antisymmetric component
in the kaon DA, are given by \cite{Fra00}~:
\begin{eqnarray}
\left\{ \begin{array}{c} A_{K^0 \Sigma^+} \\ 
B_{K^0 \Sigma^+} \\ \end{array} \right\} 
&&= \int_{-1}^1 dx \,
\left\{ \begin{array}{c} \widetilde H^{p \to \Sigma^+} \\ 
\widetilde E^{p \to \Sigma^+} \\ \end{array} \right\}  \, 
\left[ {{(1 - \eta^a_K / \eta^s_K) \, e_d}
 \over {x - \xi + i \epsilon}} 
+  {{(1 + \eta^a_K / \eta^s_K) \,  e_s }
\over {x + \xi - i \epsilon}} \right] . 
\la{eq:abko}
\end{eqnarray}
In contrast to $\pi^0$ electroproduction, $K^0$ electroproduction 
can contain a pole contribution,
which is given by~: 
\begin{equation}
B^{pole}_{K^0 \Sigma^+} \,=\, {4 \over 3} \, \eta^a_K \,
\left( {3 \over {2 \xi}} \right) \, 
{{f_K \, g_{K N \Sigma} \, ({2 m_N})} \over {-t + m_K^2}} \;,
\la{eq:kosigpole}
\end{equation}
and which vanishes when the kaon DA is symmetric (i.e. when $\eta^a_K$ = 0).
Therefore, the $K^0$ pole contribution to $B_{K^0 \Sigma^+}$, provides
a direct measure of the antisymmetric component of the kaon DA.
\newline
\indent
In Fig.~\ref{fig:pionkhyp}, we show besides the leading order 
predictions for the pion electroproduction cross sections 
also the leading order predictions for 
kaon hard electroproduction cross sections at $Q^2$ = 10 GeV$^2$. 
The charged pion and kaon channels obtain a
large contribution in the range $x_B \gtrsim 0.1$ from the pion (kaon)
pole. This largely determines the ratio between these
channels at larger $x_B$. For values of $-t$ in the range 
0.1 $\to$ 0.5 GeV$^2$, this yields 
$\pi^+ n : K^+ \Lambda \approx 7 : 1 \to 1.8 : 1$, 
using an asymptotic DA for both $\pi$ and $K$. The kaon DA is
not well known however, and the results with a CZ kaon DA yield $K^+$ cross
sections, for the pole contribution, 
larger by a factor (7/5)$^4 \approx$ 3.8.    
The ratio  $K^+ \Lambda : K^+ \Sigma^0$ at large $x_B$ 
is determined from the ratio of
the couplings : $g^2_{K N \Lambda} / g^2_{K N \Sigma} \approx 12$.
For the $K^0 \Sigma^+$ channel, the pole contribution is absent if
$\eta^a_K$ = 0 (as for $\pi^0$). 
In this case, the ratio $\pi^0 p : K^0 \Sigma^+$
is determined by the amplitude $A$ and is
very sensitive to the input valence quark distribution into $\tilde
H$. For $\Delta u_V \approx - \Delta d_V$ expected in the large $N_c$
limit, $\pi^0 : K^0 \approx 1 : 3$, while for 
$\Delta u_V \approx -2 \Delta d_V$ prefered by the global fit to DIS of 
Ref.~\cite{Lea98},  $\pi^0 : K^0 \approx 3 : 1$. The
sensitivity of this ratio to the polarized quark distributions 
might be interesting to provide cross-checks on such
global fits from DIS. In Fig.~\ref{fig:pionkhyp}, we show the results for 
$K^0 \Sigma^+$ by using the polarized distributions of \cite{Lea98} as
input for $\tilde H$ (as discussed in Sec.~\ref{chap5_2d}). 

Besides the contribution of the amplitude $A$, $K^0 \Sigma^+$ 
electroproduction has also a pole contribution, given by 
Eq.~(\ref{eq:kosigpole}), 
which is nonzero if $\eta^a_K \neq 0$. 
In the estimates shown here, 
we include the pole contribution of Eq.~(\ref{eq:kosigpole}), and use 
the CZ kaon DA with $\eta_K^a$ = 0.25. 
The resulting $K^0$ pole contribution
provides a sizeable enhancement of the $K^0 \Sigma^+$ cross section (it 
gives roughly half the value of the amplitude $A$ at the largest $x_B$). 

\begin{figure}[h]
\vspace{1.cm}
\epsfxsize=7.cm
\epsfysize=8.cm
\centerline{\epsffile{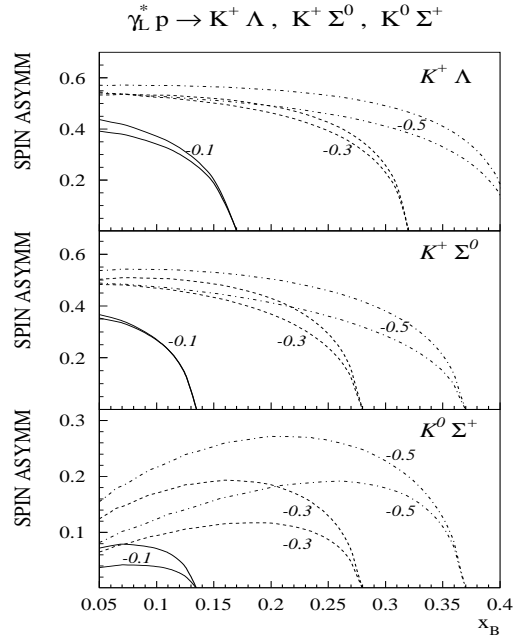}}
\caption{\small Transverse spin asymmetry, 
  for $K^+ \Lambda$, $K^+ \Sigma^0$ and $K^0 \Sigma^+$ longitudinal
  electroproduction for different values of $t$ (indicated on the
  curves in GeV$^2$). For $K^+ \Lambda$ and $K^+ \Sigma^0$, 
the thick (thin) lines are the predictions with the 
asymptotic (Chernyak-Zhitnitsky) kaon distribution amplitude respectively. 
For $K^0 \Sigma^+$, the thick (thin)  lines are the predictions with 
Chernyak-Zhitnitsky type kaon DA, with 
  antisymmetric part : $\eta^a_K$ = 0.25 (0.1).} 
\label{fig:khypasymm}
\end{figure}

After the cross section, 
we next discuss the transverse spin asmmetry ${\cal A}_{K Y}$ 
for the strangeness channels, which is given by \cite{Fra00}~: 
\begin{eqnarray}
{\cal A}_{K Y} &=& {{2 \, |\Delta_\perp|} \over {\pi}} \,
\frac{{\rm{Im}}(A B^*) \, 4 \xi m_N}{D_{K Y}} \, ,
\hspace{2cm} {\rm{with, }}\\
D_{K Y} &=& |A|^2  \, 4 \,m_N^2 (1-\xi^2) \,+\,
|B|^2 \, \xi^2 \, \left[ - t + (m_Y - m_N)^2 \right] \nonumber\\
&-& \, {\rm{Re}}(AB^*) \, 4 \, \xi \, m_N
\left[ \xi \, (m_Y + m_N) + m_Y - m_N \right] \, , \nonumber
\end{eqnarray}
where $A$ and $B$ are as defined before. 
It is interesting to note that in the case of hyperon production,
the same transverse spin asymmetry can be measured on an 
{\it unpolarized} target by measuring the polarization 
of the recoiling hyperon through its decay angular distribution.  
The transverse spin asymmetries 
${\cal A}_{K^+ \Lambda}$, ${\cal A}_{K^+ \Sigma^0}$  and 
${\cal A}_{K^0 \Sigma^+}$ are shown in Fig.~\ref{fig:khypasymm}. 
A comparison with Fig.~\ref{fig:pidelasymm} for the $\pi^+ n$ channel 
shows that the transverse spin asymmetries for the charged kaon channels are
as large as for $\pi^+ n$. One also sees that  
${\cal A}_{K^+\Sigma^0}\sim {\cal A}_{K^+\Lambda}$.
For $K^0$ production, the sensitivity to the 
$SU(3)_f$ symmetry breaking effects in the kaon DA 
is illustrated (lower panel of Fig.~\ref{fig:khypasymm}),
by plotting ${\cal A}_{K^0 \Sigma^+}$ for two values of $\eta^a_K$. 
Because ${\cal A}_{K^0 \Sigma^+}$ is directly
proportional to $\eta^a_K$, it provides a very sensitive observable
to extract the $K^0$ form factor.
\newline
\indent
As we discussed before, the hard electroproduction of strangeness 
also gives an 
unique access to the second class currents (e.g. weak electricity), 
for a first estimate see Sec.~\ref{chap3_6}.
\newline
\indent
To summarize this section, we have shown that the yields for hard exclusive 
electroproduction reactions of decuplet and octet baryons are similar. 
Strange and nonstrange 
channels can be comparable and in some cases the strange channels 
can even dominate (depending on the 
distribution amplitude and polarized parton distributions),   
in contrast to low-energy strangeness production. 
Large transverse spin asymmetries are also predicted for 
these reactions.

\subsubsection{Hard electroproduction of $\pi \Delta$ final states}
\label{chap_5_5_4}

In this section, we discuss the hard electroproduction of 
$\pi \Delta$ final states~: $\gamma ^{*}_{L} + p \rightarrow \pi + \Delta$.  
\newline
\indent
In Sec.~\ref{chap3_7}, we introduced the GPDs for the $N \to \Delta$ 
transition, and have seen how they can be expressed in terms of 
$N \to N$ GPDs through Eq.~(\ref{ndNc}), using large $N_c$ relations. 
The large $N_c$ relations make definite predictions for the ratios 
between the different cross sections for charged pion production as 
\cite{Fra00}~: 
$\sigma_L^{\gamma^* p \to \pi^+ n}:\sigma_L^{\gamma^* p \to \pi^+\Delta^0}:\sigma_L^{\gamma^* p \to \pi^-\Delta^{++}}:\sigma_L^{\gamma^* n \to \pi^- p}
\approx 1:0.5:1.25:0.8$, see also Fig.~\ref{fig:pionkhyp}.
For the production of neutral pseudoscalar mesons these estimates give~:
$\sigma^{\gamma^* p \to \eta(\eta')\Delta^+}_L  \ll
 \sigma_L^{\gamma^* p \to \eta(\eta') p}$ , 
and $\sigma^{\gamma^* p \to \pi^0\Delta^+}_L  \approx      
0.1 \; \sigma_L^{\gamma^* p \to \pi^0 p}$. 

In the $N \rightarrow \Delta$ DVCS process considered above, both
unpolarized and polarized skewed quark distributions enter, as 
shown in Eq.~(\ref{eq:dvcsampl}). 
The longitudinal electroproduction of pions 
in the reaction $\gamma_L^* N \rightarrow \pi \Delta$ at large $Q^2$. 
allows to select out the polarized quark distributions in the 
$N \rightarrow \Delta$ transition.  
\newline
\indent
For {\it charged} pions, 
the leading order amplitudes for longitudinal $\pi \Delta$ 
electroproduction on the proton are given by~\cite{Fra00}
\begin{eqnarray}
\left\{ \begin{array}{c}
{\mathcal{M}}^{L} \left(\gamma^* p \rightarrow \pi^- \Delta^{++} \right)\\
{\mathcal{M}}^{L} \left(\gamma^* p \rightarrow \pi^+ \Delta^0 \right)
\end{array}\right\}
&&=\, -ie\, {4\over 9}\, {1\over {Q}}\; \left
[ \int _{0}^{1}dz{{\Phi _{\pi }(z)}\over z}\right] \,
\left\{ \begin{array}{c}
-\sqrt{3}/2\\
-1/2
\end{array}\right\}
\nonumber \\
\times && \hspace{-.35cm} {1\over 2}\, \int _{-1}^{+1}dx\left[
\left\{ \begin{array}{c}
e_d\\
e_u
\end{array}\right\} {1\over {x-\xi +i\epsilon }}
+\left\{ \begin{array}{c}
e_u\\
e_d
\end{array}\right\} {1\over {x+\xi -i\epsilon }}\right]
\,(4\pi \alpha _{s})\nonumber \\
\times && \hspace{-.35cm}
\bar \psi^\beta(p^{'}) \,
\left\{ C_1(x,\xi ,t)\, n_\beta \,+\,
C_2(x,\xi ,t) \, \Delta_\beta \,
{{\Delta \cdot n}\over {m_N^2}} \right.\nonumber\\
&&\hspace{1.2cm}\,+\; C_3(x,\xi ,t) \, {1\over {m_N}}
 \left( {\Dirac \Delta \, n_\beta - {\Dirac n} \, \Delta_\beta}\right)
 \nonumber\\
&&\left. \hspace{1.2cm}\,+\; C_4(x,\xi ,t) \, {2\over {m_N^2}}
 \left( {\Delta \cdot \bar P}\, n_\beta - \Delta_\beta\right)
\right\} N(p)\;.
\label{eq:pichargampl}
\end{eqnarray}

For {\it neutral} pions, 
the leading order amplitude for longitudinal $\pi \Delta$ 
electroproduction on the proton is given by
\begin{eqnarray}
{\mathcal{M}}^{L} \left(\gamma^* p \rightarrow \pi^0 \Delta^+ \right)
&=&\, -ie\, {4\over 9}\, {1\over {Q}}\; \left
[ \int _{0}^{1}dz{{\Phi _{\pi }(z)}\over z}\right] \,
\left( {1 \over {\sqrt{2}}} {{e_u + e_d} \over 2}\right) \nonumber \\
&\times& \,{1\over 2}\, \int _{-1}^{+1}dx\left[
{1\over {x-\xi +i\epsilon }}
\;+\; {1\over {x+\xi -i\epsilon }}\right]
\;(4\pi \alpha _{s})\nonumber \\
&\times& \,
\bar \psi^\beta(p^{'}) \,
\left\{ C_1(x,\xi ,t)\, n_\beta \,+\,
C_2(x,\xi ,t) \, \Delta_\beta \,
{{\Delta \cdot n}\over {m_N^2}} \right.\nonumber\\
&&\hspace{1.2cm}\,+\; C_3(x,\xi ,t) \, {1\over {m_N}}
 \left( {\Dirac \Delta \, n_\beta - {\Dirac n} \, \Delta_\beta}\right)
 \nonumber\\
&&\left. \hspace{1.2cm}\,+\; C_4(x,\xi ,t) \, {2\over {m_N^2}}
 \left( {\Delta \cdot \bar P}\, n_\beta - \Delta_\beta\right)
\right\} N(p)\;.
\label{eq:pineutrampl}
\end{eqnarray}
\newline
\indent
The charged pion production amplitudes (Eq.~(\ref{eq:pichargampl}))
receive an important contribution from the pion pole contained in the
function $C_2(x,\xi ,t)$, which can again be
evaluated analytically.
Using the asymptotic distribution amplitude for the pion, 
and the pion-pole part of $h_A(t)$ given by Eq.~(\ref{eq:ha_pipo}), 
the pion-pole contribution to the $\pi^- \Delta^{++}$ amplitude 
of Eq.~(\ref{eq:pichargampl}) is given by  
\begin{equation}
{\mathcal{M}}^{L}_{\pi^- \Delta^{++}} \left( \pi - {\rm pole} \right)
=\, -ie\; Q \, F_\pi\left( Q^2 \right) 
\, {1 \over {-t + m_\pi^2}} \; {{f_{\pi N \Delta}} \over {m_\pi}} \,
\psi^\beta(p') \, \Delta_\beta \, N(p) \, ,
\label{eq:pideltapipole}
\end{equation} 
where $F_\pi\left(Q^2\right)$ represents the pion electromagnetic form
factor as given by Eq.~(\ref{eq:piasff}).
In Eq.~(\ref{eq:pideltapipole}), we have introduced the $\pi N \Delta$
coupling $f_{\pi N \Delta}$ by using the ``flavor-skewed'' 
Goldberger-Treiman relation 
\begin{equation}
{{g_A} \over {f_\pi}} \;=\; 2 \, {{\sqrt{2}} \over 3} \, 
{{f_{\pi N \Delta}} \over {m_\pi}} \;,
\end{equation}
which is equivalent to the ordinary Goldberger-Treiman relation
through $f_{\pi N \Delta} = 3/\sqrt{2} f_{\pi NN}$. Using the
phenomenological value $f^2_{\pi
  NN}/4 \pi \approx 0.08$, this yields $f_{\pi N \Delta} \approx 2.13$.
\newline
\indent
The leading order $\pi \Delta$ electroproduction cross sections are 
also shown in Fig.~\ref{fig:pionkhyp}. 
One sees that those $\pi \Delta$ channels 
which contain a charged pion pole exchange give again
large cross sections, e.g. the $\pi^- \Delta ^{++}$ cross sections are
comparable with the $\pi^+ n$ cross sections. Comparing the
$\pi^- \Delta^{++}$ and $\pi^+ \Delta^0$ channels, their ratio is
around 2.25 in our calculations, close to the ratio 3 which one would
obtain if there was only charged pion pole exchange. 
We found the non-pole
contribution to $\pi^- \Delta^{++}$ to be 
roughly about one-tenth of the total $\pi^- \Delta^{++}$ cross section. 
Also the small $\pi^0 \Delta^+$ cross
section, which receives no contribution from pion exchange confirms
the smallness of the non-pole contributions to the forward $\pi \Delta$
cross sections in the valence region. 
\newline
\indent
Similar as for the electroproduction of $\pi N$ final states,  
one can also define a transverse spin asymmetry 
${\cal A}_{\pi \Delta}$ for $\pi \Delta$ electroproduction, 
which is given by~\cite{Fra00}~: 
\begin{eqnarray}                                         
\label{eq:pidelasymm}
&&{\cal A}_{\pi \Delta} = - \, {{2 \, |\Delta_\perp|} \over {\pi}} \,
\frac{{\rm{Im}}(A B^*) \, 2 \, \xi \, m_N^2 \, m_\Delta}{D_{\pi \Delta}}\, , 
\hspace{2cm} {\rm{with, }} \\
D_{\pi \Delta} &=& |A|^2 \, m_N^4 (1-\xi)^2 
\;+\; |B|^2 \, \xi^2 \left[ t^2 - 2 \, t \, (m_\Delta^2 + m_N^2)
+ (m_\Delta^2 - m_N^2)^2\right]  \nonumber\\
&+& {\rm{Re}}(AB^*) \, 2 \, \xi \, m_N^2 
\left[ \xi \, t - \xi \, (3 m_\Delta^2 + m_N^2) -
t - m_\Delta^2 + m_N^2 \right] . \nonumber
\end{eqnarray}
For $\pi^+ \Delta^0$ electroproduction , $A$ and $B$ are given by~:
\begin{eqnarray}
\label{eq:apipdelo}
A_{\pi^+ \Delta^0} \,&=&\, \int_{-1}^1 dx \; C_1
\left\{ {{e_u} \over {x - \xi + i \epsilon}}
+ {{e_d} \over {x + \xi - i \epsilon}} \right\} , \\
B_{\pi^+ \Delta^0} \,&=&\, \int_{-1}^1 dx \; C_2
\left\{ {{e_u} \over {x - \xi + i \epsilon}}
+ {{e_d} \over {x + \xi - i \epsilon}} \right\} ,
\label{eq:bpipdelo}
\end{eqnarray}
where the functions $C_1$, $C_2$ are as defined in 
Eq.~(\ref{HND-helicity-QCD-2}).
Eq.~(\ref{eq:bpipdelo}) implies that the pion pole contribution to
$B_{\pi^+ \Delta^0}$ is given by
$B^{pole}_{\pi^+ \Delta^0} = \sqrt{3}/4 \, B^{pole}_{\pi^+ n}$. 
We use an asymptotic pion distribution amplitude (DA), for which  
$B^{pole}_{\pi^+ n}$ is given by~:
\begin{equation}
B^{pole}_{\pi^+ n} = - 3 / (2 \xi) 
\,g_A \, (2 m_N)^2 / (-t + m_\pi^2)\, .
\end{equation}

For the electroproduction of $\pi^- \Delta^{++}$ one has 
- up to a global isospin factor - analogous expressions as
Eqs.~(\ref{eq:apipdelo},\ref{eq:bpipdelo}), by making the replacement
$e_u \leftrightarrow e_d$.
\newline
\indent
In Fig.~\ref{fig:pidelasymm} (lower plot), we also show 
the transverse spin asymmetry ${\cal A}_{\pi^+\Delta^0}$ 
for $\pi^+ \Delta$ production 
at several values of $-t$, as a function of $x_B$.
We use the large $N_c$ relations Eqs.~(\ref{ndNc}) for the 
$N \to \Delta$ GPDs. 
Fig.~\ref{fig:pidelasymm} shows that ${\cal A}_{\pi^+\Delta^0}$ 
has the opposite 
sign compared with ${\cal A}_{\pi^+n}$.
Although the magnitude of ${\cal A}_{\pi^+\Delta^0}$ 
is smaller than ${\cal A}_{\pi^+n}$, 
as anticipated in \cite{Fra99}, it is still sizeable. 
Similarly, we find 
${\cal A}_{\pi^-\Delta^{++}}\approx 0.5 \, {\cal A}_{\pi^+\Delta^0}$.

\newpage
\section{CONCLUSIONS, PERSPECTIVES AND KEY EXPERIMENTS}
\label{chap6}

In this work, we outlined in detail the structure of generalized parton
distributions (GPDs), which are universal non-perturbative objects entering the
description of hard exclusive electroproduction processes.
We focussed mostly on the physics of the internal quark-gluon structure
of hadrons encoded in the GPDs. For this we presented calculations of the
GPDs within the context of the chiral quark-soliton model.
Guided by this physics we constructed
a parametrization of the GPDs $H$, $E$, $\tilde H$ and $\tilde E$,
all depending on the variable $x$, the skewedness $\xi$, and the
momentum transfer to the target nucleon, $t$.
The parameters entering this parametrization
can be related in a rather general way to such (not yet measured)
quantities as the contribution of the nucleon spin carried by the quark
total angular momentum ($J^u, J^d$, etc.),
$\bar q q$ components of the nucleon wave function
(in particular the D-term and the ``vector meson'' (VM) part of the GPD $E$),
the strength of the skewedness effects in the GPDs ($b_{val}$ and
$b_{sea}$),
the weak electricity GPD in nucleon to hyperon transitions ($E_{WE}$),
flavor $SU(3)$ breaking effects, and others.

Below we give a list of promising observables which we studied in this work
and indicate their sensitivity to the different physics aspects of the
GPDs. We also discuss for each of them the present experimental status
and perspectives. We concentrate on deeply virtual Compton scattering (DVCS)
and hard meson electroproduction (HMP).

\begin{itemize}
\vspace{.5cm}
\item
{\underline {DVCS : single spin asymmetry (SSA)}} ( see Sec.~\ref{chap5_3_5} )
\vspace{.5cm}

The DVCS SSA, using a polarized lepton beam,
accesses (in leading order) directly (i.e. without convolution)
the GPDs along the line $x = \xi$, e.g. $H^q(\xi, \xi, t)$.
We found that the DVCS SSA displays~:

\begin{itemize}
\item
a sensitivity to the strength of the skewedness effects in
GPDs (parametrized by $b_{val}$ and $b_{sea}$)

\item
a modest sensitivity to the total angular momentum contributions of
the $u$-quark ($J^u$) and $d$-quark ($J^d$) to the proton spin

\item
an important sensitivity to the $t$-dependence of the GPDs.
In particular in kinematics
where the Bethe-Heitler (BH) contribution dominates,
the SSA is basically given by a ratio of the imaginary
part of the DVCS amplitude to the BH amplitude.
Therefore it depends linearly on the GPDs and therefore the
parametrization of the $t$-dependence directly influences the size of the SSA.

\end{itemize}

On the experimental side, the first data for the DVCS SSA have
been obtained very recently, and several projects both by the HERMES
Collaboration and at JLab
are already planned in the near future.

\begin{itemize}
\item
First measurements of the SSA were performed by HERMES \cite{Air01}
(27 GeV beam energy),
and at Jlab by the CLAS Collaboration \cite{Step01,Guid01}
(around 4.2 GeV beam energy) and display already a clear
``$\sin \Phi$'' structure of the DVCS SSA, indicating that the
higher twist effects are modest for this observable at the
presently accessible scales.

\item
Further measurements of the DVCS SSA
are planned at JLab (at 6 - 11 GeV beam energy),
both in Hall-A \cite{Ber00}, and by the CLAS Collaboration
\cite{Bur01b,Bur01a}.

\item
At HERMES a dedicated large acceptance recoil detector project
has been proposed \cite{Kai01}.
This will allow a substantial improvement over existing
measurements of the DVCS SSA \cite{Air01},
in particular regarding the exclusivity of the
reaction by detecting the recoiling proton.

\end{itemize}

\vspace{.5cm}
\item {\underline {DVCS : cross section and charge asymmetry}}
( see Sec.~\ref{chap5_3_4} )
\vspace{.5cm}

One accesses the real part of the DVCS amplitude
by detecting the DVCS cross sections or the DVCS charge asymmetry
(i.e. the difference between the reactions induced by an electron
and a positron). The real part of the DVCS amplitude
contains (in a convolution) the information on the GPDs away
from the line $x = \xi$. In particular we have demonstrated the
followings points.

\begin{itemize}
\item
The DVCS charge asymmetry displays an important sensitivity
to the D-term, which is
required by the polynomiality condition of GPDs.
The D-term leads to a DVCS charge asymmetry with opposite sign as
compared to the charge asymmetry when this D-term would be absent.

\item
It was discussed how this D-term is a consequence of the
spontaneous breaking of the chiral symmetry and an estimate
of the D-term in the chiral quark-soliton model has been given.

\item
More generally, the GPDs in
the region $-\xi < x < \xi$ contain totally new information
on nucleon structure compared to the forward quark distribution.
In particular, in this region the GPDs behave like a meson distribution
amplitude and one accesses $q \bar q$ (mesonic) components
in the proton state.

\item
We have shown that the real part of the DVCS amplitude displays quite some
sensitivity to the twist-3 effects, which were estimated in the
Wandzura-Wilczek approximation.

\end{itemize}

On the experimental side, different facilities are complementary in the
measurement of the DVCS charge asymmetry and DVCS cross sections.

\begin{itemize}
\item
The measurement of the DVCS charge asymmetry provides a unique opportunity
for HERMES where both $e^+$ and $e^-$ beams are available.
A dedicated measurement of the DVCS charge asymmetry is planned in the
near future \cite{Kai01}.

\item
The measurement of the DVCS cross section is favorable in a regime where the
DVCS dominates over the Bethe-Heitler process. This is the case at high beam
energy. An experiment to measure the DVCS cross section at COMPASS
(in the 100 - 200 GeV beam energy range) has been proposed \cite{d'Ho99}.

\item
At small $x_B$, first data for DVCS cross sections have been
obtained by the H1 \cite{Adl01} and ZEUS \cite{Sau99}
Collaborations, and can map out the behavior of GPDs in the small
$x_B$ region. These first DVCS data at small $x_B$ are in
agreement with theoretical predictions \cite{Fra98b,Fra99},
implying that higher-twist effects are rather small for $Q^2 \geq
4$ GeV$^2$.
\end{itemize}

\item
{\underline {HMP : Transverse spin asymmetries for longitudinal
vector meson production}} \newline ( see Sec.~\ref{chap_5_5_1} )
\vspace{.5cm}

We showed that in
the electroproduction of longitudinally polarized vector mesons on a
transversely polarized proton target,
the GPD $E$ enters {\it linearly} in the spin asymmetry.
Therefore, this transverse spin asymmetry provides a unique
observable to extract the GPD $E$ and the quark total angular momentum
contribution $J^q$ to the proton spin.

\begin{itemize}
\item
We found large asymmetries
for longitudinally polarized vector meson electroproduction,
which display a clear sensitivity to the quark total
angular momentum contribution $J^u$ and $J^d$ to the proton spin.
Specifically, $\rho^0$ electroproduction is mainly sensitive
to the combination $2 \, J^u \,+\, J^d$,
$\rho^+$ electroproduction to the combination $J^u \,-\, J^d$ and
$\omega$ electroproduction to the combination $2 \, J^u \,-\, J^d$.

\item
To control the model dependence in the extraction of the values
$J^u$ and $J^d$, the measurement of the transverse spin
asymmetries for $\rho^0_L$, $\omega_L$ and $\rho^+_L$
electroproduction over a large kinematic range may be helpful as
an additional cross check on the modeling of the functions $E^u$
and $E^d$.

\item
Because the transverse spin asymmetry involves a ratio of cross sections,
it is expected to be a clean observable already at accessible values of
$Q^2$ (in the few GeV$^2$ range), where the precise form of the
$t$-dependence of the GPDs,
the NLO corrections and higher twist effects drop out to a large extent.

\end{itemize}

These transverse spin asymmetries for longitudinally polarized vector meson
electroproduction have not yet been addressed experimentally until now.
At HERMES, an experimental program with a transversely polarized target is
just starting \cite{Now01}.
In view of their large sensitivity to $J^u$ and $J^d$, we strongly encourage
dedicated experiments, at different facilities,
to access these transverse spin asymmetries.

\vspace{.5cm}
\item
{\underline {HMP : Transverse spin asymmetries for pseudoscalar
meson production}}
\newline ( see Secs.~\ref{chap_5_5_2}, \ref{chap_5_5_3}, \ref{chap_5_5_4} )
\vspace{.5cm}

\begin{itemize}
\item
The electroproduction of charged pseudoscalar mesons $\pi^+, K^+$ on a
transversely polarized proton target also leads to large asymmetries
due to the pion (kaon) pole contribution to the GPD $\tilde E$.

\item
Given the knowledge of this pion (kaon) pole contributions,
this asymmetry is mainly sensitive to the GPD $\tilde H(\xi,\xi,t)$,
i.e. along the line $x = \xi$. With the help of the flavor $SU(3)$
relations for the $N\to Y$ GPDs one accesses in this way polarized
strange quark and antiquark distributions in the nucleon.

\item
Transverse asymmetries for kaon production can teach us about
$SU(3)$ symmetry breaking effects. In particular, the $K^0$
production gives an access to the antisymmetric part of the kaon
distribution amplitude which is due to the flavor $SU(3)$
breaking effects. Also the transverse spin
asymmetry for kaon production is sensitive to the effects of
second class currents, especially to the weak electricity.
\end{itemize}
\end{itemize}

Besides these observables which have been discussed above, other
observables can be valuable in the study of the onset of the
scaling regime, such as e.g. cross section measurements for meson
electroproduction ( for experimental plans at JLab, see Ref.~\cite{Guid98},
and projects at COMPASS, see Ref.~\cite{Poc99} ) and double
spin asymmetries (DVCS with polarized beam and target).

\vspace{.75cm}

For a more detailed extraction of the physics content of the
GPDs from observables further theoretical work is definitely
needed. Let us just give an (incomplete) list of theoretical
problems which, in our opinion, need further clarification.

\begin{itemize}

\item
Calculations of NLO perturbative corrections to hard meson
production amplitudes. For DVCS this work is largely done.
Apart from a more accurate access to the GPDs, a
NLO analysis of the data would allow to perform a
global analysis of various data on hard reactions from small to large
$x_B$.

\item
A major open theoretical issue is to get a better understanding
of the power (higher twist) corrections to meson production
which are expected to be sizeable. For DVCS observables the
immediate task is to quantify the contributions of
the target mass corrections and the genuine twist-3 operators
related to quark-gluon correlations in the nucleon.

\item
The analysis of the higher twist effects can be useful
in identifying observables for which higher twist effects tend to
cancel, such as expected in transverse spin asymmetries
for meson production.

\item
A further effort is required in modeling the
$t-$dependence of the GPDs. In this respect the emerging chiral
theory of hard processes can be of a big help.

\item
Studying hard exclusive processes such as DVCS in the presence of
an additional soft pion.

\item
Developing the theory and phenomenology of a new resonance spectroscopy
with light-cone probes can be complementary to traditional low energy probes
and bring us a breakthrough in the
understanding of hadrons as bound states of quark and gluons.

\item
Establishing links between GPDs and other nucleon structure quantities,
such as generalized form factors, polarizabilities, etc.

\item
Studying the transition from exclusive $\to$ semi-exclusive $\to$
semi-inclusive $\to$ inclusive processes.

\end{itemize}

Summarizing, one may say that
the study of hard exclusive processes has generated a lot of activity and
excitement over the last few years, bringing together several fields
investigating low and high energy strong interaction phenomena.
One may expect new discoveries in the near future both on the theoretical and
experimental side. We hope that the underlying work can stimulate such
further efforts.

\section*{\center{Acknowledgements}}

This work was supported by the Deutsche Forschungsgemeinschaft
(SFB 443, Schwerpunktprogramm), the Russian Foundation for Basic
Researches (RFBR 00-15-96610), BMBF, COSY (J\"ulich).

We would like to express our gratitude to I. B\"ornig, L.L.
Frankfurt, P.A.M. Guichon, M. Guidal, N. Kivel, M. Penttinen, V.
Petrov, P. Pobylitsa, A. Sch\"afer, A. Shuvaev, M. Strikman, O.
Teryaev, C. Weiss in collaboration with whom some of the results,
which are reviewed in this work, were obtained.

Furthermore we would like to thank M. Amarian, A. Belitsky, P.Y.
Bertin, V. Burkert, N. d'Hose, D. Diakonov, M. Diehl, R. Kaiser,
P. Kroll, L. Mankiewicz, L. Moss\'e, D. M\"uller, A. Radyushkin,
D. Ryckbosch, F. Sabati\'e, N. Stefanis, M. Tytgat, G. Van der
Steenhoven, D. von Harrach for many useful and stimulating
discussions.

We also would like to thank Dima Kiptily for his assistance in
checking the references.




\end{document}